\documentclass[usegraphicx,usenatbib,useAMS]{mn2e}

\usepackage{times} 
\usepackage{aas_macros}     
\usepackage{amssymb}        
\usepackage{xspace}
\usepackage{ifthen}

\newcommand{\etal}{{\em et al.\ }}

\newcommand{\gsim}{\gtrsim} 
\newcommand{\lsim}{\lesssim} 

\newcommand{\Msol}{{\rm M}_\odot}
\newcommand{\hMsol}{h^{-1} {\rm M}_\odot}
\newcommand{\Lsol}{{\rm L}_\odot}

\newcommand{\Zsol}{Z_\odot}

\newcommand{\cm}{{\rm cm}}
\newcommand{\mum}{\mu{\rm m}}
\newcommand{\K}{{\rm K}}
\newcommand{\pc}{{\rm pc}}

\newcommand{\hMpc}{h^{-1}{\rm Mpc}}
\newcommand{\kms}{{\rm km\,s^{-1}}}

\newcommand{\Myr}{{\rm Myr}}
\newcommand{\Gyr}{{\rm Gyr}}

\newcommand{\mJy}{{\rm mJy}}

\newcommand{\GHz}{{\rm GHz}}

\newcommand{\GALFORM}{\textsc{galform}\xspace}
\newcommand{\galform}{\textsc{galform}\xspace}
\newcommand{\GRASIL}{\textsc{grasil}\xspace}

\newcommand{\SPITZER}{{\em Spitzer}}

\newcommand{\Herschel}{{\em Herschel}}

\newcommand{\Planck}{{\em Planck}}

\newcommand{\kB}{k_{\rm B}}

\newcommand{\Vc}{V_{\rm c}}
\newcommand{\Vcrit}{V_{\rm crit}}
\newcommand{\zreion}{z_{\rm reion}}

\newcommand{\Mhalo}{M_{\rm halo}}
\newcommand{\Jhalo}{J_{\rm halo}}
\newcommand{\rvir}{r_{\rm vir}}
\newcommand{\Vvir}{V_{\rm vir}}
\newcommand{\Tvir}{T_{\rm vir}}

\newcommand{\Mhot}{M_{\rm hot}}
\newcommand{\MZhot}{M_{\rm hot}^Z}
\newcommand{\Zhot}{Z_{\rm hot}}
\newcommand{\Mres}{M_{\rm res}}
\newcommand{\MZres}{M_{\rm res}^Z}

\newcommand{\Mcold}{M_{\rm cold}}
\newcommand{\MZcold}{M_{\rm cold}^Z}
\newcommand{\Zcold}{Z_{\rm cold}}
\newcommand{\Mstar}{M_{\star}}
\newcommand{\MZstar}{M_{\star}^Z}

\newcommand{\Zg}{Z_{\rm g}}

\newcommand{\rhohot}{\rho_{\rm hot}}
\newcommand{\tform}{t_{\rm form}}
\newcommand{\taucool}{\tau_{\rm cool}}
\newcommand{\rcool}{r_{\rm cool}}
\newcommand{\tauff}{\tau_{\rm ff}}
\newcommand{\rff}{r_{\rm ff}}
\newcommand{\racc}{r_{\rm acc}}
\newcommand{\Mdotacc}{\dot M_{\rm acc}}

\newcommand{\nuSF}{\nu_{\rm SF}}

\newcommand{\tauburstmin}{\tau_{\ast {\rm burst,min}}}
\newcommand{\fdyn}{f_{\rm dyn}}

\newcommand{\taustar}{\tau_{\ast}}
\newcommand{\taustarburst}{\tau_{\ast {\rm burst}}}
\newcommand{\taub}{\tau_{\rm b}}

\newcommand{\VSN}{V_{\rm SN}}
\newcommand{\gammaSN}{\gamma_{\rm SN}}
\newcommand{\alpharet}{\alpha_{\rm ret}}
\newcommand{\alphacool}{\alpha_{\rm cool}}
\newcommand{\fEdd}{f_{\rm Edd}}
\newcommand{\epsilonheat}{\epsilon_{\rm heat}}
\newcommand{\fBH}{f_{\rm BH}}

\newcommand{\tesc}{t_{\rm esc}}
\newcommand{\fcloud}{f_{\rm cloud}}
\newcommand{\Mcloud}{m_{\rm cloud}}
\newcommand{\rcloud}{r_{\rm cloud}}
\newcommand{\Mdust}{M_{\rm dust}}
\newcommand{\Tdust}{T_{\rm dust}}

\newcommand{\taudiffuse}{\tau_{\rm diff}}
\newcommand{\taucloud}{\tau_{\rm cloud}}

\newcommand{\Mdisk}{M_{\rm disk}}
\newcommand{\Mbulge}{M_{\rm bulge}}
\newcommand{\rdisk}{r_{\rm disk}}
\newcommand{\rbulge}{r_{\rm bulge}}
\newcommand{\rcirc}{r_{\rm circ}}

\newcommand{\fburst}{f_{\rm burst}}
\newcommand{\fellip}{f_{\rm ellip}}

\newcommand{\Fstab}{F_{\rm stab}}
\newcommand{\Fdisk}{F_{\rm disk}}

\newcommand{\forbit}{f_{\rm orbit}}
\newcommand{\fDM}{f_{\rm DM}}

\title[Unified model of galaxy formation]
{A unified multi-wavelength model of galaxy formation}

\author[Lacey  et al.]
{ \parbox{18cm}{Cedric G. Lacey\thanks{E-mail: Cedric.Lacey@durham.ac.uk (CGL)}$^1$,
Carlton M. Baugh$^1$, 
Carlos S. Frenk$^1$, 
Andrew J. Benson$^2$,
Richard G. Bower$^1$, 
Shaun Cole$^1$,
Violeta Gonzalez-Perez$^{1,3}$,  
John C. Helly$^1$,
Claudia D.P. Lagos$^{4,5}$,
Peter D. Mitchell$^{1,6}$ 
} 
\\
\\
$^{1}$Institute for Computational Cosmology, Department of Physics,
University of Durham, South Road, Durham, DH1 3LE, UK\\
$^{2}$Carnegie Observatories, 813 Santa Barbara Street, Pasadena, CA
91101, USA\\
$^{3}$Institute of Cosmology and Gravitation, Portsmouth University,
Dennis Sciama Building, Burnaby Road, Portsmouth PO1 3FX, UK\\
$^{4}$European Southern Observatory, Karl-Schwarzschild-Strasse 2,
85748 Garching, Germany\\
$^{5}$International Centre for Radio Astronomy Research, 7 Fairway,
Crawley, 6009, Perth, WA, Australia\\
$^{6}$Centre de Recherche Astrophysique de Lyon, Observatoire de Lyon,
9 Avenue Charles Andre, 69230 Saint-Genis-Laval, France}

\begin{document}


\maketitle

\begin{abstract}
  We present a new version of the \galform semi-analytical model of
  galaxy formation. This brings together several previous developments
  of \galform into a single unified model, including a different
  initial mass function (IMF) in quiescent star formation and in
  starbursts, feedback from active galactic nuclei supressing gas
  cooling in massive halos, and a new empirical star formation law in
  galaxy disks based on their molecular gas content. In addition, we
  have updated the cosmology, introduced a more accurate treatment of
  dynamical friction acting on satellite galaxies, and updated the
  stellar population model. The new model is able to simultaneously
  explain both the observed evolution of the K-band luminosity
  function and stellar mass function, and the number counts and
  redshift distribution of sub-mm galaxies selected at 850$\mum$. This
  was not previously achieved by a single physical model within the
  $\Lambda$CDM framework, but requires having an IMF in starbursts
  that is somewhat top-heavy. The new model is tested against a wide
  variety of observational data covering wavelengths from the far-UV
  to sub-mm, and redshifts from $z=0$ to $z=6$, and is found to be
  generally successful. These observations include the optical and
  near-IR luminosity functions, HI mass function, fraction of early
  type galaxies, Tully-Fisher, metallicity-luminosity and
  size-luminosity relations at $z=0$, as well as far-IR number counts,
  and far-UV luminosity functions at $z \sim 3-6$. Discrepancies are
  however found in galaxy sizes and metallicities at low luminosities,
  { and in the abundance of low mass galaxies at high-$z$,}
  suggesting the need for a more sophisticated model of supernova
  feedback.
\end{abstract}


{\em ``Everything should be as simple as it can be, but not simpler.''
  -- Albert Einstein}

\section{Introduction}


Galaxy formation is a two-stage process: structure forms in the dark
matter (DM) by hierarchical clustering and galaxies then form by
cooling and collapse of baryons in the gravitational potential wells
of dark matter halos \citep{White78}. In the standard picture, the
dark matter interacts only via gravity and its evolution in the
standard Lambda cold dark matter ($\Lambda$CDM) model is now well
understood from large N-body simulations
\citep[e.g.][]{Springel05}. On the other hand, the evolution of the
baryons involves many more physical processes, some of which (such as
star formation, and feedback effects from stars and active galactic
nuclei (AGN)) are still poorly understood in detail, and interact in
complex ways. Galaxy formation therefore is a complex problem, and
progress in understanding it relies on combining insights from
analytical models, numerical simulations, and observations.

Two main theoretical approaches have been developed for trying to
understand how the complex and non-linear physics of the baryons leads
to galaxies with the properties observed in the real Universe: (i)
semi-analytical modelling, in which simplified mathematical
descriptions are adopted for the baryonic processes, which are then
applied to evolving dark matter halos calculated from N-body
simulations or by Monte Carlo methods; (ii) gas-dynamical simulations
which follow the gas dynamics in more detail, and try to model the
physical processes in a more fine-grained way.

Both approaches have advantages and disadvantages. The semi-analytical
approach is fast and flexible, allowing large parameter spaces to be
explored, and making it easy to generate mock catalogues of galaxies
over large volumes, which on the one hand can be compared to
observational data to test the model assumptions and constrain the
model parameters, and on the other hand, can be used to interpret
large observational surveys. Gas-dynamical simulations can calculate
the anisotropic distribution and flows of gas in much more detail and
with fewer approximations, and provide detailed predictions for the
internal structure of halos and galaxies, rather than just global
properties. However, cosmological gas-dynamical simulations of galaxy
formation are still restricted to relatively small volumes and are
forced to treat many important physical processes for the baryons
(e.g. effective equation of state of the cold ISM, star formation, and
feedback from both stars and AGN) using ``subgrid'' models, in which
the effects of processes occurring at scales below the resolution
limit of the simulation are calculated using simple analytical
expressions. These subgrid models, whose form is phenomenological,
contain various free parameters, which are then adjusted so that, in
analogy to semi-analytical models, the predictions from the simulation
agree with a predetermined set of observed properties such as the
stellar mass function
\citep[e.g.][]{Vogelsberger14,Schaye15,Crain15}. The use of these
subgrid models in simulations is thus closely analogous to the
approach in semi-analytical models, albeit on a smaller spatial
scale. Given the scope to vary more easily the treatment of different
baryonic processes within semi-analytical models compared with large
gas-dynamical simulations, the former are particularly useful for
testing the relative roles of different processes.

Here, we follow the semi-analytical (SA) approach, whose origins lie
in the early work by \citet{White78}. It was developed greatly in
sophistication by \citet{Cole91}, \citet{White91} and \citet{Lacey91},
who added much more detailed treatments of processes such as gas
cooling in halos, star formation in galaxy disks, feedback from
supernova explosions, chemical enrichment, and luminosity evolution of
stellar populations, as well as updating the structure formation model
to that of cold dark matter (CDM). However, the first papers to
incorporate self-consistently the merging of both dark matter halos
and galaxies in the SA approach were those by \citet{Kauffmann93} and
\citet{Cole94}, which used halo merger histories calculated using
different Monte Carlo methods based on the extended Press-Schecter
approach \citep{Bond91,Bower91,LC93}. \citet{Kauffmann99} extended the
SA approach to use halo merger histories extracted from cosmological
N-body simulations of the evolution of the DM. Since then,
semi-analytical models based on the same general principles have been
developed by several other groups
\citep[e.g.][]{Somerville99,Nagashima99,Menci02,Hatton03,Monaco07,
Lagos08,Benson12}.

Over the last decade, SA models have continued to increase in
sophistication, both by including additional physical processes such
as formation of supermassive black holes (SMBH)
\citep{Kauffmann00,Malbon07} and consequent feedback effects from
active galactic nuclei (AGN) \citep{Croton06,Bower06}, and by
replacing very simplified treatments of processes such as star
formation with more realistic ones \citep[e.g.][]{Lagos11a}. Alongside
these developments, SA models have been compared with an ever wider
range of observational data, placing ever more stringent constraints
on the models. In parallel, gas-dynamical simulations of galaxy
formation have also developed enormously, both in terms of numerical
resolution and dynamic range, and in the sophistication of the subgrid
modelling. There have been various studies over the years comparing
the predictions of SA models with gas-dynamical simulations
\citep[e.g.][]{Benson01,Yoshida02,Helly03,Bower12}. Comparisons of
state-of-the-art SA models with the latest generation of gas-dynamical
simulations show remarkable agreement between the two approaches, when
the SA models and simulations are calibrated on observational data in
similar ways \citep{Somerville15,Guo15}.

At the same time as SA models have developed in both scope and
sophistication, some studies, motivated by a desire for simplicity
over sophistication and accuracy, have reverted to much simpler
formulations, which extract a few ideas and ingredients from SA
models, but ignore most of the physics, such as halo and galaxy
mergers, gas cooling in halos, and any physical modelling of feedback
\citep[e.g.][]{Bouche10,Dave12}\footnote{These models do however
  include halo mass growth based on mean accretion histories}. As we
show in \S\ref{sec:simplistic}, such simplistic models are extremely
limited in their applicability, and completely fail to represent the
galaxy formation process accurately over the whole range of mass and
redshift, as revealed by the panoply of current observational data.

Semi-analytical models have led to important insights into fundamental
aspects of galaxy formation, including: showing the importance of
supernova (SN) feedback for establishing both the shallow slope of the
galaxy stellar mass or luminosity function compared to the halo mass
function at low masses, and the low fraction of baryons converted into
stars overall \citep{White78,Cole91,Lacey91,White91}; showing that AGN
feedback is required to explain why galaxy formation is suppressed so
effectively in high-mass halos \citep{Benson03,Bower06,Croton06}; the
general form of the cosmic star formation history
\citep{White91,Lacey93}; the origin of the galaxy clustering bias in
terms of galaxy formation physics and the first formulation of the
``halo occupation distribution'' \citep{Benson00}; the origin of the
metallicity-mass relation \citep{Cole00}; and the dependence of galaxy
colour and specific star formation rate on environment
\citep[e.g.][]{Baldry06}.

The Durham semi-analytical model, \galform, has undergone continual
development. The original version \citep{Cole94} was based on very
simplified halo merger trees. In \citet{Cole00} the code was rewritten
to use much more accurate halo merger trees based on the extended
Press-Schechter model; the treatment of processes such as gas cooling
in halos, star formation in galaxy disks and supernova feedback was
improved; and additional physical processes were added, including
starbursts triggered by galaxy mergers and disk instabilities,
chemical enrichment of stars and gas, calculation of sizes of galactic
disk and bulge components, and the effects of dust extinction on the
light emitted by galaxies. The resulting model was found to be in
generally good agreement with a wide range of properties of galaxies
in the local Universe, including galaxy luminosity functions at
optical and near-IR wavelengths, galaxy gas contents and
metallicities, galaxy disk sizes, and the fraction of disk- or
bulge-dominated galaxies. The model was extended by \citet{Granato00}
to calculate the reprocessing of starlight by dust, predicting far-IR
luminosity functions also in good agreement with observations of the
local Universe. The same model also predicted galaxy clustering in
excellent agreement with observations, without any further fine tuning
\citep{Benson00}.

However, the good fit of the \citet{Cole00} model to the break at the
bright end of the galaxy luminosity function resulted from assuming a
cosmic baryon fraction which was later shown to be too low. Subsequent
work showed that the \citet{Cole00} model also ran into problems at
high redshifts, predicting too few rapidly star-forming galaxies at
high redshifts ($z \sim 2-3$). This deficiency was found both in the
rest-frame far-UV, comparing with observations of Lyman-break galaxies
(LBGs), and in the rest-frame far-IR, comparing with number counts and
redshift distributions of faint sub-mm galaxies (SMGs). Solving these
problems motivated the development of the \citet{Baugh05} version of
\galform, in which a new channel of feedback was posited (following
\citealt{Benson03}), with SN-driven superwinds ejecting gas from halos
and thus reducing the gas cooling rates in massive halos, and
reproducing the observed bright-end break in the galaxy luminosity
function at $z=0$. The phenomenological star formation law in galaxy
disks was modified to make galaxies more gas-rich at high redshifts,
resulting in star formation dominated by starbursts at high
redshift. Finally, and most controversially, the initial mass function
(IMF) of stars formed in starbursts was made very top-heavy (while the
IMF for disk star formation remained of solar neighbourhood form), so
boosting both the stellar luminosities and dust production in
starbursts. This change in the IMF appreared necessary to reproduce,
in particular, the number counts and median redshift ($z\sim 2$) of
the SMGs observed at mJy fluxes at 850$\mum$, and also reproduced the
far-UV luminosity function of LBGs at $z\sim 3$. The \citet{Baugh05}
model was subsequently shown, without further adjustment, to predict
far-UV LFs of LBGs in excellent agreement with observations over the
whole range $z\sim 3-10$ \citep{Lacey11}, as well as galaxy evolution
at mid- and far-IR wavelengths in reasonable agreement with
observational data from \SPITZER\ at $z \lsim 2$ \citep{Lacey08}.

However, the SN superwinds feedback mechanism used in the
\citet{Baugh05} model had the physical drawback that it required an
implausibly large energy input from supernovae in order to produce the
correct break in the galaxy LF at $z=0$. Furthermore, the model was
subsequently shown to predict an evolution in the bright end of the
rest-frame $K$-band luminosity function (which is closely related to
the stellar mass function) in conflict with observations at $z \sim
1-2$, with the model predicting too few bright galaxies, implying that
this feedback mechanism had the wrong redshift dependence
\citep{Bower06}. These two problems were solved by \citet{Bower06},
who introduced into \GALFORM a mechanism of AGN feedback to replace
the SN superwind mechanism.

In the \citet{Bower06} model, supermassive black holes at the centres
of galaxies are assumed to accrete gas from galaxy halos at highly
sub-Eddington rates, with the accretion energy powering relativistic
jets, which are assumed to deposit energy in the hot gas halo,
balancing the effect of radiative cooling. This ``radio mode'' AGN
heating was assumed to be effective only for halos where the gas is in
the ``slow'' or ``quasistatic'' cooling regime, resulting in a
characteristic halo mass $\sim 10^{12} \Msol$, above which cooling of
gas in halos is mostly suppressed. With this feedback mechanism, the
model was able to reproduce the observed $K$-band LF not only at $z=0$
but also its evolution to $z \lsim 3$. By modifying the model to allow
also the gradual return to halos of gas ejected by supernova feedback
(rather than requiring this return to happen at discrete halo
formation events, as in \citet{Cole00} and \citet{Baugh05}), the
\citet{Bower06} model was also able to reproduce qualitatively the
observed bimodal distribution of galaxy colours (although not
quantitatively, see \citet{Gonzalez09}). These were important
successes. However, in order to simplify the task of finding an
acceptable model, \citet{Bower06} set aside the observational
constraints from gas contents, metallicities and disk sizes which had
been applied when calibrating model parameters in both the
\citet{Cole00} and \citet{Baugh05} models, and the resulting
\citet{Bower06} model, in fact, violated these constraints. Likewise,
they also set aside observational constraints from SMGs and LBGs at
high redshift. The model fails to match either of these constraints,
\citep[see e.g.][]{Lacey11}, with the consequence that they did not
need to vary the IMF in their model.

In summary, the two earlier \galform models make different physical
assumptions (superwinds and a varying IMF in \citet{Baugh05} versus
AGN feedback in \citet{Bower06}), and have different successes and
failures. The aim of this paper is to develop a single unified model
which combines features from both of these earlier models and can
simultaneously satisfy all of the key observational constraints
described above. (Another version of the \galform model has recently
been released by \citet{Gonzalez-Perez14}. This uses many of the same
ingredients as in the model presented here, including the same
cosmology, but with the important difference that a single IMF is
assumed. As a consequence, the \citeauthor{Gonzalez-Perez14} model
fails to reproduce some key observations, such as the redshift
distribution of SMGs, although it does successfully match many other
observational constraints.)

The unified model which we present in this paper incorporates or uses
various theoretical and observational advances since \citet{Baugh05}
and \citet{Bower06}. (i) We now have a better observational
understanding of the relation between star formation rates (SFRs) and
gas contents in galaxy disks at low redshifts, allowing the use in our
model of an empirical star formation law based on star formation from
molecular gas, first implemented in \GALFORM by \citet{Lagos11a}. (ii)
New models of stellar population synthesis (SPS) are available which
include an improved treatment of the luminosity from the thermally
pulsing asymptotic giant branch (TP-AGB) phase of stellar evolution
\citep{Maraston05}. (iii) Thanks to more recent measurements of the
cosmic microwave background (CMB), we now have { improved} estimates
of the cosmological parameters. (iv) In addition, we have also updated
the treatments of some other physical processes, such as dynamical
friction on satellite galaxies in halos. (v) Thanks to observations by
\Herschel, we now have measurements of the evolution of the galaxy
population at far-IR wavelengths extending back to $z \sim 2$. Since
most of the star formation over the history of the Universe has been
obscured by dust, observations of the far-IR emission from dust are
crucial in constraining galaxy evolution models. (v) Finally, galaxies
have now been observed back to $z \sim 10$ \citep[e.g.][]{Oesch12a}.

An important feature of our approach is that we try to test the models
and constrain their parameters by comparing theoretical predictions
with directly observed quantities, such as galaxy luminosities at
different wavelengths (``forward modelling''), rather than with
quantities inferred from observations, such as stellar masses and SFRs
(``backwards inference''). The latter approach to testing models has
become very popular in recent years \citep[e.g.][]{Guo11}. However, it
has the drawback that stellar masses and SFRs can only be inferred
from observations by using models for stellar populations and dust
absorption and re-emission in galaxies, together with assumptions for
the IMF, for the form of the star formation history and for the
metallicity. All of these are currently uncertain, as analysed in {
  various papers
  \citep[e.g.][]{Conroy09,Gallazzi09,Zibetti09,Pforr12,Mitchell13}.}
Of course, in the approach where we forward model to predict
observable quantities, we also have to use stellar population models
and make assumptions for the IMF, but at least the star formation and
chemical enrichment histories are predicted and accounted for
self-consistently. Furthermore, in the backwards inference approach
where stellar masses and SFRs are inferred from observed SEDs, the
dust absorption is generally modelled as due to a foreground screen,
which is unrealistic. In contrast, in the forward modelling approach,
we use the model predictions for the mass and geometrical distribution
of the dust, together with a physical radiative transfer model, in
order to predict the dust absorption self-consistently. The forward
modelling approach is thus fully self-consistent, while the backwards
inference approach is not.
We therefore argue that the forwards modelling approach, where we
compare model predictions for directly observable quantities with
observational data, { in principle} provides the more robust
procedure for comparing models with observations.
{ Even more importantly,} if the galaxy formation model includes a
varying IMF, as does the model in this paper, then the only rigorous
way to compare the model with observations is in terms of the
observable quantities, since the values of stellar masses and SFRs
inferred from observational data depend strongly on the assumed IMF,
which is no longer unique. However, having constrained model
parameters by comparing with observable quantities, it is then still
of great interest to examine the model predictions for physical
quantities such as stellar masses and SFRs, and we do this later in
the paper.

Some predictions from this model have already published in other
studies
\citep{Fanidakis13a,Fanidakis13b,Mitchell13,Guo14,Lagos14,Lagos14b,
  Bethermin15,Bussmann15,Campbell15,Cowley15,Cowley15b,Cowley16,Gutcke15,
  Lagos14,Lagos15,Farrow15}\footnote{where { the model} is variously referred to as Lacey13,
  Lacey14, Lacey15 or Lacey \etal}, but this is the first paper in
which the model and its calibration are described in full.

The plan of this paper is as follows: In \S\ref{sec:foundations} we
describe the general methodology of semi-analytical models. In
\S\ref{sec:model} we describe the specific implementation of the
\GALFORM semi-analytical model used in this paper, and how it differs
in its assumptions from previous versions of \GALFORM. In
\S\ref{sec:fiducial} we describe the set of observational constraints
we use for calibrating the model, and show how our fiducial model
performs against these constraints. In \S\ref{sec:param-variations} we
explore which observables constrain which physical processes. In
\S\ref{sec:phys-predictions} we examine what the fiducial model
predicts for the evolution of key physical quantities such as the
stellar mass function and SFR density. In \S\ref{sec:simplistic} we
compare our modelling approach to that used in more simplistic
models. In \S\ref{sec:disc} we discuss our results, and in
\S\ref{sec:conc} we conclude.








\section{Principles and aims of the semi-analytical approach}
\label{sec:foundations}

In this section, we describe the general methodology of
semi-analytical models. The aim of such models is to understand how
galaxies formed, but this can be attempted at different levels,
depending on the level of detail in the modelling of physical
processes. Galaxy formation is determined by a complex interaction
between gravity, fluid dynamics and thermal and radiative
processes. In semi-analytical models, rather than calculate all of
these processes in fine-grained detail, we make simplifying
assumptions regarding geometry and timescales. This enables us to
describe galaxy formation by a set of coupled non-linear equations for
the evolution of various global properties of galaxies and their host
halos. These are a mixture of differential equations in time for
continuous processes (e.g. gas cooling, star formation) combined with
algebraic equations for processes modelled as discrete transformations
(e.g. galaxy mergers, disk instabilities). These equations for
different physical processes contain parameters whose values are
estimated by a variety of methods: from general theoretical arguments;
from targeted numerical simulations; from direct observational
measurements of the process concerned; or by comparing predictions
from the galaxy formation model with observations of the galaxy
population.

SA modelling has several aims:
\begin{enumerate}
\item By using a simplified but at the same time comprehensive
  theoretical framework, we hope to obtain a better intuitive
  understanding of the effects of different physical processes,
  something which is difficult using gas-dynamical simulations.
\item It provides a flexible way of combining a wide set of different
  physical processes together in a consistent way, and exploring what
  such combinations predict for the observable properties (and
  evolution) of the galaxy population, { including how different
    processes interplay in their effects.} By comparing such
  predictions with observational data, we can then learn about whether
  the model is complete, or whether additional physical processes need
  to be included. Examples of processes which are now regarded as
  fundamental and whose importance was revealed in this way include
  feedback from SN \citep[e.g][]{White78} and from AGN
  \citep{Benson03,Bower06,Croton06}.
\item SA modelling provides a means for interpreting observational
  data within a consistent theoretical context, and for assembling
  different types of observational data taken from different redshifts
  into a consistent evolutionary picture.
\end{enumerate}

The fact that some of the parameters in semi-analytical models need to
be calibrated by comparing predictions from the model with
observational data often leads to the criticism that such models lack
predictive power. However, this criticism is misplaced. In our
approach, we compare the model predictions with a very wide range of
observational data. We use only a subset of these observational data
to constrain the model parameters. Once we have done this, the model
is fully specified, and can be used to make genuine predictions for
other observable properties.

We emphasize that the purpose of semi-analytical modelling is not
simply to match all of the observational data, but to gain physical
understanding. In some cases, improved fits to particular
observational datasets could be obtained by fine-tuning the models by
adding {\em ad hoc} ingredients devoid of physical motivation or
meaning specifically for this purpose. However, such an approach would
be contrary to the principles of semi-analytical modelling, as it does
not lead to any improved physical understanding. Instead, when we find
discrepancies between model predictions and observational data, we use
this to try to advance our understanding, by seeking to understand
whether this points to some missing physics in the model, or the need
for improvements how some physical process is treated, or to a
possible flaw in the observational data. However, given that we aim to
construct a model which is physically realistic but still simplified,
we do not expect it to be able to reproduce all observational datasets
to arbitrary precision.


\section{Astrophysics of galaxy formation}
\label{sec:model}

In this section we first give an overview of our new model of galaxy
formation, listing the basic components of the calculation and
pointing out the similarities and differences from previous releases
of the \galform model (\S\ref{ssec:overview}). We then give a
comprehensive description of all of the components of the model. This
is intended to be self-contained. The reader who is more interested in
an executive summary of how the model presented in this paper differs
from our previous work may wish to focus on \S\ref{ssec:overview},
and omit the more detailed exposition of the model in a first
reading. Our model is discussed in the context of a simple,
reductionist view of galaxy formation in \S\ref{sec:simplistic}.

\subsection{Overview: basic processes modelled and relation to 
previous models}
\label{ssec:overview}

We carry out an {\it ab initio} calculation of the formation and
evolution of galaxies using the semi-analytical model \galform, which
is set in the context of the hierarchical growth of structure in the
dark matter (for reviews of hierarchical galaxy formation, see
\citealt{Baugh06}, \citealt{Benson10b} and \citealt{Somerville15}).
The processes included in our calculation are listed below, followed
by the subsection in which a more extensive discussion of the
implementation is given: (i) the collapse and merging of DM halos
(\S\ref{ssec:DM}); (ii) the shock-heating and radiative cooling of gas
inside DM halos, leading to the formation of galactic disks
(\S\ref{ssec:hot_gas}); (iii) star formation (SF) in galaxy disks and
in starbursts (\S\ref{ssec:SFR}); (iv) feedback from supernovae (SNe),
from AGN and from photo-ionization of the IGM (\S\ref{ssec:feedback});
(v) galaxy mergers driven by dynamical friction within common DM
halos, and bar instabilities in galaxy disks, which can trigger bursts
of SF and lead to the formation of spheroids (\S\ref{ssec:dynamics});
(vi) calculation of the sizes of disks and spheroids
(\S\ref{ssec:sizes}); (vii) chemical enrichment of stars and gas
(\S\ref{ssec:chem}).  Galaxy stellar luminosities are computed from
the predicted star formation and chemical enrichment histories using a
stellar population synthesis model (\S\ref{ssec:stars_dust}). The
reprocessing of starlight by dust, leading to both dust extinction at
UV to near-IR wavelengths, and dust emission at far-IR to sub-mm
wavelengths, is calculated self-consistently from the gas and metal
contents of each galaxy and the predicted scale lengths of the disk
and bulge components using a radiative transfer model
(\S\ref{ssec:stars_dust}).

\galform was introduced by \cite{Cole00} to model the 
processes listed above in a cold dark matter universe (see also 
\citealt{Benson10a}). This early calculation enjoyed a number of 
successes. Once the model parameters were chosen to reproduce a 
subset of the available observations of the local galaxy population 
(e.g. the observed break and faint-end slope of the optical and 
near-infrared luminosity functions), the \citeauthor{Cole00} model was 
able to match, for example, the observed scale-length distributions 
of galactic disks and the gas-to-luminosity ratio in spirals and irregulars.  

However, the \citeauthor{Cole00} model had two major problems which
motivated subsequent revisions to \galform.  The first of these
concerned the predictions for the high redshift Universe, which
disagreed significantly with observations.  The model predicted more
than an order of magnitude fewer galaxies than was observed in the
rest-frame UV at $z=3$, after taking into account a realistic
calculation of the impact of dust extinction on the predicted UV
luminosity function \citep{Granato00}. A related problem was the
number counts of galaxies detected through emission at sub-millimetre
wavelengths, due to dust heated by starlight.  At an $850 \mu$m flux
of $\sim 5$mJy, the \citeauthor{Cole00} model predicts around 30 times
fewer galaxies than are observed.  The second problem concerned the
predicted break in the local galaxy luminosity function at optical and
near-IR wavelengths. While this was reproduced in the original
\citeauthor{Cole00} model, this was dependent on the value assumed for
density parameter of baryons, $\Omega_{b0}$.
The value used by \citeauthor{Cole00}, whilst consistent with the
constraints available at the time, is around half of the best-fitting
value today.  Increasing the baryon fraction leads to more gas cooling
in massive haloes.  \citeauthor{Cole00} allowed the density profile of
hot gas to differ from that of the dark matter, with the possibility
of a constant density core in the gas distribution which grows as
radiative cooling removes lower entropy gas. Whilst this led to some
increase in the gas cooling time in higher-mass haloes, in general
this functionality did not suppress gas cooling sufficiently to
reconcile the predicted number of bright galaxies with observations,
once $\Omega_{b0}$ was increased to a value consistent with more
recent constraints.

These problems with the \citeauthor{Cole00} model illustrate the
central principle behind semi-analytical modelling. The physics of
galaxy formation is encoded, to the best of our ability, in a set of
equations which contain some parameters. The parameter values are
chosen to reproduce a subset of observations. The specified model is
then compared to other observations. If the model does not match these
observations, then either a better model lies in a different part of
parameter space or the original calculation is missing some process or
needs to be improved in some way. The two problems faced by the
\citeauthor{Cole00} model, the failure to match the high-redshift
Universe and the difficulty in reproducing the break of the present
day luminosity function with a realistic baryon density, drove two
efforts to improve the model which until now have been pursued
essentially independent of one another.

The first extension was introduced by \cite{Baugh05}. After an
extensive exploration of the model parameter space,
\citeauthor{Baugh05} concluded that the only way to reconcile the
model predictions with observations of high-redshift galaxies was to
adopt a top-heavy stellar initial mass function (IMF) in bursts of
star formation triggered by galaxy mergers.  This choice was not taken
lightly. The framework of the \galform calculations imposes
restrictions on the model parameter space that are widely
under-appreciated. By requiring that the model reproduce the local
galaxy population, a large swathe of parameter space is immediately
excluded (see \citealt{Bower10}).  Similarly, by adopting a
self-consistent calculation of the extinction of starlight by dust and
the radiation of this energy at longer wavelengths, much of the
freedom present in more simplistic calculations (e.g. to set by hand
the amount of dust extinction or the temperature of the dust) is
removed.  The \cite{Baugh05} model gave an excellent match to the
number counts and redshift distribution of galaxies observed in the
sub-millimetre and to the $z=3$ rest-frame UV luminosity
function. This model was subsequently shown to reproduce the observed
UV luminosity function out to $z=10$ \citep{Lacey11}.

The problem of reproducing the location and sharpness of the observed
break at the bright end of the local galaxy luminosity function was
investigated by \cite{Benson03}. These authors demonstrated, that, for
a realistic baryon density, it was possible to predict the observed
number of bright galaxies by invoking a wind which removed baryons
from intermediate mass haloes. This had the consequence of reducing
the gas density in massive haloes, thereby reducing the rate at which
gas cools. However, if the wind was to be driven by supernovae, the
coupling of the energy released by the supernovae to the wind would
have to be extraordinarily efficient. A more plausible energy source
was identified as the energy released by the accretion of material
onto a supermassive black hole. A few years later, several groups
introduced heating by Active Galactic Nuclei (AGN) into
semi-analytical models of galaxy formation, as a means to suppress gas
cooling in massive haloes
\citep{Bower06,Croton06,Cattaneo06,Monaco07,Lagos08}.

The objective of this paper is to combine the best features of the
models of \citeauthor{Baugh05} and \citeauthor{Bower06}, along with
other subsequent improvements in the treatment of various processes in
\galform.  This effort is motivated by the realization that both
models have attractive features that should be retained, if possible,
but they also have shortcomings to be resolved. For example, the
\citeauthor{Bower06} model gives an excellent match to the observed
evolution of the K-band luminosity function yet fails to match the
rest-frame UV luminosity function of Lyman-break alaxies (LBGs) at
high redshift \citep{Lacey11} or the number counts and redshift
distribution of sub-mm galaxies (SMGs).  The \citeauthor{Baugh05}
model does match the LBG and SMG observations, yet fails to match the
evolution of the K-band luminosity function. To our knowledge, there
is no model in the literature which is {\it simultaneously} able to
match: (i) the observed optical and near-IR luminosity functions of
$z=0$ galaxies; (ii) the evolution of the bright end of the rest-frame
K-band luminosity function; (iii) the evolution of the rest-frame UV
luminosity function; (iv) the number counts and redshift distribution
of SMGs.

Our objective is to establish whether or not these ideals can be achieved 
with a single model. A related question we address, is that, given the 
improvements to \galform since \citeauthor{Baugh05}, do we still need to 
invoke a top-heavy IMF to explain observations of the high-redshift 
Universe? And, if the answer is ``yes'', do we need such an extreme 
IMF as the one used by \citeauthor{Baugh05}? 

In summary the new features of the model introduced in this paper,
compared to the models of \citeauthor{Baugh05} and
\citeauthor{Bower06}, are:

\begin{enumerate}
\item The adoption of the best-fitting cosmological 
parameters of the cold dark matter (CDM) model based on recent data. 
\item A new treatment of star formation in galactic disks, 
which follows the atomic and molecular hydrogen content of 
the ISM, as implemented in \galform by \cite{Lagos11a,Lagos11b,Lagos12} 
\item A more accurate description of the dynamical friction timescale 
for galaxy mergers, calibrated against numerical simulations \citep{Jiang08}. 
\item The use of a stellar population synthesis model which
includes the contribution from stars in the thermally-pulsing
asymptotic giant branch (TP-AGB) stage of stellar evolution
\citep{Maraston05}.
\end{enumerate}

Various consequences of using the new treatment of disk star formation
in \galform have previously been explored by
\cite{Lagos11a,Lagos11b,Lagos12}, but without retuning most of the
model parameters from their values in
\citet{Bower06}. \citet{Gonzalez-Perez14} presented a new \galform
model using the same cosmology and star formation prescription as in
this paper, which was retuned to match a range of observational data,
but still with the assumption of a universal Solar neighbourhood IMF
for all star formation, with the consequence that it is unable to
reproduce some observational constraints, such as the redshift
distribution of SMGs.

\subsection{Dark matter halos}
\label{ssec:DM}
The basic framework for our galaxy formation model is provided by the
assembly histories, density profiles and angular momenta of the dark
matter halos in which gas collapses and cools to form galaxies. We
require both a halo mass function, specifying the number density of
halos as a function of mass and redshift, and also merger trees
describing how these halos are hierarchically assembled by mergers of
smaller objects. We have two approaches for obtaining these quantities
in \galform: (i) use an analytical expression for the halo mass
function, and halo merger trees generated using a Monte Carlo method
based on the Extended Press-Schechter (EPS) model \citep{Cole00}, with
improvements by \citet{Parkinson08}; or (ii) use halos and halo merger
trees extracted from an N-body simulation of the dark matter
\citep{Helly03,Bower06}. The two approaches give very similar results
for statistical quantities such as galaxy mass or luminosity
functions. In this paper, we will mainly use the second approach,
based on N-body simulations, since it allows us to also predict the
spatial distribution of galaxies. 

The halo merger trees are constructed using the method described in
\citet{Merson13} and \citet{Jiang14}.  For this paper, we use the
Millennium-WMAP7 (or MR7) N-body simulation of dark matter in a flat
$\Lambda$CDM universe
which assumes cosmological parameters based on the WMAP-7 dataset
\citep{Komatsu11}, with $\Omega_{\rm m0}=0.272$, $\Omega_{\rm v0}=0.728$,
$\Omega_{\rm b0}=0.0455$ and $h=H_0/(100~\kms)=0.704$, and an initial power
spectrum with slope $n_{\rm s}=0.967$ and normalization
$\sigma_8=0.810$.\footnote{This is the simulation referred to as MS-W7
  in \citet{Guo13} and \citet{Gonzalez-Perez14} and as MW7 in
  \citet{Jenkins13}. }
The simulation has a boxsize $500 \hMpc$ and a particle mass $9.364
\times 10^8 \hMsol$, corresponding to a minimum resolvable halo mass
$1.87 \times 10^{10}\hMsol$. Merger trees are constructed from outputs
at 61 different redshifts.

{ The halo mass resolution in the N-body simulation used here has
  some effects on the properties calculated for the galaxy population,
  especially at low galaxy masses. We plan to make a detailed study of
  the convergence of the \GALFORM predictions with respect to the halo
  mass resolution and the redshift spacing of the N-body outputs used
  for constructing the merger trees in a future paper. For this paper,
  we have made only a limited investigation of the effects of the halo
  mass resolution, by comparing with results obtained using an N-body
  simulation having the same cosmology but much { higher} mass resolution and
  smaller volume (the DOVE simulation described in \citet{Jiang15}),
  as well as results obtained using Monte Carlo merger trees with
  higher mass resolution. Based on these comparisons, we have
  estimated down to what galaxy mass or luminosity the resulting mass
  or luminosity functions are insensitive to the minimum halo mass in
  the Millennium-WMAP7 simulation. We have indicated these resolution
  limits in the relevant plots.}

When they form, halos are assumed to have virial radii $\rvir =
(3\Mhalo/(4\pi \Delta_{\rm vir} \overline\rho))^{1/3}$, where $\Mhalo$ is the
halo mass, $\overline\rho$ is the cosmological mean density at that
redshift, and the overdensity $\Delta_{\rm vir}(\Omega_m,\Omega_v)$ is
calculated from the spherical top-hat collapse model
\citep[e.g.][]{Eke96}. The DM density profiles of halos are assumed to
have the NFW form \citep{Navarro97}:
\begin{equation}
\rho_{\rm DM}(r) \propto \frac{1}{(r/r_{\rm s})(1+r/r_{\rm s})^2} ,
\end{equation}
where $r_{\rm s}$ is the scale radius, related to the virial radius by
the concentration, $r_{\rm s} = \rvir/c_{\rm NFW}$. We calculate
$c_{\rm NFW}$ using the analytical prescription of
\citet{Navarro97}\footnote{Modified to account for the slightly
  different definition of virial radius used there.} (see also
\citealt{Gao08}). Halos grow by merging with other halos and by
accretion. When a halo has grown by a factor 2 in mass, we treat this
as a new ``halo formation'' event, and update the density profile
according to the mass and redshift at this formation event (see
\citealt{Cole00} for more details). Between such halo formation
events, the halo mass and radius continue to grow, but the circular
velocity $\Vvir$ and halo concentration $c$ are assumed to remain
constant.

Halos have angular momentum, acquired through tidal torques. Based on
the results of N-body simulations \citep[e.g.][]{Cole96}, we calculate
the halo angular momentum in the model by randomly drawing a value of
the dimensionless spin parameter $\lambda = \Jhalo |E_{\rm
  halo}|^{1/2}/(G\Mhalo^{5/2})$ from a lognormal distribution having a
median $\lambda_{\rm med}=0.039$ and a dispersion
$\sigma_{\lambda}=0.53$ in $\ln\lambda$\footnote{These values are also
  very close to the best-fit lognormal parameters from
  \citet{Bett07}}. The halo spin is calculated anew at each halo
formation event. We do not keep track of the direction of this angular
momentum, only its magnitude.

\begin{figure}

\begin{center}

\includegraphics[width=8cm]{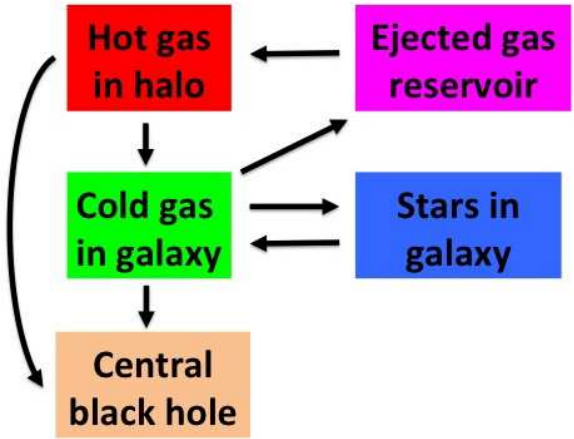}

\end{center}

\caption{Flow chart showing the different baryonic components in a
  halo, and transfers between them.}

\label{fig:baryons}
\end{figure}

\subsection{Gas in halos} 
\label{ssec:hot_gas}
The \galform model assumes that the baryons associated with a DM halo
are in five different components: hot gas in halos (available for
cooling), a reservoir of gas ejected from the halo by feedback (not
yet available for cooling), cold gas in galaxies, stars in galaxies,
{ and central black holes (BH) in galaxies.} The physical processes
causing mass transfers between these different components are shown
schematically in Fig.~\ref{fig:baryons}, and discussed in the
following subsections.

Gas falling into halos must dissipate its energy through radiative
cooling in order to condense into a galaxy and form stars. We assume
that gas falling into halos is all shock-heated to the virial
temperature $\Tvir = (\mu m_{\rm H}/2\kB) \Vvir^2$, where $\Vvir =
(GM/\rvir)^{1/2}$, and $\mu$ is the mean molecular weight, 
and then settles into a spherically symmetric distribution with
density profile
\begin{equation}
\rhohot(r) \propto \frac{1}{(r^2 + r_{\rm c}^2)} ,
\label{eq:rhohot}
\end{equation}
with gas core radius $r_{\rm c} = 0.1 \rvir$. The hot gas then loses its
thermal energy by radiative cooling due to atomic processes, at a rate
per unit volume $\rhohot^2 \Lambda(\Tvir,\Zhot)$ (assuming collisional
ionization equilibrium), where $\Zhot$ is metallicity of this
gas. The local cooling time, defined as the timescale for the gas to
radiate its thermal energy, is then
\begin{equation}
\taucool(r) = \frac{3}{2} \frac{\kB}{\mu m_{\rm H}}
\frac{\Tvir}{\rhohot(r) \Lambda(\Tvir,\Zhot)} .
\label{eq:taucool}
\end{equation}
We use the metallicity-dependent cooling function $\Lambda(T,Z)$
tabulated by \citet{Sutherland93}. Given the gas density profile
$\rhohot(r)$ and the formation time $\tform$ for the halo, we
calculate the radius $\rcool(t)$ at which the cooling time equals the
time since halo formation by solving $\taucool(\rcool) = t-\tform$.

Given the density profile of the dark matter $\rho_{\rm DM}(r)$, we can
also calculate the free-fall timescale $\tauff(r)$, defined as the
time for a particle to fall from radius $r$ to the halo centre under
the force of gravity alone. The corresponding free-fall radius
$\rff(t)$ is then defined through $\tauff(\rff) = t-\tform$. We
then define an accretion radius for the halo gas as
\begin{equation}
\racc(t) = \min[\rcool(t),\rff(t)] .
\label{eq:racc}
\end{equation}
This is the radius within which halo gas both has time to cool and
time to fall to the centre. If the calculated $\racc$ exceeds the
virial radius, then we set $\racc=\rvir$. We assume that the rate at
which gas drops out of the halo and accretes onto the galaxy at the
centre of the halo is
\begin{equation}
\Mdotacc = 4\pi \, \racc^2 \, \rhohot(\racc) \, \frac{d\racc}{dt} .
\label{eq:Mdot_acc}
\end{equation}
We assume that gas can only accrete onto the {\em central} galaxy in a
halo, and not onto any {\em satellite} galaxies.  Note that $\tform$
and $\Tvir$ are reset after each halo formation event, but $\Mhalo$,
$\rvir$ and $r_{\rm c}$ are all updated continually as the halo grows,
and the normalization of the gas density profile in
eqn.(\ref{eq:rhohot}) is also updated continually to account for
accretion of gas due to halo growth.

As mentioned, satellite galaxies are treated differently from central
galaxies in \galform. All galaxies are assumed to originate as central
galaxies, but when the halo of a galaxy merges with another more
massive halo, that galaxy is assumed to become a satellite in the new
larger halo. In the present model, as in most previous versions of
\galform, we assume that the hot gas halo of the satellite is
instantly stripped away by the ram pressure of the hot gas in the main
halo as soon as the galaxy becomes a satellite, and added to the main
hot gas halo. Consequently, no gas is able to cool onto satellite
galaxies. (This assumption of instantaneous ram pressure stripping has
been relaxed in the \galform model of \citet{Font08}, who considered
the effects of gradual ram pressure stripping, and also in
\citet{Lagos14b}, who consider a variant of the Lacey16 model in the
context of predicting the gas contents of early-type galaxies.)

This model for accretion of gas from the halo (which is essentially
identical to that in \citeauthor{Cole00}) predicts two different
accretion modes: {\em hot accretion} when $\taucool > \tauff$, for
which gas accretes in a quasi-static cooling flow, and {\em cold
  accretion} when $\taucool < \tauff$, for which gas cools rapidly and
then falls in at the free-fall speed. There has been much debate in
recent years about whether or not most of the gas accreted by galaxies
was ever shock-heated close to the virial temperature and radius of
the host halo. \citet{Birnboim03} used a combination of analytical
calculations and 1D hydrodynamical simulations to argue that
shock-heating was only effective in more massive halos ($M \gsim
10^{11}-10^{12} \Msol$). Subsequent studies using 3D hydrodynamical
simulations with both SPH and fixed-mesh Eulerian codes seemed to
support the picture that most gas was accreted onto galaxies through
cold flows \citep[e.g.][]{Keres05,Ocvirk08}. However, recent
simulations using the new moving-mesh hydrodynamic code AREPO imply
that the earlier simulation results suffered from numerical
inaccuracies, and that most of the gas forming galaxies does in fact
get shock-heated to the halo virial temperature as it falls into
halos, and then radiatively cools \citep{Nelson13}, although this
shock-heating may occur well within the virial radius. Whichever of
these viewpoints about shock heating turns out to be more correct, the
consequences for the rate of accretion of gas onto galaxies are
probably modest. As discussed in \citet{Benson11}, the
\citet{Birnboim03} criterion for gas to be shock-heated near the
virial radius is similar to the condition for $\taucool > \tauff$ at
this radius. As such, gas accreting onto the halo which avoids
shock-heating according to the \citeauthor{Birnboim03} criterion, in
the \galform model will typically have $\taucool < \tauff$, and so
will in any case fall in from the virial radius on the free-fall
timescale, leading to a very similar mass accretion rate onto the
central galaxy in either case. Furthermore, as shown by
\citet{Benson11}, once the reheating of gas by supernova feedback is
included, the differences for predictions of galaxy formation between
these two approaches become even smaller.

The gas accreted from the halo has angular momentum, and so forms a
disk at the halo centre.  We assume that at the time the halo forms,
the gas in the halo has the same specific angular momentum as the dark
matter, which in turn is related to the halo spin parameter
$\lambda$. The halo gas is assumed to have a constant rotation speed
around a fixed axis. The specific angular momentum of the gas
accreting onto the central galaxy at time $t$ is then equal to that in
a spherical shell of radius $\racc$ (see \citealt{Cole00} for more
details). We assume that the disk of the central galaxy always has its
angular momentum aligned with that of the current halo, so that the
angular momentum of the accreted gas adds linearly to that already
there. (See \citet{Lagos15} for a different approach within the \galform
framework, which relaxes this assumption.)

\subsection{Star formation in galaxies}
\label{ssec:SFR}
Cold gas in galaxies is able to form stars. Galaxies are assumed to
contain separate disk and spheroid components, each of which can
contain stars and gas. We assume two separate modes of star formation,
the {\em quiescent} mode (in the disk) and the {\em starburst} mode
(associated with the spheroid). Gas accreted from the halo is assumed
to add to the disk. Galaxy mergers and disk instabilities can transfer
this gas to a starburst component associated with the spheroid.

\subsubsection{Star formation in disks}
We calculate the star formation rate (SFR) in the disk using the
empirical \citet{Blitz06} law (as implemented in \galform in
\citealt{Lagos11a}), which is based on observations of nearby
star-forming disk galaxies (see also \citealt{Leroy08,Bigiel08}). In
this formulation, the cold gas in the disk is divided into atomic and
molecular phases, with the local ratio of surface densities
$\Sigma_{\rm atom}$ and $\Sigma_{\rm mol}$ at each radius in the disk
depending on the gas pressure, $P$, in the midplane as
\begin{equation}
R_{\rm mol} = \frac{\Sigma_{\rm mol}}{\Sigma_{\rm atom}} 
= \left( \frac{P}{P_0} \right)^{\alpha_P} .
\label{eq:at_mol}
\end{equation}
We use $\alpha_P=0.8$ and $P_0/\kB = 1700~\cm^{-3} \K$ based on
observations \citep{Leroy08}. We calculate the pressure from the
surface densities of gas and stars, as described in
\citet{Lagos11a}. The SFR is then assumed to be proportional to the
mass in the molecular component only; integrated over the whole disk,
this gives an SFR
\begin{equation}
\psi_{\rm disk} = \nuSF M_{\rm mol,disk} 
= \nuSF f_{\rm mol} M_{\rm cold,disk} ,
\label{eq:SFR_mol}
\end{equation}
where $f_{\rm mol} = R_{\rm mol}/(1+R_{\rm mol})$. \citet{Bigiel11}
find a best-fit value $\nuSF = 0.43 \Gyr^{-1}$ for a sample of local
galaxies, with a $1\sigma$ range of 0.24~dex around this. We treat
$\nuSF$ as being an adjustable parameter in the model, but only within
the $1\sigma$ range described. The disk SFR law (\ref{eq:SFR_mol}) has
a non-linear dependence on the total cold gas mass through the
dependence on $f_{\rm mol}$. As discussed in more detail in
\citet{Lagos11a}, at low gas surface densities, $f_{\rm mol} \ll 1$,
resulting in a steeper than linear dependence of SFR on cold gas mass,
while at high gas surface density, $f_{\rm mol} \approx 1$, resulting
in a linear dependence. { The effects of using an SFR law based on
  molecular gas have been investigated in \GALFORM by
  \citet{Lagos11b,Lagos12,Lagos14} and \citet{Gonzalez-Perez14}, and
  in other SA models by \citet{Fu10,Fu12}, \citet{Berry14} and
  \citet{Popping14}.}

\subsubsection{Starbursts}
For star formation in bursts, we assume that $f_{\rm mol} \approx 1$, but
with a dependence of the SFR timescale on the dynamical timescale in
the host spheroid
\begin{equation}
\psi_{\rm burst} = \nu_{\rm SF,burst} M_{\rm cold,burst} 
= \frac{M_{\rm cold,burst}}{\taustarburst} ,
\label{eq:SFR_burst}
\end{equation}
where 
\begin{equation}
\taustarburst = \max [ \fdyn\tau_{\rm dyn,bulge},\tauburstmin ] ,
\label{eq:taustar_burst}
\end{equation}
and the bulge dynamical time is defined in terms of the half-mass
radius and circular velocity as $\tau_{\rm dyn,bulge} =
\rbulge/\Vc(\rbulge)$.  Eqn.(\ref{eq:taustar_burst}) has the behaviour
that $\taustarburst \propto \tau_{\rm dyn,bulge}$ when the dynamical
time is large, but has a floor value when it is small. A scaling of
the SFR timescale in bursts with the dynamical time was suggested by
\citet{Kennicutt98}, based on observations of galaxies in the local
Universe, with a value $\fdyn \sim 50-100$ in our notation.

\subsection{Feedback}
\label{ssec:feedback}
\galform includes three modes of feedback by stars and AGN on the
galaxy formation process.
\subsubsection{Photoionization feedback}
The IGM is reionized and photo-heated by ionizing photons produced by
stars and AGN. This inhibits subsequent galaxy formation in two ways:
(i) the increased IGM pressure inhibits the collapse of gas into dark
matter halos; (ii) continued photo-heating of gas inside halos by the
ionizing UV background inhibits the cooling of gas. We model these
effects by assuming that after the IGM is reionized at a redshift
$z=\zreion$, no cooling of gas occurs in halos with circular
velocities $\Vvir < \Vcrit$. This simple model of photoionization
feedback has been shown to reproduce more detailed treatments quite
well \citep{Font11}. We adopt the standard value $\zreion=10$
\citep[e.g][]{Dunkley09}, and $\Vcrit = 30~\kms$, based on gas
dynamical simulations \citep{Hoeft06,Okamoto08}. The latter value
corresponds to a virial temperature $\Tvir = 3.3 \times 10^4\K$, and
to a halo mass $\Mhalo = 9.0 \times 10^9~\hMsol$ at $z=0$.

\subsubsection{Supernova feedback}
Supernova explosions inject energy into the ISM, which causes gas to
be ejected from galaxies. The energy injection is typically dominated
by Type~II supernovae due to short lived, massive stars, and so is
approximately proportional to the SFR. We make the standard assumption
that the rate of gas ejection due to supernova feedback is
proportional to the instantaneous SFR { $\psi$}, with a ``mass loading'' factor
$\beta$ that depends as a power law on the galaxy circular velocity $\Vc$:
\begin{equation}
{\dot M}_{\rm eject} = \beta(\Vc) \psi = \left( \frac{\Vc}{\VSN} 
\right)^{-\gammaSN} \psi .
\label{eq:Meject}
\end{equation}
{ This is calculated separately for star formation in disks and
  starbursts, and the results added to get the total ejection rate.}
The circular velocity used is that at the half-mass radius of the disk
for disk star formation, and of the spheroid for starbursts.  This
formulation involves two adjustable parameters, $\gammaSN$, which
specifies the dependence of $\beta$ on circular velocity, and $\VSN$
which specifies the normalization.\footnote{Note that in our previous
  papers, \citep[e.g.][]{Cole00} these parameters were called
  $\alpha_{\rm hot}$ and $V_{\rm hot}$.}  { We assume that cold gas
  is ejected from galaxies at the rate ${\dot M}_{\rm eject}$} to
beyond the virial radius of the host DM halo. The motivation for this
form for $\beta$ is that, for a given SN energy injection rate, the
efficiency of mass ejection into the halo should decrease with
increasing depth of the gravitational potential well, which is related
to $\Vc$. Unlike in \citet{Baugh05}, there is no ``superwind'' term in
the SN feedback.

Gas which has been ejected from the galaxy by SN feedback is assumed
to accumulate in a reservoir of mass $\Mres$ beyond the virial radius,
from where it gradually returns to the hot gas reservoir within the
virial radius, at a rate
\begin{equation}
{\dot M}_{\rm return} = \alpharet \frac{\Mres}{\tau_{\rm dyn,halo}} ,
\label{eq:Mreturn}
\end{equation}
where $\tau_{\rm dyn,halo} = \rvir/\Vvir$ is the halo dynamical
time.\footnote{The parameter $\alpharet$ and $\Mres$ were previously
  called $\alpha_{\rm reheat}$ and $M_{\rm reheat}$ respectively
  \citep[e.g.][]{Bower06}.} This assumption of gradual return of
ejected gas to the hot halo is the same as in \citet{Bower06}, but
differs from the model in \citet{Cole00} and \citet{Baugh05}, where it
was assumed that ejected gas returned only after the host halo mass
doubled. (\citet{Bower12} proposed a modified version of this SN
feedback scheme, in which some fraction of the ejected gas returns on
a longer timescale than in eqn.(\ref{eq:Mreturn}), controlled by the
growth of the DM halo, but we do not use this here.)

\subsubsection{AGN feedback}
Supermassive black holes (SMBHs) release energy through accretion of
gas, making them visible as AGN, and producing feedback. In \galform,
SMBHs grow in three ways \citep{Malbon07,Bower06,Fanidakis11}: (i)
accretion of gas during starbursts triggered by galaxy mergers or disk
instabilities ({\em starburst mode}); (ii) accretion of gas from the
hot halo ({\em hot halo mode}); (iii) BH-BH mergers. The mass accreted
onto the SMBH in a starburst is assumed to be a constant fraction
$f_{\rm BH}$ of the mass formed into stars, where $f_{\rm BH}$ is an
adjustable parameter. We assume that AGN feedback occurs in the {\em
radio mode} \citep{Croton06,Bower06}: energy released by direct
accretion of hot gas from the halo onto the SMBH powers relativistic
jets which propagate into the halo and deposit thermal energy in the
hot gas which can balance energy losses by radiative cooling. In
\galform, we assume that this radio-mode feedback sets up a steady
state in which energy released by the SMBH accretion exactly balances
the radiative cooling, if both of the following conditions are
satisfied: (a) the cooling time of halo gas is sufficiently long
compared to the free-fall time
\begin{equation}
\taucool(\rcool)/\tauff(\rcool) > 1/\alphacool ,
\label{eq:alpha_cool}
\end{equation}
where $\alphacool \sim 1$ is an adjustable parameter (with larger
values causing more galaxies to be affected by AGN feedback); and (b)
the AGN power required to balance the radiative cooling luminosity
$L_{\rm cool}$ is below a fraction $\fEdd$ \footnote{This parameter
was called $\epsilon_{\rm SMBH}$ in \citet{Bower06}.}  of the
Eddington luminosity $L_{\rm Edd}$ of the SMBH of mass $M_{\rm BH}$
\begin{equation}
L_{\rm cool} < \fEdd L_{\rm Edd}(M_{\rm BH}) .
\label{eq:f_Edd}
\end{equation}
The physical motivations for these two conditions are that: (a) the
halo gas needs to be in the quasi-hydrostatic rather than rapid cooling
regime for relativistic jets to be able to heat it effectively; and
(b) accretion disks around BHs are efficient at producing relativistic
jets only for very sub-Eddington accretion rates (see the discussion
in \citealt{Fanidakis11}). We assume that accretion of hot gas onto
the SMBH takes place only when these radio-mode feedback conditions
are satisfied, and when it does, the efficiency of converting mass
into energy in relativistic jets is $\epsilonheat$, causing the SMBH
mass to grow at a rate given by $\epsilonheat c^2 {\dot M}_{\rm BH} =
L_{\rm cool}$. We adopt values $\fEdd = 0.01$ and $\epsilonheat=0.02$
for these parameters, based on \citet{Fanidakis15}. The results in
this paper are not very sensitive to these values. 
(\citet{Bower08} considered an alternative AGN feedback scheme in
\galform, in which energy input from AGN is able to expel most of the
hot gas halo, rather than just balance radiative cooling. However, in
this paper we retain the simpler \citet{Bower06} AGN feedback scheme,
in line with other \galform papers.)

\subsection{Dynamical processes}
\label{ssec:dynamics}
Galaxies evolve according to a variety of dynamical processes, as we
now describe.
\subsubsection{Galaxy mergers}
We classify galaxies into {\em central} galaxies, which sit at the
centres of their DM halos and can grow by accreting gas which cools in
that halo, and {\em satellite} galaxies which orbit within the DM
halo, and are assumed not to accrete any gas from the hot gas
halo. When DM halos merge, we assume that the central galaxy in the
most massive progenitor halo becomes the new central galaxy, while all
other galaxies are left as satellites in the new halo. Satellite
galaxies merge with the central galaxy in their host DM halo on a
timescale set by {\em dynamical friction}. In \citet{Cole00}, we used
a dynamical friction timescale which was calculated analytically from
the Chandraskhar dynamical friction formula, but did not include the
effects on the dynamical friction rate of tidal stripping of the DM
subhalo hosting the satellite galaxy. In the new model, we replace
this with a modified expression obtained by fitting to the results of
cosmological N-body/hydrodynamical simulations of galaxy formation
\citep{Jiang08,Jiang10}, which automatically incorporates the effects
of this tidal stripping:
\begin{eqnarray}
\lefteqn{\tau_{\rm merge} = }  \nonumber \\
& & \frac{f(\epsilon)}{2 C} 
\frac{M_{\rm pri}}{M_{\rm sat}} \,
\frac{1}{\ln\left( 1 + M_{\rm pri}/M_{\rm sat} \right)} \, 
\left( \frac{\rcirc}{\rvir} \right)^{1/2} \,  \tau_{\rm dyn,halo} .
\label{eq:taumerge}
\end{eqnarray}
This gives the time for a satellite to merge with the central galaxy
from when it falls in through the virial radius of the main halo, in
terms of the masses $M_{\rm sat}$ of the satellite system (galaxy plus
DM halo) and $M_{\rm pri}$ (galaxies plus DM halo) of the primary at
infall, and {\em circularity} $\epsilon$ of the satellite orbit (at
infall), defined as the ratio of the orbital angular momentum to that
of a circular orbit of the same energy in the same potential, and
$\rcirc$, the radius of this equivalent circular orbit. The constant
$C=0.43$, while $f(\epsilon) = 0.90 \epsilon^{0.47} + 0.60$ is the fit
found by \citeauthor{Jiang08}. In applying this formula, we draw a
random value of $\epsilon$ for each satellite from the probability
distribution of orbital parameters of infalling satellite halos
measured by \citet{Benson05} from cosmological N-body simulations. The
merger time is calculated when the satellite first falls into the main
halo. If the satellite has not merged with the central galaxy by the
time of the next halo mass doubling event, then the merger time is
recalculated for the new halo, drawing a new value of $\epsilon$ from
the distribution. { (Note that, since we calculate the galaxy
  merger timescale analytically, rather than using the orbit of the
  satellite galaxy subhalo measured from the N-body simulation used in
  constructing the halo merger trees, our galaxy merger timescales are
  not affected by galaxies becoming ``orphaned'', i.e. losing their
  subhalos due to effects of limited numerical resolution.)}

The result of a galaxy merger depends on the ratio of baryonic mass
(including both stars and cold gas) of the satellite, $M_{\rm b,sat}$,
to that of the central galaxy, $M_{\rm b,cen}$. We define two
different thresholds, $\fburst \leq \fellip \leq 1$. (a) Mergers with
$M_{\rm b,sat}/M_{\rm b,cen} > \fellip$ are classed as {\em major}. We
assume that any stellar disks are destroyed and transformed into a
stellar spheroid, while all of the cold gas collapses into the newly
formed spheroid. Other mergers are classed as {\em minor}. In minor
mergers, stars from the satellite are added to the spheroid of the
central galaxy, but the cold gas is added to the disk of the central
galaxy, without changing the specific angular momentum of the
latter. (b) Mergers with $M_{\rm b,sat}/M_{\rm b,cen} > \fburst$
(which includes all major mergers) trigger starbursts, in which all of
the cold gas from the merging galaxies is transferred to the spheroid
and then consumed by star formation or ejected by the resulting SN
feedback. Numerical simulations of galaxy mergers imply $\fellip \sim
0.3$ and $\fburst \sim 0.1$
\citep[e.g.][]{Mihos94,Barnes98,Hopkins09}. We treat $\fellip$ and
$\fburst$ as adjustable parameters, but only in small ranges around
these values.

\subsubsection{Disk instabilities} 
Galaxies can also undergo morphological transformations and trigger
starbursts due to disk instabilities. Galaxy disks which are dominated
by rotational motions are unstable to bar formation when they are
sufficently self-gravitating. Based on the work of
\citet{Efstathiou82}, we assume that disks are dynamically
unstable to bar formation if
\begin{equation}
  \Fdisk \equiv \frac{\Vc(\rdisk)}{\left( 1.68 \, G\Mdisk/\rdisk
    \right)^{1/2}} 
  <\Fstab ,
\label{eq:disk_stability}
\end{equation}
where $\Mdisk$ is the total disk mass (stars plus gas), $\rdisk$ is
the disk half-mass radius, and the factor $1.68$ relates this to the
disk exponential scalelength.\footnote{Note that the original
  \citeauthor{Efstathiou82} criterion used the maximum disk circular
  velocity in place of the circular velocity at the disk half-mass
  radius.} The quantity $\Fdisk$ measures the contribution of disk
self-gravity to its circular velocity, with larger values
corresponding to less self-gravity and so greater disk
stability.\footnote{The parameters $\Fdisk$ and $\Fstab$ were
  previously called $\epsilon$ and $\epsilon_{\rm disc}$ in
  \citet{Bower06}.}  \citet{Efstathiou82} found a stability threshold
$\Fstab \approx 1.1$ for a family of exponential stellar disk models,
while \citet{Christodoulou95} found $\Fstab \approx 0.9$ for a family
of gaseous disks. We treat $\Fstab$ as an adjustable parameter in the
range $0.9 \lsim \Fstab \lsim 1.1$, with larger values resulting in
more disks becoming unstable. Note that a completely self-gravitating
stellar disk would have $\Fdisk=0.61$.

If the disk satisfies the instability condition $\Fdisk<\Fstab$ at any
timestep, then we assume that the disk forms a bar, { which then
  thickens due to vertical buckling instabilities and evolves into a
  spheroid \citep{Combes90,Debattista06}. We assume that this newly
  formed spheroid incorporates all of the stellar mass of the
  pre-existing disk and of any pre-existing spheroid. We also assume
  that bar formation triggers a starburst that consumes any cold gas
  present.} While in reality the timescale for growth of the bar and
its evolution into a bulge is likely to be at least several disk
dynamical times, in the model we approximate this whole process as
happening instantaneously, as soon as the disk instability condition
is met.

\subsection{Galaxy sizes}
\label{ssec:sizes}
Our model for galaxy sizes is identical to that in
\citet{Cole00}. Galaxies consist of a disk and a spheroid embedded in
a DM halo. These three components interact with each other
gravitationally. (a) We assume that the disk has an exponential
surface density profile, with a half-mass radius $\rdisk$ that is set
by angular momentum conservation and by centrifugal equilibrium in the
combined gravitational potential of disk, bulge and halo. When the
disk accretes gas by cooling from the halo, it is assumed to gain
angular momentum equal to that which this gas had in the halo before
it cooled. When the disk loses gas through SN feedback, this is
assumed to leave the specific angular momentum of the disk
unchanged. Apart from this, the disk angular momentum remains
constant, but the disk radius adiabatically adjusts in response to
changes in the gravitational potential. (b) We assume that the
spheroid is spherical and has an $r^{1/4}$ law surface density
profile, with 3D half-mass radius $\rbulge$. The initial size of the
spheroid formed in a galaxy merger or disk instability is set by a
combination of energy conservation and virial equilibrium (see
below). The bulge size subsequently evolves adiabatically in response
to changes in the gravitational potential, based on conservation of an
approximate radial action. (c) The DM halo is assumed to initially
have an NFW profile, but then to deform adiabatically in response to
the gravity of the disk and spheroid, assuming that each spherical
shell adiabatically conserves its value of $r \Vc(r)$
\citep{Barnes84,Blumenthal86}. The DM halo here means the main halo
for a central galaxy, but for a satellite galaxy it means the subhalo
which hosts the satellite galaxy. For the purpose of calculating the
galaxy size, the subhalo is assumed to have the same properties as it
had when it was last a separate halo. The disk and bulge sizes and
halo profile are updated to their new equilibrium values at each
timestep.

The details of how we calculate disk and spheroid sizes and halo
contraction are all given in \citet{Cole00}. Here we just remind the
reader of our procedure for calculating the sizes of spheroids formed
in mergers and disk instabilities:

\noindent{\em Galaxy mergers:} Dynamical friction causes the satellite
galaxy orbit to shrink as it loses energy to the host DM halo, until
the separation of the satellite and central galaxies becomes
comparable to the sum of their half-mass radii, at which point the
galaxies merge. We assume that the internal energy (kinetic plus
gravitational binding energy) of the spheroidal merger remnant just
after the merger is equal to the sum of the internal and relative
orbital energies of the two merging galaxies just before the merger:
\begin{equation}
E_{\rm int,remnant} = E_{\rm int,1} + E_{\rm int,2} + E_{\rm orbit} .
\end{equation}
This equation neglects any energy dissipation by gas or any energy
transfer to the dark matter during the merger. We also neglect any
mass loss from the galaxies during the merger. Using the virial
theorem, the internal energy of a galaxy is related to its
gravitational binding energy, which in turn depends on its mass
$M_{\rm gal}$ and half-mass radius $r_{\rm gal}$ as
\begin{equation}
E_{\rm int} = - \frac{1}{2} E_{\rm bind} = - \frac{c_{\rm gal}}{2}
\frac{G M_{\rm gal}^2}{r_{\rm gal}} .
\end{equation}
Here the dimensionless form factor $c_{\rm gal}$ depends (weakly) on
the galaxy density profile. Since $c_{\rm disk}=0.49$ for a pure
exponential disk and $c_{\rm bulge}=0.45$ for an $r^{1/4}$-law
spheroid, and galaxies in general contain both a disk and a spheroid,
we adopt a fixed value $c_{\rm gal}=0.5$ for simplicity. We can write
the energy of the relative orbital motion of the two galaxies at the
point they merge as
\begin{equation}
E_{\rm orbit} = - \frac{\forbit}{2} \, 
\frac{G M_{\rm gal,1} M_{\rm gal,2}}{r_{\rm gal,1} + r_{\rm gal,2}} ,
\end{equation}
where $\forbit$ is another dimensionless parameter, which would have a
value $\forbit=1$ for two point masses in a circular orbit with
separation $r_{\rm gal,1} + r_{\rm gal,2}$. We treat $\forbit$ as an
adjustable parameter in the range $0 \leq \forbit \lsim 1$. Putting
these equations together, we obtain
\begin{eqnarray}
\frac{(M_{\rm gal,1} + M_{\rm gal,2})^2}{r_{\rm remnant}} &=& 
\frac{M_{\rm gal,1}^2}{r_{\rm gal,1}} + \frac{M_{\rm gal,2}^2}{r_{\rm gal,2}} \nonumber \\ 
&+& \frac{\forbit}{c_{\rm gal}} \, \frac{M_{\rm gal,1} M_{\rm gal,2}}{r_{\rm gal,1} +
  r_{\rm gal,2}} ,
\label{eq:r_remnant}
\end{eqnarray}
which can be solved for the radius $r_{\rm remnant}$ of the remnant
spheroid. Finally, we note that the effective galaxy masses appearing
in eqn(\ref{eq:r_remnant}) include not only the stars and cold gas in
the merging galaxies, but also some part of the dark matter, since the
DM in the centre of the halo will have similar dynamics to the stars
during the merger. We therefore write the effective galaxy mass as
$M_{\rm gal,eff} = M_{\rm gal,b} + \fDM M_{\rm halo}(r_{\rm gal})$, where
$\fDM$ is another parameter. We choose $\fDM=2$, which would mean that
if the DM had the same spatial distribution as the baryons, then the
effective galaxy mass would be simply $M_{\rm gal,eff} = M_{\rm gal,b}
+ M_{\rm halo}$. 

In the case of a minor merger, we use the same equations, except that
now $M_{\rm gal,1}$ and $r_{\rm gal,1}$ for the primary galaxy are replaced by
the mass and half-mass radius of the primary spheroid.

\noindent{\em Disk instabililities:} We follow a similar approach to
calculating the size of the spheroid formed by a disk instability as
for a galaxy merger. In this case, the input system is the disk and
spheroid of the galaxy before the instability occured, with masses and
radii $\Mdisk$, $\Mbulge$, $\rdisk$ and $\rbulge$
respectively, and the output system is a new spheroid with half-mass
radius $r_{\rm new}$ containing all of the mass previously in the disk and
spheroid. Applying energy conservation and the virial theorem leads to
the relation
\begin{eqnarray}
c_{\rm bulge} \frac{(\Mdisk + \Mbulge)^2}{r_{\rm new}} &=& 
c_{\rm bulge} \frac{\Mbulge^2}{\rbulge} 
+ c_{\rm disk} \frac{\Mdisk^2}{\rdisk} \nonumber \\ 
&+& f_{\rm int} \, \frac{\Mdisk \Mbulge}{\rdisk + \rbulge} .
\label{eq:r_new}
\end{eqnarray}
Here $c_{\rm disk}$ and $c_{\rm bulge}$ have the same meanings as above. The
last term represents the gravitational interaction energy of the disk
and bulge, which is reasonably well approximated for a range of
$\rbulge/\rdisk$ by this form with $f_{\rm int}=2.0$. The disk and
bulge masses in this formula include stars and cold gas only.

\subsection{Chemical evolution and IMF}
\label{ssec:chem}
\subsubsection{Evolution equations for mass and metals}
We now combine the processes described above into a set of evolution
equations for the mass and metals in different components. We have
four different baryonic components: hot gas in halos, the
reservoir of ejected gas outside halos, cold gas in galaxies, and
stars in galaxies. These components have masses $\Mhot$, $\Mres$,
$\Mcold$ and $\Mstar$ respectively, which evolve according to the
following differential equations between halo formation and galaxy
merger events:
\begin{eqnarray}
\dot\Mhot &=& -\Mdotacc + \alpharet \frac{\Mres}{\tau_{\rm dyn,halo}} \\
\dot\Mcold &=& \Mdotacc - (1-R+\beta)\psi \\
\dot\Mstar &=& (1-R)\psi \\
\dot\Mres &=& \beta\psi - \alpharet \frac{\Mres}{\tau_{\rm dyn,halo}}
\end{eqnarray}
In the above, $\Mdotacc$ is the rate at which gas is added to the disk
by cooling and accretion from the halo (eqn(\ref{eq:Mdot_acc})),
$\psi$ is the SFR (eqn(\ref{eq:SFR_mol} or (\ref{eq:SFR_burst})), and
$\beta\psi$ is the rate of ejection of gas from the cold component
into the halo reservoir by SN feedback (eqn(\ref{eq:Meject})). We use
the instantaneous recycling approximation, meaning that we negelect
the time delay between when stars form and when they die and eject gas
and metals, so that the rate of gas ejection by dying stars into the
cold component is $R\psi$. The value of the {\em returned fraction},
$R$, depends on the IMF, as described below. (The effects of relaxing
the instantaneous recycling approximation in \galform are described in
\citet{Nagashima05a,Nagashima05b,Li15}.)  As discussed in
\S\ref{ssec:hot_gas}, the hot gas content is continually updated for
the effects of DM halo growth by mergers and accretion. We note that
the stellar mass is split between disk and bulge components, but for
simplicity we do not show this explicitly in the above equations.

The masses of heavy elements (``metals'') in the different components
obey a similar set of equations. We define $\MZhot$ as the mass of
metals in the hot component and $\Zhot = \MZhot/\Mhot$ as its
metallicity, and similarly for the other components. The evolution
equations are then, again using the instantaneous recycling
approximation:
\begin{eqnarray}
\dot\MZhot &=& -\Zhot\Mdotacc + \alpharet \frac{\MZres}{\tau_{\rm dyn,halo}} \\
\dot\MZcold &=& \Zhot\Mdotacc + \left[ p- (1-R+\beta)\Zcold\right] \psi \\
\dot\MZstar &=& (1-R)\Zcold\psi \\
\dot\MZres &=& \beta\Zcold\psi - \alpharet \frac{\MZres}{\tau_{\rm dyn,halo}}
\end{eqnarray}
The term $p\psi$ in the above equations is the rate of ejection of
newly synthesized metals into the ISM by dying stars. The value of the
{\em yield} $p$ also depends on the IMF, as detailed below. We assume
that metals ejected from stars are instantaneously mixed into the cold
gas component. Ejection of metals from the galaxy by SN feedback
therefore occurs via the cold gas.

\begin{figure}

\begin{center}

\includegraphics[width=8cm]{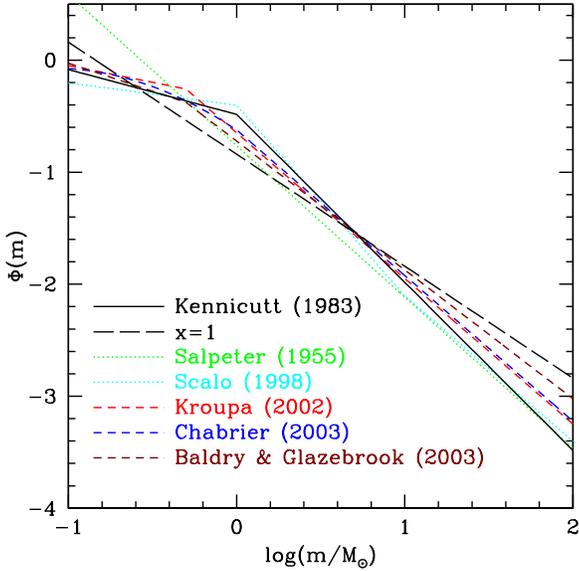}

\end{center}

\caption{{ The IMFs used in this work (black lines) compared to
    some other IMFs in the literature (coloured lines). The IMF is
    defined as $\Phi(m) = dN/d\ln m$ and normalized according to
    eqn.(\ref{eq:IMF_norm}) for $0.1<m<100 \Msol$. The solid black line
    shows the \citet{Kennicutt83} IMF which is assumed for quiescent
    star formation, and the long-dashed black line the $x=1$ power-law
    IMF assumed for starbursts in the standard model. The coloured
    lines show estimates for the solar neighbourhood IMF from
    \citet{Salpeter55}, \citet{Scalo98}, \citet{Kroupa02} and
    \citet{Chabrier03}, and also the galaxy IMF from \citet{Baldry03},
    as labelled.}  }

\label{fig:IMF}
\end{figure}

\subsubsection{Initial mass function}
The evolution of the gas, star and metal contents of galaxies, as well
as their luminosity evolution, depends on the stellar initial mass
function (IMF). The IMF is defined as the distribution of stars in
mass $m$ at the time of formation of a stellar
population. Specifically, we define $\Phi(m)$ such that $dN = \Phi(m)
\, d\ln m$ is the number of stars formed with masses in the range
$m,m+dm$ per unit total mass of stars formed. $\Phi$ is therefore
normalized as
\begin{equation}
\int_{m_L}^{m_U} m \, \Phi(m)\, d\ln m = 1 ,
\label{eq:IMF_norm}
\end{equation}
where $m_L$ and $m_U$ are respectively the lower and upper mass limits
on the IMF.

The returned fraction $R$ is the fraction of the initial mass of a stellar
population that is returned to the ISM by mass loss from dying
stars. In the instantaneous recycling approximation, it is given by
the integral
\begin{equation}
R = \int_{1 \Msol}^{m_U} (m - m_{\rm rem}(m)) \, \Phi(m) \, d\ln m ,
\label{eq:R_integral}
\end{equation}
where $m_{rem}(m)$ is the mass of the remnant (white dwarf, neutron star
or black hole) left by a star of initial mass $m$, obtained from
stellar evolution calculations.

The yield $p$ is the fraction of the initial mass of a stellar
population that is synthesized into new metals and then ejected, and
is given by
\begin{equation}
p = \int_{1 \Msol}^{m_U} p_Z(m) m \, \Phi(m) \, d\ln m 
\label{eq:p_integral}
\end{equation}
where $p_Z(m)$ is the corresponding fraction for a single star of
initial mass $m$, also obtained from stellar evolution calculations.

We will assume IMFs that are power laws or piecewise
power laws in mass, i.e. 
\begin{equation}
\Phi(m) = \frac{dN}{d\ln m} \propto m^{-x} ,
\label{eq:IMF_slope}
\end{equation}
where $x$ is the IMF slope. For a \citet{Salpeter55} IMF, $x=1.35$. We
assume that stars form with different IMFs in the {\em quiescent}
(disk) and {\em starburst} modes. {\em Quiescent mode:} We assume an
IMF similar to that measured in the Solar neighbourhood and in the
disks of nearby galaxies, specifically that of \citet{Kennicutt83},
which has $x=0.4$ for $m < \Msol$ and $x=1.5$ for $m > \Msol$. 
{\em Starburst mode:} We
assume an IMF that is a single power law, with slope $0\leq x \leq 1$,
i.e. having a shallower slope compared to the Solar neighbourhood for
$m > \Msol$, and so being {\em top heavy}. We treat this IMF slope $x$
in starbursts as an adjustable parameter. We adopt lower and upper
mass limits $m_L = 0.1 \Msol$ and $m_U = 100 \Msol$ for both quiescent
and burst IMFs, in order to be consistent with the IMFs assumed in the
stellar population models which we use (see below).

For any choice of IMF in our model, we use self-consistent values of
the recycled fraction and yield, based on integrating
eqns.(\ref{eq:R_integral}) and (\ref{eq:p_integral}) over the assumed
IMF. We use remnant masses $m_{\rm rem}(m)$ and stellar yields $p_Z(m)$
from the stellar evolution calculations of \citet{Marigo96} for
intermediate mass stars and \citet{Portinari98} for high mass
stars. We calculate $R$ and $p$ for Solar metallicity, neglecting the
metallicity dependence of these quantities. We obtain the following
values for the IMFs listed above: (a) \citet{Kennicutt83} IMF:
$R=0.44$, $p=0.021$; (b) tilted $x=1$ IMF: $R=0.54$, $p=0.048$; (c)
tilted $x=0$ IMF: $R=0.91$, $p=0.13$. It can be seen that $R$ and $p$
both have a strong dependence on the form of the IMF.

{ We plot the two IMFs used in our standard model in
  Fig.~\ref{fig:IMF} (solid and dashed black lines for the quiescent
  and $x=1$ burst IMFs respectively), where we also compare them with
  some other widely used IMFs from the literature (shown as coloured
  lines). The \citet{Kennicutt83} IMF that we use as our normal galaxy
  IMF was originally proposed to fit the H$\alpha$ equivalent widths
  and colours of nearby star-forming galaxies. It is very close to the
  \citet{Scalo98} IMF estimated for the solar neighbourhood. Compared
  to the \citet{Kroupa02} and \citet{Chabrier03} IMFs that were also
  estimated for the solar neighbourhood, it is slightly higher around
  $m \sim 1~\Msol$, but slightly lower for $m \gsim 10~\Msol$. We also
  show the \cite{Baldry03} IMF, which was an estimate of the {\em
    average} galaxy IMF, obtained by fitting the galaxy luminosity
  density at $z\sim 0$, and is significantly flatter at high masses,
  with a slope $x = 1.15$ that is closer to our starburst IMF.}

Our assumption of a top-heavy IMF in starbursts is a controversial
one. Indeed, the whole issue of whether the IMF varies with
environment or has varied over cosmic history remains hugely
controversial, with a large literature, but arriving at conflicting
conclusions (see recent reviews by \citet{Bastian10} and
\citet{Krumholz14}). In their review of observational studies,
\citet{Bastian10} argued against significant IMF variations in the
nearby Universe, { but a number of other recent studies have
  reached different conclusions, as discussed below.  Direct
  observational constraints on the IMF in starbursts remain weak, in
  large part because of the large dust extinctions typical of such
  systems.}

{ Many recent observational studies provide evidence for IMF variations,
but paint a complex picture of the nature of these variations. From a
study of the spectra of nearby star-forming galaxies,
\citet{Gunawardhana11} infer an IMF that becomes more top-heavy with
increasing SFR, with the IMF slope flattening to $x \approx 0.9$
(similar to our starburst IMF) in the most actively star-forming
galaxies in their sample. \citet{Finkelstein11} infer a similarly flat
IMF in a star-forming galaxy at $z \sim 3$. For early-type galaxies, a
number of studies measuring stellar mass-to-light ratios from stellar
dynamics \citep[e.g.][]{Cappellari12} or gravitational lensing
\citep[e.g.][]{Treu10} find $M/L$ increasing with stellar mass, implying
an IMF in massive early-type galaxies that is either top-heavy or
bottom-heavy compared to the Solar neighbourhood (\citet{Cappellari12}
infer IMF slopes $x=0.5$ or $x=1.8$ for these two cases, assuming a
single power-law IMF).  An independent constraint on the low mass ($m
\lsim 1~\Msol$) IMF in early-type galaxies comes from studying
spectral features sensitive to low-mass stars. Several such studies
\citep[e.g.][]{Conroy12,LaBarbera13} find evidence for a bottom-heavy
  IMF in high mass galaxies. However, the overall picture for
  early-type galaxies is currently unclear, with different methods in
  some cases giving conflicting results for the IMF when applied to
  the same galaxy \citep{Smith14,Smith15}. \citet{Weidner13} argue
  that, in any case, an IMF in early-type galaxies that is
  bottom-heavy at all times is incompatable with their observed
  metallicities, and propose instead a time-dependent IMF that is
  top-heavy at early times but bottom-heavy at later times.}

On the theoretical side, there is also a lack of consensus { about
  variations in the IMF}. \citet{Larson05} argued that the
characteristic mass in the IMF should scale with the Jeans mass in the
star-forming cloud, and that the latter should be larger in more
actively star-forming regions, due to heating by the radiation from
massive young stars. \citet{Krumholz10} proposed a modified version of
this idea, in which the characteristic mass increases in star-forming
regions of higher gas surface density, due to the effects of radiation
trapping. Either of these scenarios could plausibly lead to a more
top-heavy IMF in starbursts, where the gas densities are higher.  {
  On the other hand, \citet{Hopkins13} recently argued that the IMF in
  starbursts should be bottom-heavy due to increased turbulence.}
To conclude, we would argue that the issue of IMF variations is still
an open one, which makes the possibility of such variations worth
exploring in galaxy formation models { (see \citet{Fontanot14} for
  another recent study of the effects of IMF variations in SA
  models)}. We note that the exact form of the top-heaviness for the
starburst IMF is not critical for our results. We have chosen a tilted
power-law IMF for convenience, but an IMF truncated below some mass
would give very similar results { for the quantities that we
  predict in this paper}.

\subsection{Stellar populations and dust}
\label{ssec:stars_dust}
For each galaxy, the model calculates a complete star formation and
metallicity history. We combine this with a stellar population
synthesis (SPS) model based on stellar evolution models to calculate
the luminosity and spectral energy distribution (SED) of the stellar
population. We then apply a physical model for absorption and emission
of radiation by dust, in order to calculate the effects of dust
extinction on the stellar SED and also the luminosity and SED of the
IR/sub-mm emission by the dust.

\subsubsection{Stellar population synthesis}
The SED at time $t$ of a stellar population with a mixture of ages and
metallicities but a single IMF can be written as 
\begin{equation}
L_{\lambda}(t) = \int_0^t dt' \, \int_0^{\infty} dZ' \, \Psi(t',Z') \,
L^{\rm (SSP)}_{\lambda}(t-t',Z';\Phi) ,
\label{eq:L_tot}
\end{equation}
where $\Psi(t',Z')\, dt' dZ'$ is the mass (at birth) of stars which
formed in the time interval $t',t'+dt'$ and metallicity range
$Z',Z'+dZ'$, and $L^{\rm (SSP)}_{\lambda}(t,Z;\Phi)$ is the SED of a
single stellar population (SSP) of unit mass with age $t$ and
metallicity $Z$, formed with an IMF $\Phi(m)$. $\Psi(t,Z)$ is obtained
by summing over the star formation histories of all the progenitor
galaxies which merged to form the final galaxy. The SSP luminosity is
related to the luminosity of a single star $L^{\rm (star)}(t,Z,m)$ by
\begin{equation}
L^{\rm (SSP)}_{\lambda}(t,Z;\Phi) = \int_{m_L}^{m_U} \, L^{\rm (star)}_{\lambda}(t,Z,m) \, 
 \Phi(m) \, d\ln m .
\end{equation}
{ Since we have two IMFs in our model, we apply eqn.(\ref{eq:L_tot})
  separately (for both disk and spheroid) to the stars formed in the
  disk and starburst modes, and then add these to get the total
  luminosities of the disk and spheroid in each galaxy.}

There are several libraries available which provide $L^{\rm
  (SSP)}_{\lambda}(t,Z;\Phi)$ for different ages, metallicities and
IMFs
\citep[e.g.][]{Bruzual93,Bressan98,Bruzual03,Maraston05,Conroy09,Vazdekis15}. These
are based on theoretical stellar evolution tracks and either
theoretical or observed stellar spectra. Here we use the
\citet{Maraston05} SPS, since it incorporates what appears to be
currently the most accurate treatment of the light produced by stars
on the thermally-pulsing asymptotic giant branch (TP-AGB), which is
important for the rest-frame near-IR luminosities of stellar
populations with ages $\sim 0.1-1 \Gyr$.  The contribution to the SED
from the TP-AGB phase is difficult to model accurately from
theoretical stellar evolution models alone, so \citet{Maraston05}
calibrate this using observations of star clusters. The
\citeauthor{Maraston05} models are computed for a large grid of ages,
but only a coarse grid of metallicities: $Z=0.001$, $Z=0.01$, $Z=0.02$
and $Z=0.04$. We therefore interpolate $L^{\rm
  (SSP)}_{\lambda}(t,Z;\Phi)$ in both $t$ and $Z$ as needed. We use
the blue horizontal branch models for $Z=0.001$, and red horizontal
branch models for higher metallicities. { The impact of having a
  strong TP-AGB contribution has previously been investigated in SA
  models by
  \citet{Tonini09,Tonini10,Fontanot10,Henriques11,Henriques12}.}
\citet{Gonzalez-Perez14} have made a comparison of \galform results
using different SPS models.

To calculate broad-band luminosities and magnitudes from the stellar
SEDs of galaxies, we multiply $L_{\lambda}$ by the suitably normalized
filter response function and integrate. In the case of observer-frame
bands, we first shift the SED by a factor $(1+z)$ in wavelength before
doing this, to account for the k-correction \citep[e.g.][]{Hogg02}. We
calculate absolute magnitudes with zeropoints on either the Vega or AB
systems, depending on the observational data with which we are
comparing.

\subsubsection{Absorption and emission by dust}
\label{sec:dust}
\galform includes a self-consistent model for the reprocessing of
starlight by dust, with UV, optical and near-IR light being absorbed
by stars and the energy then reradiated at IR and sub-mm
wavelengths. We calculate the dust absorption using radiative
transfer, and we solve for the temperature of the dust emission based
on energy balance. This model, which is the same as that used in
\citet{Lacey11,Gonzalez11,Gonzalez12,
  Gonzalez-Perez09,Gonzalez-Perez13,Gonzalez-Perez14,
Lagos11b,Lagos12,Lagos14,Mitchell13,Cowley15,Cowley15b},
is described in more detail in Appendix~\ref{app:dust_model}, so we
give only an overview of the main features here. The model shares
features with
the \GRASIL spectrophotometric model \citep{Silva98}, which we
combined with \galform in several previous papers
\citep[e.g.][]{Granato00,Baugh05,Lacey08,Lacey10}, but with a number of
important simplifying approximations relative to \GRASIL, especially
for the dust emission, which are designed to speed up the
calculations.

We assume a two-phase dust medium, with molecular clouds embedded in a
diffuse dust medium having an exponential radial and vertical
distribution. For quiescent galaxies, with stars forming in the disk,
this dust medium is co-extensive with the stellar disk, with the same
half-mass radius, while for bursts, the dust is co-extensive with the
starburst stellar population, which is assumed to have the same
half-mass radius as the stellar bulge.
Stars are assumed to form inside the molecular clouds, and then to
leak out on a timescale $\tesc$.

The mass and radius of the dust medium are directly predicted by
\galform (unlike many other models where they are treated as
adjustable functions of galaxy mass and redshift). We calculate the
total dust mass $\Mdust$ from the mass and metallicity of the cold gas
component, assuming a dust-to-gas ratio that scales linearly with
metallicity, { equivalent to assuming that a constant fraction
$\delta_{\rm dust}$ of metals in the cold gas component are in dust grains,
\begin{equation}
{\Mdust} = \delta_{\rm dust} \, {\Zcold} \, {\Mcold},
\label{eq:delta_dust}
\end{equation}
where we choose $\delta_{\rm dust}=0.334$ to match the Solar neighbourhood
dust-to-gas ratio $6.7 \times 10^{-3}$ for $\Zsol=0.02$
\citep{Silva98}.
}
The dust is assumed to always have the same extinction curve shape
$k_{\lambda}$ and albedo as in the Solar neighbourhood, so that the
(extinction) optical depth of the dust for light passing through gas
with surface density $\Sigma_{\rm gas}$ is
\begin{equation}
\tau_{{\rm dust},\lambda} = 0.043 \left( \frac{k_{\lambda}}{k_V}\right)
\left( \frac{\Sigma_{\rm gas}}{\Msol\pc^{-2}}\right)
\left( \frac{\Zcold}{0.02} \right) ,
\label{eq:tau_dust}
\end{equation}
again normalized to match the local ISM for $\Zcold=0.02$ (see
\citealt{Cole00} for more details).

We assume that a fraction $\fcloud$ of the dust is in clouds of mass
$\Mcloud$ and radius $\rcloud$, and the remainder in the diffuse
medium. $\fcloud$ and $\tesc$ are treated as adjustable parameters,
while $\Mcloud$ and $\rcloud$ are kept fixed, based on observations of
nearby galaxies \citep{Granato00}. (In fact, only the combination
$\Mcloud/\rcloud^2$ affects the model predictions, since this
determines the optical depth through a cloud. In practice, the model
predictions presented in this paper are very insensitive to the
value of $\Mcloud/\rcloud^2$, provided it is large enough to make the
optical depth through a cloud large at UV wavelengths, as is the case
for our standard parameter choice.)

The calculation of the absorption of starlight by dust is in two
parts. (a) We first calculate the fraction of the galaxy luminosity at
each wavelength that is emitted by stars still inside their birth
clouds, based on the star formation history and stellar population
model. We then apply dust attenuation by clouds to this fraction,
assuming that the emission occurs from the centres of clouds. The dust
optical depth of a single cloud scales as $\taucloud \propto \Zcold
\Mcloud/\rcloud^2$. (b) The starlight emerging from molecular clouds
together with the light from stars outside clouds are then attenuated
by the diffuse dust component. The optical depth through the centre of
this component scales as $\taudiffuse \propto (1-\fcloud) \Mcold
\Zcold/r_{\rm diff}^2$, where $r_{\rm diff}=\rdisk$ or $\rbulge$ for
quiescent or starburst components respectively. We calculate the
attenuation by the diffuse dust by interpolating the tabulated
radiative transfer models of \citet{Ferrara99}, which assume that the
stars are distributed in an exponential disk and a bulge, and the dust
is distributed in an exponential disk. The tables provide the dust
attenuations of the disk and bulge luminosities as functions of
wavelength, disk inclination, central dust optical depth and ratio of
disk to bulge half-light radii. (The inclinations of galaxy disks to
the line of sight are chosen randomly.) By combining (a) and (b), we
predict the SEDs of the disk and bulge (including any starburst) after
attenuation by dust.

We calculate the IR/sub-mm emission by dust as follows. From the
difference between the stellar SEDs with and without dust attenuation,
we can calculate the luminosity absorbed by dust at each
wavelength. Integrating over wavelength gives the total stellar
luminosity absorbed by dust in a galaxy. We calculate this separately
for the molecular clouds and diffuse dust. We then assume that each
dust component radiates as a modified blackbody:
\begin{equation}
L_{\lambda}^{\rm dust} \propto \Mdust \, \kappa_{\rm d}(\lambda) \,
B_{\lambda}(\Tdust) ,
\label{eq:L_lambda,dust}
\end{equation}
where $\kappa_{\rm d}(\lambda)$ is the dust opacity per unit mass,
$\Mdust$ and $\Tdust$ are the mass and temperature of that dust
component (clouds or diffuse), and $B_{\lambda}(T)$ is the Planck
function. By integrating this over wavelength, we obtain the total
dust luminosity of that component, and by equating this to the
absorbed luminosity we can then solve for the dust temperature
$\Tdust$. In general, the clouds and diffuse dust have different
temperatures. The total SED of dust emission is then the sum of the
SEDs of the two components. We approximate the opacity at IR
wavelengths as a broken power law:
\begin{equation}
\kappa_{\rm d}(\lambda) \propto \left\{ \begin{array}{ll}
                             \lambda^{-2} & \lambda<\lambda_{\rm b} \\
                             \lambda^{-\beta_{\rm b}} &
                             \lambda>\lambda_{\rm b} .
		  \end{array} \right .
\label{eq:kappa_d}
\end{equation}
The normalization and slope of $\kappa_{\rm d}(\lambda)$ at
$\lambda<\lambda_{\rm b}$ are chosen to match the Solar neighbourhood.
We allow a break in this power law at $\lambda>\lambda_{\rm b}$ in
starbursts. We fix $\lambda_{\rm b}=100\mum$, but allow the long-wavelength
slope to be adjustable in the range $1.5<\beta_{\rm b}<2$, motivated by the
results of \citeauthor{Silva98} on fitting the sub-mm SED of
Arp220. For quiescent galaxies, we assume an unbroken power law
(i.e. $\beta_{\rm b}=2$).

Our model for dust emission thus has a number of approximations: (i)
single dust temperature for each component; (ii) no temperature
fluctuations for small grains; (iii) power-law opacity, so no PAH
features. These approximations break down in the mid-IR, but seem to
work reasonably well at far-IR and sub-mm wavelengths. Comparisons
with more detailed calculations using \GRASIL indicate that our
approximate method is reasonably accurate for rest-frame wavelengths
$\lambda \gsim 70\mum$, for which the emission is dominated by fairly
large dust grains in thermal equlibrium in the general interstellar
radiation field \citep{Cowley16}.

{ We note that most published SA models do not include a detailed
  model for IR/sub-mm emission from dust. Some exceptions to this
  include \citet{Fontanot07}, who coupled their SA model to the
  \GRASIL spectrophotometric model, similar to what we had done for
  \GALFORM in some earlier papers (as described above), and
  \citet{Devriendt99} and \citet{Somerville12}, who combined simpler
  geometrical models for absorption of starlight by dust with
  templates for the SED of the IR/sub-mm emission from dust. The
  disadvantages of the template approach are: (i) the templates are
  derived from or calibrated on observed SEDs of galaxies in the
  nearby Universe; and (ii) it is assumed that the template SED shape
  depends only on the total IR luminosity. Both of these assumptions
  may break down for galaxies at higher redshifts.}

\begin{table*}
\centering

\caption{
Values of input parameters for standard model. Parameters labelled F
were kept fixed when searching for the parameter set which produces
the best fit to the observational constraints described in the
text. Parameters which were varied are labelled as primary (P) or
secondary (S) in terms of how strongly they affect these predictions.
} 

\begin{tabular}{clccll}
\hline
parameter  & value  & range & type=F/P/S & description  &  Eqn/paper \\
\hline
Cosmology  &        &        &          & \cite{Komatsu11}\\
$\Omega_{\rm m0}$ & $0.272$  & -  & F & matter density &   \\
$\Omega_{\rm b0}$ & $0.0455$ & -  & F & baryon density  &  \\
$h $             & $0.704$  & -  & F & Hubble parameter  &  \\
$\sigma_{8}$     & $0.81 $  & -  & F & fluctuation amplitude & \\
$n_{s}$          & $0.967$    & -  & F & scalar spectral index & \\
\hline
Stellar population &        &   &    & & \cite{Maraston05} \\
\hline 
IMF : quiescent   \\
$x $  &  Kennicutt   & - & F & IMF & Eqn.~\ref{eq:IMF_slope} \\
$p $  &  0.021       & - & F & yield & Eqn.~\ref{eq:p_integral} \\
$R $  &  0.44       & - & F & recyled fraction & Eqn.~\ref{eq:R_integral} \\
IMF : starburst   \\
$x $  &  1   & 0-1 & P & IMF slope & Eqn.~\ref{eq:IMF_slope} \\
$p $  &  0.048       & - & P & yield & Eqn.~\ref{eq:p_integral} \\
$R $  &  0.54       & - & P & recyled fraction & Eqn.~\ref{eq:R_integral} \\
\hline
Star formation: quiescent &                       &  &   & & \citet{Lagos11a}\\
$\nuSF$  & $0.74~{\rm Gyr}^{-1}$ &  $0.25-0.74~{\rm Gyr}^{-1}$
& P & efficiency factor for molecular gas & Eqn.~\ref{eq:SFR_mol} \\ 
$ P_{0}       $  & $1.7\times10^4$     & -  &  
F & normalisation of pressure relation & Eqn.~\ref{eq:at_mol} \\ 
$ \alpha_P      $  & 0.8        & -  &  F & slope of pressure  relation    &
Eqn.~\ref{eq:at_mol} \\
\hline
Star formation: bursts &                       & & & & \citet{Baugh05} \\
$\fdyn$ & 20   &  0 - 100  & P & multiplier for dynamical time & Eqn.~\ref{eq:taustar_burst} \\
$\tauburstmin $ & 0.1 Gyr  &   0-1.0  & P & minimum burst timescale & Eqn.~\ref{eq:taustar_burst} \\
\hline 
Photoionization feedback  & &   &   & & \citet{Benson03} \\
$\zreion$ & 10 & - & F & reionization redshift & \\
$\Vcrit$ & $30 ~\kms$ & - & F & threshold circular velocity & \\
\hline 
SNe feedback  & &   &   & & \citet{Cole00} \\
$\VSN$  & $320 ~\kms$ & anything  & P & pivot velocity & Eqn.~\ref{eq:Meject} \\
$\gammaSN$ & 3.2 & 0-5.5 & P & slope on velocity scaling &
Eqn.~\ref{eq:Meject} \\
$\alpharet$ & 0.64 & 0.3-3 & P & reincorporation timescale multiplier &
Eqn.~\ref{eq:Mreturn} \\ 
\hline 
AGN feedback \& SMBH growth & & & & & \citet{Bower06} \\
$\fBH$ & 0.005 & 0.001-0.01 & S & fraction of mass accreted onto BH in
starburst & \cite{Malbon07} \\
$\alphacool$ & 0.8 & 0-2 & P & ratio of cooling/free-fall time &
Eqn.~\ref{eq:alpha_cool} \\
$\fEdd$ & 0.01 & - & S & controls maximum BH heating rate & Eqn.~\ref{eq:f_Edd} \\
$\epsilonheat$ & 0.02 & - & S & BH heating efficiency & \\
\hline 
Disk stability & & & & & \citet{Cole00} \\
$\Fstab$  & 0.9 & 0.9-1.1 & P & threshold for instability & Eqn.~\ref{eq:disk_stability} \\
\hline  
Galaxy mergers & & & & & \citet{Jiang08} \\
\hline  
Size of merger remnants & & & & & \citet{Cole00} \\
$\forbit$ & 0 & 0 - 1 & S & orbital energy contribution &
Eqn.~\ref{eq:r_remnant} \\
$\fDM$ & 2 & - & S & dark matter fraction in galaxy mergers & \\
\hline 
Starburst triggering in mergers    &  &  &  & & \citet{Baugh05} \\
$\fellip$ & 0.3 & 0.2 - 0.5 & P & threshold on mass ratio for major merger & \\
$\fburst$ & 0.05 & 0.05 - 0.3 & S & threshold on mass ratio for burst & \\
\hline 
Dust model     &  &  &  & & \citet{Granato00} \\
$\fcloud$ & 0.5 & 0.2 - 0.8 & P & fraction of dust in clouds & \\
$\tesc$ & $1~\Myr$ & $1 - 10~\Myr$ & P & escape time of stars from
clouds & Eqn.~\ref{eq:tesc} \\
$\beta_{\rm b}$ & 1.5 & 1.5 - 2 & S & sub-mm emissivity slope in starbursts &
Eqn.~\ref{eq:kappa_d} \\
\hline 
\multicolumn{6}{l}{Note: $P_0$ in units $k_B\cm^{-3}\K$} \\
\hline 

\end{tabular}


\label{table:param}
\end{table*}



\section{Results from the new model}
\label{sec:fiducial}





\subsection{Fitting the model parameters}
In this section, we introduce the key observational constraints which
we use when choosing what are the best values for the adjustable
parameters in the model, and show how the predictions from the
fiducial version of our model compare to these observational data. We
also discuss how the predictions from our new model compare with the
earlier \citet{Baugh05} and \citet{Bower06} models. We discuss in the
following section (\S\ref{sec:param-variations}) which observational
constraints are sensitive to which physical parameters in the
model. We find there that there are some tensions between fitting the
different observational constraints, in the sense that some of the
constraints can be fit well by the model only at the expense of
fitting others poorly. For this reason, we do not give all constraints
equal weight when finding the best values of the model parameters, but
instead choose to give some constraints higher priority than others.
We therefore divide the observational constraints into {\em primary}
and {\em secondary}. We insist that our fiducial model reproduces our
primary constraints to a good approximation. We regard reproducing the
secondary constraints as desirable, but only if that does not
significantly degrade the fit to our primary constraints.

The input parameters for the standard version of our new model are
presented in Table~\ref{table:param}. Some of these parameters,
labelled as F in the table, have been kept fixed throughout. These
include parameters for the cosmology and CDM power spectrum, for the
IMF in quiescent star formation, for the \citet{Blitz06} pressure law
controlling the molecular gas fraction, and for the photoionization
feedback. Other parameters were allowed to vary, and are labelled as
either P (primary) or S (secondary) in the table according to how
strongly they affect the model predictions presented in this paper. We
note that the 3 parameters relating to the IMF in starbursts are not
independent, in that once the IMF slope, $x$, is chosen, the yield,
$p$, and recycled fraction, $R$, are completely determined by
integrals over the IMF. For the variable (P/S) input parameters, we
chose the standard values given in the table by trying to find the
best fit to the observational constraints presented in
\S\ref{sec:primary_constraints} and \S\ref{sec:secondary_constraints},
giving more weight to primary than secondary constraints, as discussed
above. Note that additional observational comparisons shown later in
this paper (in \S\ref{sec:phys-predictions} and
\S\ref{sec:simplistic}) were {\em not} used in calibrating the model
parameters. The search for the best-fitting parameters was performed
by running grids of models and visually comparing the results, rather
than by any automated procedure, such as Monte Carlo Markov
Chain. When performing this search, the input parameters were allowed
to vary only over the ranges given in the table. Some of these ranges
were set according to theoretical considerations, and others according
to independent observational constraints, as discussed in
\S\ref{sec:model}. The AGN feedback parameters $\fEdd$ and
$\epsilonheat$ are in principle variable, but were calibrated in a
companion study by \citet{Fanidakis15}, and so were not varied here.

\subsection{Primary observational constraints}
\label{sec:primary_constraints}
\subsubsection{Optical and near-IR luminosity functions at $z=0$}
\label{ssec:optNIRLFs}
We require our model to give a good fit to the observed $b_J$- and
$K$-band galaxy luminosity functions (LFs) in the local Universe. The
LFs in these bands at the present day mainly depend on the galaxy
stellar mass function (SMF), with some dependence also on the ages and
metallicities of the stellar populations, and also (for the $b_J$-band)
the dust extinction. 

While some other recent papers on galaxy formation models constrain
their model parameters by comparing their predicted SMFs directly with
SMFs inferred from observational data by SED fitting
\citep[e.g.][]{Guo11}, we prefer to use the observed LFs
instead. There are several reasons for this, { of which the first
  is most critical in any model with a variable IMF} (see
\citealt{Mitchell13} for more details): (1) When stellar masses are
inferred by fitting stellar population models to observed galaxy SEDs,
an IMF must be assumed. In our model, stars form with different IMFs
in disks and in starbursts. This means that any direct comparison
between predicted and observationally-inferred SMFs would be
meaningless in this case. (2) Inferring stellar masses by SED fitting
also requires using a stellar population model. There are differences
in the SEDs predicted by different stellar population models, so if a
different stellar population model is used in the estimation of
stellar masses from that used in the galaxy formation model for
predicting other observed properties, then this will lead to
inconsistencies. (3) The stellar masses inferred using SED fitting
also depend on assumptions about the star formation histories and
metallicity distributions in galaxies, and on how dust extinction is
modelled. The assumptions made in the SED fitting may be inconsistent
with what is assumed in the galaxy formation model. In particular,
\citet{Mitchell13} showed that differences between the empirical dust
attenuation laws typically used in SED fitting and the more physical
dust attenuation calculation used in \GALFORM (see \S\ref{sec:dust})
can lead to large systematic differences between the true stellar
masses in the model and what would be inferred by SED fitting. We
explore these issues further in \S\ref{sec:phys-predictions}, where we
show the evolution of the SMF predicted by our model.

\begin{figure}

\begin{center}

\includegraphics[width=7cm, bb= 20 55 275 510]{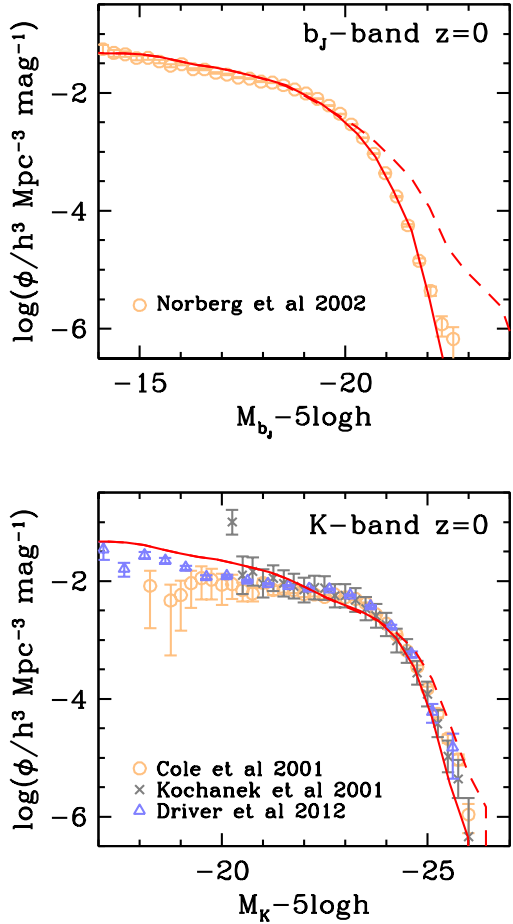}

\end{center}

\caption{Predictions of the default model for the $b_J$- and $K$-band
  LFs at $z=0$. { $\phi$ is defined as $dn/dM_X$, where $M_X$ is
    the absolute magnitude in the relevant band.} The dashed lines
  show the predicted LFs without dust extinction, and the solid lines
  show the predictions including dust extinction. Observational data
  are from \citet{Norberg02,Cole01,Kochanek01,Driver12}.}

\label{fig:lf_default}
\end{figure}

Fig.~\ref{fig:lf_default} compares the predictions from our fiducial
model with observational data on the $b_J$- and $K$-band LFs in the
local Universe. The dashed lines show the predicted LFs when the
effects of dust extinction in the model are ignored, while the solid
lines show the predictions including dust extinction. The fiducial
model is in good agreement with observations over the whole range of
luminosity. In particular, the predicted faint-end slope in the
$K$-band agrees much better with recent, deeper, observational data
than with older data, which gave a very shallow faint-end slope {
  (although the agreement at the faint end is still not perfect)}.

\subsubsection{HI mass function at $z=0$}
\begin{figure}

\begin{center}

\includegraphics[width=7cm, bb= 20 520 280 750]{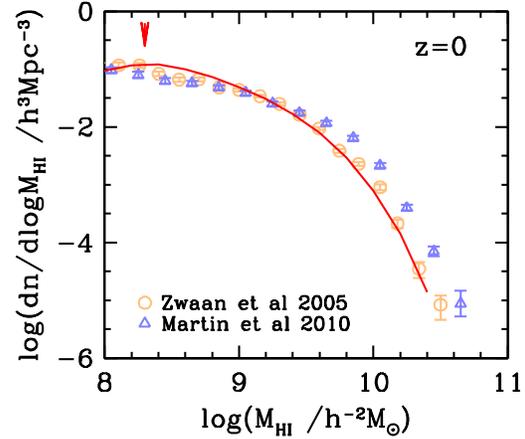}

\end{center}

\caption{Predictions of the default model for the HI mass function at
  $z=0$ (solid line). { The vertical arrow at the top of the panel
  indicates the HI mass below which the results are affected by the
  halo mass resolution.} Observational data are from
  \citet{Zwaan05,Martin10}.}

\label{fig:mfHI_default}
\end{figure}

It is important that the model agrees with the observed gas contents
of galaxies. Our model predicts both the total cold gas masses in
galaxies, and how this is partitioned between the atomic ($HI$) and
molecular ($H_2$) components of the ISM (\S\ref{ssec:SFR}). We use the
$HI$ mass function of galaxies in the local Universe as our primary
constraint on cold gas contents, since this has been quite accurately
measured from large 21~cm surveys (although there is still a factor 2
difference at high $HI$ mass between the two surveys which we
plot). The comparison of our fiducial model with observations is shown
in Fig.~\ref{fig:mfHI_default}. In contrast, the $H_2$ mass function
has not yet been measured as accurately from $CO$ surveys, and in
addition there are still uncertainties in relating $CO$ observations
to $H_2$ masses. The new model is seen to be in very good agreement
with the observed $HI$ mass function for $M_{HI}> 10^8 \hMsol$. The
dip in the $HI$ mass function for $M_{HI} \lsim 10^8 \hMsol$ is
produced by the transition from being dominated by central galaxies at
higher $M_{HI}$ to being dominated by satellite galaxies at lower
$M_{HI}$. However, the location of this transition is affected by the
halo mass resolution, which is around $2 \times 10^{10}~\hMsol$ for
the N-body simulation used here, and it would shift to somewhat lower
$M_{HI}$ if the minimum halo mass were reduced (see \citealt{Lagos11b}
for more details). 

\subsubsection{Morphological fractions at $z=0$}
\begin{figure}

\begin{center}

\includegraphics[width=7cm, bb= 20 520 275 750]{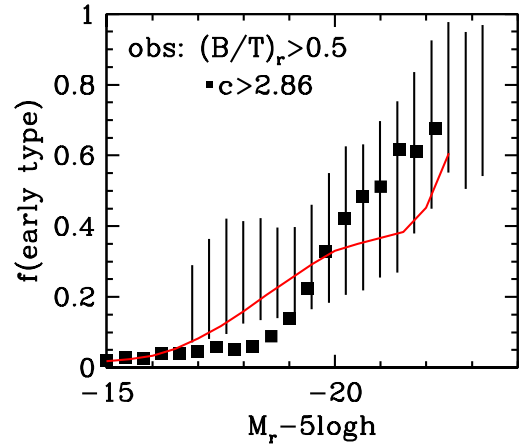}

\end{center}

\caption{Predictions of the default model for the fraction of
  early-type galaxies as a function of $r$-band luminosity at $z=0$
  (solid line). Model galaxies are classified as early-type if the
  bulge-to-total luminosity ratio in the $r$-band
  $(B/T)_r>0.5$. Observational data are from \citet{Benson07}
  (vertical hatched region), based on the bulge-to-total luminosity
  ratio in the $r$-band $(B/T)_r$ from fitting disk+bulge models {
    (with the vertical hatching indicating the range of systematic
    uncertainty in the fits),} and from \citet{Gonzalez09} (filled
  squares), based on the Petrosian concentration index, $c$, in the
  $r$-band.}

\label{fig:morph_default}
\end{figure}

In our model, stars are split between disk and spheroidal components,
and we morphologically classify galaxies as {\em late-} or {\em
  early-}type depending on which component dominates. We require that
our model broadly reproduces the trend of early vs late-type fractions
with luminosity that is observed in the local Universe. We compare the
fraction of early-type galaxies vs luminosity with observational data
from the SDSS survey in Fig.~\ref{fig:morph_default}. Since the SDSS
results are based on $r$-band imaging data, we classify model galaxies
as early-type for this plot if their bulge-to-total luminosity ratio
in the $r$-band $(B/T)_r > 0.5$. We compare with two different
observational estimates of the early-type fraction, one based on
$(B/T)_r$ estimated from fitting disk+bulge models to galaxy images,
and the other based on the Petrosian concentration index $c$. These
two methods of classifying galaxies have been shown previously to be
in reasonable agreement (see \citealt{Gonzalez09} for more
details). The fraction of early-type galaxies in the model is in
reasonable agreement with the observations. (See \citealt{Lagos14b}
for a more detailed comparison between the model presented here and
observations of the gas contents of early-type galaxies.)

\subsubsection{Black hole - bulge mass relation at $z=0$}
\begin{figure}

\begin{center}

\includegraphics[width=7cm, clip=true, bb= 20 520 280 750]{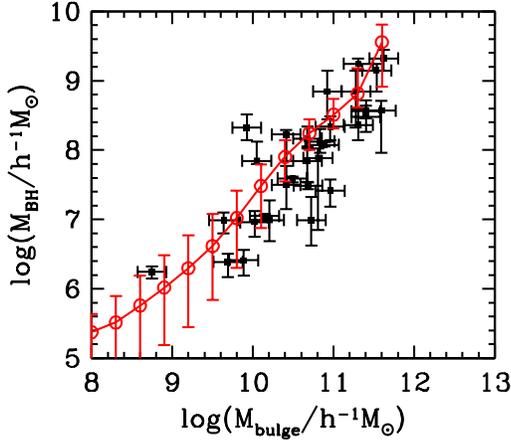}

\end{center}

\caption{Predictions of the default model for the black hole mass vs
  bulge mass relation at $z=0$. The solid line shows the predicted
  median relation, with the error bars on it showing the 10--90\%
  range in the distribution. Observational data are from
  \citet{Haering04}. In order to match the bias towards early-type
  galaxies in the observational sample, only model galaxies with
  $(B/T)_B >0.3$ are included. }

\label{fig:SMBH_default}
\end{figure}

Our final primary observational constraint from the local Universe is
the relation between the mass of the central supermassive black hole
(SMBH) and the mass of the bulge. This is plotted in
Fig.~\ref{fig:SMBH_default}. In this plot, we have chosen to include
only model galaxies with $B$-band bulge-to-total luminosity ratios
$(B/T)_B >0.3$, so as to roughly match the bias towards early-type
galaxies in the observational sample which we compare with. However,
the predicted relation is in fact highly insensitive to the cut
chosen, over the whole range $0<(B/T)_B <0.9$.

\subsubsection{Evolution of near-IR luminosity function}
\begin{figure*}

\begin{center}

\begin{minipage}{5.4cm}
\includegraphics[width=5.4cm, clip=true, bb= 280 525 530 750]{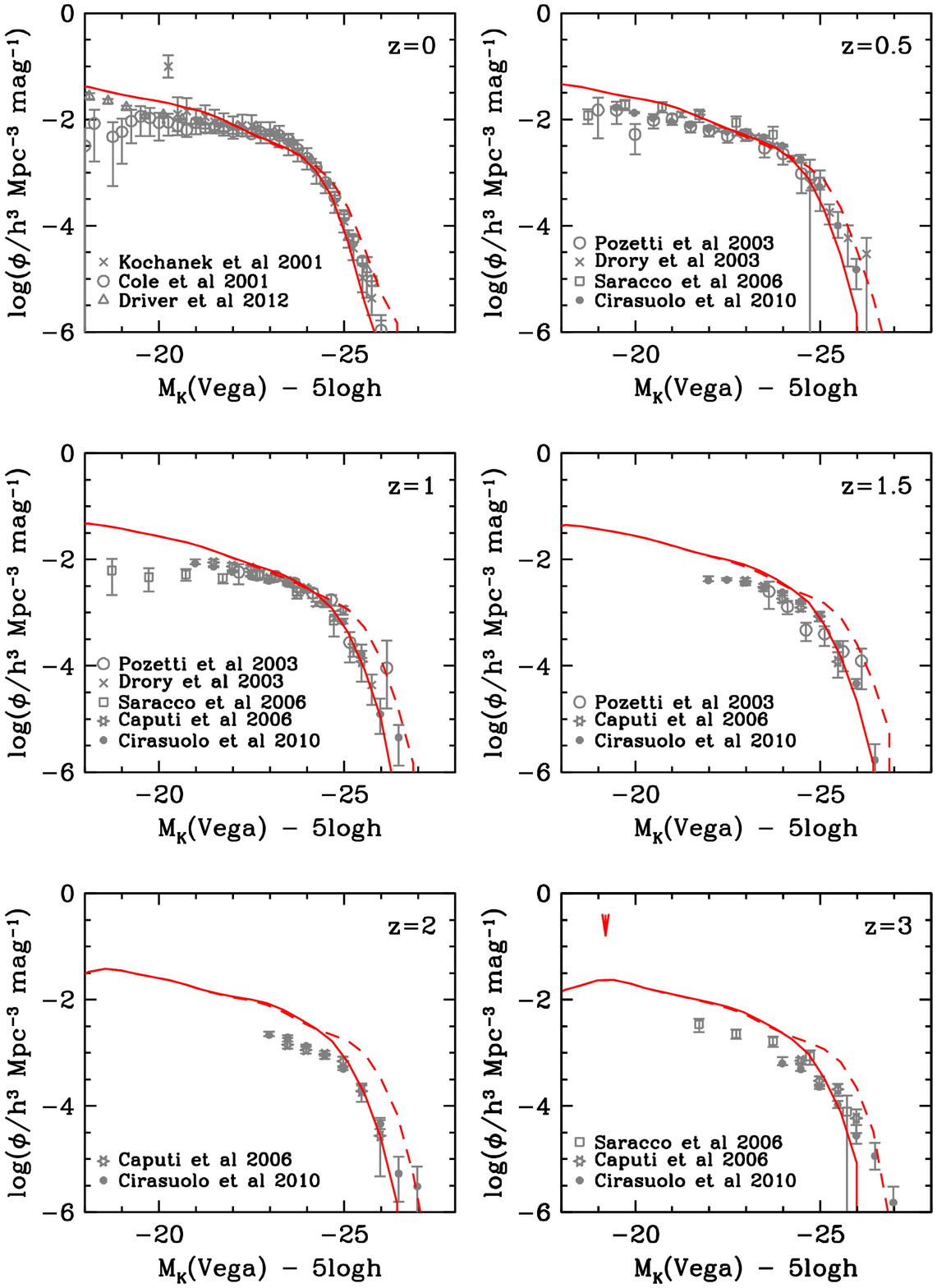}
\end{minipage}
\hspace{0.4cm}
\begin{minipage}{5.4cm}
\includegraphics[width=5.4cm, clip=true, bb= 24 288 275 514]{figs/Klfz_default.ps}
\end{minipage}
\hspace{0.4cm}
\begin{minipage}{5.4cm}
\includegraphics[width=5.4cm, clip=true, bb= 280 51 530 277]{figs/Klfz_default.ps}
\end{minipage}

\end{center}

\caption{Predictions of the default model for the evolution of the
  rest-frame K-band luminosity function. The solid and dashed dashed
  lines show the model LFs with and without dust extinction. We
  compare the model with observational data at $z=0.5$, $z=1$ and
  $z=3$, as labelled in each panel. { The vertical arrow at the top
    of the $z=3$ panel indicates the K-band luminosity below which the
    results are affected by the halo mass resolution.} Observational
  data are from
  \citet{Pozzetti03,Drory03,Saracco06,Caputi06,Cirasuolo10}. }

\label{fig:lfKz_default}
\end{figure*}

Our next set of primary observational constraints tests the evolution
of galaxies at different wavelengths. We start with the evolution of
the rest-frame $K$-band LF in the range $z=0-3$. This depends mostly
on the evolution of the SMF, but we prefer to use the $K$-band LF
rather than the SMF to constrain our model for the reasons given in
\S\ref{ssec:optNIRLFs}. As shown by \citet{Mitchell13}, errors in SMFs
inferred from observations by SED fitting are expected to increase
with redshift due to both increases in dust attenuation and (in the
present model) due to the larger fraction of stars formed in
starbursts with a top-heavy IMF. We compare the fiducial model with
observational data on the $K$-band LF at $z=0.5$, $z=1$ and $z=3$ in
Fig.~\ref{fig:lfKz_default}. The predicted $K$-band LF is in
fair agreement with the observational data up to $z=3$, although it
appears somewhat high at the faint end at $z \sim 1-3$. { Previous
studies of the evolution of the K-band LF using SA models include
\citet{Bower06,Kitzbichler07,Henriques11,Somerville12}. }

\subsubsection{Sub-mm galaxy number counts and redshift distributions}
\begin{figure}

\begin{center}

\includegraphics[width=7cm, bb= 20 295 275 750]{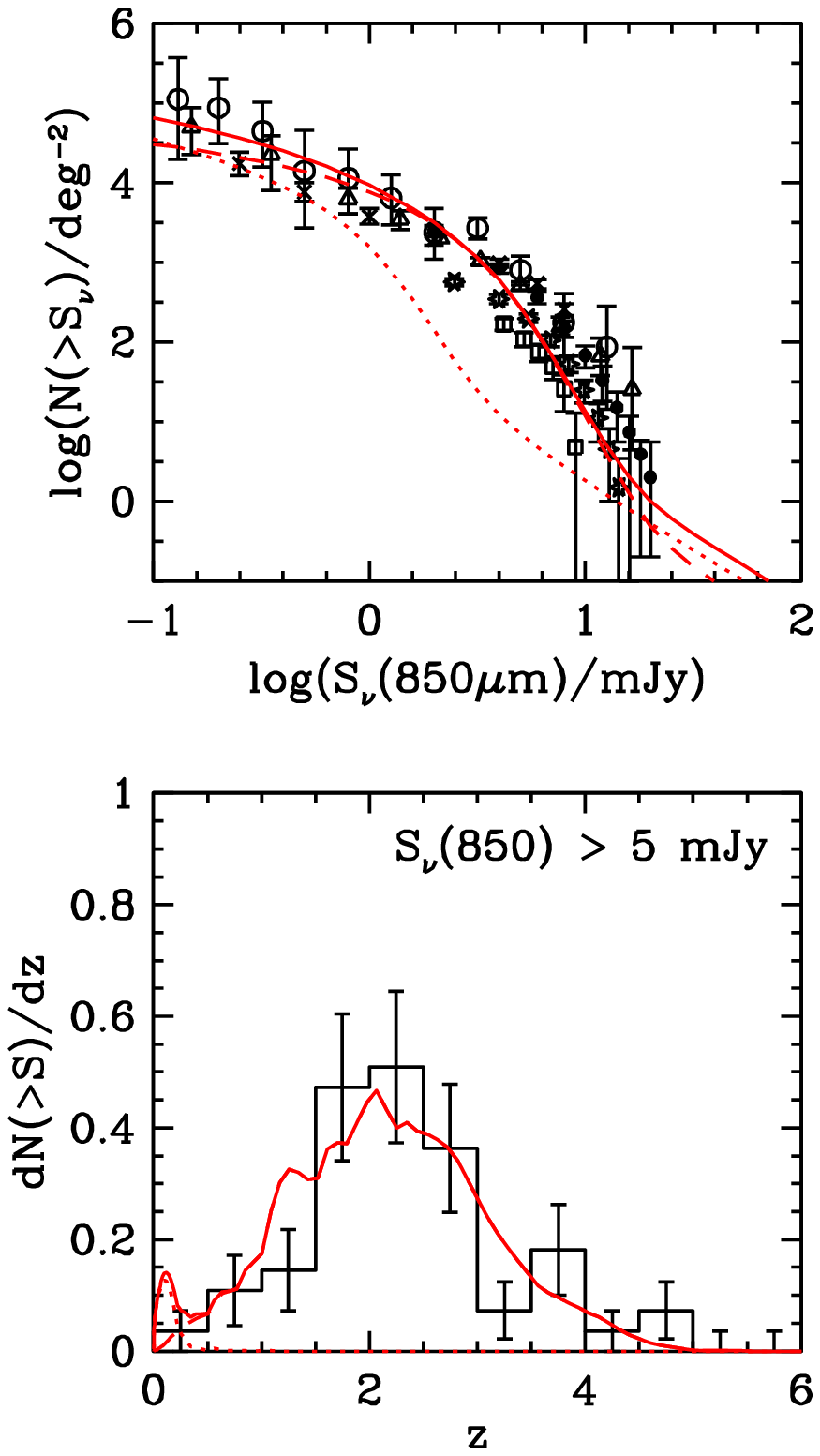}

\end{center}

\caption{Predictions of the default model for the $850\mum$ number
  counts and redshift distribution. Upper panel shows cumulative
  number counts at $850\mum$, compared to observational data from
  \citet{Coppin06} (filled circles), \citet{Knudsen08} (open circles),
  \citet{Weiss09} (stars), \citet{Zemcov10} (crosses), \citet{Karim13}
  (open squares), \citet{Chen13b} (open triangles). Lower panel shows
  redshift distribution of sources brighter than $S(850\mum) > 5
  \mJy$, compared to observational data from \citet{Wardlow11}. Both
  predicted and observed redshift distributions have been normalized
  to unit area. The dotted and dashed lines show the contributions to
  the total from quiescent and starburst galaxies respectively. Note
  that the distribution in the lower panel is dominated by starbursts
  for $z>0.3$.}

\label{fig:SMGs_default}
\end{figure}

One of the most important constraints on our model comes from the
observed number counts and redshift distribution of galaxies detected
in deep surveys at 850~$\mum$, the so-called sub-mm galaxies
(SMGs). Observations of these constrain the properties of dusty
star-forming galaxies at high redshifts. We compare the fiducial model
with observed cumulative number counts in the upper panel of
Fig.~\ref{fig:SMGs_default}, while in the lower panel we show the
redshift distribution for galaxies brighter than $S>5 \mJy$ at
850~$\mum$. We show here observations from single-dish surveys. Recent
work \citep{Karim13,Chen13a} using sub-mm interferometers has shown
that some SMGs which appear as single sources when observed at low
angular resolution split up into multiple sources when observed at
higher angular resolution. The implications of this for comparing our
model with observed counts and redshift distributions are discussed in
\citet{Cowley15}.

\subsubsection{Far-IR number counts}
\begin{figure*}

\begin{center}
\begin{minipage}{5.4cm}
\includegraphics[width=5.4cm, clip=true, bb= 24 525 275 750]{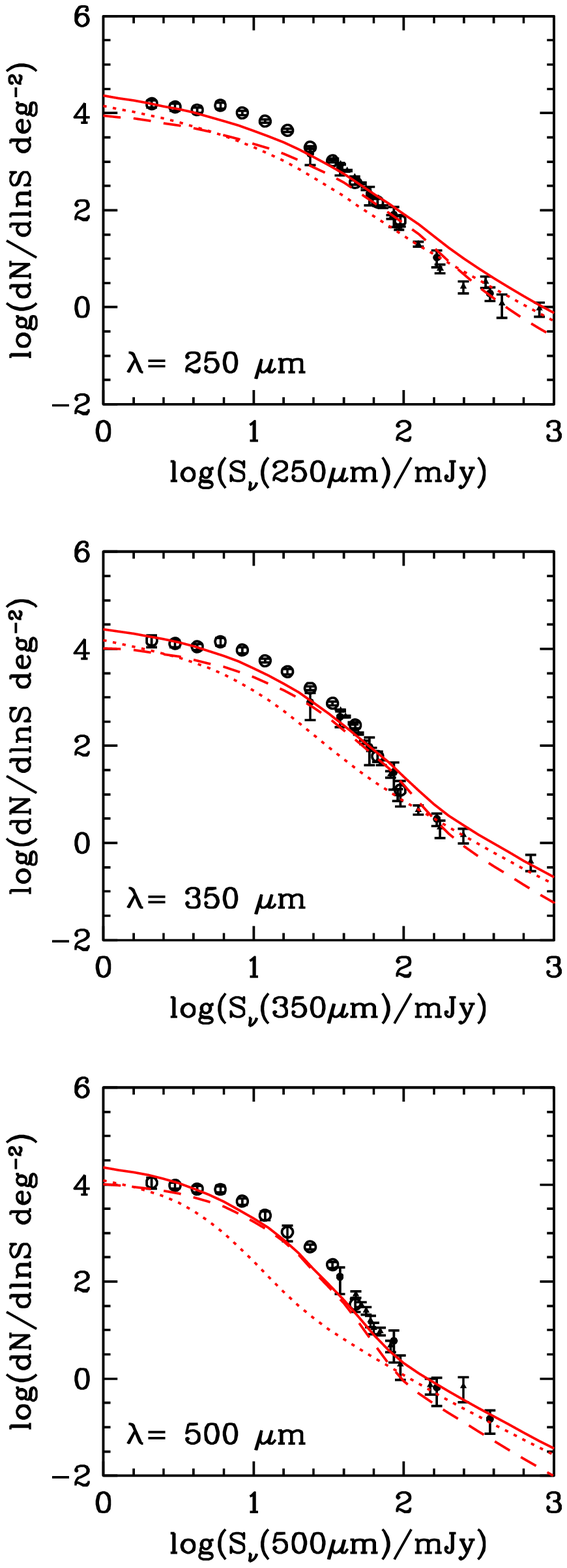}
\end{minipage}
\hspace{0.4cm}
\begin{minipage}{5.4cm}
\includegraphics[width=5.4cm, clip=true, bb= 24 288 275 514]{figs/FIRcounts_default.ps}
\end{minipage}
\hspace{0.4cm}
\begin{minipage}{5.4cm}
\includegraphics[width=5.4cm, clip=true, bb= 24 51 275 277]{figs/FIRcounts_default.ps}
\end{minipage}
\end{center}

\caption{Predictions of the default model for the far-IR differential
  number counts at (a) 250, (b) 350 and (c) 500~$\mum$. The dotted and
  dashed lines show the contributions to the total from quiescent and
  starburst galaxies respectively. Observational data are shown from
  \citet{Clements10} (open triangles), \citet{Oliver10} (open
  squares), \citet{Bethermin12} (open circles). }

\label{fig:FIRcounts_default}
\end{figure*}

An independent constraint on the population of dusty star-forming
galaxies comes from galaxy number counts at far-IR wavelengths
measured by \Herschel\ and also by \Planck, which probe this
population at lower redshifts. We only use the far-IR counts at 250,
350 and 500~$\mum$ to constrain our model, since far-IR counts at
shorter wavelengths are affected by inaccuracies in our model of dust
emission (\S\ref{sec:dust}). The comparison of our fiducial model with
observations is shown in Fig.~\ref{fig:FIRcounts_default}.

\subsubsection{Far-UV luminosity functions of Lyman-break galaxies}
\begin{figure}

\begin{center}

\includegraphics[width=7cm, bb= 20 295 275 750]{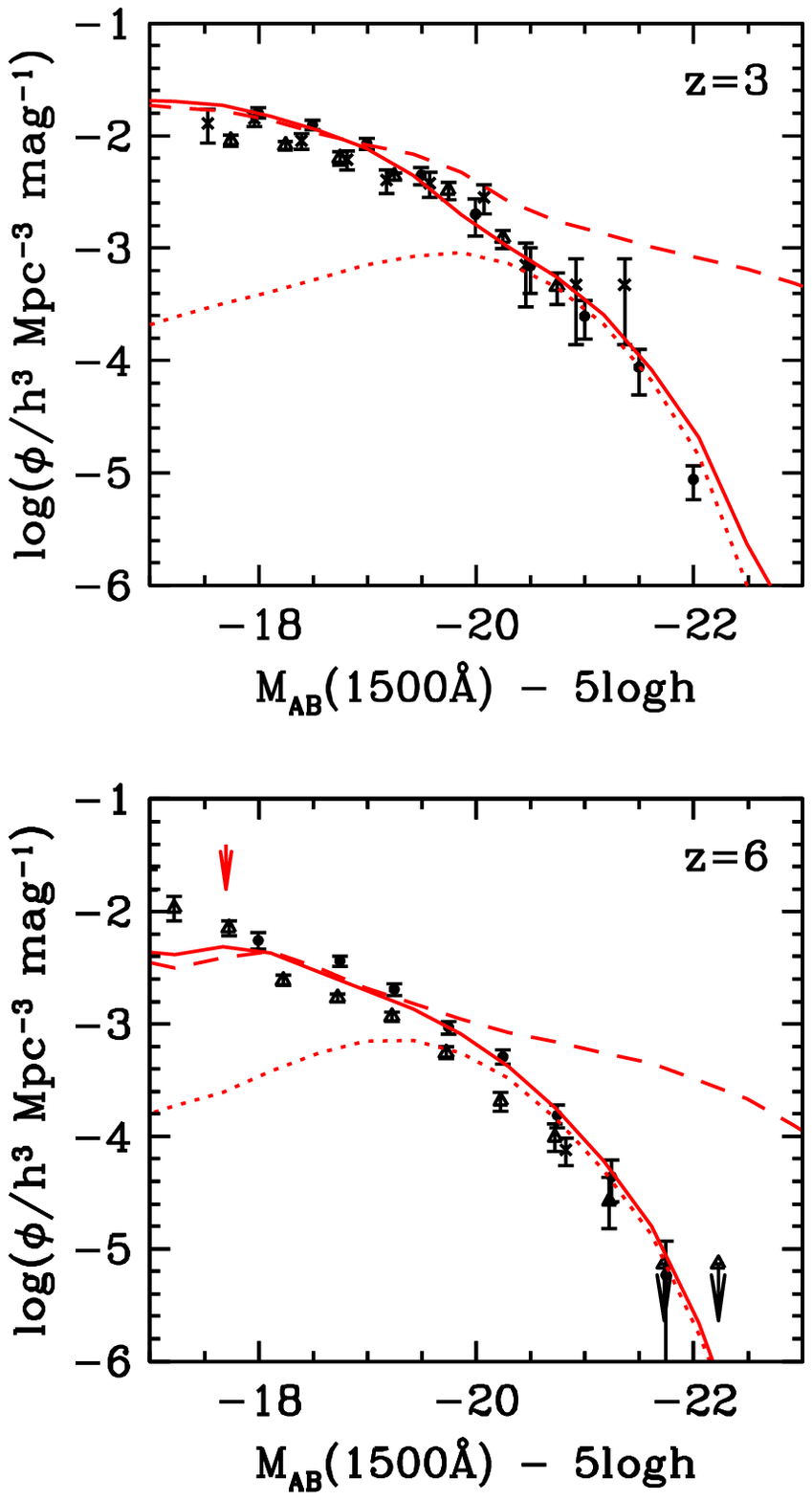}

\end{center}

\caption{Predictions of the default model for the rest-frame far-UV
  (1500\AA) LF at $z=3$ (top panel) and $z=6$ (bottom panel). The
  solid lines show the predictions including dust attenuation, and the
  dashed lines the predictions without dust attenuation. The dotted
  lines show the contribution from bursts including dust
  attenuation. { The vertical arrow at the top of the $z=6$ panel
    indicates the UV luminosity below which the results are affected
    by the halo mass resolution.} Observational data are from
  \citet{Arnouts05} (crosses),\citet{Reddy09} (filled circles) and
  \citet{Sawicki06} (open triangles) at $z=3$, and \citet{Bouwens15}
  (filled circles), \citet{Finkelstein15} (open triangles) and
  \citet{Shimasaku05} (crosses) at $z=6$.  }

\label{fig:LBGs_default}
\end{figure}

Our final primary observational constraint is the rest-frame far-UV
luminosity function of galaxies at high redshifts. This probes the
star-forming galaxy population at very high redshifts, though only the
part of it that is not obscured by dust. Observationally, this is
typically measured from samples of galaxies selected by the
Lyman-break technique, the so-called Lyman-break galaxies
(LBGs). Fig.~\ref{fig:LBGs_default} compares the fiducial model with
the observed far-UV LF at $z=3$ and $z=6$. The effects of dust
extinction on the far-UV LF are predicted to be very large, as can be
seen by comparing the solid and dashed lines (see
\citealt{Gonzalez-Perez13} for a detailed study of the effects of dust
extinction on properties of UV-selected galaxies in \GALFORM models.)

\subsection{Secondary observational constraints}
\label{sec:secondary_constraints}

We remind the reader that when trying to find the best values for the
model parameters, we first try to fit the {\em primary} observational
constraints described in the previous subsection. Only then do we try
to fit the {\em secondary} observational constraints described in this
subsection. When trying to fit these secondary constraints, we only
allow parameter variations which do not degrade the fits to the
primary constraints.

\subsubsection{Tully-Fisher relation at $z=0$}
\begin{figure}

\begin{center}

\includegraphics[width=7cm, bb= 270 290 530 510]{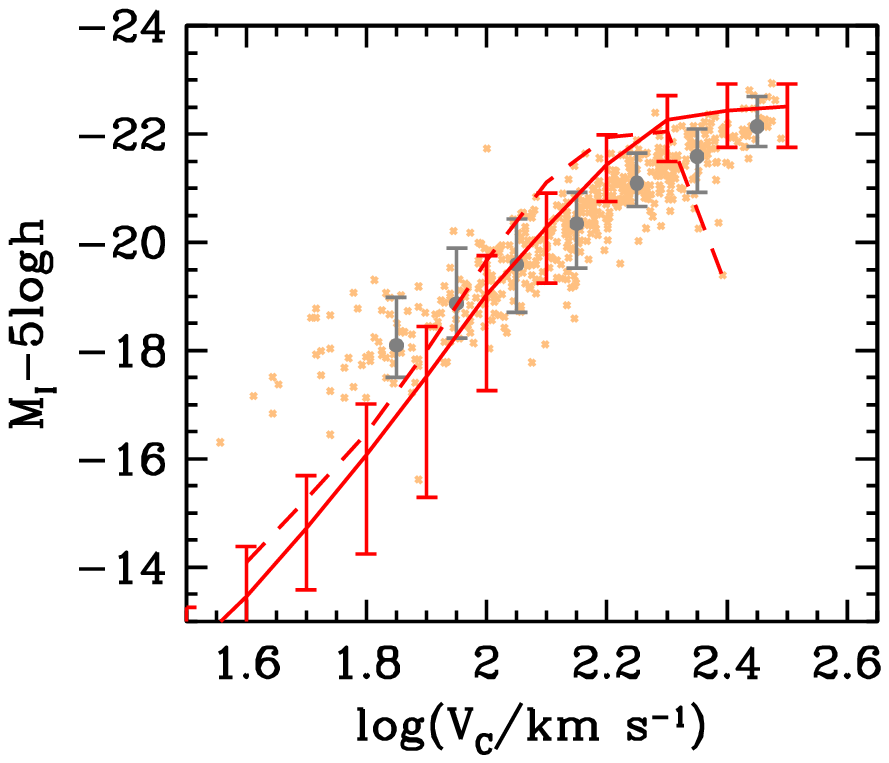}

\end{center}

\caption{Predictions of the default model for the $I$-band
  Tully-Fisher relation at $z=0$. The solid line shows the predicted
  median $I$-band magnitude as a function of circular velocity for the
  model, and the error bars show the 10 and 90 percentiles of the
  distribution. The magnitudes are face-on values, including the
  effects of dust extinction. The circular velocities are measured at
  the half-mass radius of the disk. Model galaxies have been selected
  with $B$-band bulge-to-total luminosity ratios $(B/T)_B<0.2$ and gas
  fractions $\Mcold/\Mstar>0.1$, to try to replicate the selection in
  the observational sample. The dashed line shows the model prediction
  using the circular velocity at the halo virial radius instead of the
  disk half-mass radius. The points show the observed distribution for
  a subsample of Sb-Sd galaxies selected by \citet{deJong00} from the
  \citet{Mathewson92} catalogue, and again all magnitudes have been
  converted to face-on values. The points with error bars show the
  medians and 10-90 percentile ranges for this observational data.}

\label{fig:TF_default}
\end{figure}

Another observational relation which has been widely used in previous
work to constrain galaxy formation models is the Tully-Fisher (TF)
relation between the luminosities and circular velocities of spiral
galaxies. We show this relation in Fig.~\ref{fig:TF_default}. We note
that in our work we distinguish between the circular velocity of the
galaxy disk (measured at its half-mass radius) and the circular
velocity at the virial radius of the host halo. These differ due to
several effects: (i) the rotation curve of the halo is not flat, but
instead follows an NFW profile; (ii) the disk circular velocity is
increased by the self gravity of the galaxy; and (iii) the halo
density profile undergoes adiabatic contraction due to the gravity of
the baryons in the galaxy. The net effect is that the circular
velocity of the disk is generally somewhat higher than that at the
halo virial radius (by around $\sim 10\%$ for $L_{\star}$ spiral
galaxies in our fiducial model). { Note that predictions for the TF
  relation from other SA models in the literature have often used some
  measure of the DM halo circular velocity as a proxy for the disk
  circular velocity, either the halo circular velocity at the virial
  radius \citep[e.g.][]{Somerville99}, or the peak value
    \citep[e.g.][]{Guo11}.} In the past, it has proved challenging for
  galaxy formation models to reproduce both the optical LF and TF
  relation at $z=0$ \citep[e.g.][]{Cole00} { (although more recent
    models have been more successful, e.g. \citet{Guo11})}, but our current
  model is seen to agree quite well with the observed TF relation in
  Fig.~\ref{fig:TF_default}. In this figure, we have chosen to include
  only model galaxies with $B$-band bulge-to-total luminosity ratios
  $(B/T)_B<0.2$ and gas fractions $\Mcold/\Mstar>0.1$, to roughly
  replicate the selection in the observational sample. However, the
  model predictions for the disk circular velocity in fact depend only
  weakly on these cuts in the range $0.1<(B/T)_B<0.5$ and
  $0<\Mcold/\Mstar<0.2$.

\subsubsection{Sizes of early- and late-type galaxies at $z=0$}
\begin{figure}

\begin{center}

\includegraphics[width=7cm, bb= 10 285 280 750]{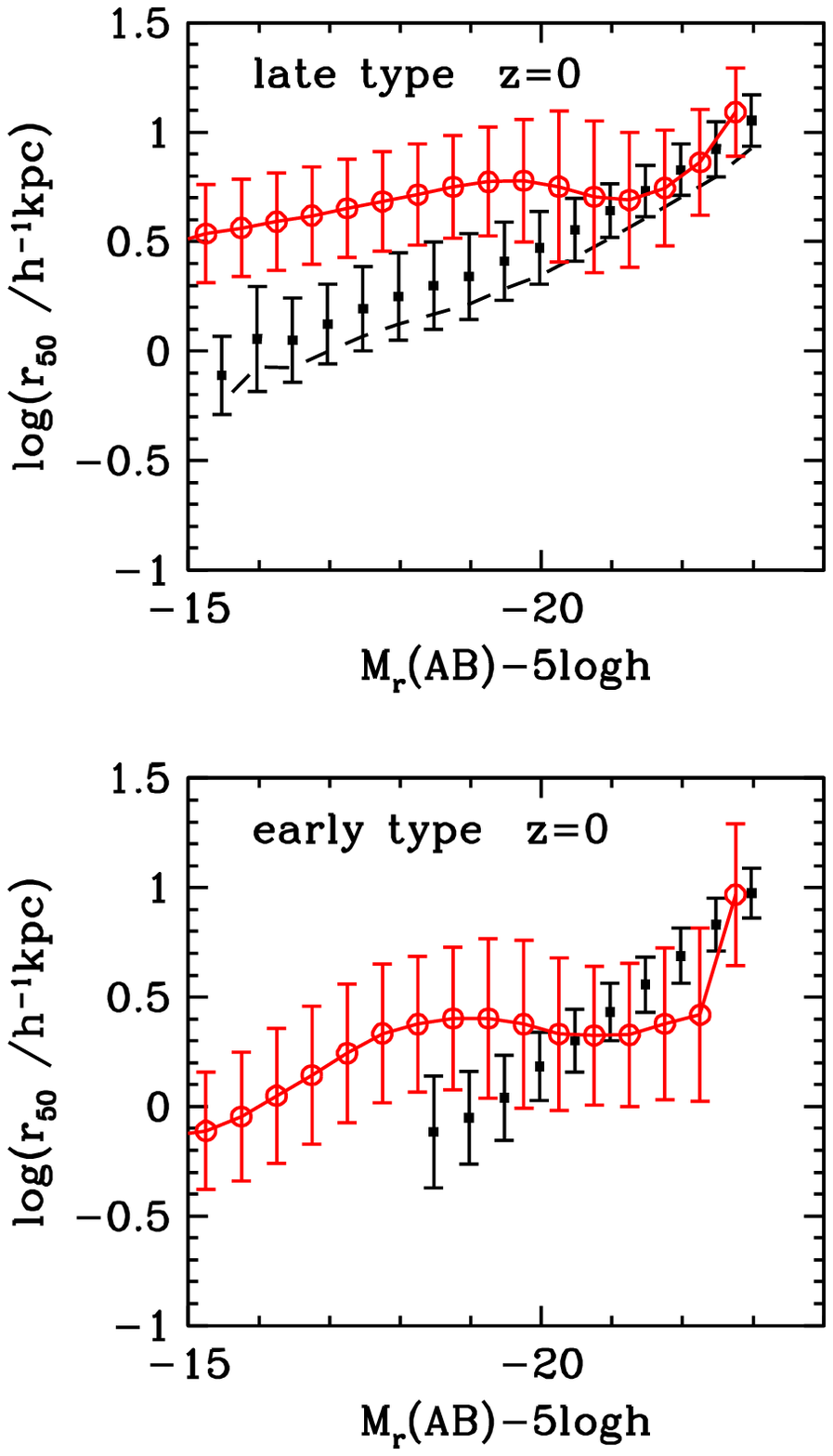}

\end{center}

\caption{Predictions of the default model for the half-light radii of
  late-type (top panel) and early-type (bottom panel) galaxies at
  $z=0$. In each panel, the solid line joining the open circles shows
  the predicted median face-on projected half-light radius in the
  $r$-band as a function of (dust extincted) $r$-band absolute
  magnitude, with the error bars showing the 10-90 percentile
  range. In this figure, model galaxies are classified as late- or
  early-type according to whether $(B/T)_r<0.5$ or $>0.5$
  respectively. The filled squares show measurements of the median and
  10-90\% size ranges from \citet{Shen03}, based on SDSS
  data. \citeauthor{Shen03} measured half-light radii by fitting
  Sersic profiles to galaxy images, and classified galaxies as late-
  or early-type according to whether the Sersic index $n<2.5$ or
  $>2.5$ respectively. For the \citeauthor{Shen03} data on late-type
  galaxies, we have multiplied the median sizes by a factor $1.34$, to
  correct them to face-on values (see text for details). The dashed
  line in the upper panel shows the median observed size relation if
  this correction is not applied.}

\label{fig:sizes_default}
\end{figure}

Another important property of galaxies is the relation between galaxy
size and luminosity (or stellar mass). We explore this in
Fig.~\ref{fig:sizes_default}, which shows the relation between
galaxy half-light radius and luminosity, for galaxies split into
late-type (i.e. disk-dominated, upper panel) and early-type
(i.e. bulge-dominated, lower panel). We compare the fiducial model
with measurements from the SDSS by \citet{Shen03}. Since
\citeauthor{Shen03} measured half-light radii in circular apertures
projected on the sky, we multiply their median sizes for late-type
galaxies by a factor 1.34, to correct them to face-on values. (The
factor 1.34 is the median correction from the projected to the face-on
half-light radius, for thin exponential disks having random
inclinations.) After applying this correction, the \citeauthor{Shen03}
median sizes for late-type galaxies are in good agreement with the
measurements by \citet{Dutton11}, who measured sizes by fitting
disk+bulge models to 2D galaxy images, over the range of overlap in
luminosity. 

Previous galaxy formation models have generally struggled to produce
the correct sizes for both disks and spheroids at $z=0$ { (although
  some recent SA models have been more successful
  \citep[e.g.][]{Guo11,Porter14}).} Our fiducial model is seen to
predict roughly correct sizes for brighter ($L \gsim L_{\star}$)
galaxies, but to predict sizes for both disks and spheroids which are
too large for fainter galaxies. This discrepancy is explored further
in \S\ref{sec:param-variations}.

\subsubsection{Stellar metallicities of early-type galaxies at $z=0$}
\begin{figure}

\begin{center}

\includegraphics[width=7cm, bb= 10 520 275 750]{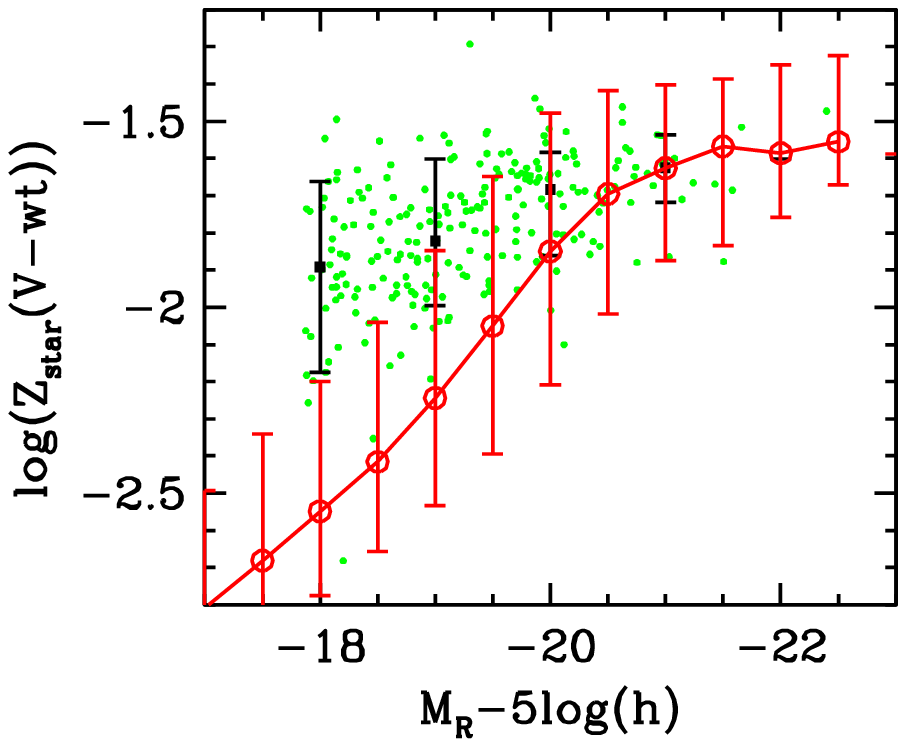}

\end{center}

\caption{Predictions of the default model for the stellar metallicity
  in early-type galaxies at $z=0$. The solid line shows the median
  stellar metallicity as a function of $R$-band luminosity for
  early-type galaxies in galaxy clusters, and the error bars show the
  10-90 percentile range. The stellar metallicities for individual
  model galaxies are mean values weighted by $V$-band luminosity. In
  order to replicate the galaxy selection in the observational sample,
  model galaxies are selected to be in dark matter halos with $\Mhalo
  > 10^{14} \hMsol$, and to have equivalent widths for $H\alpha$
  emission $EW(H\alpha) < 0.5$\AA. The green points show metallicities
  of individual galaxies from the sample of \citet{Smith09}, estimated
  from stellar absorption line strengths. Since the original spectra
  were measured in 1~arcsec radius fibre apertures, we correct the
  observed metallicity for each galaxy to a global value assuming a
  uniform metallicity gradient $d\log Z/d\log r = -0.15$, based on
  \citet{Rawle10}, and that the light profile in each galaxy follows
  an $r^{1/4}$ law. Observed metallicities relative to Solar are
  converted to absolute metallicities assuming $\Zsol=0.017$, to be
  consistent with the stellar population models used in
  \citeauthor{Smith09}. The black points with error bars show medians
  and 10-90\% ranges for the observational data in bins of $R$-band
  absolute magnitude.}

\label{fig:Zstar_default}
\end{figure}

The final secondary observational constraint which we consider here is
the metallicity-luminosity relation for galaxies. Observationally,
there are two main versions of this: (i) the stellar metallicity vs
luminosity relation for passive or early-type galaxies; and (ii) the
gas metallicity vs luminosity relation for star-forming or late-type
galaxies. We prefer to use the first of these as our constraint on
metallicities, for two reasons: (a) the predicted gas metallicities in
the models are coupled to the gas fractions; and (b) the observed gas
metallicities are generally not corrected for metallicity gradients,
and so do not represent global mean values for the cold gas
component. We compare predictions from our fiducial model with
observations of the stellar metallicty vs luminosity relation of
passive galaxies in galaxy clusters in
Fig.~\ref{fig:Zstar_default}. Since the observed metallicities are
inferred from absorption line features in the optical wavelength
range, we compare them with $V$-band luminosity-weighted mean
metallicities for model galaxies. In addition, we correct the observed
metallicities from aperture values to mean global values assuming a
fixed metallicity gradient - this results in a median correction of
$-0.10$~dex to the observed metallicities.

\subsection{Comparison with previous models}
We now compare the predictions from our new model with the predictions
from the earlier \galform models by \citet{Baugh05} and
\citet{Bower06}. We focus here on comparing with these \galform models
because they have been used in many previous papers, and because the
current model grew out of the desire to overcome various problems with
both earlier models, while retaining their respective strengths. We
note that there has also been significant work using the
\citet{Lagos12} and \citet{Gonzalez-Perez14} \galform models, which
developed out of the \citet{Bower06} model. Some comparisons of the
present model with the \citeauthor{Lagos12} and
\citeauthor{Gonzalez-Perez14} models have already been presented in
\citet{Lagos14,Lagos14b}.

The \citet{Baugh05} and \citet{Bower06} models differ from the current
model in various respects, as mentioned in the Introduction. We here
summarize the main differences:
\begin{enumerate}
\item The \citeauthor{Baugh05} and \citeauthor{Bower06} models and the
  new model all used different star formation laws for the quiescent
  (or disk) mode. Both of the earlier models assumed a quiescent SFR
  linearly proportional to the total cold gas mass in the galaxy
  disk. In \citeauthor{Baugh05}, the SFR timescale depended mildly on
  circular velocity, leading to SFR timescales that varied only weakly
  with redshift. In \citeauthor{Bower06}, the SFR timescale was also
  proportional to the disk dynamical time, leading to much shorter SFR
  timescales at high redshifts. In contrast, in the new model the
  quiescent SFR depends non-linearly on the total cold gas mass
  through the dependence on molecular gas fraction and hence on
  surface densities of gas and stars. This leads to typical quiescent
  SFR timescales that at first decrease with increasing redshift, but
  then tend to a constant value when most of the cold gas is
  molecular. (These differences and their effects are discussed in
  more detail in \citet{Lagos11a,Lagos11b}.)
\item All three models include star formation in bursts triggered by
  major and minor galaxy mergers, and transformation of stellar disks
  into spheroids in major mergers. In addition, the
  \citeauthor{Bower06} and new models include triggering of
  starbursts and transformation of stellar disks into spheroids by
  disk instabilities, with similar values for the stability threshold
  $\Fstab$. In both of the latter models, most of the star formation
  in bursts over the history of the universe is triggered by disk
  instabilities rather than galaxy mergers. All three models adopt the
  same dependence of starburst timescale on bulge dynamical time, but
  with different parameters. The starburst timescales in the new model
  are a factor $\sim 10-20$ larger than in \citeauthor{Bower06}, but a
  factor $\sim 2$ smaller than in \citeauthor{Baugh05}.  
\item All three models adopt the same formulation for the ejection of
  gas from galaxies and halos by supernova feedback, but with
  different parameters. The \citeauthor{Bower06} and the new model both
  adopt the same slope $\gammaSN=3.2$ for the dependence of the mass
  ejection rate on circular velocity, which is steeper than the slope
  $\gammaSN=2$ assumed in \citeauthor{Baugh05}, leading to much
  stronger SN feedback in low mass galaxies in the former two
  models. However, the normalization of the mass-loading factor is
  different between the models, leading to much larger mass ejection
  rates for a given SFR and circular velocity (by a factor $\sim 4$)
  in the \citeauthor{Bower06} model compared to the new model. Another
  difference is that both \citeauthor{Bower06} and new models include
  gradual return of ejected gas to the hot halo (at rates which are a
  factor $\sim 2$ higher in \citeauthor{Bower06}), while in the
  \citeauthor{Baugh05} model, ejected gas is only returned to the hot
  gas halo at halo mass doubling events.
\item The new model uses the same formulation for SMBH growth and AGN
  feedback as in \citeauthor{Bower06}, with similar values for the
  parameters. The \citeauthor{Baugh05} model did not include AGN
  feedback, but instead included an additional ``superwind'' mode of
  SN feedback, in which gas was ejected from halos and never
  reincorporated. The AGN and superwind feedback mechanisms both
  produce a high-mass break in the stellar mass function, but predict
  different dependences of this break mass on redshift.
\item All of the models assume identical Solar neighbourhood IMFs for
  quiescent star formation, but the \citeauthor{Baugh05} and new
  models both assume a top-heavy IMF in starbursts, while in
  \citeauthor{Bower06}, the starburst IMF is the same as the quiescent
  one. However, the starburst IMF in \citeauthor{Baugh05} is much more
  top-heavy than in the new model (with slopes $x=0$ and $x=1$
  respectively). 
\item The \citeauthor{Baugh05} and \citeauthor{Bower06} models used
  similar stellar population models (\citealt{Bressan98}, and an
  updated version of \citealt{Bruzual93} respectively), while the new
  model uses the \citet{Maraston05} models, which predict a larger
  contribution to the luminosity from TP-AGB stars.
\item The three models assume somewhat different cosmologies.
\end{enumerate}

\begin{figure}
\begin{center}
\includegraphics[width=7cm, bb= 20 55 275 510]{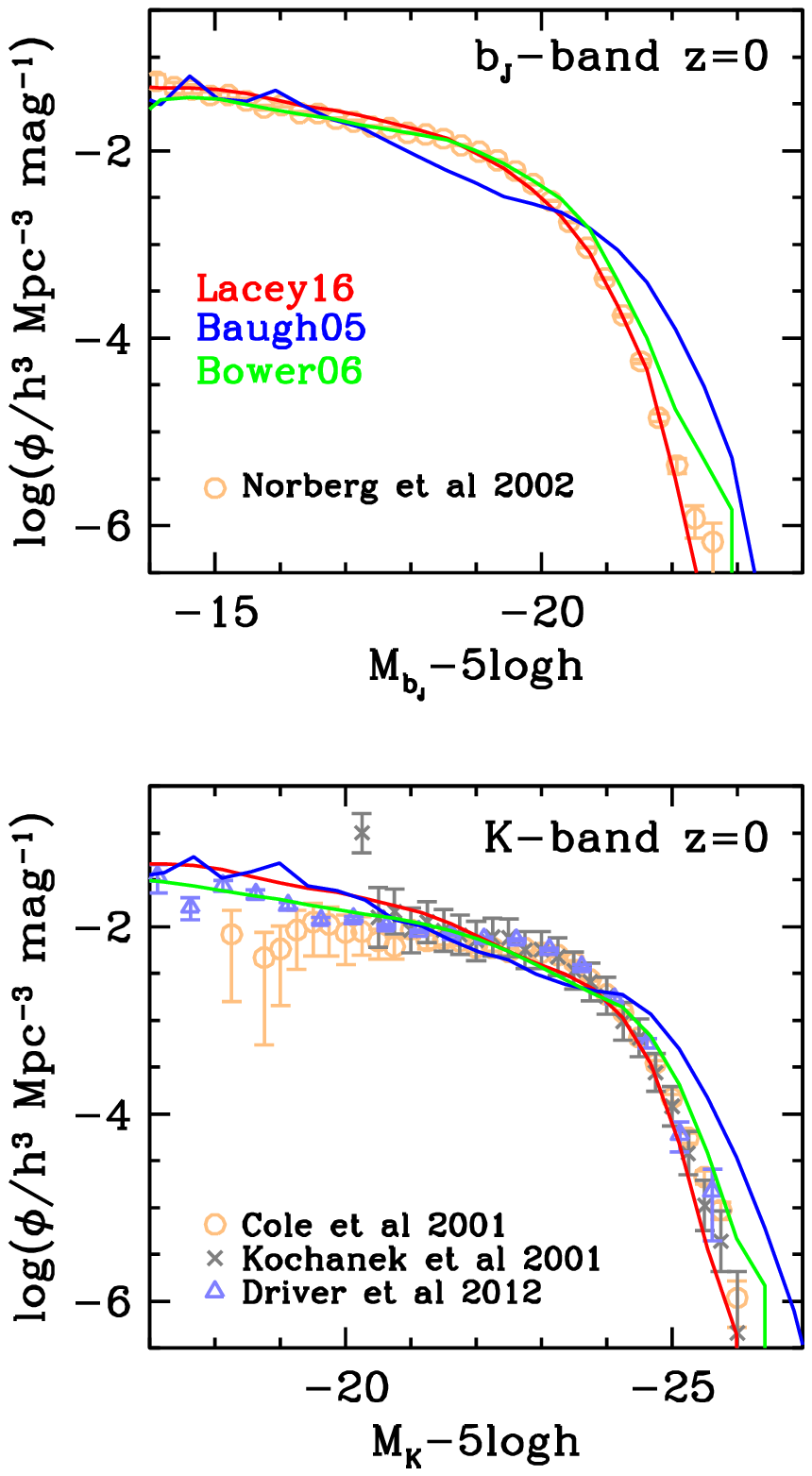}
\end{center}

\caption{Predictions for the $b_J$- and $K$-band LFs at $z=0$,
  comparing the new model with the \citet{Baugh05} and \citet{Bower06}
  \galform models. See Fig.~\ref{fig:lf_default} for more details
  about the curves and the observational data.}

\label{fig:lf_old_new}
\end{figure}

We see in Fig~\ref{fig:lf_old_new} that the $b_J$ and $K$-band LFs at
$z=0$ are similar in all 3 models.  The reason for this is that the
parameters in all 3 models were calibrated to approximately reproduce
these observational data. The \citeauthor{Baugh05} model is a poorer
fit than the other 2 models, having been calibrated to match a much
wider range of other observational data than for the
\citeauthor{Bower06} model.

\begin{figure}
\begin{center}
\includegraphics[width=7cm, bb= 20 520 280 750]{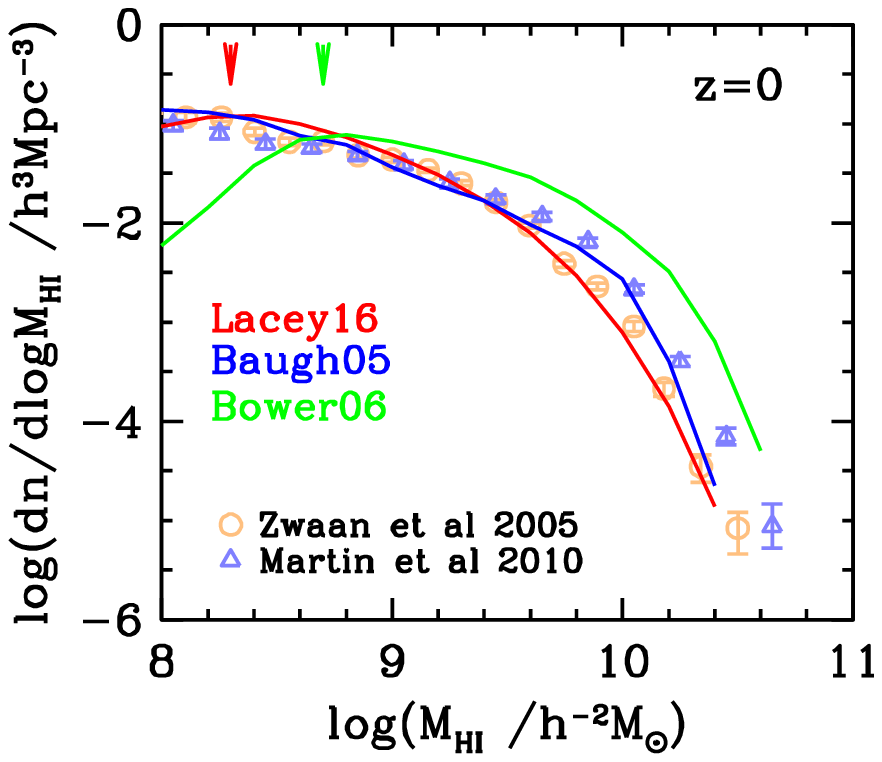}
\end{center}

\caption{Predictions for the HI mass function at $z=0$, comparing the
  new model with the \citet{Baugh05} and \citet{Bower06} \galform
  models. For the old models, we assumed a constant ratio 0.38 of
  molecular to atomic hydrogen masses. { The vertical arrows at the
    top of the panel indicate the HI mass below which the results for
    the corresponding model are affected by the halo mass resolution.}
  See Fig.~\ref{fig:mfHI_default} for more details about the curves
  and the observational data.}
\label{fig:mfHI_old_new}
\end{figure}

Fig.~\ref{fig:mfHI_old_new} compares the $HI$ mass functions at
$z=0$. The \citeauthor{Baugh05} model was calibrated to match
observational data on the $M_H/L_B$ vs. $L_B$ relation for late-type
galaxies (see \citealt{Cole00} for more details of the observational
data), while for the \citeauthor{Bower06} model, no calibration
against observed gas fractions or gas masses was performed.
The \citeauthor{Baugh05} model is seen to fit the observed $HI$ MF well
down to the $HI$ mass $\sim 10^8 \Msol$ at which halo mass resolution
effects set in. On the other hand, the \citeauthor{Bower06} model
predicts too many objects at high gas masses, and too few at low gas
masses. This seems to be a consequence mainly of the disk SF law
assumed in this model. Note that in the older \galform models, we
had to assume a constant ratio $M_{H_2}/M_{HI}$ in order to relate the
theoretically predicted cold gas masses to $HI$ masses. In contrast,
in the new model the $M_{H_2}/M_{HI}$ is predicted, and in fact has a
wide range of values.

\begin{figure}
\begin{center}
\includegraphics[width=7cm, bb= 20 520 275 750]{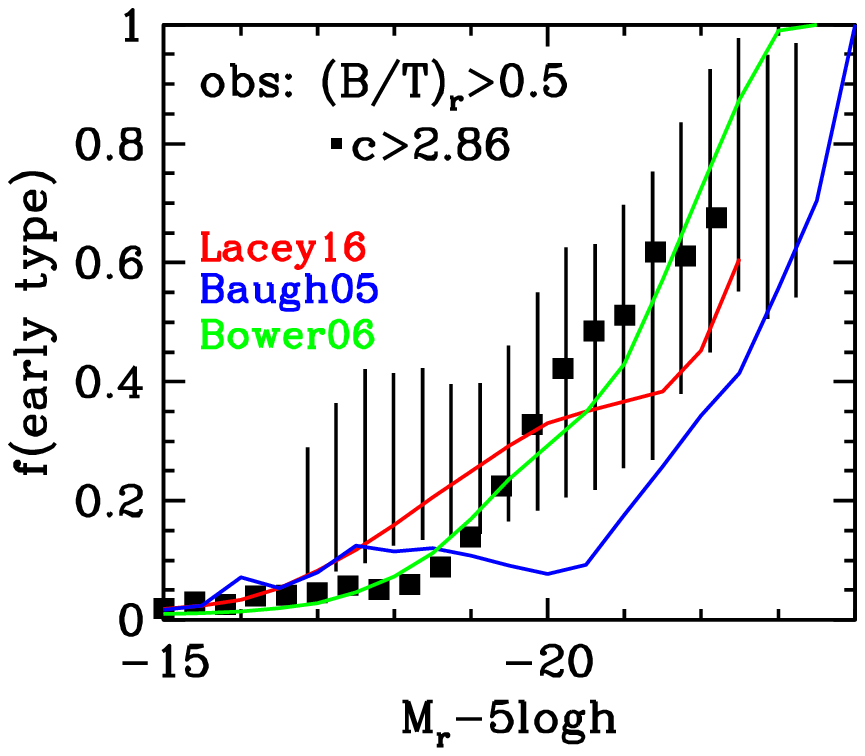}
\end{center}

\caption{Predictions for the fraction of early-type galaxies at $z=0$,
  comparing the new model with the \citet{Baugh05} and \citet{Bower06}
  \galform models. See Fig.~\ref{fig:morph_default} for more details
  about the curves and the observational data.}
\label{fig:morph_old_new}
\end{figure}

Fig~\ref{fig:morph_old_new} compares the fraction of early-type
galaxies as a function of luminosity at $z=0$. The
\citeauthor{Bower06} model is seen to be in good agreement with the
observed relation, like the new model, while the \citeauthor{Baugh05}
model predicts too low a fraction of early-type galaxies at high
luminosities (as was found earlier by \citealt{Gonzalez09}).



The BH vs bulge mass relation at $z=0$ for the new model is very
similar to that for the \citeauthor{Bower06} model, with both being in
good agreement with observations.

\begin{figure*}
\begin{center}
\begin{minipage}{5.4cm}
\includegraphics[width=5.4cm, clip=true, bb= 280 525 530 750]{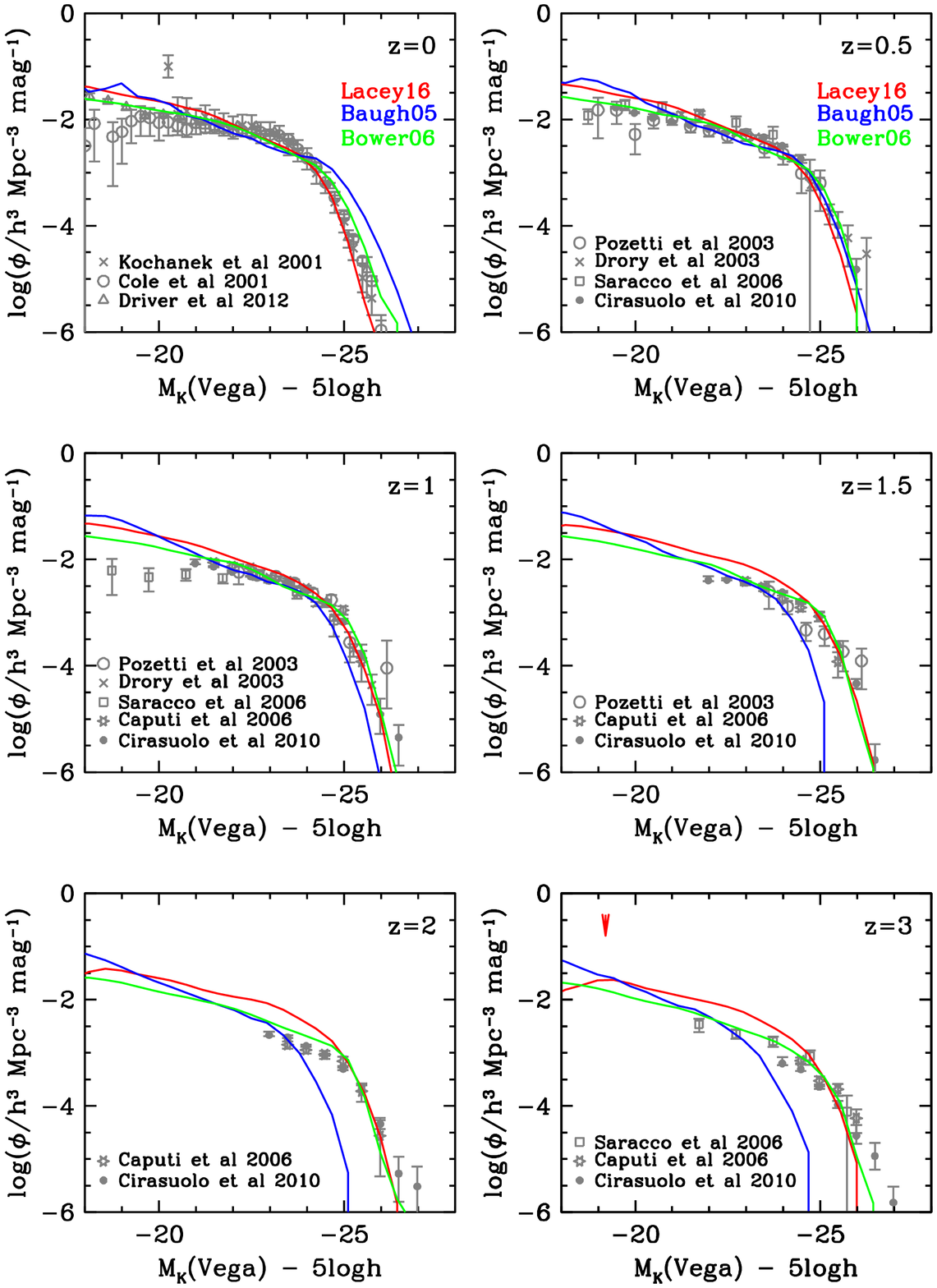}
\end{minipage}
\hspace{0.4cm}
\begin{minipage}{5.4cm}
\includegraphics[width=5.4cm, clip=true, bb= 24 288 275 514]{figs/Klfz_old_new.ps}
\end{minipage}
\hspace{0.4cm}
\begin{minipage}{5.4cm}
\includegraphics[width=5.4cm, clip=true, bb= 280 51 530 277]{figs/Klfz_old_new.ps}
\end{minipage}
\end{center}

\caption{Predictions for the evolution of the rest-frame K-band
  luminosity function, comparing the new model with the
  \citet{Baugh05} and \citet{Bower06} \galform models. { The
    vertical arrows at the top of the panels indicate the K-band
    luminosity below which the results for the corresponding model are
    affected by the halo mass resolution.}  See
  Fig.~\ref{fig:lfKz_default} for more details about the curves and
  the observational data. }
\label{fig:lfKz_old_new}
\end{figure*}

In Fig.~\ref{fig:lfKz_old_new} we compare the evolution of the
rest-frame $K$-band LF between the 3 models. The \citeauthor{Bower06}
model reproduced the observed evolution very well. On the other hand, the
\citeauthor{Baugh05} model underpredicted the number of high luminosity
galaxies at high redshift, which was one of the main failings of that
model, resulting from the too slow buildup of stellar mass in massive
galaxies, due to the long SF timescales in disks and the lack of disk
instabilities. In contrast, the new model is in good agreement with
the observed $K$-band LFs even at high redshift, { apart from being
  somewhat high at the faint end}.

\begin{figure}
\begin{center}
\includegraphics[width=7cm, bb= 20 295 275 750]{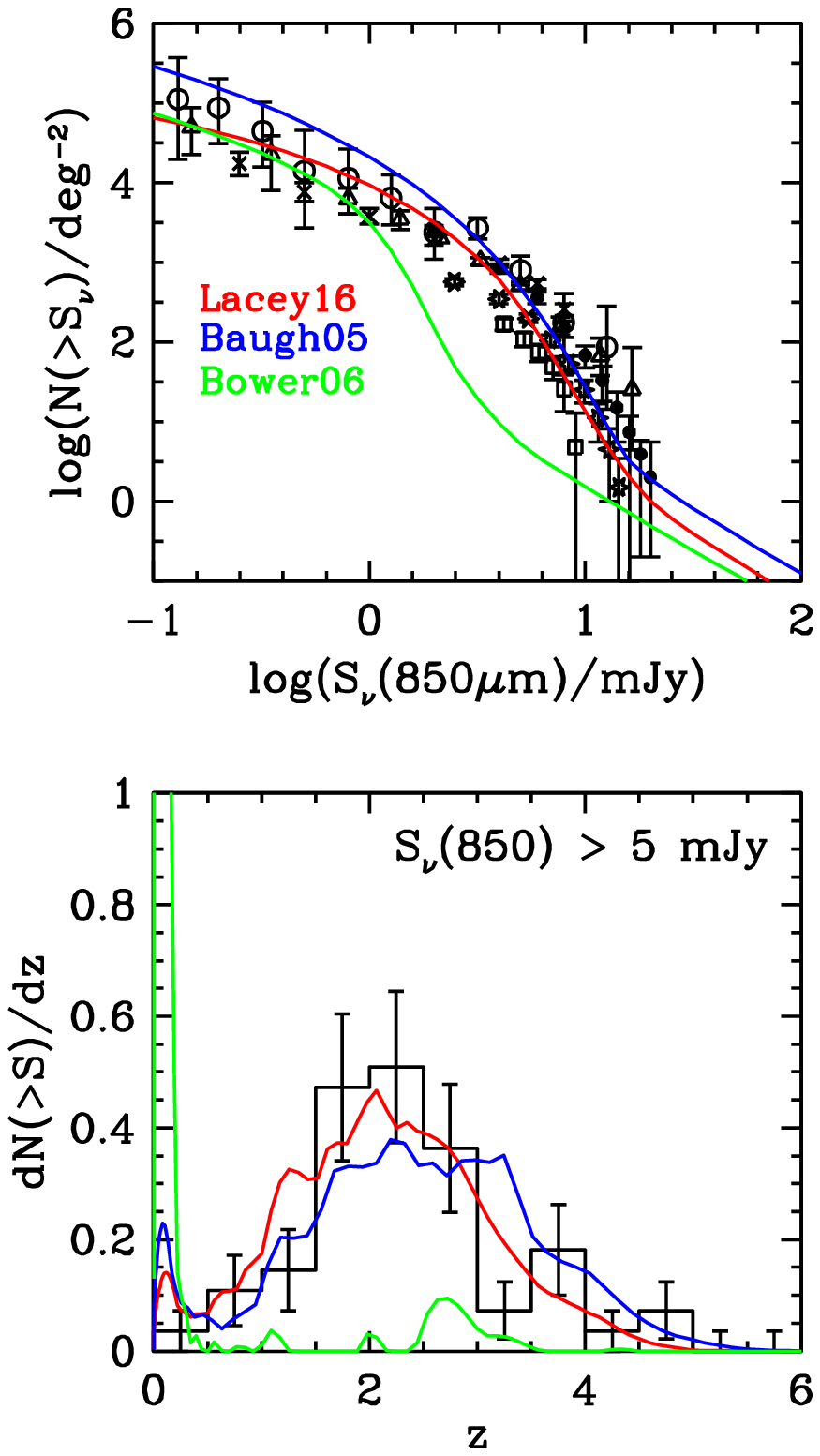}
\end{center}

\caption{Predictions for the $850\mum$ number counts and redshift
  distribution, comparing the new model with the \citet{Baugh05} and
  \citet{Bower06} \galform models. See Fig.~\ref{fig:SMGs_default} for
  more details about the curves and the observational data.}
\label{fig:SMGs_old_new}
\end{figure}

In Fig.~\ref{fig:SMGs_old_new} we compare predictions for number counts
and redshift distributions at 850~$\mum$ between the 3 models. The
\citeauthor{Baugh05} model is in good agreement with the observational
data, due to the top-heavy IMF in starbursts which was introduced for
that purpose. In contrast, the \citeauthor{Bower06} model, which
assumed a normal IMF in starbursts, predicts number counts which are
far too low in the 1-10~mJy flux range, by more than a factor 10. In
addition, this model predicts that 850~$\mum$ sources at these fluxes
should be at very low redshifts, $z \sim 0.1$, in complete
contradiction with observational measurements which put them at $z
\sim 2$. The new model, which also assumes a top-heavy IMF in bursts,
though with a less extreme slope, is also in very good agreement with
the observed counts and redshifts.

\begin{figure*}

\begin{center}
\begin{minipage}{5.4cm}
\includegraphics[width=5.4cm, clip=true, bb= 24 525 275 750]{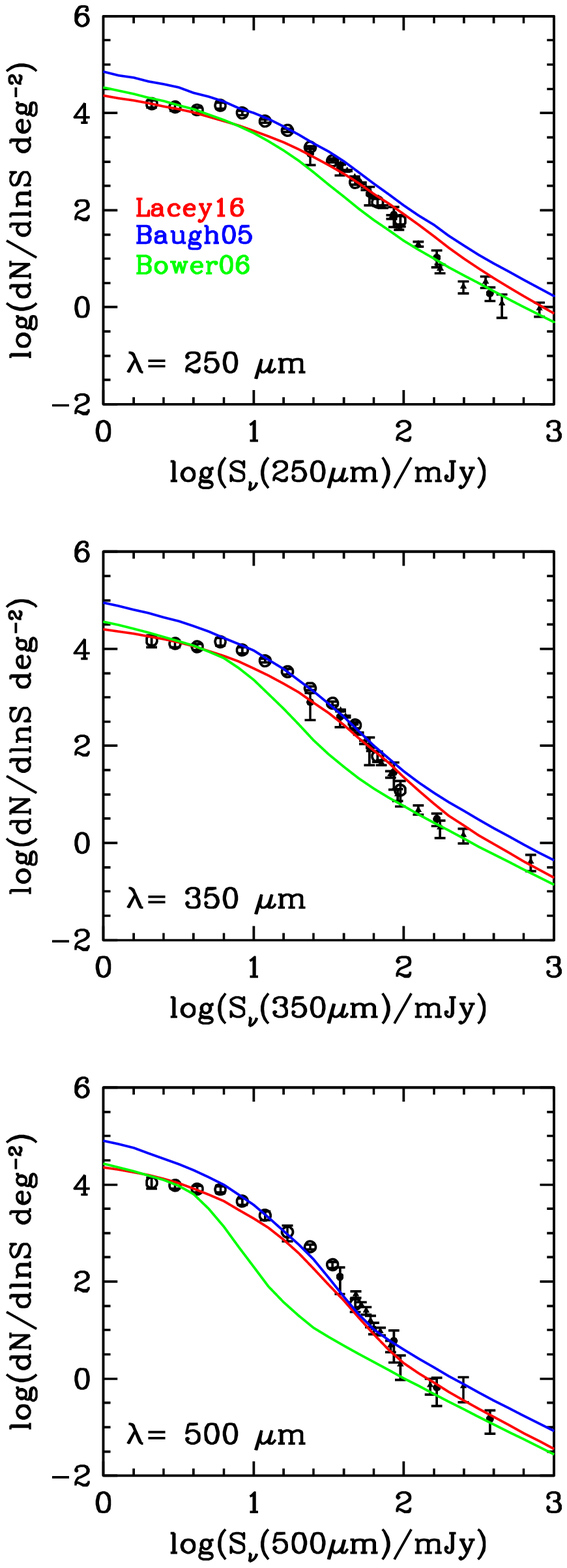}
\end{minipage}
\hspace{0.4cm}
\begin{minipage}{5.4cm}
\includegraphics[width=5.4cm, clip=true, bb= 24 288 275 514]{figs/FIRcounts_old_new.ps}
\end{minipage}
\hspace{0.4cm}
\begin{minipage}{5.4cm}
\includegraphics[width=5.4cm, clip=true, bb= 24 51 275 277]{figs/FIRcounts_old_new.ps}
\end{minipage}
\end{center}

\caption{Predictions for the far-IR differential number counts at (a)
  250, (b) 350 and (c) 500~$\mum$, comparing the new model with the
  \citet{Baugh05} and \citet{Bower06} \galform models. See
  Fig.~\ref{fig:FIRcounts_default} for more details about the curves
  and the observational data.}
\label{fig:FIRcounts_old_new}
\end{figure*}

Fig.~\ref{fig:FIRcounts_old_new} compares predictions for the far-IR
number counts at 250--500~$\mum$. The \citeauthor{Baugh05} model
predicted far-IR counts which were in reasonable agreement with
observations at faint fluxes, but which were too high at bright
fluxes, due to predicting too many far-IR luminous galaxies in the
nearby Universe. On the other hand, the \citeauthor{Bower06} model,
while in better agreement for bright fluxes, predicted counts which
were far too low at faint fluxes, especially at longer far-IR
wavelengths. The new model, though still not a perfect match to the
observed counts at intermediate fluxes, is now in much better
agreement at bright and faint fluxes, especially for the longer
wavelengths. The improvement appears to be mainly due to the IMF in
starbursts, which is top-heavy, unlike in \citeauthor{Bower06}, but
less top-heavy than in \citeauthor{Baugh05}.

\begin{figure}
\begin{center}
\includegraphics[width=7cm, bb= 20 295 275 750]{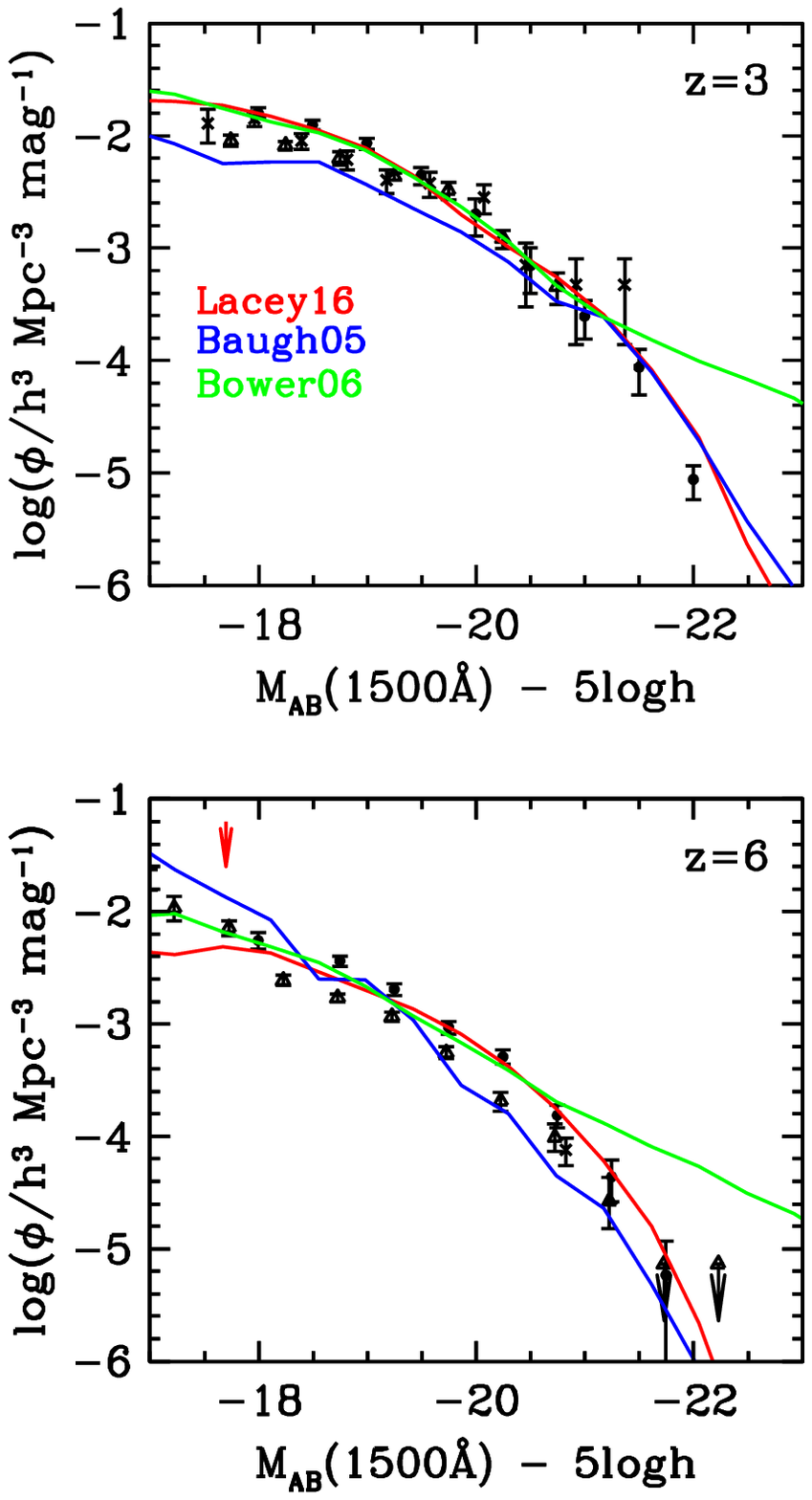}
\end{center}

\caption{Predictions for the rest-frame far-UV (1500\AA) LF at (a)
  $z=3$ and (b) $z=6$, comparing the new model with the
  \citet{Baugh05} and \citet{Bower06} \galform models.  { The
    vertical arrows at the top of the panels indicate the UV
    luminosity below which the results for the corresponding model are
    affected by the halo mass resolution.} See
  Fig.~\ref{fig:LBGs_default} for more details about the curves and
  the observational data. }
\label{fig:LBGs_old_new}
\end{figure}

In Fig.~\ref{fig:LBGs_old_new}, we compare predictions from the three
models for the rest-frame far-UV LFs at $z=3$ and $z=6$. The
\citeauthor{Baugh05} model fits the observed LF at both redshifts well
--- for $z=3$, this was because the model parameters were calibrated to
do this. On the other hand, the \citeauthor{Bower06} model is in
serious disagreement with the observed LFs, mainly due to the very
short timescales it assumed for star formation in bursts (this
discrepancy was previously noted in \citealt{Lacey11}). The new model
is in good agreement with the observations at $z=3$, and slightly
poorer at $z=6$.



Both the \citeauthor{Baugh05} and \citeauthor{Bower06} models predict
$I$-band Tully-Fisher relations at $z=0$ which are slightly too low in
normalization. In contrast, the new model predicts a TF relation in
better overall agreement with observational data, although the slope
is somewhat steeper than implied by observations.

\begin{figure}
\begin{center}
\includegraphics[width=7cm, bb= 10 285 280 750]{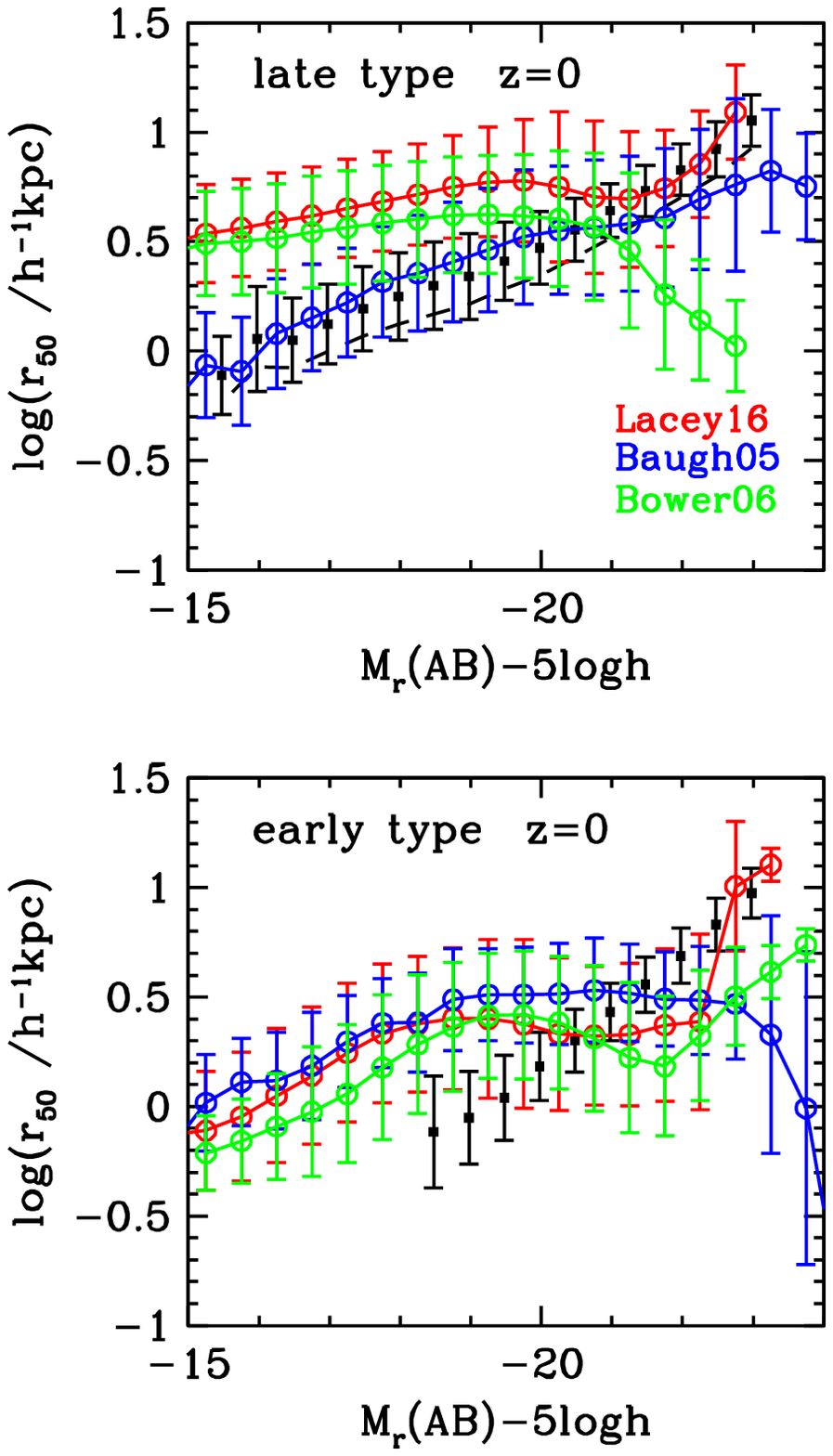}
\end{center}

\caption{Predictions for the half-light radii of late-type (top panel)
  and early-type (bottom panel) galaxies at $z=0$, comparing the new
  model with the \citet{Baugh05} and \citet{Bower06} \galform
  models. See Fig.~\ref{fig:sizes_default} for more details about the
  curves and the observational data.}
\label{fig:sizes_old_new}
\end{figure}

Fig.~\ref{fig:sizes_old_new} compares the size vs luminosity relations
at $z=0$ for the three models. The \citeauthor{Baugh05} model predicts
sizes which agree very well with observations for late-type galaxies,
but very poorly for early-type galaxies. The \citeauthor{Bower06}
model instead predicted sizes in very poor agreement with observations
for both late- and early-type galaxies. The new model predicts sizes
for both late and early types in quite good agreement with
observations for brighter ($L \gsim L_{\star}$) galaxies, but which
are too large for lower luminosity galaxies. The larger sizes of $L
\lsim L_{\star}$ late-type galaxies in both the new and
\citeauthor{Bower06} models compared to the \citeauthor{Baugh05} model
are primarily due to the stronger SN feedback adopted in the former
models, which results in galaxies of a given stellar mass forming in
larger halos. This is explored further in \S\ref{sec:param-variations}.

\begin{figure}
\begin{center}
\includegraphics[width=7cm, bb= 10 520 275 750]{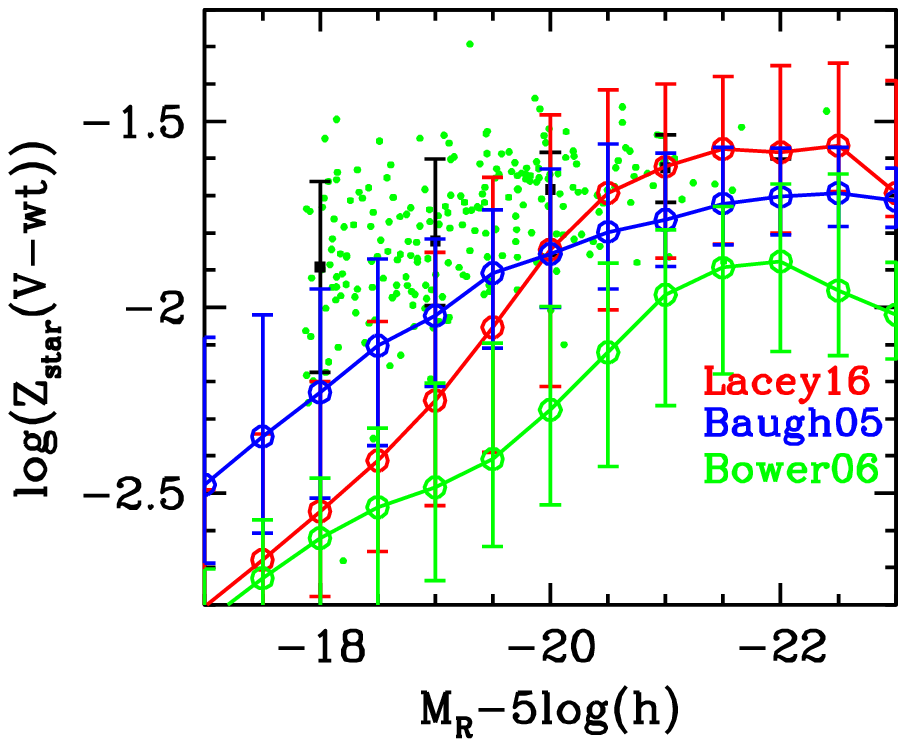}
\end{center}

\caption{Predictions for the stellar metallicity in early-type
  galaxies at $z=0$, comparing the new model with the \citet{Baugh05}
  and \citet{Bower06} \galform models. See
  Fig.~\ref{fig:Zstar_default} for more details about the curves and
  the observational data.}

\label{fig:Zstar_old_new}
\end{figure}

Finally, Fig.~\ref{fig:Zstar_old_new} compares the stellar metallicity
vs luminosity relation for early-type galaxies for the 3 models. The
\citeauthor{Baugh05} model predicts a slope for this relation in very
good agreement with the observational data in
Fig.~\ref{fig:Zstar_old_new}, but with normalization that is somewhat
too low. In contrast, the \citeauthor{Bower06} model predicts
metallicities which are too low by a factor $\sim 4$ at all
luminosities. The new model predicts metallicities which are in good
agreement with observations at higher luminosities ($L \gsim
L_{\star}$), but which fall below observed values at lower
luminosities. The lower metallicities at $L \lsim L_{\star}$ in the
new model compared to \citeauthor{Baugh05} are mainly due to the
stronger SN feedback in the new model. The higher metallicities
compared to \citeauthor{Bower06} are mainly result from the top-heavy
starburst IMF in the new model, which results in a higher yield of
metals.  These issues are explored in \S\ref{sec:param-variations}.



\section{Parameter space of galaxy formation 
and effects of different physical processes on observable quantities}
\label{sec:param-variations}


In this section, we examine in more detail how the predictions of the
\galform model for the key observational constraints identified in the
previous section depend on different physical processes and the
parameters describing them. The text below summarizes the effects of
different physical processes and of varying the associated parameters
in the model. The plots showing the effects of varying different model
parameters on different predicted properties are collected in
Appendix~\ref{sec:param_var_plots}. Each plot shows the effect of
varying one parameter around its standard value given in
Table~\ref{table:param} (as indicated by the red curve and
corresponding label in each plot), while keeping the other parameters
fixed at their standard values. The plots are grouped together
according to the observational constraint which the model is being
compared to.

\subsection{Supernova feedback}




Supernova feedback plays a crucial role in galaxy formation. In
\galform it depends on 3 parameters: $\gammaSN$ and $\VSN$ which
control respectively the circular velocity dependence and
normalization of the mass-loading (eqn.\ref{eq:Meject}), and
$\alpharet$, which controls the timescale for ejected gas to return to
the hot gas halo and so become available for cooling
(eqn.\ref{eq:Mreturn}). Of these, $\gammaSN$ and $\VSN$ have dramatic
effects on a very wide range of properties, while $\alpharet$ has a
somewhat more modest effect on the observational properties which we
compare with here. 
Variations in $\VSN$ and $\alpharet$ are somewhat degenerate in their
effects, in that decreasing $\VSN$ (stronger SN feedback) has effects
in the same sense as increasing $\alpharet$ (faster return of ejected
gas), for example in their effects on the galaxy luminosity function
(Fig.~\ref{fig:lf_SNfeedback}) and on far-IR and sub-mm number counts
(Figs.~\ref{fig:FIRcounts_SNfeedback} and
\ref{fig:SMGs_SNfeedback}). However, this degeneracy is reduced by
also considering other properties, for example stellar metallicities
(Fig.~\ref{fig:Zstar_SNfeedback}). 

Fig.~\ref{fig:lf_SNfeedback} shows that we need both a steep dependence
of SN feedback on circular velocity ($\gammaSN \gsim 3$) and a high
normalization ($\VSN \gsim 300\kms$) in order for the faint end of the
$b_J$ and $K$-band LFs at $z=0$ to agree with observations. With weaker
feedback, the model predicts far too many low luminosity
galaxies. (Note that the turnover at low luminosities seen for the
models with weaker feedback is a result of the DM N-body simulation
only resolving halos more massive than $\sim 2\times 10^{10} \hMsol$.)
The same result is found from the evolution of the $K$-band LF up to
$z=3$ (Fig.~\ref{fig:lfKz_SNfeedback}). We use these data as our
primary constraint on $\gammaSN$ and $\VSN$.

The SN feedback (especially $\VSN$) likewise has strong effects on the
far-IR and sub-mm number counts, with weaker feedback leading to
higher counts (Figs.~\ref{fig:FIRcounts_SNfeedback} and
\ref{fig:SMGs_SNfeedback}). However, the SN feedback has important
effects on several other of the observational datasets which we use to
constrain the model, some of which are in tension with the constraints
from the LFs. The size-luminosity relations for disks and spheroids
are much better fit at lower luminosities with much weaker feedback
(lower $\gammaSN$ and/or lower $\VSN$ - see
Fig.~\ref{fig:sizes_SNfeedback}), because galaxy sizes tend to scale
with the radii of the DM halos in which they formed, and weaker
feedback results in galaxies of the same mass forming in smaller
halos. However, having much weaker feedback tends to boost the
fraction of spheroid-dominated galaxies far above observed values at
low luminosities (see Fig.~\ref{fig:morph_SNfeedback}). { This is
  because in our model, low mass spheroidal galaxies are produced
  mainly by disk instabilities rather than by galaxy mergers (see
  \S\ref{ssec:effects_disk_instab}, and weaker SN feedback results in
  disks being more massive, and so more more self-gravitating and
  hence more bar unstable.}  The stellar metallicity vs luminosity
relation is also best fit with somewhat weaker feedback than in our
fiducial model (see Fig.~\ref{fig:Zstar_SNfeedback}). Finally, the
slope of the TF relation is also fit better with $\gammaSN=2$ than
with our fiducial $\gammaSN=3.2$ (see
Fig.~\ref{fig:TF_SNfeedback}). The HI MF varies quite weakly with SN
feedback parameters, except for the lowest values of $\VSN$, which
cause a large decrease in the MF at high HI masses and a modest
increase at intermediate masses
(Fig.~\ref{fig:mfHI_Vhot_SF_noAGN}). (See \citet{Kim13} for a detailed
study of the effects of SN and AGN feedback on the HI MF.)

On the other hand, the gas return timescale parameter $\alpharet$ has
a modest effect on the bright end of the $b_J$ and $K$-band LFs, as
well as on the far-UV LF at $z=3$, with faster gas return (i.e. larger
$\alpharet$) resulting in a higher number density of brighter galaxies
(Figs.\ref{fig:lf_SNfeedback}, \ref{fig:lfKz_SNfeedback} and
\ref{fig:LBGs_SNfeedback}). There are also appreciable effects on the
far-IR and sub-mm number counts (Figs.\ref{fig:FIRcounts_SNfeedback}
and \ref{fig:SMGs_SNfeedback}). { However, there is less effect on
  the 850~$\mum$ redshift distribution
  (Fig.~\ref{fig:SMGs_SNfeedback}), because varying $\alpharet$ tends
  to shift the whole bright end of the far-IR/sub-mm luminosity
  function up or down.}  Larger $\alpharet$ also results in a higher
fraction of early-type galaxies at high luminosities at $z=0$
(Fig.~\ref{fig:morph_SNfeedback}). The effects of $\alpharet$ on other
properties considered here are quite small. However, it does have a
significant effect on galaxy colours. If there is no gradual return of
ejected gas (i.e. $\alpharet=0$, so that gas only returns to the hot
halo after the halo mass has doubled), then the fraction of blue
galaxies at low luminosities is much lower at $z=0$ than
observed. Predictions for galaxy colours will be discussed in more
detail in \S\ref{sec:phys-predictions}.

\subsection{AGN feedback and SMBH growth}




AGN feedback plays a very important role in the model. The most
important factors controlling the strength of the AGN feedback are the
masses of the SMBHs hosted by galaxies, and the parameter $\alphacool$
(eqn.\ref{eq:alpha_cool}). SMBHs are assembled mostly during
starbursts, and in the fiducial model, starbursts are triggered mainly
by disk instabilities, as discussed in
\S\ref{ssec:effects_disk_instab}. The amount of mass accreted onto
SMBHs during starbursts is controlled by the parameter $\fBH$, but we
always adjust this to reproduce the normalization of the SMBH vs bulge
mass relation at $z=0$.
This leaves $\alphacool$ as the main parameter to be considered
here. $\alphacool=0$ corresponds to turning off AGN feedback, while
increasing $\alphacool$ reduces the halo masses at which AGN feedback
turns on and so results in larger effects from AGN feedback
overall. Fig.~\ref{fig:lf_AGNfeedback} shows that for no AGN feedback
(equivalent to $\alphacool=0$), there are far too many bright galaxies
in the $b_J$ and $K$ LFs at $z=0$. The value of $\alphacool$ is then
calibrated to reproduce the $z=0$ LFs. Fig.~\ref{fig:lf_AGNfeedback}
also shows that increasing $\alphacool$ results in a modest decrease
in the bright end of the $z=0$ LFs. Similar effects are seen in the
$K$-band LF at higher redshifts, though the effects of $\alphacool$
become less pronounced for $z \gsim 3$
(Fig.~\ref{fig:lfKz_AGNfeedback}). The value of $\alphacool$ also has
a quite significant effect on the far-IR and sub-mm counts
(Figs.\ref{fig:FIRcounts_AGNfeedback_IMF} and
\ref{fig:SMGs_AGNfeedback}), but again not on the sub-mm redshift
distribution, nor on the far-UV LFs at $z=3-6$
(Fig.~\ref{fig:LBGs_AGNfeedback}). There is a noticeable effect on the
morphological fractions at $z=0$, where stronger AGN feedback results
in lower fractions of early-type galaxies at higher luminosities
(Fig.~\ref{fig:morph_AGNfeedback}). The effects of $\alphacool$ on
other properties considered here are quite small.


\subsection{Disk instabilities}
\label{ssec:effects_disk_instab}



Disk instabilities play a key role in our model. They play a direct
role in triggering starbursts and causing the morphological
transformation of disks into spheroids.  They are also the main
mechanism triggering the growth of SMBHs in our fiducial model
(through accretion in starbursts), and hence play a large role in AGN
feedback. The parameter in our model which modulates the effects of
disk instabilities is the stability threshold $\Fstab$
(eqn.\ref{eq:disk_stability}). For $\Fstab<0.61$, all disks are
stable, but as $\Fstab$ is increased, more disks become unstable.

Examining the direct effects first, we see in
Fig.~\ref{fig:morph_AGNfeedback} that with no disk instabilities, the
fraction of early-type (i.e. spheroid-dominated) galaxies at $z=0$ is
far too low at all luminosities. In that case, spheroids are assembled
only through galaxy mergers. Increasing $\Fstab$ increases the
fraction of early types at all luminosities
(Fig.~\ref{fig:morph_AGNfeedback}). { We also see from
  Fig.~\ref{fig:morph_SF_nomerge} that if galaxy mergers are turned
  off, the fraction of early-type galaxies at low luminosities
  ($M_r-5\log h \lsim -19$) is almost identical to the fiducial model,
  while at higher luminosities, the fraction is appreciably lower. In
  our fiducial model, disk instabilities therefore play the dominant
  role in building up stellar spheroids at low galaxy masses, and make
  an important contribution even at high masses
  \citep[c.f.][]{Parry09}. The buildup of SMBHs is closely linked to
  the buildup of spheroids in our model. Even though the fraction of
  spheroid-dominated galaxies is sensitive to the parameters for disk
  instabilities and mergers, the SMBH vs. bulge mass relation is only
  weakly dependent on these
  (Fig.~\ref{fig:SMBH_AGNfeedback_nomerge}).}

Disk instabilities also have large effects on galaxy sizes. In
Fig.~\ref{fig:sizes_AGNfeedback}, we see that in the absence of disk
instabilities, the sizes of early-type galaxies are far too large at
low luminosities compared to observations, and far too small at high
luminosities, while turning on disk instabilities brings these sizes
into much closer agreement with observations. For late-type galaxies,
in the absence of disk instabilities, the average sizes are much too
small at high luminosities, but turning on the instabilities converts
these compact disks into spheroids (since smaller disks are more self
gravitating and so more unstable), so bringing the average sizes of
disk-dominated galaxies into good agreement with observations.

The direct effects of disk instabilities in triggering starbursts can
be seen in Fig.~\ref{fig:LBGs_AGNfeedback}, showing the far-UV LF at
$z=3-6$. The bright part of the far-UV LF is dominated by starbursts,
and the number of these increases when more disks become unstable,
especially at $z=6$. 

The indirect effects of disk instabilities through their impact on AGN
feedback are shown in the $b_J$ and $K$-band LFs in
Figs.\ref{fig:lf_AGNfeedback} and \ref{fig:lfKz_AGNfeedback}. In the
absence of disk instabilities, the LFs at $z=0$ look very close to the
case of no AGN feedback, producing a large excess of bright galaxies
(Fig.~\ref{fig:lf_AGNfeedback}). In the $K$-band LF, this effect
reduces with increasing redshift, until at $z=3$ the LF is insensitive
to disk instabilities (Fig.~\ref{fig:lfKz_AGNfeedback}). The far-IR
and sub-mm number counts are much higher when disk instabilities are
turned off, due to the absence of AGN feedback, but are only mildly
sensitive to $\Fstab$ for $\Fstab>0.61$ (see
Figs.\ref{fig:FIRcounts_AGNfeedback_IMF} and
\ref{fig:SMGs_AGNfeedback}). The excess sub-mm counts when disk
instabilities are turned off are dominated by galaxies at lower
redshifts than the observed peak at $z\sim 2$
(Fig.~\ref{fig:SMGs_AGNfeedback}).

\subsection{Galaxy mergers}




Galaxy mergers have two consequences in the model: major mergers (with
mass ratio $M_2/M_1 > \fellip$) cause stellar disks to be transformed
into spheroids, and major and minor mergers with $M_2/M_1 > \fburst$
trigger starbursts.
However, in the current model, most of the properties we have been
examining are almost unchanged if either starbursts in galaxy mergers
are turned off, or if galaxy mergers are turned off completely. The
main exceptions to this are for the fractions of early-type galaxies
and their sizes at $z=0$. Fig.~\ref{fig:morph_SF_nomerge} shows that
when mergers are turned off, the fraction of early-type galaxies at
high luminosities is much lower, while
the sizes of high-luminosity early-type galaxies are also smaller. On
the other hand, the galaxy luminosity functions and number counts are
almost identical whether galaxy mergers and their associated
starbursts are turned on or not (Figs.\ref{fig:lf_IMF_nomerge},
\ref{fig:SMGs_IMF_tauburst_nomerge} and \ref{fig:LBGs_nomerge}). Most
starbursts are triggered by disk instabilities in this model, and disk
instabilities also dominate the morphological transformation of disks
into spheroids at low galaxy masses. The SMBH vs. bulge mass relation
at $z=0$ is also insensitive to whether galaxy mergers are included
(Fig.~\ref{fig:SMBH_AGNfeedback_nomerge}). The main importance of
mergers in this model is therefore in building up stellar spheroids at
high masses at the present day.

\subsection{Disk star formation timescale}



The value of the star formation rate coefficient $\nuSF$, which
controls the rate of conversion of molecular gas into stars in
quiescent galaxy disks (eqn.(\ref{eq:SFR_mol})), has only a small effect
on most of the observable properties we compare to here, when it is
varied over the range $\nuSF = 0.25-1.2 \Gyr^{-1}$ allowed by direct
observational constraints at $z=0$. The $b_J$-band LF at $z=0$ and also
the $K$-band LFs over the whole range $z=0-3$ are extremely
insensitive to this parameter (a similar result was found earlier by
\citealt{Lagos11a}). Somewhat more surprisingly, the HI mass function at
$z=0$ also depends only very weakly on $\nuSF$ in the allowed range
(Fig.~\ref{fig:mfHI_Vhot_SF_noAGN}). This insensitivity is due to the
non-linear dependence of SFR and HI mass on total cold gas mass in
this version of \galform, which contrasts with the simpler linear
dependence assumed in earlier versions of \galform. On the other hand,
this means that the HI mass function is a robust prediction of the
model, and the fact that it agrees so well with observational data is
a significant success. The morphological fractions at $z=0$ do depend
significantly on $\nuSF$, with higher values leading to a higher
fraction of late-type galaxies at high luminosities (see
Fig.~\ref{fig:morph_SF_nomerge}).  The other main effect is on the
far-IR and sub-mm counts (Fig.~\ref{fig:FIRcounts_SNfeedback}), where
the amplitude of the counts at bright fluxes decreases as $\nuSF$
increases.


\subsection{Starburst timescale}




The timescales for starbursts due to both galaxy mergers and disk
instabilities are controlled by the parameters $\fdyn$ and
$\tauburstmin$ (eqn.\ref{eq:taustar_burst}). Varying $\fdyn$ over the
range 2-40 has almost negligible effect on any of the properties
considered here, apart from the bright end of the far-UV LF at $z=3$
and $z=6$ (Fig.~\ref{fig:LBGs_IMF_tauburst}). On the other hand,
varying $\tauburstmin$, the minimum star formation timescale in
bursts, has more noticeable effects. There is no effect on the $b_J$-
and $K$-band LFs at $z=0$, but the effect on the $K$-band LF increases
with redshift, with values of $\tauburstmin$ larger than our fiducial
value causing a large drop in the number of bright galaxies by $z=3$
(Fig.~\ref{fig:lfKz_IMF_SPS}). The same effect is seen in the far-UV
LF at $z=3$ and $z=6$ (Fig.~\ref{fig:LBGs_IMF_tauburst}). The effects
on the far-IR and sub-mm number counts are quite modest, but the
effects on SMG redshifts are large, with larger values of
$\tauburstmin$ shifting the distribution to much lower redshifts
(Fig.~\ref{fig:SMGs_IMF_tauburst_nomerge}). Finally, larger values of
$\tauburstmin$ also cause a reduction in the fraction of early-type
galaxies at high luminosities at $z=0$.  These observations therefore
constrain starburst timescales in the model to be not too large.

\subsection{IMF in starbursts}



The slope $x$ of the IMF in starbursts is a very important parameter
in our model. Comparing first our fiducial model with a starburst IMF
slope of $x=1$ with a model having the same Kennicutt IMF in
starbursts as in quiescent disks, we find that the largest effect is
on the sub-mm counts and redshift distribution
(Fig.~\ref{fig:SMGs_IMF_tauburst_nomerge}), where for a Kennicutt IMF,
the counts are too low by a factor up to 100 at intermediate fluxes,
and in addition the predicted redshifts are far lower than
observed. There are also important effects on the far-IR number counts
(Fig.~\ref{fig:FIRcounts_AGNfeedback_IMF}). The $b_J$- and $K$-band
LFs at $z=0$ are only slightly different for the two different
starburst IMFs (Fig.~\ref{fig:lf_IMF_nomerge}), as is the evolution of
the $K$-band LF (Fig.~\ref{fig:lfKz_IMF_SPS}). However, the stellar
metallicity in early-type galaxies at $z=0$ is too low at high
luminosities for a Kennicutt IMF (Fig.~\ref{fig:Zstar_IMF}).

We also show in the same plots the effects of varying the slope of the
starburst IMF over the range $x=0-1.2$. The results for $x=1.2$ are
quite close to those for a Kennicutt IMF in bursts, because these two
IMFs have similar fractions of mass in high-mass ($m \gsim 10\Msol$)
stars, despite having different shapes in detail. Much flatter
starburst IMF slopes (i.e. $x<1$) than our fiducial slope of $x=1$
result in sub-mm counts which are much too high
(Fig.~\ref{fig:SMGs_IMF_tauburst_nomerge}), as well as overpredicting
the bright end of the $K$-band LF at $z=0$
(Fig.~\ref{fig:lf_IMF_nomerge}) and of the far-UV LF at high redshift
(Fig.~\ref{fig:LBGs_IMF_tauburst}), and also stellar metallicities of
luminous early-type galaxies (Fig.~\ref{fig:Zstar_IMF}). These results
are in contrast to the \citet{Baugh05} model, which obtained similar
fits to many of the same observational constraints assuming $x=0$ for
the starburst IMF. The most important reason for the difference is
that the \citet{Baugh05} model had a different model for feedback in
high mass galaxies, which produced much stronger suppression of
high-mass galaxies at high redshifts than in the AGN feedback model we
use here.

\subsection{Stellar population model}

As described in \S\ref{ssec:stars_dust}, in our fiducial model we use
the stellar population synthesis (SPS) models of \citet{Maraston05},
which include an enhanced contribution from TP-AGB stars compared to
earlier models \citep[e.g.][]{Bruzual03}. The enhanced TP-AGB results
in increased near-IR luminosities for stellar populations with ages
$\sim 0.1-1 \Gyr$. We here investigate the effect this has on our
predictions for galaxy evolution by comparing with predictions using
two other SPS models, PEGASE-2 \citep{Fioc97,Fioc99} and FSPS (version
2.4) \citep{Conroy09}. (See Table~2 in \citet{Gonzalez-Perez14} for a
summary of the differences between these SPS models.) Previous work
\citep[e.g.][]{Gonzalez-Perez14} has shown that PEGASE-2 predicts very
similar broad-band SEDs to the \citet{Bruzual03} SPS models, while
FSPS uses an alternative calibration of the contribution of TP-AGB
stars. As expected, the main differences between these SPS models are
seen in the rest-frame $K$-band LF. Fig.~\ref{fig:lfKz_IMF_SPS} shows
that at $z=3$, the bright end of the $K$-band LF with the Maraston SPS
is around 0.4~mag brighter than with PEGASE, and around 0.7~mag
brighter than with FSPS, but this difference shrinks with decreasing
redshift due to the change in the typical ages of the stellar
populations in bright galaxies. { Similar effects have been found
  in previous studies \citep[e.g.]{Tonini09,Henriques11}.} For $z=0$,
the $b_J$-band LFs are esssentially identical for the 3 SPS models,
while the $K$-band LFs differ only slightly. There are also very small
differences in the far-IR and sub-mm counts and redshift distributions
between the 3 SPS models. We conclude that the choice of SPS model has
only a modest effect on our model predictions.


\section{Exploring physical predictions}
\label{sec:phys-predictions}

In this section, we explore the predictions of our new model for basic
physical properties such as stellar and gas masses and SFRs. We make
some limited comparisons with observational data, but defer detailed
comparisons to future papers.


\subsection{Global evolution of densities and metallicities}

\begin{figure}

\begin{center}
\includegraphics[width=7cm]{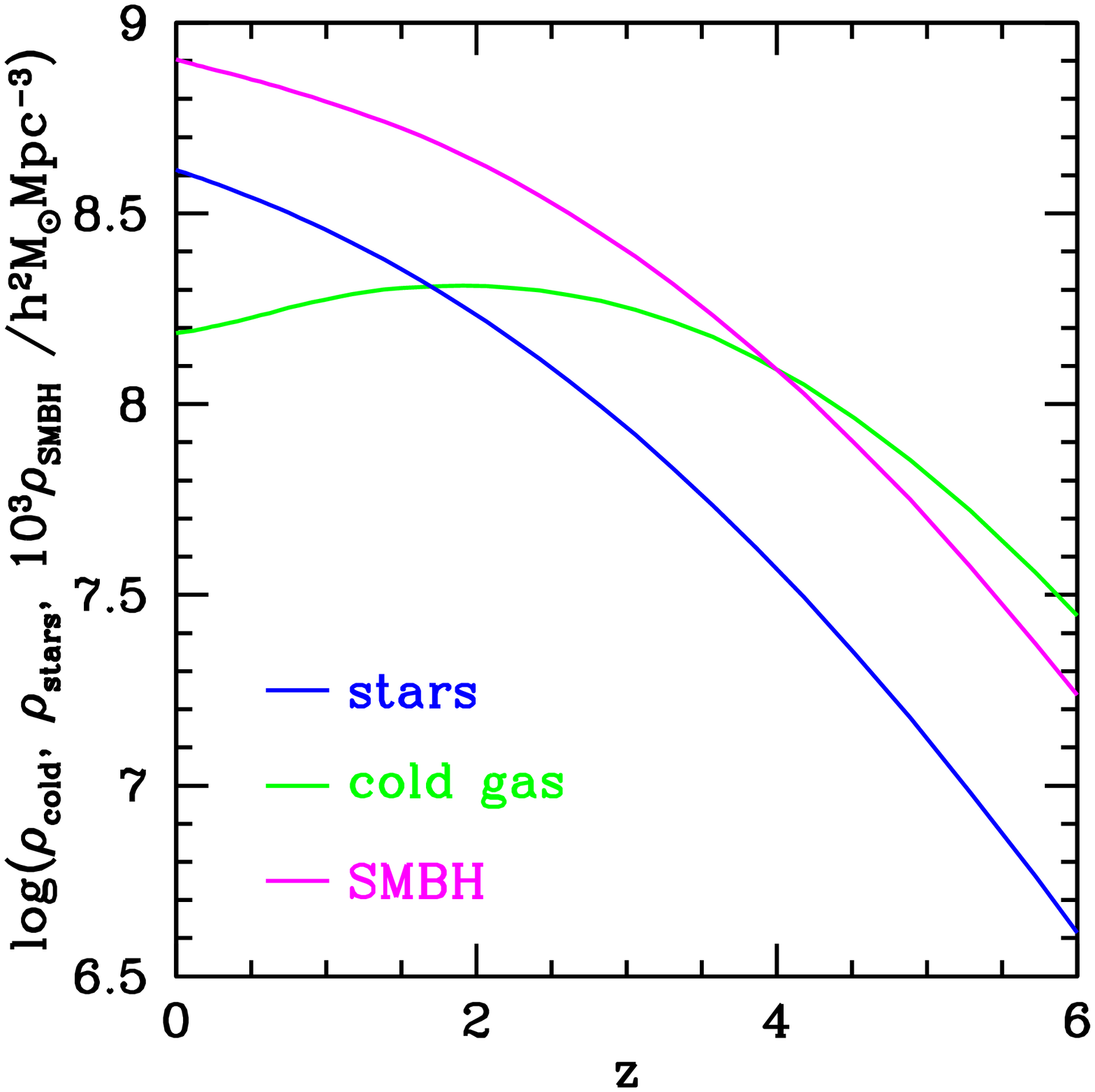}
\includegraphics[width=7cm]{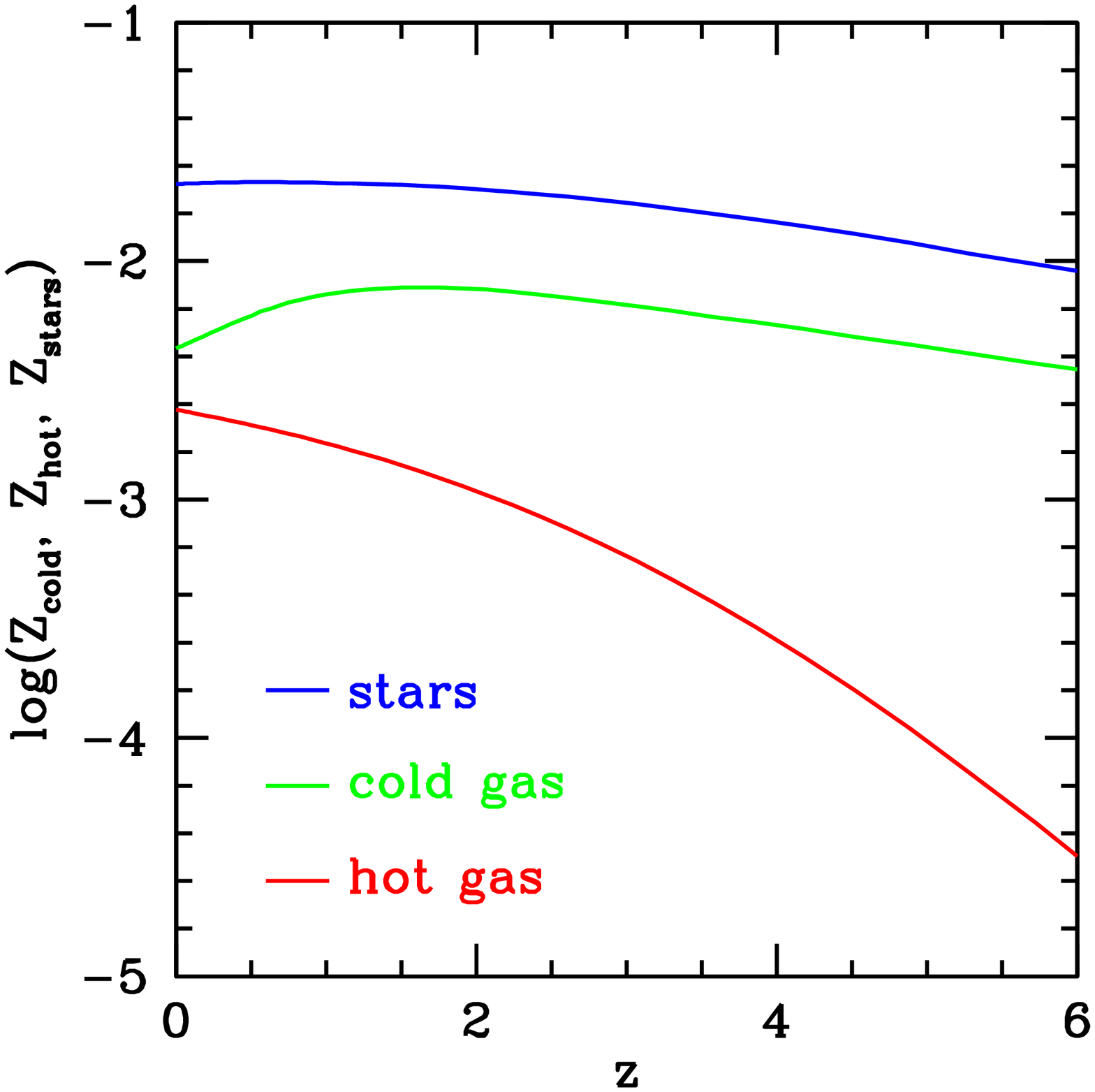}
\end{center}

\caption{Top: Evolution of mean comoving densities in stars (blue),
  cold gas (green) and SMBHs (magenta). The SMBH density has been
  mulitiplied by $10^3$ for plotting purposes. Bottom: Evolution of
  mean metallicities in stars (blue), cold gas (green) and hot gas
  (red).}

\label{fig:global_dens_met}
\end{figure}

In the top panel of Fig.~\ref{fig:global_dens_met} we show the evolution
with redshift of the global mean densities in cold gas, stars and
SMBHs. The cold gas density rises at early times, reaches a peak at
$z\sim 2$, and then declines by about 30\% up to the present day. The
stellar mass density increases monotonically, with 50\% of the current
mass in stars having formed since $z=1.7$. The SMBH density roughly
tracks the growth in stellar mass since $z=6$, as both grow by a
factor $\sim 10^2$, but with the SMBH mass growing somewhat more
slowly at $z \lsim 2$. 

The lower panel of Fig.~\ref{fig:global_dens_met} shows the evolution
of the mass-weighted mean metallicities of the hot gas, cold gas and
stars. Note that in this plot, the ``hot gas'' component for each halo
includes the ejected gas reservoir (with mass $\Mres$) as well as the
gas cooling in the halo (with mass $\Mhot$). The mean stellar
metallicity is seen to reach values not greatly different from Solar
at quite early times, and then to increase by only a factor of 2 from
$z=6$ up to the present day. This is due to galaxies enriching
themselves in metals through star formation. The cold gas metallicity
similarly evolves only modestly (by a factor $\sim 2$) over the same
redshift range $0<z<6$, increasing from $z=6$ to $z \sim 1.5$, and
then decreasing again to $z=0$. On the other hand, the mean
metallicity of the hot gas increases steadily, by almost a factor
$10^2$ over the range $0<z<6$, but is still a factor $\sim 10$ below
the mean stellar metallicity at the present day. This enrichment of
the hot gas is due to gas ejected from galaxies by SN feedback.

One concern with these plots is that at high redshifts they may be
affected by the mass resolution of the halo merger trees, which is set
by the N-body simulation used. However, we have checked that this
effect is small for this model by running \galform on halo merger
trees with different mass resolutions { (for example, at $z=6$, the
  stellar mass density changes by only $\sim 20\%$ when the halo mass
  resolution is increased by a factor $\sim 100$).} The basic reason
for this insensitivity to halo mass resolution is the very strong SN
feedback in low mass halos in this model, which prevents significant
star formation and accumulation of cold gas in the low mass halos
below the resolution limit of the N-body simulation.

\begin{figure*}

\begin{center}

\includegraphics[width=15cm]{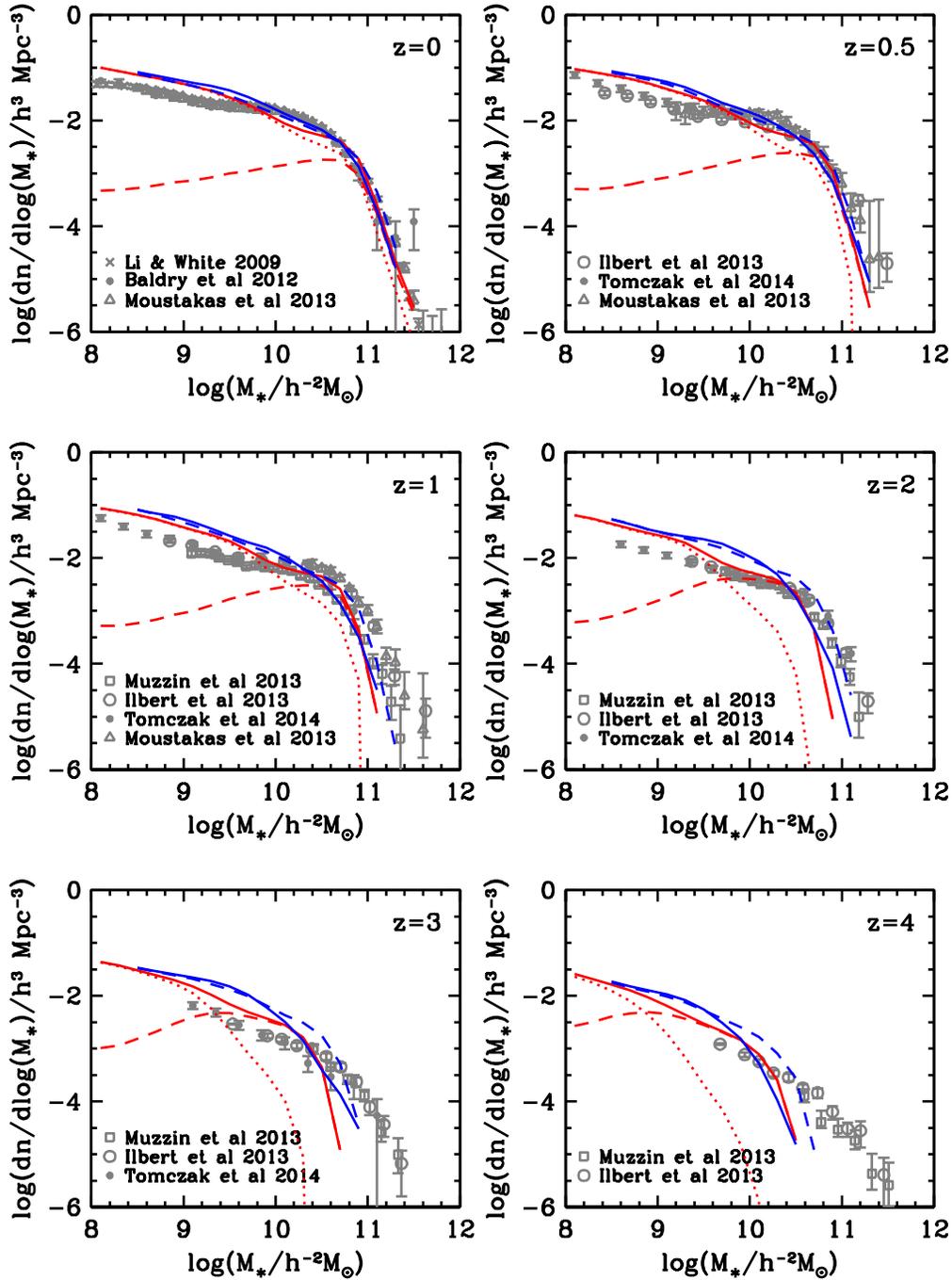}
\end{center}

\caption{Evolution of stellar mass function. The red lines show the
  model prediction for the stellar mass function using the true
  stellar masses, for the redshifts indicated in each panel. The solid
  red lines show the total stellar mass function, while the dotted and
  dashed red lines respectively show the contributions to this from
  galaxies in which most of the stellar mass present at that redshift
  was formed either quiescently or in starbursts respectively. The
  blue lines instead show the predicted stellar mass function when
  using stellar masses estimated from fitting model galaxy SEDs in a
  similar way to what is done for observations, but without allowing
  for photometric errors. The solid blue line shows the stellar mass
  function when effects of dust are included, and the dashed blue line
  when effects of dust are excluded. The grey points with error bars
  show observational data, which have all been corrected to a
  Kennicutt IMF, as described in the text. { The observational data
    are from \citet{Li09} (including the correction described by
    \citet{Guo10}); \citet{Baldry12}; \citet{Moustakas13};
    \citet{Ilbert13}; \citet{Tomczak14}.} Note that the redshifts for
  the observational SMFs are close to but do not exactly coincide with
  the redshifts for the model predictions.}

\label{fig:SMFz}
\end{figure*}

\subsection{Stellar mass function evolution}

Fig.~\ref{fig:SMFz} shows the model prediction for the evolution of the
stellar mass function (SMF), compared to observational estimates. In
this figure, the red lines show the predicted SMF using the true
stellar masses in the model. The dotted red line shows the
contribution to this from galaxies in which most of the stellar mass
at the redshift has been formed by quiescent star formation in disks,
while the red dashed line instead shows the contribution from galaxies
in which most of the mass has formed in starbursts. It can be seen
that the quiescent component dominates at low stellar masses at all
redshifts, while the starburst component dominates at high stellar
masses. The latter effect is marginal at $z=0$, but becomes strong for
$z \gsim 1$.

For comparison, Fig.~\ref{fig:SMFz} also shows recent observational
estimates of the SMF at redshifts $z=0-4$. For all of these, the
stellar masses have been estimated by fitting galaxy SEDs measured
from broad-band photometry with stellar population synthesis (SPS)
models. The results depend on the SPS model, on assumptions about
galaxy star formation histories and metallicity distributions, on the
model for dust attenuation, and on the assumed IMF. They also depend
on the set of photometric bands used, and are affected by errors in
the photometry. As analysed in detail in \citet{Mitchell13}, all of
these effects can cause the observationally inferred SMF to differ
from the true one. \citeauthor{Mitchell13} found that the effects of
dust attenuation and the assumed IMF had particularly large effects on
the inferred SMF. However, our theoretical model has different IMFs
for the quiescent and starburst modes of star formation, while the
SED-fitting method always assumes a single IMF, so it is impossible
for SED-fitting to recover the correct SMF from observational data,
even if that data is perfect. The observational SMFs shown in
Fig.~\ref{fig:SMFz} were originally derived with a variety of assumed
IMFs. We have applied approximate corrections to the observed stellar
masses to convert all of the observed SMFs to what would have been
measured if a \citet{Kennicutt83} IMF had been assumed in the SED
fitting. This is the IMF for the quiescent mode of star formation in
our model. The correction factors used are listed in
Table~\ref{table:mstar_conv_IMF}, and discussed further in
Appendix~\ref{sec:IMF_conv}. We emphasize that the observational SMFs
corrected in this way are not expected to agree with the model SMFs in
ranges of stellar mass and redshift for which stars formed in the
starburst mode make an important contribution, i.e. at higher stellar
masses and redshifts.

In order to understand better the effects on the comparison between
predicted and observed SMFs of inferring stellar masses from
observations using SED fitting, we have applied the SED-fitting
procedure to broad-band SEDs of model galaxies, as described in
\citep{Mitchell13}. For this exercise, we used the \citet{Bruzual03}
(BC03) SPS with a \citet{Chabrier03} IMF in the SED fitting, since
this is what was typically used in deriving the observed SMFs. To be
consistent with what is done in observational analyses, we also used
the \citet{Calzetti00} empirical dust attenuation law in the SED
fitting, even though the effect of dust attenuation on \galform model
galaxies is calculated using a physically-based radiative transfer
model (see \S\ref{sec:dust}). We use a fixed set of photometric bands
in the SED fitting, $B,V,R,i,z,J,H,K$ and the \SPITZER\ IRAC
$3.6,4.5,5.8,8 \mum$ bands, and assume zero photometric errors. Both
of these assumptions are optimistic compared to the actual
observational data, which use often a more restricted set of bands.

The results from estimating SMFs by applying SED fitting to model
galaxies are shown by blue lines in Fig.~\ref{fig:SMFz}, with the
solid blue lines showing the results when the effects of dust are
included as described, and the dashed blue lines showing the results
if dust attenuation is ignored (both in \galform and in the SED
fitting). We have applied the same correction factors to the stellar
masses estimated by SED fitting to convert them from the
\citet{Chabrier03} to the \citet{Kennicutt83} IMF as we apply to the
observational data. The differences between the SMFs based on true and
estimated stellar masses are seen to increase with redshift. There are
two main effects: (i) At higher redshifts, the contribution to the SMF
from stars formed in starbursts is larger. Such stars form with a
top-heavy IMF, and SED fitting assuming a Solar neighbourhood IMF
tends to overestimate the stellar masses of galaxies in which such
stars dominate. This causes the blue dashed line (showing the SMF from
SED fitting with no dust attenuation) to be offset to higher masses
than the solid red line (showing the SMF based on true stellar masses)
at high redshifts. (ii) On the other hand, the model predicts that
high-mass galaxies at high redshifts are typically heavily dust
extincted, and SED fitting tends to underestimate the stellar masses
in such cases. This partly offsets the effect of the top-heavy IMF, as
shown by the shift of the solid blue line (showing the SMF from SED
fitting including dust attenuation) relative to the dashed blue
line. These effects are discussed in more detail in
\citet{Mitchell13}.

Using the predicted SMF based on SED fitting is seen to bring the
model into closer agreement with observational data at higher masses
($\Mstar \gsim 10^{10} \hMsol$) and lower redshifts ($z \lsim
0.5$). This is mainly due to errors in stellar masses inferred from
SED fitting smoothing out the dip in the true SMF around $\Mstar \sim
10^{10} ~\hMsol$. However, at lower masses, the model predicts
somewhat too many galaxies compared to most observational estimates,
{ at both low and high redshifts. (A similar discrepancy has been
  found previously in other SA models \citep[e.g.][]{Fontanot09} and
  in gas-dynamical simulations \citep[e.g.][]{Weinmann12}.)} At high
redshifts ($z \gsim 3$), the model predicts too few high-mass galaxies
compared to recent observational estimates. However, this comparison
could be affected by photometric errors and also errors in photometric
redshifts, both of which are expected to become more significant at
higher redshifts, and which would be expected to broaden the
observationally inferred SMF. Neither of these effects was included
when calculating the blue curves in Fig.~\ref{fig:SMFz}.

\begin{figure}

\begin{center}
\includegraphics[width=8cm]{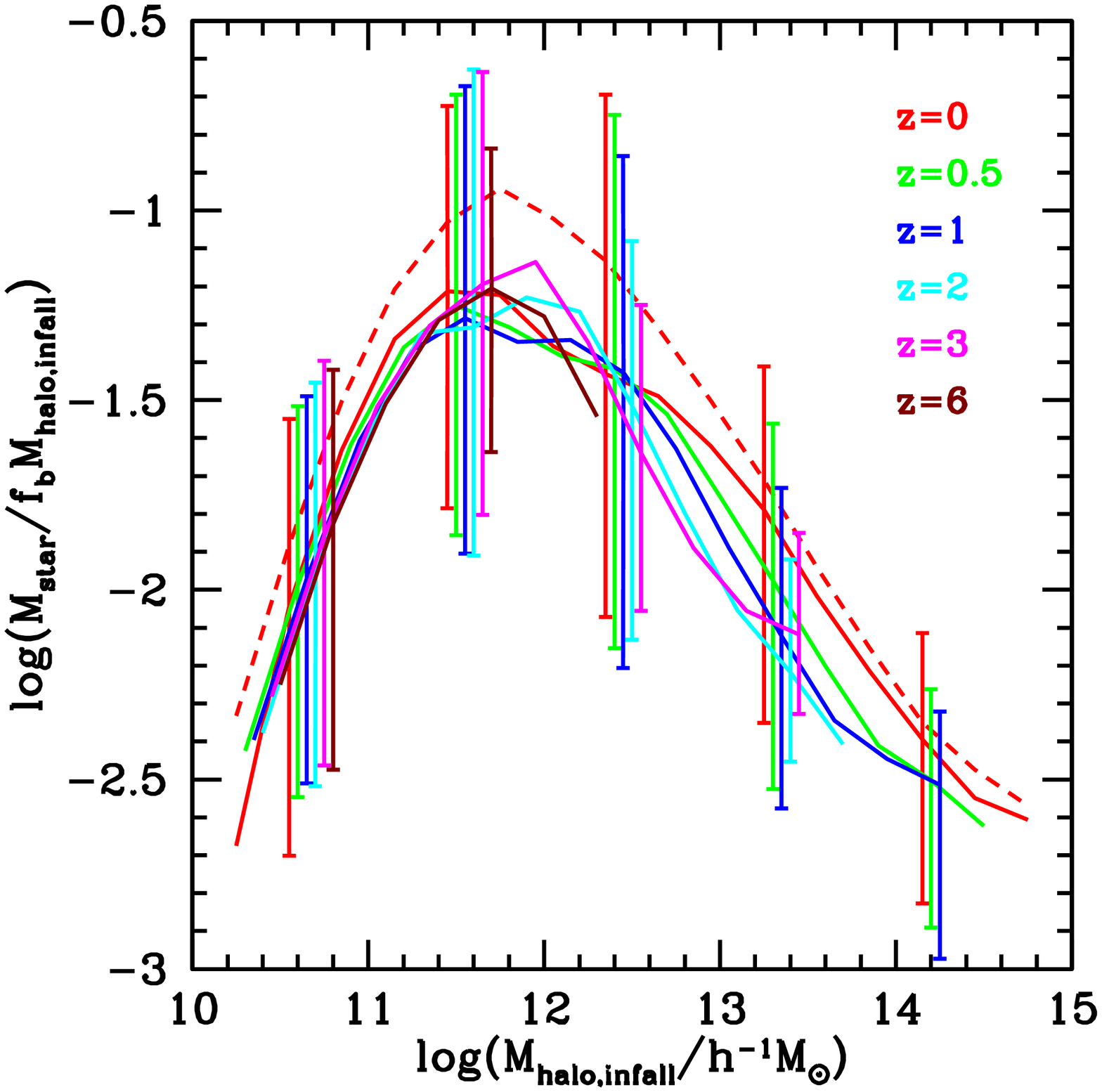}  
\end{center}

\caption{Fraction of baryons associated with a halo in the form of
  stars. The halo mass used in this plot is the host halo mass for
  central galaxies, and the subhalo mass at infall for satellite
  galaxies. The different colour lines are for different redshifts, as
  labelled in the key in each panel. $f_{\rm b}$ is the universal
  baryon fraction. Solid lines show the median, while the error bars
  show the 10-90\% range. The dashed line shows the mean for $z=0$.}

\label{fig:mstar_mchalo}
\end{figure}

In Fig.~\ref{fig:mstar_mchalo} we plot the fraction of baryons
associated with a halo in the form of stars (or baryon conversion
efficiency) as a function of halo mass. The halo mass
$M_{\rm halo,infall}$ used here is the current host halo mass for central
galaxies, and the host subhalo mass at infall into the main halo for
satellite galaxies. This is related to the SMF via the halo mass
function for main+satellite halos (expressed in terms of
$M_{\rm halo,infall}$). The baryon conversion efficiency is seen to peak
for halo masses around $10^{12}\hMsol$, which is a result of SN
feedback being more effective at low masses, and AGN feedback being
more effective at high masses. The conversion efficiency (and hence
also the $\Mstar$ vs $M_{\rm halo,infall}$ relation) is seen to evolve
little with redshift. However, the scatter at a given halo mass is
quite large, so the mean value of $\Mstar/M_{\rm halo,infall}$ as a
function of $M_{\rm halo,infall}$ is significantly different from the
median. The stellar mass vs. halo mass relation is often estimated
from observational samples using the abundance matching technique,
discussed further in \S\ref{sec:bathtub_vs_galform} { (see
  Fig.~\ref{fig:SMHM_bathtub})}. We note that if
the scatter in the stellar mass vs. halo mass relation is large, as
predicted here, then the relation inferred from abundance matching may
be significantly biased compared to the true relation (see
\citealt{Mitchell15} for more discussion of this point).

\subsection{SFR density evolution}

\begin{figure}

\begin{center}

\includegraphics[width=8cm]{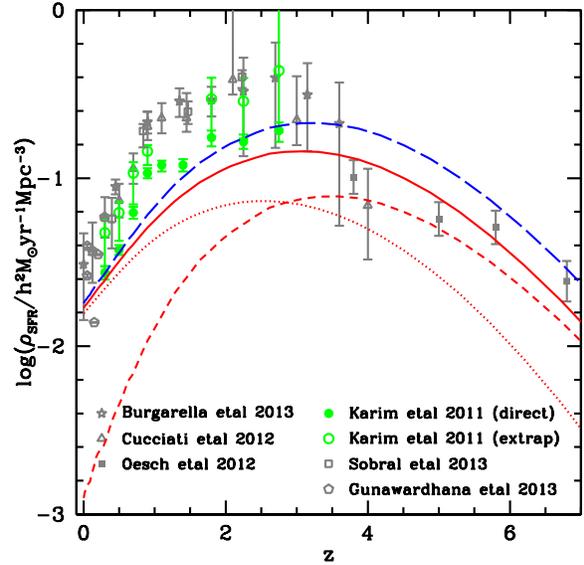}

\end{center}

\caption{Comoving SFR density (SFRD) as a function of redshift for the
  standard model. The solid red line shows the total SFRD, while the
  dotted and dashed red lines show the separate contributions to this
  from quiescent star formation and starbursts respectively. The long
  dashed blue line shows an estimate of the { ``apparent'' SFRD in
    the model} that would be inferred from observations of UV, IR or
  radio SFR tracers assuming a Kennicutt IMF. The black and green
  symbols show observational estimates, with the solid symbols showing
  direct estimates, and the open symbols based on extrapolating an
  analytic fit to the observed SFR distribution. The SFR tracers used
  are: UV \citep{Cucciati12,Oesch12b}; UV+IR \citep{Burgarella13};
  radio continuum \citep{Karim11}; $H\alpha$
  \citep{Sobral13,Gunawardhana13}. All of the observational data have
  been corrected to a Kennicutt IMF, as described in the text.}

\label{fig:SFRD}
\end{figure}

Another basic physical quantity in galaxy formation models is the
evolution of the comoving SFR density. The model predictions for this
are shown in Fig.~\ref{fig:SFRD}, together with a selection of recent
observational estimates. In this plot, the solid red line shows the
true total SFR density in the model, while the dotted and dashed red
lines show the contributions to this from quiescent SF (in disks) and
starbursts respectively. The quiescent SF mode dominates the SFR
density at $z \lsim 3$, while the starburst mode dominates at higher
redshifts. More than 90\% of the SF in the starburst mode at all
redshifts is triggered by disk instabilities, rather than by galaxy
mergers. 

The observational estimates of the SFR density plotted in
Fig.~\ref{fig:SFRD} are based on a variety of SFR tracers: far-UV
light \citep{Cucciati12,Oesch12b}, far-IR + far-UV luminosity
\citep{Burgarella13}, $H\alpha$ emission
\citep{Sobral13,Gunawardhana13}, and non-thermal radio emission
\citep{Karim11}. These tracers are all sensitive to high-mass star
formation only, although the stellar mass range depends on the tracer,
varying from $m \gsim 5 \Msol$ for far-UV and far-IR, to $m \gsim 8
\Msol$ for non-thermal radio, and $m \gsim 20 \Msol$ for $H\alpha$
(see \citealt{Kennicutt12} for a recent review). Since the SFRs in
these papers were derived assuming different IMFs, we convert all SFRs
to a \citet{Kennicutt83} IMF using the conversion factors in
Table~\ref{table:SFR_conv_IMF}, and discussed further in
Appendix~\ref{sec:IMF_conv}. Note that the conversion factors depend
on the SFR tracer. However, while this allows a fair comparison with
SFRs in model galaxies when quiescent star formation dominates, that
is not the case when the starburst mode dominates, due to the
top-heavy $x=1$ IMF adopted for the latter. We have therefore made an
approximate correction for this in Fig.~\ref{fig:SFRD} by plotting the
{ blue} dashed line, in which the starburst SFR is weighted by a
factor 1.9 before adding to the quiescent SFR. This factor is
calculated as the ratio of the fractions of mass in stars with
$m>5\Msol$ for the $x=1$ compared to the Kennicutt IMF. The { blue}
curve thus approximately represents the { ``apparent''} SFR density
that would be inferred for this model if SFRs were derived from SFR
tracers assuming a Kennicutt IMF. Note that applying a single
correction factor for the starburst IMF is only an approximation,
since in detail different SFR tracers are sensitive to different
ranges of stellar mass, so the correction factor should depend on the
SFR tracer used.

We see that although the predicted SFR density evolution has a
generally similar shape to the observed relations, the predicted SFR
density still lies below most of the observational estimates at $z<3$,
by a factor $\sim 2$, even after allowing for the top-heavy IMF in
starbursts { (although the discrepancy is smaller for $z \sim 0$).}
We note { however} that most of the observational estimates plotted
in Fig.~\ref{fig:SFRD} involve extrapolating the measured distribution
of luminosities or SFRs down to low values, to account for the
low-luminosity galaxies that are missed in the observational samples
(an exception is the data by \citet{Oesch12b} at $z\gsim 4$). The
effect of this extrapolation can be quite large. We show { an
  example of} this in Fig.~\ref{fig:SFRD} by plotting the data from
\citet{Karim11} with and without this extrapolation. {
  \citeauthor{Karim11} estimated mean SFRs in bins of stellar mass
  from a radio stacking analysis. They then obtained ``directly
  observed'' estimates of the SFR density, shown in
  Fig.~\ref{fig:SFRD} as filled green circles, by summing over the
  stellar mass bins for which they had measurements. They also
  obtained ``extrapolated'' estimates, shown by open green circles, by
  fitting Schechter functions to their measurements and then
  integrating down to much lower stellar masses than were directly
  observed (we plot their extrapolation for the case of no upper limit
  on the specific star formation rate).} For the \citeauthor{Karim11}
observational dataset, this extrapolation increases the estimated SFR
density by up to a factor $\sim 2$. { We also note that
  \citet{Madau14} find a discrepancy of a similar size when comparing
  direct and indirect observational estimates of the SFR density
  evolution - when they integrate over the observationally estimated
  SFR density evolution} to obtain the corresponding stellar mass
density evolution, the answer they obtain is higher than direct
observational estimates of the stellar mass density by around $\sim
0.2$~dex at all redshifts. This might imply that current observational
estimates of the SFR density are affected by some bias that makes them
too large.

\begin{figure*}

\begin{center}

\begin{minipage}{8cm}
\includegraphics[width=8cm]{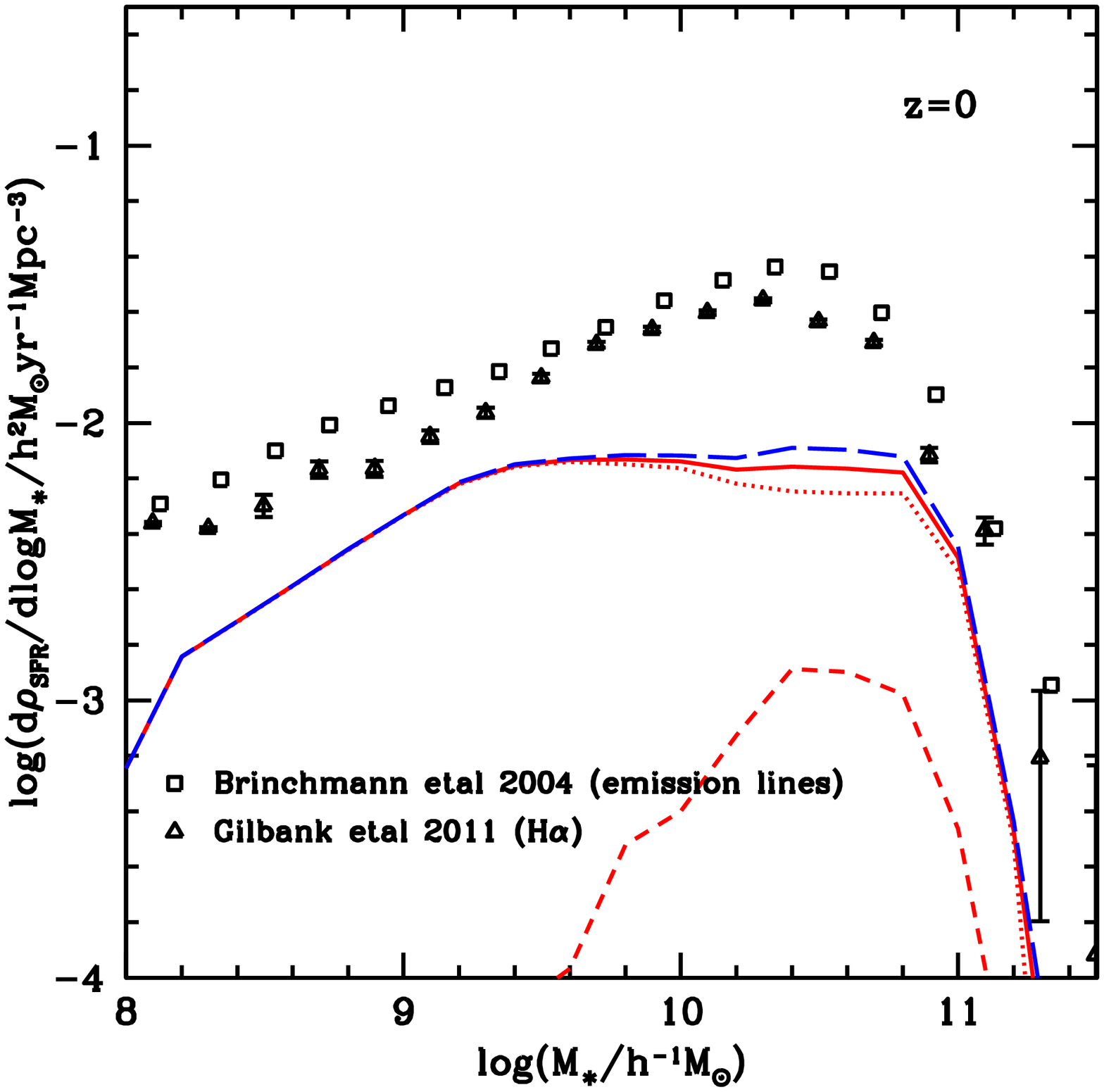}
\end{minipage}
\hspace{0.5cm}
\begin{minipage}{8cm}
\includegraphics[width=8cm]{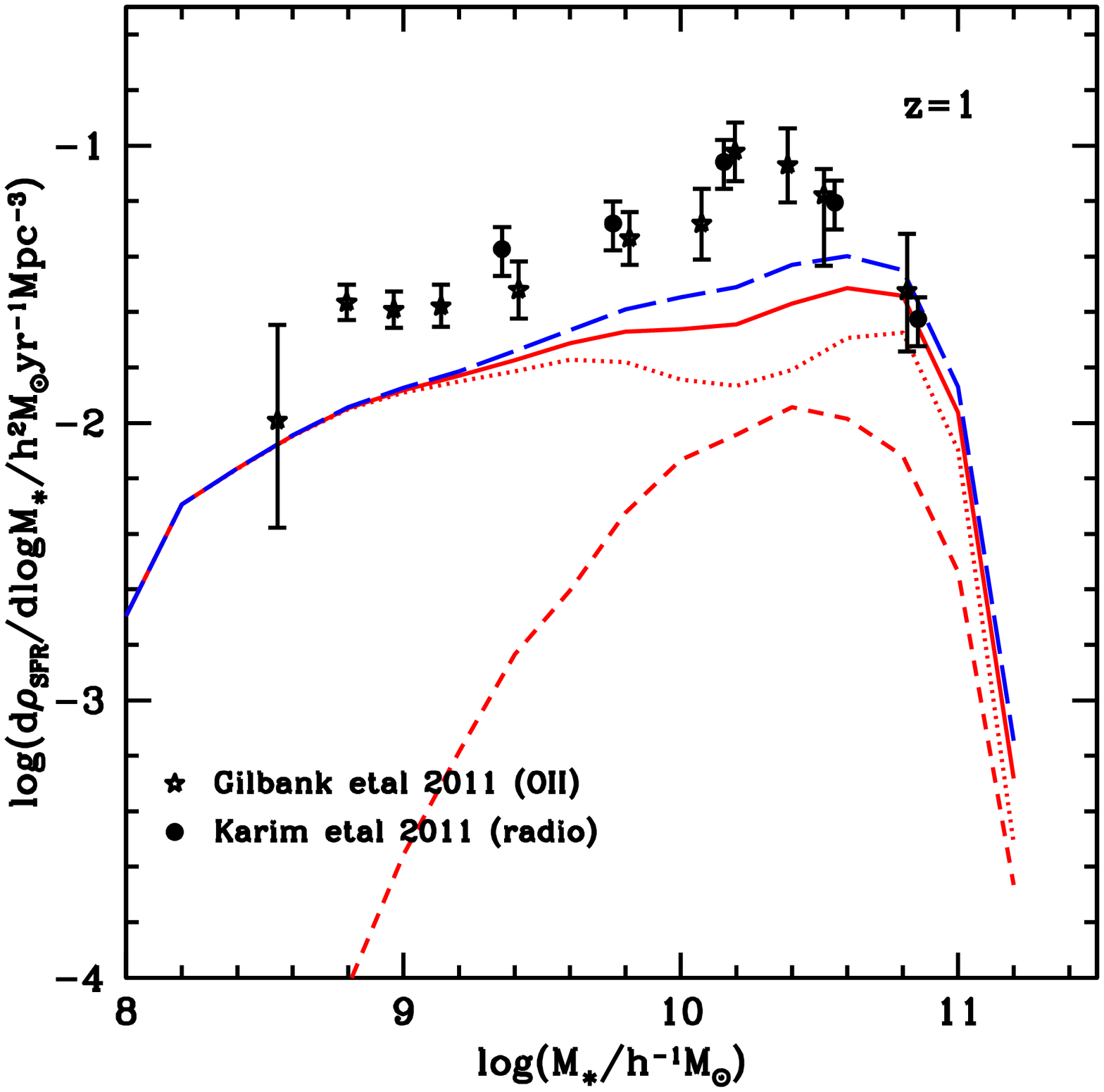}
\end{minipage}

\begin{minipage}{8cm}
\includegraphics[width=8cm]{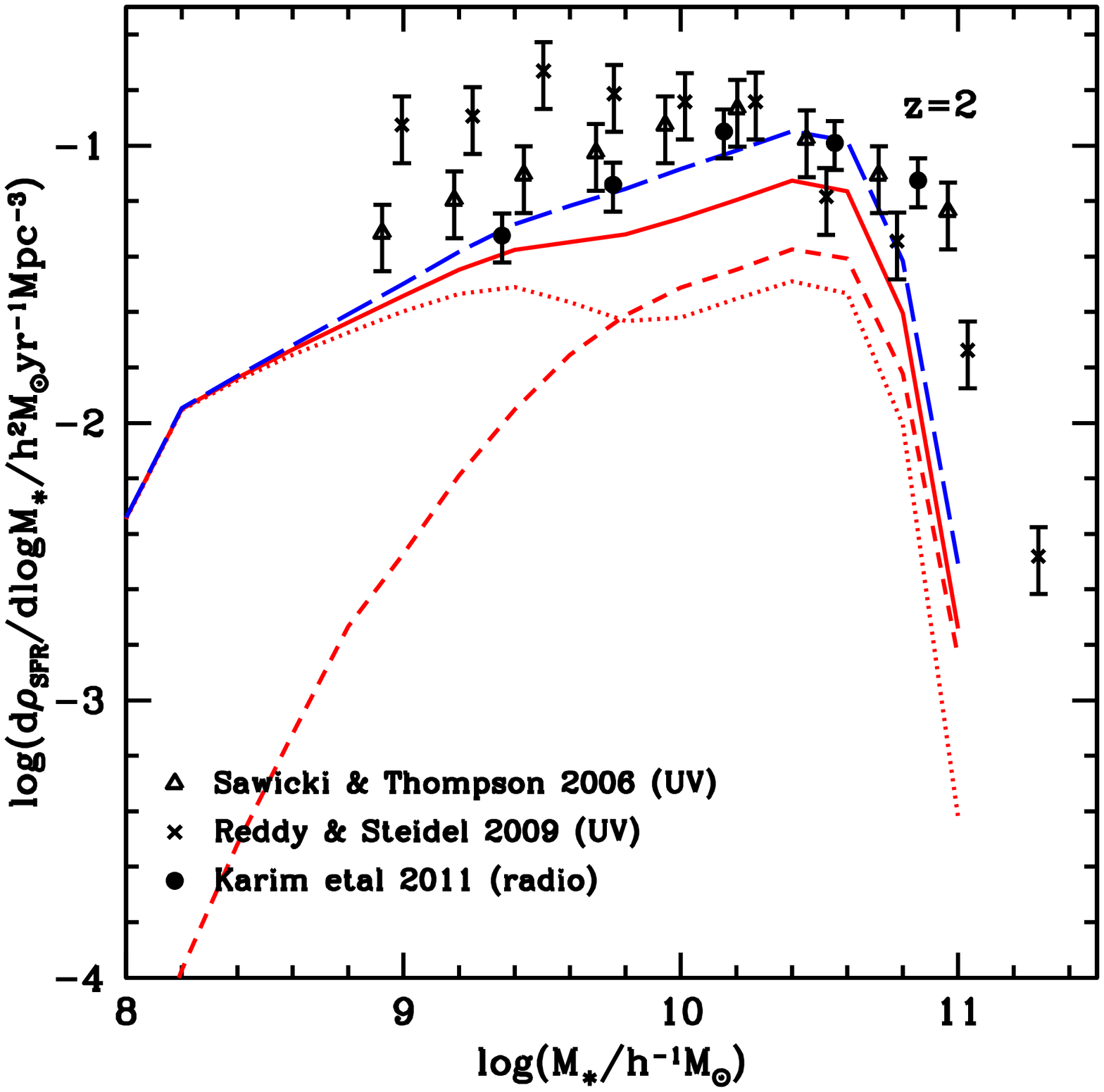}
\end{minipage}
\hspace{0.5cm}
\begin{minipage}{8cm}
\includegraphics[width=8cm]{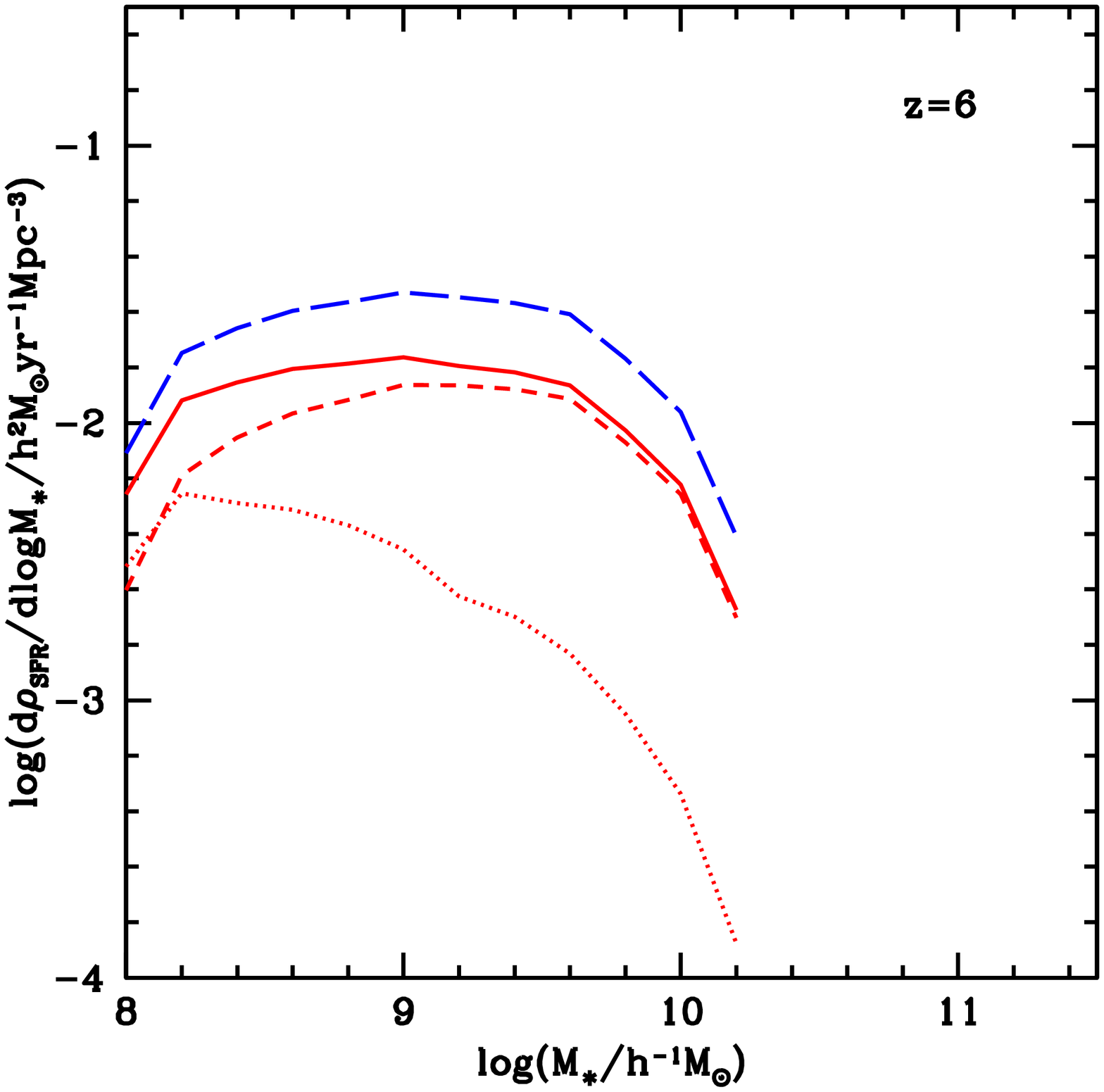}
\end{minipage}

\end{center}

\caption{Contribution to the SFR density as a function of stellar mass
  at redshifts $z=0,1,2,6$. The solid red lines show the total SFR
  density per logarithmic range in stellar mass, while the dotted and
  dashed red lines show the separate contributions to this from
  quiescent SF and starbursts respectively. The long dashed blue lines
  show an estimate of the total SFR density that would be obtained
  from observations of UV, IR or radio tracers assuming a Kennicutt
  IMF. { Observational data are from \citet{Brinchmann04};
    \citet{Gilbank11}; \citet{Karim11}; \citet{Sawicki06};
    \citet{Reddy09}.}  }

\label{fig:SFRD_Mstar}
\end{figure*}

We investigate this discrepancy in SFR densities further by plotting
in Fig.~\ref{fig:SFRD_Mstar} the differential distribution of SFR
density with stellar mass for various redshifts.  The different lines
have the same meaning as in Fig.~\ref{fig:SFRD}.  We include
observational data for $z=0,1,2$. The observational estimates have
been converted to a Kennicutt IMF as in Fig.~\ref{fig:SFRD}.  We see
that the predicted $d\rho_{\rm SFR}/d\log\Mstar$ vs $\Mstar$ relation
has a quite similar shape to that implied by observations, but is too
low by factors $\sim 2-3$ at $z\lsim 1$.  We also see that that the
starburst mode is predicted to make a larger contribution to the SFR
density at higher stellar masses and higher redshifts. { The
  differential SFR density $d\rho_{\rm SFR}/d\log\Mstar$ in the model
  is seen to peak at a roughly constant stellar mass $\Mstar \sim
  10^{10.5}~\hMsol$ for $z \sim 0 - 2$, similar to what is implied by the
  observations plotted here. At higher redshifts, the peak in
  $d\rho_{\rm SFR}/d\log\Mstar$ in the model gradually shifts to lower
  stellar masses. As in Fig.~\ref{fig:SFRD}, red curves show
  predictions for the true SFR density, while the long dashed blue
  lines show model predictions for the ``apparent'' total SFR density
  that would be inferred from observations of UV, IR or radio
  luminosities assuming a universal Kennicutt IMF.  The difference
  between the predictions for the ``true'' and ``apparent''
  $d\rho_{\rm SFR}/d\log\Mstar$ is seen to be small at low redshifts,
  but is appreciable at $z \gsim 2$. At $z \approx 2$, the correction
  from ``true'' to ``apparent'' SFRs is seen to bring the model into
  significantly closer agreement with the observational data. In
  particular, the model then agrees quite closely with the
  \citet{Karim11} data at this redshift.  The discrepancy seen in
  Fig.~\ref{fig:SFRD} between the model prediction for the
  ``apparent'' SFR density and the \citeauthor{Karim11} extrapolated
  value is seen to be caused mainly by the extrapolation to lower
  masses used by the latter.}

\begin{figure*}

\begin{center}

\begin{minipage}{8cm}
\includegraphics[width=8cm]{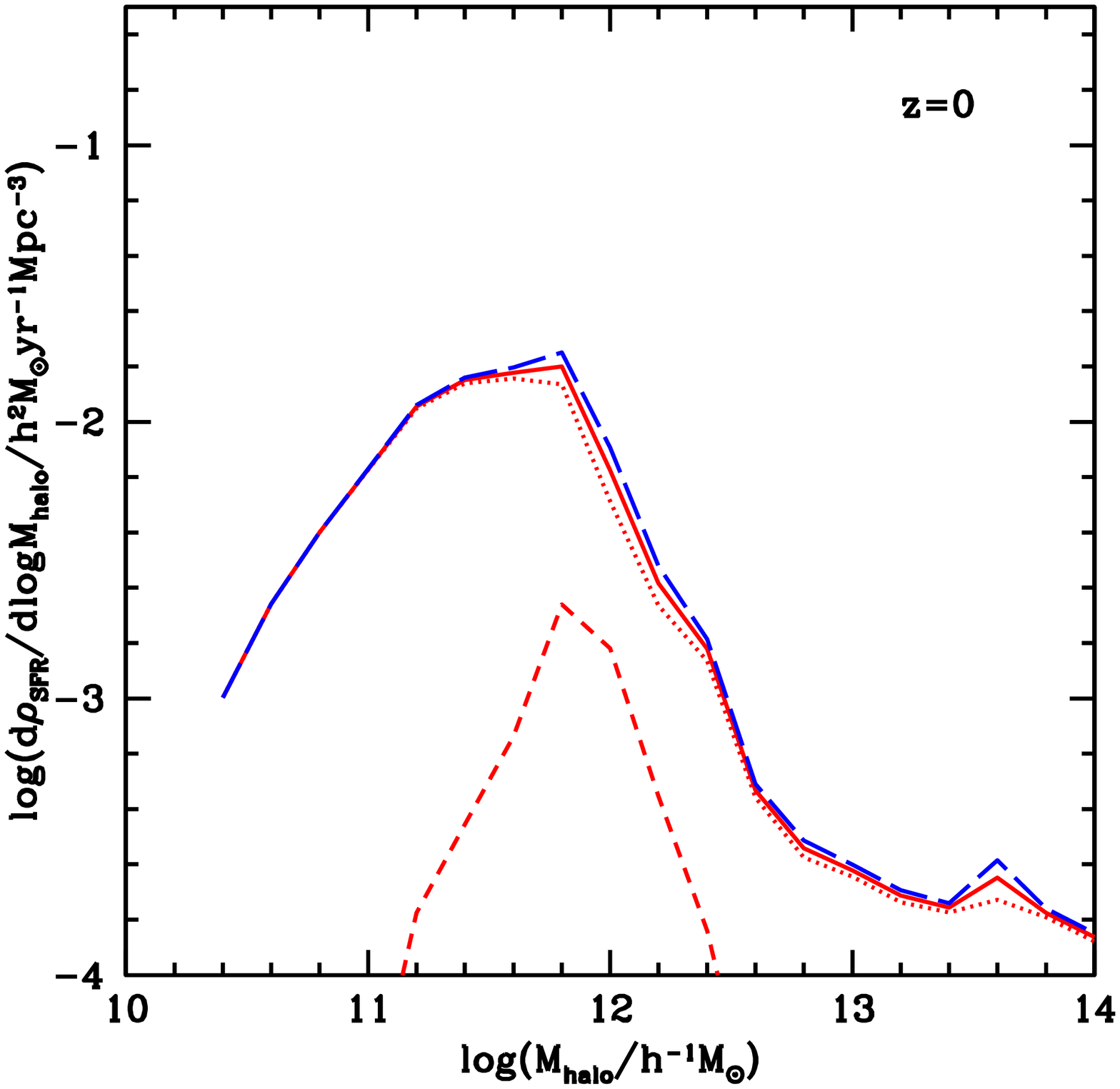}
\end{minipage}
\hspace{0.5cm}
\begin{minipage}{8cm}
\includegraphics[width=8cm]{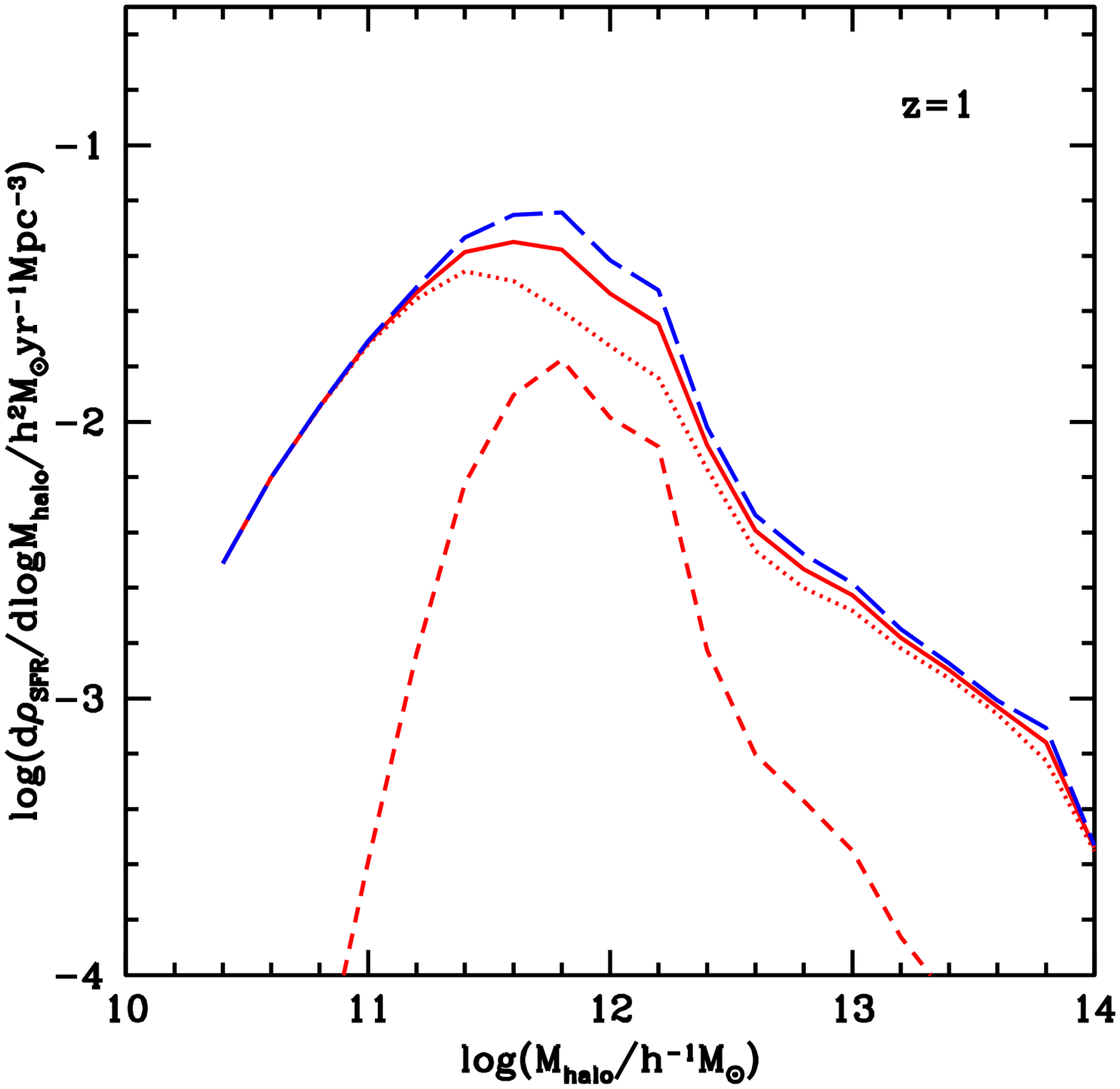}
\end{minipage}

\begin{minipage}{8cm}
\includegraphics[width=8cm]{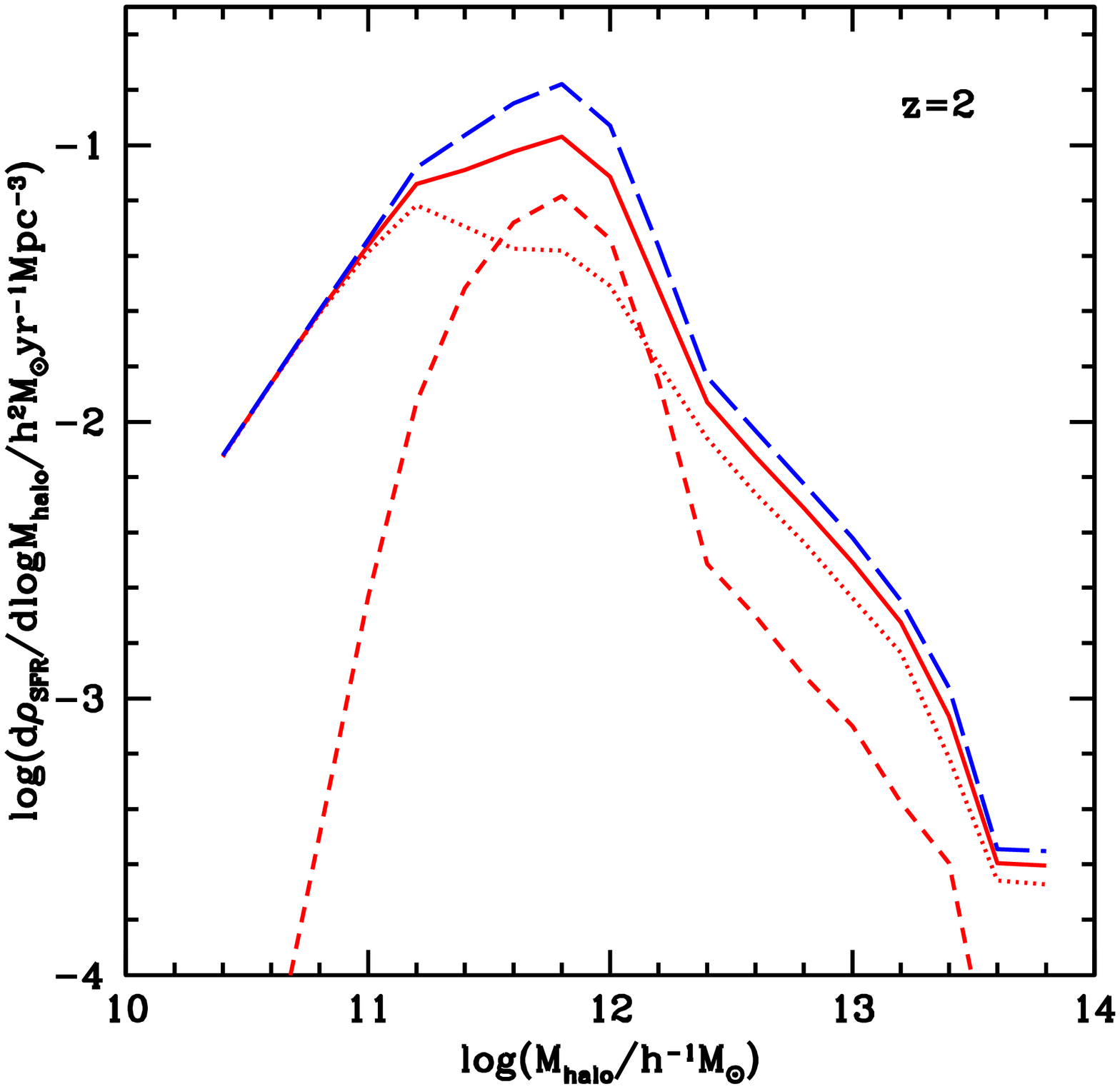}
\end{minipage}
\hspace{0.5cm}
\begin{minipage}{8cm}
\includegraphics[width=8cm]{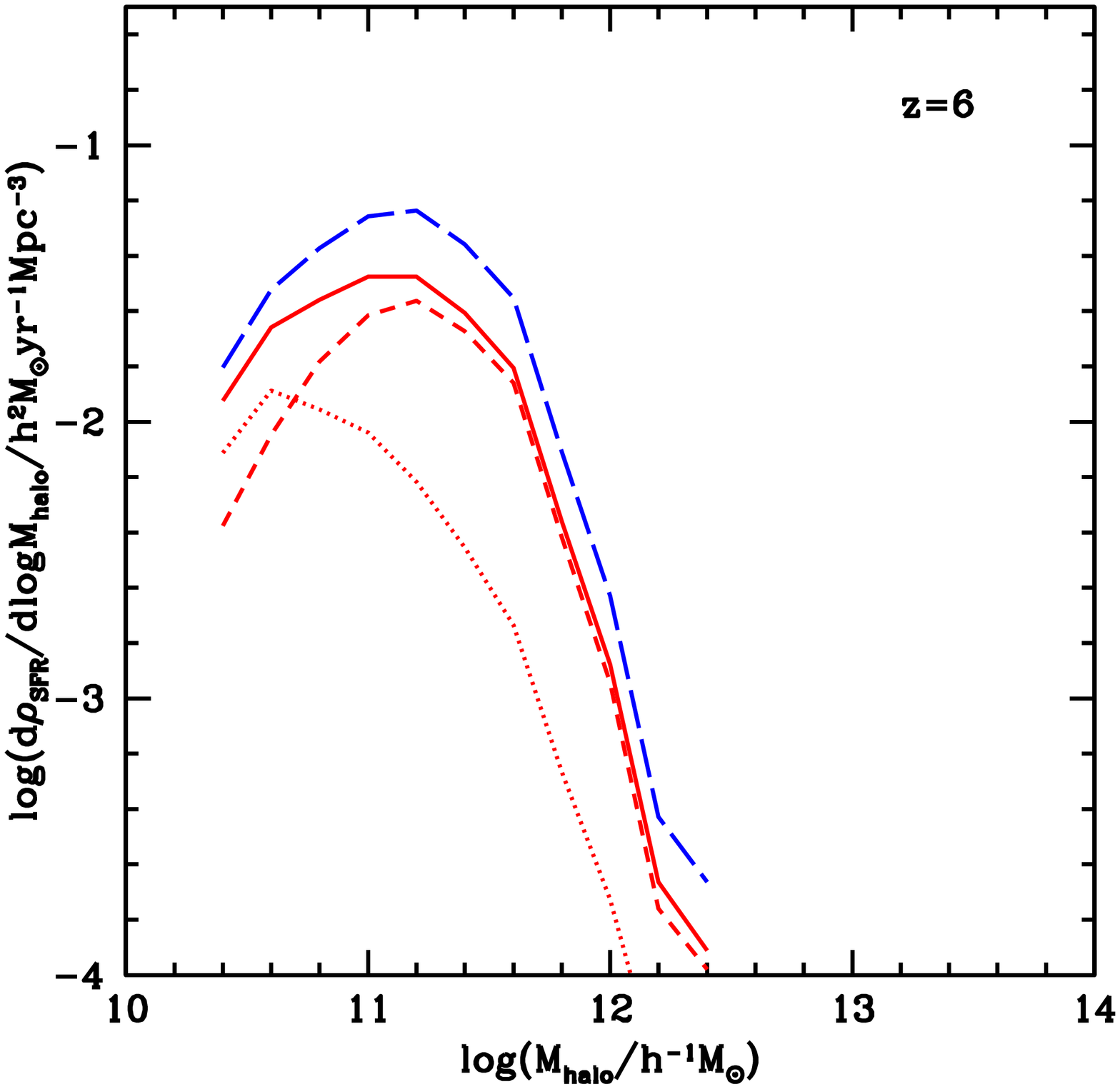}
\end{minipage}

\end{center}

\caption{Contribution to the SFR density as a function of halo mass at
  redshifts $z=0,1,2,6$. The solid lines show the total SFR density
  per logarithmic range in halo mass, while the dotted and dashed
  lines show the separate contributions to this from quiescent SF and
  starbursts respectively. { The long dashed blue lines
  show an estimate of the SFR density that would be obtained from
  observations of UV, IR or radio tracers assuming a Kennicutt
  IMF.}}

\label{fig:SFRD_Mhalo}
\end{figure*}

Finally, we show in Fig.~\ref{fig:SFRD_Mhalo} the distribution of SFR
density over halo mass for the same redshifts as in
Fig.~\ref{fig:SFRD_Mstar}. The differential SFR density
$d\rho_{\rm SFR}/d\log\Mhalo$ is seen to peak at $\Mhalo \sim
10^{12}\hMsol$ for $z \sim 0-2$, almost independent of redshift within
this range. The position of the peak reflects the effects of SN and
AGN feedback and also gas cooling, as discussed in relation to
Fig.~\ref{fig:mstar_mchalo}. At even higher redshifts, the peak shifts
to somewhat lower masses, reflecting the buildup of the halo mass
function. For all redshifts in the range plotted, the contribution to
the SFR density from very high mass halos ($\Mhalo \gsim (2-4) \times
10^{12} \hMsol$, depending on redshift) is dominated by satellite rather
than central galaxies, while at all lower halo masses (including the
peak), central galaxies dominate.

\subsection{Evolution of gas fractions and specific star formation
  rates}

\begin{figure}

\begin{center}
\includegraphics[width=8cm]{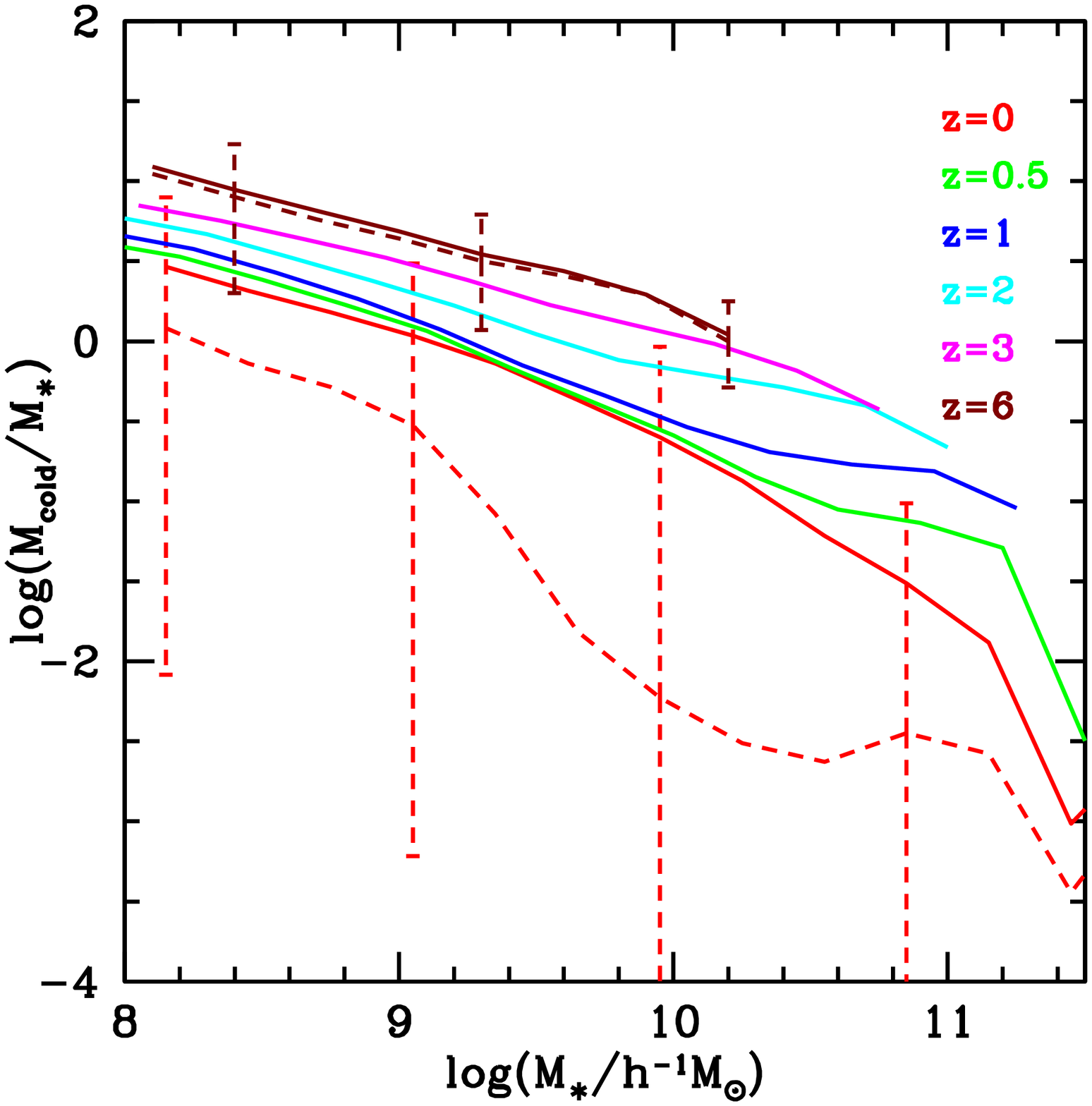}  
\includegraphics[width=8cm]{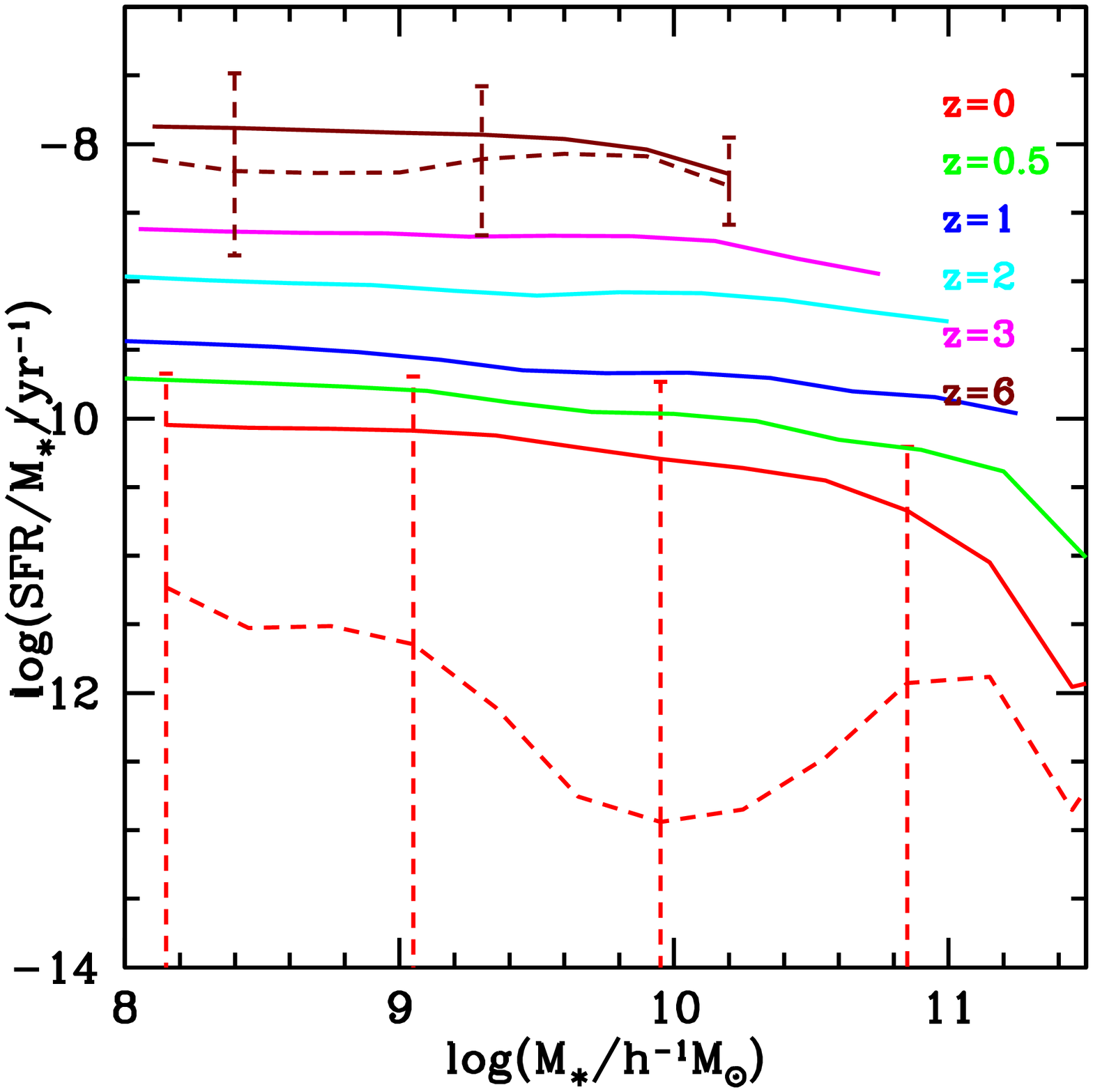}  
\end{center}

\caption{Top: mean ratio of cold gas mass to stellar mass as a
  function of stellar mass. Bottom: mean specific star formation rate
  (sSFR) as a function of stellar mass. The different colour lines are
  for different redshifts, as labelled in the key in each panel. The
  dashed lines show the median relations at $z=0$ and $z=6$, with the
  error bars showing the 10-90\% range.  }

\label{fig:fgas_SSFR_mstar}
\end{figure}

In Fig.~\ref{fig:fgas_SSFR_mstar} we show the evolution of the
average ratios of gas to stars (top panel) and specific star formation
rates (bottom panel) in galaxies. The top panel shows that the
gas-to-star ratio has a strong dependence on stellar mass, with
low-mass galaxies being more gas rich. The relation evolves with
redshift, with galaxies at a given stellar mass becoming more gas-rich
with increasing redshift. However, the amount of evolution depends
strongly on whether the mean or median gas-to-star ratio is used
(solid and dashed lines in Fig.~\ref{fig:fgas_SSFR_mstar}). The mean
and median relations are quite close at high redshift, when most
galaxies are star forming and contain significant cold gas, but the
median is much lower than the mean at low redshift, when a significant
fraction of galaxies have become passive, with low SFRs and gas
contents. The dependence of gas fractions on galaxy mass results
mostly from the assumed SFR law for disks (\S\ref{ssec:SFR}):
higher-mass galaxies typically have higher surface densities,
resulting in higher gas pressures, which causes a larger fraction of
their gas to be in the molecular star-forming phase. The efficiency of
converting cold gas into stars is therefore higher in high mass
galaxies. The increase in gas fractions with redshift results from the
fact that the adopted timescale for converting molecular gas into
stars is constant with redshift, while the time available (the age of
the universe) shrinks. This effect is only partly offset by the
increase with redshift of the fraction of gas in molecular form, again
driven by the increase in gas pressure.
These dependencies of gas contents on mass and redshift are analysed
in more detail in \citet{Lagos11b,Lagos14}.

The lower panel in Fig.~\ref{fig:fgas_SSFR_mstar} shows that the
specific star formation rate $sSFR = SFR/\Mstar$ has only a weak
dependence on stellar mass, but the average $sSFR$ increases strongly
with redshift. As for the gas-to-star ratios, the mean and median
sSFRs (solid and dashed lines) are similar at high redshift, but the
median is much lower at low redshift, due to a significant fraction of
galaxies being passive. The mean $sSFR$ increases by a factor $\sim
10^2$ between $z=0$ and $z=6$. The behaviour of the $sSFR$ vs $\Mstar$
relation is analysed in more detail in \citet{Mitchell14}. It is shown
there that the dependence of $sSFR$ on both stellar mass and redshift
in the model is controlled mainly by the timescale for dark matter
halos to grow by mergers and accretion, which depends weakly on halo
mass but strongly on redshift.

\subsection{Galaxy colours}

\begin{figure*}

\begin{center}
\begin{minipage}{5.4cm}
\includegraphics[width=5.4cm, clip=true, bb= 24 525 275 750]{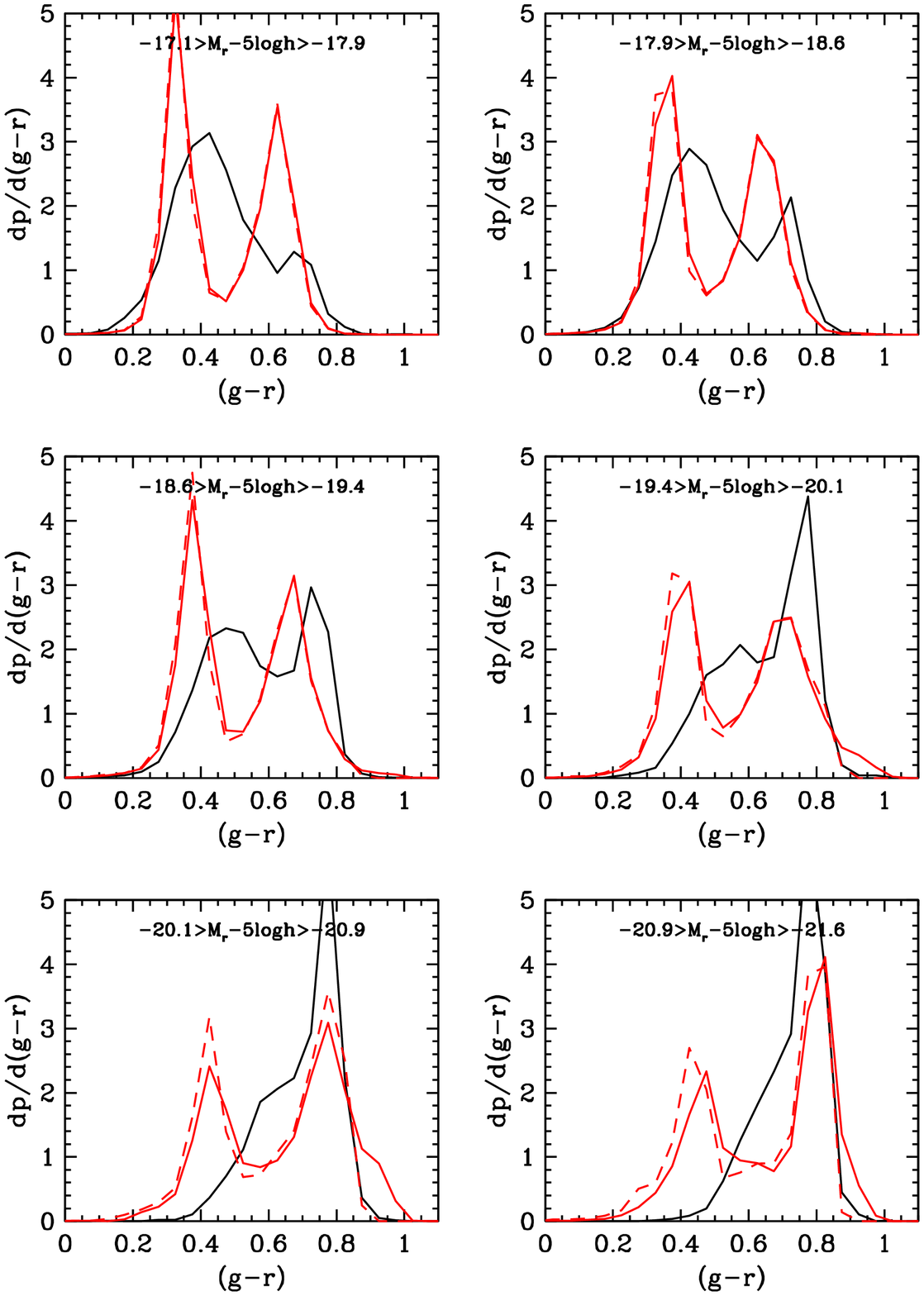}
\end{minipage}
\hspace{0.4cm}
\begin{minipage}{5.4cm}
\includegraphics[width=5.4cm, clip=true, bb= 24 288 275 514]{figs/gr_colours_default.ps}
\end{minipage}
\hspace{0.4cm}
\begin{minipage}{5.4cm}
\includegraphics[width=5.4cm, clip=true, bb= 24 51 275 277]{figs/gr_colours_default.ps}
\end{minipage}
\end{center}

\caption{Galaxy $g-r$ colours at $z=0$. The normalized distribution of
  $g-r$ colours are plotted for three different ranges of absolute
  $r$-band magnitude, as indicated. The red lines show model
  predictions (solid including dust extinction, and dashed without
  dust extinction). The black lines show the colour distributions
  measured from SDSS by \citet{Gonzalez09}.}

\label{fig:gr_colours_default}
\end{figure*}

A further interesting prediction from the models is for galaxy
colours. In Fig.~\ref{fig:gr_colours_default} we show the predicted
$g-r$ colour distributions at $z=0$ for galaxies selected in different
ranges of absolute $r$-band magnitude, compared to observational data
from SDSS. The rest-frame $g-r$ colour depends mainly on the star
formation history of a galaxy, as well as its metallicity and dust
extinction. Comparing the solid and dashed red lines in
Fig.~\ref{fig:gr_colours_default}, we see that the effects of dust
extinction on $g-r$ colours are predicted to be quite small in the
models, except at the highest luminosities. The models show a clear
bimodal colour distribution, corresponding to ``star forming'' and
``passive'' galaxies. The observations show a similar bimodality, but
the bimodality in the models is stronger. In particular, the models
show a stronger blue peak at high luminosities, and a stronger red
peak at low luminosities, when compared to observations.
Reproducing the detailed colour distributions of galaxies (as opposed
to their qualitative form) has been a longstanding problem for
semi-analytical models
{ \citep[e.g.][]{Gonzalez09,Font08,Guo11,Gonzalez-Perez14,Henriques15}.}

\section{Comparison of a simplistic galaxy formation model with physical models}
\label{sec:simplistic}



The model presented in this paper attempts to make as complete a
representation as possible of the interplay between the main processes
thought to be important in shaping the formation and evolution of
galaxies. These processes are dealt with under certain approximations
and assumptions, as set out in Section~\ref{sec:model}.  We have
demonstrated how this model can make an extremely wide range of
predictions for observables. Furthermore, we have shown how the model
responds to perturbations to the parameters which are built into the
descriptions of various phenomena.

Recently there has been some interest in the literature in simplified
models of galaxy formation \citep{Bouche10,Dave12,Dekel13,Dekel14}.
These ``toy'' models focus on solving a small number of the equations
presented in Section~\ref{sec:model} in isolation, focusing on the
balance between inflows and outflows of gas in a halo. These
calculations have the attraction of simplicity but, as we have argued
above, galaxy formation is a complex phenomenon which requires many
processes to be modelled simultaneously. In this section we outline
one of these simple calculations and compare it with the more complete
calculation which is the focus of this paper.

For this exercise, we focus on the ``reservoir'' or ``bathtub'' model
that was introduced by \cite{Bouche10}. This model follows the growth
of a single galaxy inside a dark matter halo. The galaxy consists of
baryons in the form of cold gas and stars. The cold gas component is
modelled as a reservoir with sources and sinks. The ``source'' of the
cold gas is the accretion of new material as mass is added onto the
host dark matter halo.  The ``sinks'' of cold gas are star formation
and the ejection of gas through supernova-driven winds.

\subsection{The bathtub model equation}

The basic equation of the model is a differential equation expressing
the conservation of mass outlined above:

\begin{equation}
\dot{M}_{\rm gas} = \dot{M}_{\rm gas, in} - 
\left( 1 - R \right) \psi - \dot{M}_{\rm gas, out}, 
\end{equation}
where $\dot{M}_{\rm gas}$ is the overall rate of change of the cold
gas mass in the galaxy, $\dot{M}_{\rm gas, in}$ is the rate at which
cold gas is accreted onto the galaxy, $\psi$ is the star formation
rate, $R$ is the fraction of the material turned into stars that is
recycled into the ISM, and $\dot{M}_{\rm gas, out}$ is the gas outflow
rate from the galaxy. The recycled fraction is assumed to be fixed
and is determined by the choice of stellar initial mass function
(IMF). Note the choice of IMF does not have any other influence over
the model predictions as the luminosity of the galaxy is not computed.
The gas outflow rate is assumed to be proportional to the star
formation rate:
\begin{equation} 
\dot{M}_{\rm gas, out} = a \psi, 
\end{equation} 
where $a$ is a model parameter. With the outflow rate written in this
way, the bathtub equation simplifies to
\begin{equation} 
\dot{M}_{\rm gas} = \dot{M}_{\rm gas, in} - \alpha \psi, 
\label{eq:bathtub_gas}
\end{equation} 
where $\alpha = \left( 1 - R \right) + a$ and the second term on the
right-hand side of eqn.(\ref{eq:bathtub_gas}) gives the net mass loss
rate due to star formation and outflows.

The gas accretion rate is obtained from the rate at which the host
dark matter halo grows, modulated by efficiency
factors. \cite{Bouche10} quote the halo growth rate as
\begin{equation}
\dot{M}_{\rm halo} = 39.5 \left( 
\frac{\Mhalo}{10^{12}M_{\odot}} \right)^{1.1} 
\left(1 + z \right)^{2.2} M_{\odot} {\rm yr}^{-1}.   
\label{eq:halo}
\end{equation}
where $\Mhalo$ is the halo mass.  The origin of this expression is
\cite{Genel08}, who give a fit to the mean mass accretion rate
measured for haloes in the Millennium simulation of \cite{Springel05};
however, the numerical coefficient given by \cite{Genel08} is $35$
rather than $39.5$, a reduction of 12\%.

The baryon accretion rate onto the galaxy is taken to be a fraction of
the dark matter accretion rate onto the halo
\begin{equation}
\dot{M}_{\rm gas, in} = \epsilon_{\rm in} f_{\rm b} 
\dot{M}_{\rm halo}, 
\end{equation}
where $\epsilon_{\rm in}$ is an ``efficiency'' factor for the
accretion and $f_{\rm b}$ is the universal baryon fraction. The gas
accretion efficiency factor is defined by:
\begin{eqnarray}
 \epsilon_{\rm in} & = &  0 \quad \quad \quad \textrm{ if } \Mhalo < M_{\rm min} \\ \nonumber
                   & = & f(z) \epsilon_0 \quad \textrm{ if } \Mhalo >
                   M_{\rm min}  \textrm{ and } \Mhalo < M_{\rm max} \\ \nonumber
                   & = &  0 \quad \quad \quad \textrm{ if } \Mhalo > M_{\rm max}, 
\end{eqnarray}
where the range of halo masses which are allowed to accrete baryons is
set by the model parameters $M_{\rm min}$ and $M_{\rm max}$.  The
parameter $\epsilon_{0}$ is set to 0.7. The efficiency factor is
assumed to be redshift dependent for redshifts below $z=2$, with the
redshift dependence given by the factor $f$.  $f$ is assumed to vary
linearly in time between values of $f(z=2.2)=1$ (note the boundary
condition is specified at $z=2.2$ and not $z=2$) and $f(z=0)=0.5$.

The star formation rate is modelled as 
$$
\psi = \epsilon_{\rm sfr} M_{\rm gas} / t_{\rm dyn},
$$ where $\epsilon_{\rm sfr}$ is a model parameter that is set to
$\epsilon_{\rm sfr} = 0.02$. The dynamical time, $t_{\rm dyn}$, is
parametrized as
$$
t_{\rm dyn} = 2 \times 10^{7} \left( \frac{1+z}{3.2} 
\right)^{-1.5} {\rm yr}. 
$$ 

\subsection{How many parameters?} 

After setting the background cosmology, the bathtub model requires the
following additional parameters to be specified:
$$
\left( M_{\rm min}, M_{\rm max}, \epsilon_{0}, 
f(z), \epsilon_{\rm sfr}, a, R \right).  
$$ 
(where in principle $R$ depends on the IMF). At first sight this list
contains 7 parameters, but in fact really requires more than this
because of the form adopted for the redshift modulation of the
accretion efficiency factor, $f(z)$. This parameter requires four
numbers to specify its form; the redshift below which $f$ is assumed
to vary ($z=2$), the boundary conditions $f(z=2.2)$, $f(z=0)$, and the
rate of change of $f(z)$ between these boundary conditions (which is
assumed to be a linear variation in time; hence, given the boundary
conditions, this translates into an additional number, a gradient).
This gives 10 parameters. In practice, $\epsilon_{\rm sfr}$,
$\epsilon_{0}$, $R$ and $f(z)$ are not varied in the models presented
in \cite{Bouche10}.

\subsection{How many outputs?} 

The bathtub model assumes that there is one galaxy per halo and tracks
the stellar mass ($\Mstar$) and cold gas mass ($M_{\rm gas}$) of the
galaxy, along with the mass of the host halo ($\Mhalo$). The model
also gives the star formation rate in the galaxy ($\psi$).  The
bathtub model therefore produces 4 outputs ($\Mstar,M_{\rm gas},
\psi,\Mhalo$) as a function of time.

\subsection{How good are the assumptions in the bathtub model?}

We now review the key assumptions behind the reservoir 
model. 

\noindent {\it 1. Halo growth rate.} Eqn.(\ref{eq:halo}) says that all
haloes of a given mass accrete mass at precisely the same rate.  The
motivation for universal mass accretion histories comes from the
extended Press-Schechter theory \citep{LC93, vandenBosch02}.  Halo
mass accretion histories extracted from N-body simulations show
considerable scatter \citep{Fakhouri10}, suggesting more variety than
is implied by eqn.(\ref{eq:halo}).  \citet{McBride09} found that a two
parameter fit could describe halo formation histories measured from
the Millennium Simulation of \citet{Springel05}, provided that the
halos are divided into four different classes for which different
values of the two parameters are adopted.

{\it 2. Baryon accretion rate} The accretion of baryonic material 
is assumed to be proportional to the rate at which mass is added 
to the dark matter halo, with the modulation encoded in an efficiency 
factor. The accreted baryonic material is assumed to be in the form 
of cold gas, as it is made available immediately to be turned into 
stars. In practice, no cold gas is accreted for haloes less massive 
than $M_{\rm min}$. Also, no cold gas is added to the galaxy on the 
addition of mass to haloes more massive than $M_{\rm max}$. 
The latter cut-off is justified as the upper halo mass for 
which ``cold accretion'' operates. 

{\it 3. Outflows} Cold baryons leave the galaxy ``reservoir'' in the
form of outflows, which are described as due to supernovae, or as gas
that was involved in star formation. Note that this assumes that the
gas ejected by outflows leaves the halo forever and is not returned to
the cold gas component. This assumption seems physically rather
unreasonable, { particularly for more massive halos.}

\subsection{How does the bathtub model compare with \galform?}
\label{sec:bathtub_vs_galform}

\begin{figure}
\begin{center}
\includegraphics[width=8.2cm, bb= 60 200 570 700]{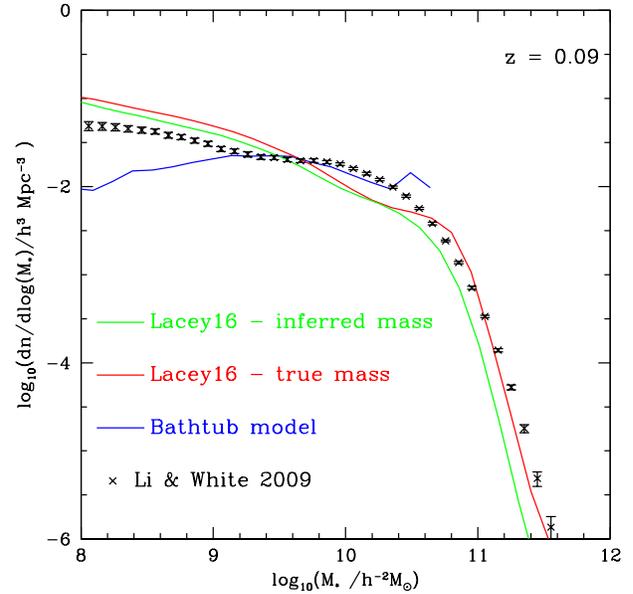}
\end{center}
\caption{ The local stellar mass function. The points show the stellar
  mass function inferred from the SDSS by \citet{Li09} (as updated by
  \citealt{Guo10}. The lines show the model predictions at the median
  redshift of SDSS, $z=0.09$.  The red line shows the mass function
  for the stellar masses predicted directly by \galform, while the
  green line shows the mass function inferred from SED fitting to the
  SDSS photometry of the model galaxies.  The blue line shows the
  stellar mass function predicted by the bathtub model. Both the
  observed SMF and the inferred one have been corrected to a
  \citet{Kennicutt83} IMF.}
\label{fig:SMF_bathtub}
\end{figure}

\begin{figure}
\begin{center}
\includegraphics[width=8.2cm, bb= 60 200 570 700]{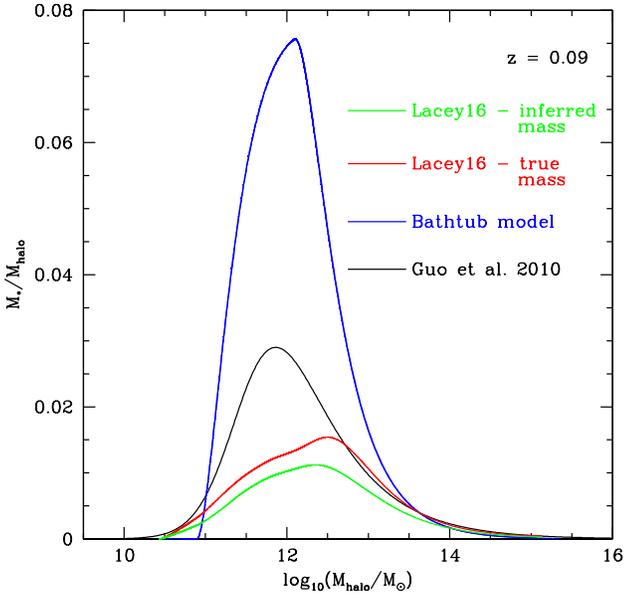}
\end{center}
\caption{The ratio of stellar mass to halo mass. The black line shows
  an estimate based on abundance matching, using the observationally
  inferred stellar mass function and the subhalo mass function
  obtained from the Millennium simulations by \citet{Guo10}. The red
  and green lines show the predicted relation for the fiducial model
  in this paper, as inferred by abundance matching, using the directly
  predicted stellar masses (red) and the stellar masses estimated from
  SED fitting to SDSS photometry (green). The blue curve shows the
  stellar mass - halo mass relation predicted by the bathtub model.  }
\label{fig:SMHM_bathtub}
\end{figure}

The parameters of the bathtub model were fixed to reproduce the
inferred specific star formation rates of galaxies at $z=2$
\citep{Bouche10}. The model also makes a limited number of other
predictions which we can compare to those of \galform.

Fig.~\ref{fig:SMF_bathtub} compares the present day stellar mass
functions in the bathtub and \galform models. For reference, we also
plot the observationally inferred mass function from \cite{Li09}. Two
predictions are shown for \galform: one using the stellar masses
output directly in the model (red line) and the other ({ green}
line) which shows the masses inferred by fitting SEDs to the SDSS
photometry of the model galaxies. These predictions are within a
factor of two of the observational estimate (which was based on SED
fitting).  { The bathtub model predictions agree with the stellar
  mass function inferred from observations over the mass range
  $10^{9.1} < \Mstar < 10^{10.3} h^{-2}M_{\odot}$.  However, beyond
  this mass range the predictions from the bathtub model vary little
  with stellar mass and disagree with the measurements.}  The
disagreement between the bathtub predictions and the observational
estimate at low masses could be blamed on the neglect of satellite
galaxies in the bathtub, with all of the stellar mass within a dark
matter halo being concentrated in one object.  However, this might
lead one to expect an excess of massive galaxies, whereas in fact none
are predicted beyond $\Mstar = 10^{10.8} h^{-2}M_{\odot}$. This is due
to the way in which cooling is suppressed by hand in halos above a
certain mass and to the neglect of galaxy mergers. Hence the bathtub
model does not predict a smooth break in the stellar mass function,
but instead a sharp cutoff.

Next we compare the stellar mass of galaxies to their host dark matter
halo masses. This has implications for the clustering of galaxies as a
function of their stellar mass. The black curve in
Fig.~\ref{fig:SMHM_bathtub} shows a prediction made by combining the
observationally inferred stellar mass function with the subhalo mass
function prediction by the Millennium simulations by
\cite{Guo10}. \citeauthor{Guo10} generated a list of stellar masses
and subhalo mass from these respective mass functions, ranked each
list in order of decreasing mass and then paired up the most massive
galaxy in terms of stellar mass with the most massive subhalo, and
then worked down each list, matching galaxies to subhaloes.  This
technique is called subhalo abundance matching \citep{Vale:2004}. The
black curve is therefore made up of a point for each galaxy from the
ranked list and the subhalo it is paired up with. We have applied the
same process to generate the \galform predictions. The subhalo mass is
derived from the galaxy merger tree. For satellites this mass is the
mass of the host halo at infall into a larger structure when the
galaxy became a satellite. For central galaxies, we use the mass of
the host halo. This is standard practice in subhalo abundance matching
\citep{Simha12}. Note that by using the mass stored in the halo merger
tree we avoid the complication of trying to find the subhalo in the
simulation output at $z=0$ and finding a proxy for its mass at infall.
As for the stellar mass function, there are two curves for the
\galform predictions corresponding to the true stellar masses (red)
and the masses inferred from SDSS photometry (green). The green line
agrees well with the observationally inferred curve at high and low
masses, { but underpredicts the peak by a factor of two.}  Note
that we have not attempted to account for the difference in the
subhalo mass functions between \galform and \cite{Guo10} due to
differences in the cosmology and halo mass definition (these
differences are discussed in more detail in \citealt{Mitchell15}).
The blue curve shows the bathtub prediction. In this case there is one
galaxy per halo, so stellar mass is plotted against host halo mass
(i.e. there are no subhaloes). The bathtub model does not reproduce
the tail at low halo mass and overpredicts the peak by a factor of
$\approx 3.5$.


\section{Discussion}
\label{sec:disc}

In this section, we discuss various issues raised by the approach to
modelling galaxy formation presented in this paper.

Firstly, we review the purpose of physical models of galaxy formation
in general (including gas-dynamical simulations), and of
semi-analytical (SA) models in particular. Galaxy formation is a very
complex process, involving many different physical mechanisms, acting
over an enormous range of scales, from the growth of cosmic structure
down to the formation of individual stars and supermassive black holes
and their interactions with the ISM. Even the best current numerical
simulations are not able to model all of these scales using only the
fundamental equations of physics. Instead, the effects of physical
processes acting below the resolution scale of the calculation
(whether a semi-analytical model or gas dynamical simulation) must be
included via ``effective'' or ``subgrid'' models. While in principle
such subgrid models could be fully specified, with no free parameters,
this is generally not the case in practice. Instead, our current
theoretical uncertainty about the details of many of the processes
acting on small scales within galaxies means that these subgrid models
contain parameters whose values can only be estimated by reference to
observations of the real Universe. (An analogous situation arises in
stellar evolution theory, where various types of convective mixing
processes, which crucially influence the evolution of stars, are
calculated using effective models depending on parameters whose values
are estimated by comparing the results of the calculations with
observations of real stellar populations.)  Consequently, it is
currently not possible (and likely will not be possible for many
decades) to construct a definitive {\em a priori} physical model for
galaxy formation starting from linear density perturbations in the
early Universe. This is in stark contrast to calculation of structure
formation in the dark matter, where, at least for the case of standard
CDM, highly detailed {\em a priori} physical predictions are possible
using large N-body simulations.

Given the current impossibility of making {\em a priori} physical
predictions for galaxy formation, independently of any calibration of
subgrid model parameters on observational data, the goals of physical
models of galaxy formation are instead to: (i) explore how the different
physical processes in galaxy formation interact to determine the
stellar and gaseous properties of the galaxy population; (ii) 
understand how this is reflected in the various observed properties of
galaxies and their evolution; and (iii) help interpret, and place in
context, observational data, for example by suggesting evolutionary
pathways and connections between galaxies observed at different
redshifts. 

While our theoretical understanding of galaxy formation will always be
incomplete as long as it rests on subgrid models containing free
parameters, we can still use theoretical models to increase enormously
our understanding of the roles of different physical processes in
determining galaxy properties. SA models effectively apply the subgrid
modelling approach at the level of an entire galaxy, and so are able
to predict only global properties of galaxies (masses, luminosities,
metallicities, gas contents, colours etc). (Examples exist of SA
models which resolve annuli within galaxy disks
\citep[e.g.][]{Kauffmann96,Stringer07,Fu10}, but these are computationally
expensive.)  However, at the level of such global galaxy properties,
SA models are still an ideal tool to carry out this exploration of
galaxy formation physics, and still have great advantages in terms of
speed and flexibility over gas-dynamical simulations. To conclude, the
aim of SA modelling is to improve our physical understanding of galaxy
formation, not simply to provide a parametric fit to observational
data. Indeed, phenomenological models based on arbitrary assumptions
lacking in physical motivation
\citep[e.g.][]{Peng10} may be able to provide better descriptions of
particular observational datasets than SA models, { but, in our
  opinion, such phenomenological models have less to teach us about
  the physics behind the observations because they consider special or
  contrived situations in which a single process is studied in
  isolation.  It is not clear if the lessons learnt from empirical
  model models hold when a more realistic interplay between processes
  is considered.}  

A second issue concerns the ``complexity'' of SA models, and the
number of input parameters. It is often claimed that SA models are
very complicated, and also that they have so many free parameters that
they can be tuned to reproduce any observational dataset, and
consequently lack any predictive power. However, galaxy formation is
intrinsically complex, due to the number of different physical
processes involved, which interact in a highly non-linear way. SA
models aim to model the individual processes with the minimum
complexity necessary, so the apparent complexity of the final model is
due only to the significant number of processes that must be included
for a realistic model of galaxy formation, able to predict a wide
range of galaxy properties. Likewise, the claim that SA models have
``many'' free parameters ignores the large number of different
physical processes being modelled, the numerous constraints on the
values of the parameters, either from physical considerations or from
observations, and the very wide range of observational data that the
model predictions can then be compared with. Our current model has
around 14 input parameters that are significant (the ``primary''
parameters in Table~\ref{table:param}), once we take into account that
many of the input parameters (e.g. the cosmological parameters, the
form of the IMF for quiescent star formation, and the parameters for
photoionization feedback) are fixed directly from observations or from
detailed simulations. (Note also that only the slope of the IMF in
starbursts is a free parameter, while the yield and recycled fraction
are then predicted from stellar evolution.) Even then, some of the 14
parameters, although significant in principle, actually have little
effect in the current model.

A third issue concerns how we find the ``best fit'' values for the
adjustable input parameters. In this study, we have done this fitting
by the traditional ``trial and error'' approach, in which parameters
are varied and the results then examined. { Based on our previous
  experience, we do not think that we have missed any other equally
  well-fitting models in some other corner of parameter space.}
However, more automated statistical techniques have the potential to
be both more objective and faster (in terms of human time), {
  although even automated schemes might miss some solutions.}
Important progress has already been made in applying statistical
methods such as emulation \citep{Bower10}, Monte Carlo Markov Chain
(MCMC)
\citep{Henriques09,Henriques13,Henriques15,Lu11,Lu14,Mutch13,Benson14}
or Particle Swarm Optimization \citep{Ruiz15} to parameter estimation
in SA models, and this will be a useful direction for future work.

What is the future of SA models in the era of large cosmological
hydrodynamical simulations such as Illustris \citep{Vogelsberger14}
and EAGLE \citep{Schaye15}? Simulations have recently made big
breakthroughs in producing galaxy populations with global properties
which agree much better with observations than previously. This
progress results from a combination of factors: (i) faster computers;
(ii) a better understanding of how to construct subgrid physics models
to enable the simulations to match key observational constraints such
as the galaxy stellar mass function; (iii) adopting the same
methodology as in SA models of running grids of simulations with
different subgrid parameters, to find the parameter set that best
reproduces a set of observational constraints. However, these
successes of hydrodynamical simulations do not make SA models
irrelevant. The subgrid models employed in these recent simulations
make a lot of assumptions and often have as many, or even more,
parameters than SA models, e.g. the EAGLE subgrid model for SN
feedback invokes a phenomenological dependence of feedback effeciency
on both metallicity and gas density, with 6 adjustable parameters. It
is not feasible to fully explore the effects of different subgrid
modelling assumptions and parameters using simulations, a task which
is currently possible only by using SA modelling. Furthermore, SA
models are still the only practical means to generate from a physical
model the very large volume galaxy catalogues needed for designing and
interpreting future large galaxy surveys, such as DESI and PAU, as
well as surveys to be carried out on next-generation telescopes such
as Euclid, LSST and SKA.

The SA model presented in this paper has a number of important
successes. As in previous \galform models, it reproduces the observed
$B$- and $K$-band luminosity functions at $z=0$, but unlike the
earlier \citet{Baugh05} model, it also matches the observed evolution
of the { bright part of the} $K$-band luminosity function up to
$z=3$, and unlike the earlier \citet{Bower06} model, it reproduces the
number counts and redshift distribution of the faint
850$\mum$-selected sub-mm galaxies. In addition, at $z=0$ it predicts
the correct $HI$ mass function, Tully-Fisher relation, fraction of
early-type galaxies vs luminosity, and black hole mass vs bulge mass
relation. At higher redshifts, it predicts the correct evolution of
the rest-frame far-UV luminosity function at $z \sim 3-6$, and far-IR
number counts of galaxies at wavelengths of 250--500$\mum$. As far as
we are aware, the model is unique in matching observations over such a
wide range of wavelength and redshift.

{ We note that \citet{Somerville12} presented a study with somewhat
  similar aims and methods to the current paper, namely predicting the
  multi-wavelength evolution of galaxies using a SA model combined
  with a model for absorption and emission of radiation by dust. As
  mentioned in \S\ref{sec:dust}, they use templates for the shape of
  the IR/sub-mm SED, instead of a self-consistent calculation of the
  dust temperature based on energy balance as in \GALFORM. They also
  assume a single universal IMF, unlike in our model. The
  \citeauthor{Somerville12} model predicts galaxy luminosity functions
  and number counts at UV, optical and near-IR wavelengths in
  reasonable agreement with observational data (although they somewhat
  overpredict the $K$-band luminosity function at the faint end,
  similar to our model). However, the galaxy number counts at far-IR
  and sub-mm wavelengths are underpredicted by large factors compared
  to observations (up to a factor $\sim 5$ at 250~$\mum$ and $\sim 50$
  at 850~$\mum$). An earlier multi-wavelength SA model by
  \citet{Fontanot07} was able to able to match the sub-mm galaxy
  number counts at 850~$\mum$ while assuming a universal Salpeter
  IMF. However, that model predicts the wrong redshift distribution
  for galaxies having 850~$\mum$ fluxes $\sim 1-10$~mJy, placing them
  at low redshift, in contradiction with observations, which show a
  redshift distribution peaked at $z \sim 2$. This means that the
  bright part of the IR/sub-mm galaxy population at $z \sim 2$
  is missing in the \citet{Fontanot07} model. A further drawback of
  the \citeauthor{Fontanot07} model is that it overpredicts the bright
  end of the $K$-band luminosity function at $z=0$.}

However, the { SA model presented here} has problems in matching
two important observed relations at $z=0$: the metallicity vs
luminosity relation for early-type galaxies, and the size vs
luminosity relation for both early- and late-type galaxies. The
metallicity-luminosity relation is too steep at low luminosity, while
the size-luminosity relation for late-type (i.e. disk-dominated)
galaxies is too flat. Both of these problems could be solved if the
mass-loading factor for SN feedback varied less strongly with circular
velocity than in our standard model, specifically if $\beta \propto
\Vc^{-2}$ rather than $\Vc^{-3.2}$. However, this change would cause
the faint end of the galaxy luminosity function at $z=0$ to be too
steep compared to observations. Within our current model framework,
there seems to be no way simultaneously to match both these sets of
observational constraints. This points to the need for an improved
treatment of SN feedback, and various work is underway to develop this
using both analytical methods and numerical simulations
\citep[e.g.][]{Creasey13,Lagos13,Muratov15}. The problem with the
sizes of early-type (i.e. bulge-dominated) galaxies seems to have a
different origin, since even if the SN feedback is adjusted to give
the correct sizes for late-type galaxies, the size-luminosity relation
for early-type galaxies is still too flat at low luminosities. Since
low-luminosity spheroids form mainly by disk instabilities in our
model, this suggests that the treatment of disk instabilities needs to
be improved.

{ We note that in our model, disk instabilities play an extremely
  important role, both in building up stellar spheroids, and in
  building up SMBHs through gas accretion triggered by
  starbursts. Through SMBH growth, they also impact strongly on the
  effectiveness of AGN feedback. However, there are currently
  significant uncertainties in the treatment of disk instabilities in
  SA models, not only in the criterion for a disk to undergo a bar
  instability, but also in what fraction of disk stars are transferred
  to the spheroid, what is the size of the resulting spheroid, and
  what fraction of gas is consumed in a starburst. \citet{Parry09} and
  \citet{DeLucia11} have compared results between SA models that make
  different assumptions about disk instabilities. They found that all
  of the models predicted a larger contribution of disk instabilities
  to spheroid formation at lower masses, but disagreed about whether
  this (rather than galaxy mergers) made the dominant
  contribution. This emphasizes the need for a better understanding of
  disk instabilities. The role of galaxy mergers in assembling the
  total stellar mass of galaxies has now been analysed in detail in
  cosmological gas-dynamical simulations of galaxy formation
  \citep[e.g.][]{Oser10,Rodriguez16}, but such studies have not yet
  yielded robust measurements of the contribution of different
  channels (mergers vs. disk instabilities) to spheroid formation at
  different galaxy masses.}

{ Other areas where the model is in some disagreement with current
  observational data are galaxy colours at $z=0$, and the evolution of
  the cosmic SFR density. The model predicts a bimodal colour
  distribution at $z=0$, qualitatively consistent with observations,
  but disagrees in detail. In particular, it predicts too large a
  fraction of red galaxies at low luminosities. This may be due to
  stripping of hot gas halos from satellite galaxies being too
  efficient in the current model
  \citep[c.f.][]{Font08,Henriques15}. The cosmic SFR density in the
  model is lower than most current observational estimates at $z<3$,
  even after allowing for the effect of the varying IMF in the
  model. However, the observational estimates still have significant
  uncertainties, in particular, at higher redshifts they typically
  involve extrapolating galaxy luminosity functions or SFR
  distributions down to much lower values than are directly observed.}

{ Finally, we return to the issue of the IMF, and whether a
  top-heavy IMF is really necessary. In our model, we assume an IMF in
  starbursts that is mildly top-heavy compared to that in normal disk
  galaxies, resulting in starbursts producing roughly twice as much UV
  light and twice as much mass in metals as they would for the same
  amount of star formation with a normal IMF. The need for the
  top-heavy IMF is driven primarily by trying to simultaneously match
  both the number counts and redshift distribution of sub-mm galaxies
  at 850~$\mum$, as well as the $K$-band lumininosity function of
  galaxies at $z=0$. The former constrains the number of dusty, far-IR
  luminous galaxies with $L_{\rm IR} \sim 10^{12} \Lsol$ present at $z
  \sim 2$, while the latter constrains the stellar mass function at
  $z=0$. Simply adopting a normal IMF in bursts, while keeping all
  other model parameters the same, results in a huge underprediction
  of the 850~$\mum$ number counts at fluxes $\sim 1-10$~mJy, and also
  shifts their redshift distribution down to much lower values than
  observed (see Fig.~\ref{fig:SMGs_IMF_tauburst_nomerge}). These
  effects at 850~$\mum$ could be compensated in part by reducing the
  strength of SN feedback or assuming a faster return of gas ejected
  from halos by SN feedback (see Fig.~\ref{fig:SMGs_SNfeedback}), or
  by reducing the strength of AGN feedback (see
  Fig.~\ref{fig:SMGs_AGNfeedback}), but such changes then result in a
  $K$-band lumininosity function at $z=0$ (and also at higher
  redshifts) that has too many galaxies at higher luminosities (see
  Figs.~\ref{fig:lf_SNfeedback} and \ref{fig:lf_AGNfeedback}, and also
  Figs.~\ref{fig:lfKz_SNfeedback} and
  \ref{fig:lfKz_AGNfeedback}). (Similar effects can also be seen in
  the far-IR number counts as for the 850~$\mum$ counts, see
  Figs.~\ref{fig:FIRcounts_SNfeedback} and
  \ref{fig:FIRcounts_AGNfeedback_IMF}.) We also note that the earlier
  \GALFORM model by \citet{Gonzalez-Perez14}, which adopts a similar
  model framework to that used here, but with a single IMF, and tries
  to fit a similar set of observational constraints, underpredicts the
  850~$\mum$ number counts by only a modest factor, but greatly
  underpredicts the typical redshifts of the 850~$\mum$ sources.  We
  conclude that, within our current modelling framework, the top-heavy
  IMF in starbursts is needed in order to match the abovementioned
  observational constraints. However, the possibility remains that
  some future modification to this framework, affecting, for example
  AGN or SN feedback, might allow the formation of a larger number of
  dusty galaxies with intrinsically high SFRs at high redshifts, and
  so ease or remove the need for a top-heavy IMF in these objects.}


\section{Conclusions}
\label{sec:conc}




We present a new multi-wavelength semi-analytical model of galaxy
formation. This extends previous versions of the \galform model
\citep{Cole00,Baugh05,Bower06,Font08,Lagos12,Gonzalez-Perez14} by
including important improvements in the input physics, and by
calibrating the model against an unprecedentedly wide range of
observational constraints. For the first time in the development of
\galform, it combines a treatment of AGN feedback with a varying IMF
in starbursts, together with a detailed modelling of the absorption
and emission of radiation by dust, enabling predictions of observable
galaxy properties from far-UV to sub-mm wavelengths. The model
includes the following physical processes: (i) assembly of dark matter
halos, calculated from the Millennium-WMAP7 cosmological N-body
simulation; (ii) shock heating and radiative cooling of gas in DM
halos; (iii) collapse of cooled gas to a rotationally-supported disk,
with the disk size calculated self-consistently based on angular
momentum and the gravity of the disk, spheroid and halo; (iv)
formation of stars in the disk, calculated using an empirical star
formation law related to the molecular gas content; (v) ejection of
gas from galaxies by supernova feedback, and gradual return of this
gas to galaxy halos; (vi) mergers of galaxies within common DM halos
driven by dynamical friction on satellite galaxies, with a timescale
calibrated on simulations; (vii) bar instabilities in galaxy disks;
(viii) galaxy mergers and bar instabilities transforming stellar disks
into spheroids, and triggering starbursts in the remaining cold gas;
(ix) starbursts triggering accretion of gas onto supermassive black
holes at the centres of galaxies; (x) AGN feedback acting in halos in
the hydrostatic cooling regime, with energy released by accretion of
gas from the hot halo onto the central black hole balancing radiative
cooling in the halo, and hence shutting down accretion of gas from the
halo; (xi) chemical evolution, tracking metal production by supernovae
and the chemical enrichment of gas and stars in the galaxies and halo;
(xii) stellar luminosity of galaxies calculated from a population
synthesis model including a { strong contribution from} TP-AGB
stars; (xiii) absorption of starlight by dust calculated by radiative
transfer, with the dust mass and optical depth calculated
self-consistently based on gas mass, metallicity and galaxy size;
(xiv) FIR/sub-mm emission from dust, with the dust temperature
calculated self-consistently from energy balance.

The most important success of the new model is that, due to the
combination of a top-heavy IMF in starbursts boosting their total
luminosities, and AGN feedback suppressing the growth of galaxies in
massive halos, it is able to reconcile the number counts and redshift
distribution of the 850$\mum$-selected SMG population at $z \sim 1-3$
with the evolution of the { bright end of the} rest-frame $K$-band
luminosity function of galaxies at $z \sim 0 - 3$. This is the first
time this has been achieved in a physical model of galaxy formation
based on $\Lambda$CDM, with a physical model of dust emission. In
contrast to the earlier \citet{Baugh05} model, the starburst IMF is
only required to be mildly top-heavy (with slope $x=1$, compared to
$x=1.35$ for Salpeter), while the IMF for star formation in disks is
assumed to have the form observed in the Solar neighbourhood. The
top-heavy IMF in starbursts also boosts the total metallicities in
high-mass elliptical galaxies, bringing them into better agreement
with observations. The cosmic SFR density in the model is dominated by
quiescent star formation in disks at $z \lsim 3$, but by starbursts at
higher redshifts. At $z=3$, 54\% of the stellar mass has formed with
the top-heavy IMF, but by $z=0$, this fraction has dropped to 30\%. In
contrast to the \cite{Baugh05} model, these starbursts are triggered
mainly by bar instabilities in disks, rather than by galaxy mergers.

The main successes of our current model, as well as its weaknesses and
possible avenues for future improvements, are discussed in the
previous section. Our new model builds upon developments of the
semi-analytic approach to modelling galaxy formation over the past 25
years, which has led to important advances in our understanding of the
physical processes at work during galaxy formation. Perhaps the most
notable successes of this approach so far have been the identification
of the two key processes that explain why the observed stellar mass
function has such a different shape to the $\Lambda$CDM dark matter
halo mass function: supernova feedback at the faint end
\citep{White78, Cole91,Lacey91,White91} and AGN feedback at the bright
end \citep{Benson03,Bower06,Croton06,Cattaneo06}. These insights led
to a prediction for the evolution of cosmic star formation from high
redshift to the present \citep{Lacey93,Cole94} and to the development
of the now widely-used approach of the ``halo occupation
distribution'' to characterize the clustering of galaxies
\citep{Benson00,Seljak00,Peacock00}.


A concern of the model presented in this paper is that in order
simultaneously to match the properties of submillimeter galaxies at
high redshift and the galaxy luminosity function today, it requires
that the IMF in starbursts should be different from the IMF in
quiescent star formation. This is a feature of our models that was
already present in \citet{Baugh05} and \citet{Lacey08,Lacey11}. At the
current time there is little direct observational evidence for such a
variation in the IMF. Should such evidence be forthcoming, however, it
would highlight the role that semi-analytical modelling plays in
helping reveal the nature of processes involved in galaxy
formation. Alternatively, should a varying IMF be conclusively ruled
out, we would be forced to revise other aspects of our model in ways
that have eluded us so far.

In spite of the tremendous progress in understanding galaxy formation
over the past three decades, several fundamental problems, some
highlighted in this paper, remain unsolved. Progress to date has
resulted from a close interaction between ever more precise
theoretical modelling and observations of ever increasing reach and
accuracy. The traditional theoretical tool of semi-analytical
modelling has now been augmented by the recently developed ability to
carry out large cosmological hydrodynamical simulations that produce
credible model galaxy populations. These two theoretical techniques,
semi-analytical modelling and hydrodynamical simulations, are
complementary and will continue to play an important role in the
continuing search for a physical understanding of galaxy formation and
evolution.

\section*{Acknowledgements} 
We thank Dave Campbell, Will Cowley and Peder Norberg for their
contributions to this work. We thank Claudia Maraston for supplying
versions of her stellar population model for different IMFs. { We thank
Rychard Bouwens, Ignacio Ferreras and Ian Smail for comments on the
paper. Finally, we thank the referee Bruno Henriques for a detailed
and constructive report that helped improve the paper.}

This work was supported by the Science and Technology Facilities
Council grants ST/F001166/1 and ST/L00075X/1, and by European Research
Council grant GA 267291 (Cosmiway). This work used the DiRAC Data
Centric system at Durham University, operated by the Institute for
Computational Cosmology on behalf of the STFC DiRAC HPC Facility
(www.dirac.ac.uk). This equipment was funded by BIS National
E-infrastructure capital grant ST/K00042X/1, STFC capital grant
ST/H008519/1, and STFC DiRAC Operations grant ST/K003267/1 and Durham
University. DiRAC is part of the National E-Infrastructure. CMB
acknowledges receipt of a Research Fellowship from the Leverhulme
Trust.

Galaxy catalogues calculated using the model described in this paper
will be made available on a relational database accessible from
{ http://virgodb.dur.ac.uk/.}

\bibliographystyle{mn2e}
\bibliography{paper}


\appendix

\section{Simplified two-temperature model for dust absorption and
  emission}
\label{app:dust_model}

In order to calculate the sub-mm luminosities and fluxes of model
galaxies, we need a model for calculating the amount of stellar
radiation absorbed by dust and for the spectral energy distribution
(SED) of the radiation emitted by the dust grains. In our previous
papers on the dust emission from galaxies
\citep{Granato00,Baugh05,Lacey08,Lacey10}, we calculated the dust emission by
coupling the \galform code with the \GRASIL
spectrophotometric code \citep{Silva98}, which incorporates a detailed
calculation of radiative transfer through the dust distribution and of
the heating and cooling of dust grains of different sizes and
compositions at different locations within each galaxy. A drawback of
the \GRASIL code is that it typically takes several minutes of
CPU time to compute the SED of a single galaxy. For the present paper,
it was necessary to calculate the dust emission for very large numbers
of \galform galaxies, for which the computational cost would have
been prohibitive if we had used \GRASIL directly. We therefore
devised a simplified approximate model for dust emission at sub-mm
wavelengths, which retains some of the main assumptions of {\tt
GRASIL}, but is much faster computationally.

We retain the \GRASIL assumptions about the geometry of the stars
and dust. Stars are in general distributed in two components: a
spherical bulge with an $r^{1/4}$-law profile, and a flattened component,
either a quiescent disk or a burst component, with an
exponential radial and vertical profile. We assume that the young
stars and dust are found only in the flattened component. We retain
the assumption made in \GRASIL that the dust and gas are in a
two-phase medium consisting of dense molecular clouds and a diffuse
intercloud medium. Stars are assumed to form inside the molecular
clouds, and then to escape into the diffuse medium on a timescale of a
few Myr. The calculation of the emission from the dust then has two
parts, calculating the amount of energy absorbed in the molecular
cloud and diffuse dust components, and then calculating the wavelength
distribution of the energy re-emitted by the dust.

\subsection{Energy absorbed by dust}

The unattenuated SED of a stellar population at time $t$ (measured
from the big bang) and with a specified IMF is given by an integral
over the star formation and metal enrichment history:
\begin{equation}
L^{\rm unatten}_{\lambda}(t) = \int_0^t L^{\rm (SSP)}_{\lambda}(\tau,Z) \Psi(t-\tau,Z) 
\, {\rm d}\tau {\rm d}Z,
\label{eq:SED_nodust}
\end{equation}
where $L_{\lambda}$ is the luminosity per unit wavelength for the
composite stellar population, $L^{\rm (SSP)}_{\lambda}(\tau,Z)$ is the
luminosity per unit wavelength for a simple stellar population (SSP)
with the specified IMF of age $\tau$ and metallicity $Z$ and unit
initial mass, and $\Psi(t,Z)\,{\rm d}t\,{\rm d}Z$ is the mass of stars
formed in the time interval $t,t+{\rm d}t$ and metallicity range
$Z+{\rm d}Z$. The SED including dust attenuation is then
\begin{eqnarray}
\lefteqn{ L^{\rm atten}_{\lambda}(t) = }  \nonumber \\
& & A^{({\rm diff})}_\lambda \int_0^t
  A^{\rm (MC)}_\lambda(\tau) L^{\rm (SSP)}_{\lambda}(\tau,Z) \Psi(t-\tau,Z)
\, {\rm d}\tau {\rm d}Z,
\label{eq:SED_dust}
\end{eqnarray}
where $A^{({\rm diff})}_\lambda$ is the dust attenuation factor at wavelength
$\lambda$ due to the diffuse dust component, and
$A^{\rm (MC)}_\lambda(\tau)$ is the mean attenuation due to molecular
clouds, which depends on stellar age. The attenuation by diffuse dust
is independent of stellar age, since we assume that the spatial
distribution of stars after they escape from their parent molecular
clouds is independent of stellar age.

\subsubsection{Dust attenuation by clouds}

Following the assumptions made in \GRASIL, we assume that a
fraction $\fcloud$ of the total gas mass is in molecular clouds, which are
modelled as uniform density spheres of gas mass $\Mcloud$ and radius
$\rcloud$. The effective absorption optical depth for the stars in each
cloud is approximated as
\begin{equation}
\tau_{\lambda,{\rm eff}} = (1-a_\lambda)^{1/2} \tau_{\lambda,{\rm
    ext}} ,
\end{equation}
\citep[e.g.][]{Silva98}, where $a_\lambda$ is the albedo, and
$\tau_{\lambda,{\rm ext}}$ is the extinction optical depth from the
centre of a cloud to its edge. The extinction optical depth is
calculated from the column density of gas through a cloud and its
metallicity using eqn(\ref{eq:tau_dust}).
The dust attenuation factor for light from stars in a single cloud is
then $e^{-\tau_{\lambda,{\rm eff}}}$, while the mean attenuation for
all stars of age $\tau$ due to clouds is given by
\begin{equation}
A^{\rm (MC)}_\lambda(\tau) = \eta(\tau)e^{-\tau_{\lambda,{\rm eff}}} +
(1-\eta(\tau)),
\end{equation}
where $\eta(\tau)$ is defined as the fraction of stars of age $\tau$
which are still in the clouds where they formed. For this fraction, we
adopt the same dependence as assumed in \GRASIL,
\begin{equation}
\eta(\tau) = \left\{ \begin{array}{ll}
                  1 & \tau < t_{\rm esc} \\
                  2-t/t_{\rm esc} & t_{\rm esc} < \tau < 2t_{\rm esc} \\
                  0 & \tau > 2t_{\rm esc} ,
		  \end{array} \right .
\label{eq:tesc}
\end{equation}
so that stars begin to escape a time $t_{\rm esc}$ after they form, and
have all escaped after time $2t_{\rm esc}$.

The dust-attenuated SED can therefore be rewritten as 
\begin{equation}
L^{\rm atten}_{\lambda}(t) = A^{({\rm diff})}_\lambda \langle \,
A^{\rm (MC)}_\lambda(\tau)
\rangle \, L^{\rm unatten}_{\lambda}(t) ,
\end{equation}
where $A^{\rm (MC)}_\lambda(\tau) \rangle$ is the dust attenuation by
clouds averaged over all stellar ages, given by 
\begin{equation}
\langle A^{\rm (MC)}_\lambda(\tau) \rangle = 1 - \langle \eta(\tau)
\rangle_{\lambda} (1-e^{-\tau_{\lambda,{\rm eff}}}) ,
\end{equation}
and $\langle \eta(\tau) \rangle_{\lambda}$ is the fraction of starlight
at wavelength $\lambda$ emitted by stars inside molecular clouds. This
is in turn given by a luminosity-weighted average
\begin{equation}
\langle \eta(\tau) \rangle_{\lambda} = \frac{\int_0^t
  \eta(\tau) L^{\rm (SSP)}_{\lambda}(\tau,Z) \Psi(t-\tau,Z) \, {\rm d}\tau {\rm d}Z }
 {\int_0^t L^{\rm (SSP)}_{\lambda}(\tau,Z) \Psi(t-\tau,Z) \, {\rm d}\tau {\rm d}Z } .
\label{eq:eta_av}
\end{equation}
In principle, in order to calculate $\langle \eta(\tau)
\rangle_{\lambda}$ we need to know the entire star formation and
chemical enrichment history for a galaxy, specified by
$\Psi(\tau,Z)$. However, we now make a number of simplifying
approximations. The absorption of starlight by dust in clouds is
important mostly for the UV light, which is emitted mainly by young
stars, which have metallicities close to the current ISM value $\Zg$. We can
therefore approximate the integral in eqn.(\ref{eq:eta_av}) as
\begin{equation}
\langle \eta(\tau) \rangle_{\lambda} \approx \frac{\int_0^T
  \eta(\tau) L^{\rm (SSP)}_{\lambda}(\tau,\Zg) \psi(t-\tau) \, {\rm d}\tau}
 {\int_0^T L^{\rm (SSP)}_{\lambda}(\tau,\Zg) \psi(t-\tau) \, {\rm d}\tau} ,
\label{eq:eta_av2}
\end{equation}
where $\psi(t)$ is now the total SFR at time $t$, integrated over all
stellar metallicities, and $T$ is a fixed upper cutoff in the integral
over stellar age. We adopt $T=10~\Gyr$, but our results are not
sensitive to this value.

We evaluate eqn.(\ref{eq:eta_av2}) separately for star
formation in disks and in bursts. For disks, the SFR typically varies
on a timescale long compared to the lifetimes of the stars responsible
for most of the UV radiation which dominates the dust heating, so we
approximate the recent SFR as constant, $\psi^{\rm disk}(t-\tau) \approx
\psi^{\rm disk}(t)$, leading to 
\begin{equation}
\langle \eta(\tau) \rangle_{\lambda}^{\rm disk} \approx \frac{\int_0^T
  \eta(\tau) L^{\rm (SSP,disk)}_{\lambda}(\tau,\Zg) \, {\rm d}\tau}
 {\int_0^T L^{\rm (SSP,disk)}_{\lambda}(\tau,\Zg) \, {\rm d}\tau},
\label{eq:eta_av_disk}
\end{equation}
where the SSPs $L^{\rm (SSP,disk)}_{\lambda}(\tau,\Zg)$ use the IMF for
quiescent star formation. In the case of a burst starting at time
$t_b$, with e-folding timescale $\taustar$ the SFR varies as
\begin{equation}
\psi^{\rm burst}(t) = \left\{ \begin{array}{ll}
                        0 & t<t_{\rm b} \\
			\psi^{\rm burst}_0 \exp(-(t-t_{\rm b})/\taustar) & t>t_{\rm b}
		  \end{array} \right .
\label{eq:psi_burst}
\end{equation}
so that eqn.(\ref{eq:eta_av2}) can be rewritten as
\begin{equation}
\langle \eta(\tau) \rangle_{\lambda}^{\rm burst} \approx \frac{\int_0^{\taub}
  \eta(\tau) L^{\rm (SSP,burst)}_{\lambda}(\tau,\Zg) e^{\tau/\taustar} \, {\rm d}\tau}
{\int_0^{\taub}
  L^{\rm (SSP,burst)}_{\lambda}(\tau,\Zg)  e^{\tau/\taustar} \, {\rm d}\tau} ,
\label{eq:eta_av_burst}
\end{equation}
where the SSPs $L^{\rm (SSP,burst)}_{\lambda}(\tau,\Zg)$ use the IMF for
bursts, and we define $\taub = t-t_{\rm b}$ as the age at which the burst
started. In practice, we tabulate both functions $\langle \eta(\tau)
\rangle_{\lambda}^{\rm disk}$ and $\langle \eta(\tau)
\rangle_{\lambda}^{\rm burst}$ as functions of $\Zg$ and
$(\Zg,\taustar,\taub)$ respectively.

Finally, we calculate the luminosity absorbed by dust in molecular
clouds as
\begin{equation}
L_{\rm abs}^{\rm MC} = \int_0^\infty (1-\langle A^{\rm (MC)}_\lambda \rangle)
L^{\rm unatten}_{\lambda} {\rm d}\lambda .
\label{eq:Labs_MC}
\end{equation}
The parameters we use for the molecular clouds are identical to those
which we use in \GRASIL. For the current model, they are:
$\fcloud=0.25$, $\Mcloud=10^6~\Msol$, $\rcloud=16~\pc$, $t_{\rm
  esc}=1~\Myr$ for both disks and bursts \citep{Baugh05}. (In fact,
$\Mcloud$ and $\rcloud$ only enter in the combination $\Mcloud/r_c^2$,
which determines the optical depth of the molecular clouds. As shown
by \citet{Vega05}, in \GRASIL the main effect of varying
$\Mcloud/r_c^2$ is on the mid-IR dust emission, which we do not
calculate in our simple model.) We note that the \GRASIL code does not
make any of the above approximations, but instead does an exact
radiative transfer calculation for the escape of starlight from
molecular clouds.

\subsubsection{Dust attenuation by diffuse medium}

We calculate the attenuation of starlight by dust in the diffuse
medium using the tabulated radiative transfer models of
\citet{Ferrara99}, as described in \citet{Cole00}.
\citeauthor{Ferrara99} calculated dust attenuation factors using a
Monte Carlo radiative transfer code, including both absorption and
scattering, for galaxies containing stars in both a disk with
exponential radial and vertical distributions, and a spherical bulge
with a Jaffe density profile (which closely approximates an $r^{1/4}$
law), with the dust smoothly distributed in an exponential disk. They
tabulated their results as functions of wavelength, disk inclination
angle, central ($r=0$) dust optical depth, and ratio of disk to bulge
scalelengths. We use their models for a Milky Way extinction curve,
equal scaleheights for dust and gas, and ratio of vertical to radial
disk scalelengths equal to 0.1. We compute the central optical depth
for our model galaxies from the mass and metallicity of the gas and
the radial scalelength of the disk, assuming that the dust-to-gas
ratio is proportional to the gas metallicity, and then interpolate in
the \citeauthor{Ferrara99} tables to get the total attenuation as a
function of wavelength. The only difference from \citet{Cole00} is
that in the present case the diffuse medium contains only a fraction
$1-\fcloud$ of the total gas mass.

The luminosity absorbed by dust in the diffuse medium is then
calculated as
\begin{equation}
L_{\rm abs}^{\rm diff} = \int_0^\infty (1-A^{({\rm diff})}_{\lambda}) 
\langle A^{\rm (MC)}_\lambda \rangle L^{\rm unatten}_{\lambda} {\rm
  d}\lambda .
\label{eq:Labs_diff}
\end{equation}

\subsection{SED of dust emission}

The dust is assumed to be in thermal equilibrium, so the total
luminosity emitted by dust is equal to the luminosity absorbed from
starlight. To calculate the wavelength distribution of the dust
emission, we approximate the dust temperature as being constant within
each of the dust components, i.e. for each galaxy, we have a single
temperature $\Tdust^{\rm MC}$ for the dust in molecular clouds, and a
single (but different) temperature $\Tdust^{\rm diff}$ for dust in the
diffuse medium. This is a major simplification compared to what is
done in \GRASIL, where the dust temperature varies with location in
the galaxy according to the strength of the stellar radiation field,
and also depending on the size and composition of each dust
grain. (\GRASIL assumes a distribution of grain sizes, and also two
compositions, carbonaceous and silicate, and in addition includes
Polycyclic Aromatic Hydrocarbon (PAH) molecules.) Furthermore, \GRASIL
includes the effects of fluctuating temperatures in small grains and
PAH molecules (due to finite heat capacities), unlike our simplified
model. For a medium in thermal equilibrium at temperature $T$, the
emissivity $\epsilon_{\lambda}$ (defined as the luminosity emitted per
unit wavelength per unit mass) can be written as
\citep[e.g.][]{Rybicki79}
\begin{equation}
\epsilon_{\lambda} = 4\pi \kappa_{\rm d}(\lambda) B_{\lambda}(\Tdust) ,
\label{eq:emissivity}
\end{equation}
where $\kappa_{\rm d}(\lambda)$ is the absorption opacity (absorption
cross-section per unit mass), and $B_{\lambda}(\Tdust)$ is the Planck
blackbody function $B_{\lambda}(\Tdust) =
(2hc^2/\lambda^5)/(\exp(hc/\lambda \kB\Tdust)-1)$. Since we assume
throughout that the dust-to-gas ratio is proportional to the gas
metallicity { (see eqn.(\ref{eq:delta_dust}))}, it is convenient to
  define the opacity relative to the total mass of metals in the gas
  (whether in dust grains or not). Assuming that the galaxy is
  optically thin at the wavelengths at which the dust emits, we can
  then write the luminosity per unit wavelength emitted by dust as
\begin{equation}
L^{\rm dust}_{\lambda} = 4\pi  Z_{\rm gas} M_{\rm gas} \,
\kappa_{\rm d}(\lambda) B_{\lambda}(\Tdust).
\label{eq:dust_em}
\end{equation}
This equation must be applied separately to the dust in the molecular
clouds and in the diffuse medium, since they have different
temperatures. (In contrast, in \GRASIL, the calculation of dust
emission from clouds includes optical depth effects.) We calculate the
dust temperatures for the clouds and diffuse medium by equating the
luminosity of dust emission (integrated over all wavelengths) to the
luminosity absorbed from starlight. 

In order to calculate eqn.(\ref{eq:dust_em}), we need to know the dust
opacity $\kappa_{\rm d}$ as a function of wavelength. We assume the same
values as for the dust model used in \GRASIL, but since the dust
emission is at long wavelengths, we approximate this by a power-law
when we calculate the emission. We find that in the \GRASIL dust
model for the local ISM (with metallicity $Z=0.02$), the absorption
opacity per unit mass of metals at $\lambda > 30~\mum$ can be
approximated as $\kappa_{\rm d} = 140~{\rm cm}^2{\rm g}^{-1}
(\lambda/30~\mum)^{-2}$. However, \citet{Silva98} found that for the
ultraluminous starburst galaxy Arp~220, the observed sub-mm SED was
reproduced better by \GRASIL if the dust emissivity at very long
wavelengths was modified by introducing a break to a $\lambda^{-1.6}$
power-law at $\lambda > 100~\mum$, and the same modification was
adopted by \citet{Baugh05} when modelling SMGs using \GRASIL. We
therefore describe the dust emissivity in our model by a broken
power-law:
\begin{equation}
\kappa_{\rm d}(\lambda) = \left\{ \begin{array}{ll}
                            \kappa_1
                            \left(\frac{\lambda}{\lambda_1}\right)^{-2}
                            & \lambda<\lambda_{\rm b} \\
			    \kappa_1
                            \left(\frac{\lambda_{\rm b}}{\lambda_1}\right)^{-2} 
			    \left(\frac{\lambda}{\lambda_{\rm b}}\right)^{-\beta_{\rm b}}
                            & \lambda>\lambda_{\rm b} 
		  \end{array} \right .
\label{eq:kappa_d_app}
\end{equation}
where $\kappa_1=140~{\rm cm}^2{\rm g}^{-1}$ at the reference
wavelength of $\lambda_1=30~\mum$, and the power-law breaks to a slope
$\beta_{\rm b}$ longwards of wavelength $\lambda_{\rm b}$. We adopt
$\lambda_{\rm b}=100~\mum$ and $\beta_{\rm b}=1.6$ in bursts, and
$\lambda_{\rm b}=\infty$ (i.e. an unbroken power-law) in quiescent disks.

\section{IMF conversion factors for observed stellar masses and SFRs}
\label{sec:IMF_conv}

In this Appendix we list the conversion factors which we apply to
observationally-inferred stellar masses and SFRs to account for
differences in assumed IMFs between observations and
models. Observational estimates of SFRs derived from luminosities in
different bands, and also stellar masses inferred from fitting stellar
population models to the broad-band SEDs of galaxies, rely on assuming
an IMF. However, different observational studies assume different
IMFs, and generally these differ from the IMFs assumed in our galaxy
formation model. To allow a fairer comparison of our models with
observational data, we apply conversion factors to observed stellar
masses and SFRs to the estimate the values that would have been
inferred if a \citet{Kennicutt83} IMF had been assumed for analysing
the observational data. We choose the \citet{Kennicutt83} IMF as our
reference IMF because this is what our model assumes for quiescent
star formation. The conversion factors presented here update those
given in \citet{Lagos14} and \citet{Gonzalez-Perez14}.


\begin{table}
\centering

\caption{Table of IMF conversion factors for stellar masses estimated
  from SED fitting. The conversion factors are given as
  $\Mstar^{\rm (Kenn)} = corr \cdot \Mstar^{\rm (IMF)}$, where
  $\Mstar^{\rm (Kenn)}$ is the stellar mass inferred assuming a
  \citet{Kennicutt83} IMF, and $\Mstar^{\rm (IMF)}$ is the stellar mass
  inferred from the same observations assuming a different IMF. We
  assume that the IMF covers the mass range $0.1 < m < 100 ~\Msol$ in
  all cases.}

\begin{tabular}{l c}
\hline
IMF  & corr \\
\hline
\citet{Salpeter55} & 0.47 \\
\citet{Kroupa01} (eqn.2) & 0.74 \\
\citet{Chabrier03} & 0.81 \\
\citet{Baldry03} & 0.85 \\
\hline

\end{tabular}

\label{table:mstar_conv_IMF}
\end{table}


The conversion factors for stellar mass are given in
Table~\ref{table:mstar_conv_IMF}. These have been obtained by
combining conversion factors between different IMFs from the
literature, for studies of stellar masses inferred from SED
fitting. Specifically, we find the conversion factors to a
\citet{Salpeter55} IMF (with $dN/d\ln m \propto m^{-1.35}$) using the
results of \citet{Ilbert10} and \citet{Santini12} for a
\citet{Chabrier03} IMF, \citet{Marchesini09} and \citet{Muzzin13} for
a \citet{Kroupa01} IMF and \citet{Glazebrook04} for a \citet{Baldry03}
and then use \citet{Mitchell13} to convert masses from Salpeter to
\citet{Kennicutt83} IMFs.

As shown by \citet{Mitchell13}, using a single conversion factor
between two IMFs is only an approximation. In reality, the ratio of
the stellar masses inferred from fitting the same SED with two
different IMFs depends on the age and star formation history (hence
also on redshift), as well as on the set of bands used to measure the
SED, the metallicity distribution, and the treatment of dust
extinction.


\begin{table}
\centering

\caption{Table of IMF conversion factors for SFRs estimated from
  different tracers. The conversion factors are given as $SFR^{\rm (Kenn)} =
  corr \cdot SFR^{\rm (IMF)}$, where $SFR^{\rm (Kenn)}$ is the SFR inferred
  assuming a \citet{Kennicutt83} IMF, and $SFR^{\rm (IMF)}$ is the SFR
  inferred from the same observations assuming a different IMF. Each
  column shows a different SFR tracer. The H$\alpha$ conversion factor
  is used also for other optical emission lines. FIR here means the $8-1000
  ~\mum$ luminosity. We assume that the IMF covers the mass range $0.1
  < m < 100 ~\Msol$ in all cases.}

\begin{tabular}{l c c c c c}
\hline
IMF  & H$\alpha$ & 1500\AA & 2500\AA & FIR & 1.4 GHz \\
\hline
\citet{Salpeter55} &  0.94 & 0.79 & 0.76 & 0.81 & 0.77 \\
\citet{Kroupa01} (eqn.2) & 1.49 & 1.19 & 1.14 & 1.22 & 1.15 \\
\citet{Chabrier03} & 1.57 & 1.26 & 1.20 & 1.29 & 1.22 \\
\citet{Baldry03} & 2.26 & 1.56 & 1.45 & 1.64 & 1.46 \\
Top-heavy IMF ($x=1$) & 3.13 & 1.89 & 1.71 & 2.02 & 1.68 \\
\hline

\end{tabular}

\label{table:SFR_conv_IMF}
\end{table}

The conversion factors for SFR are given in
Table~\ref{table:SFR_conv_IMF}, and were calculated as in
\citet{Gonzalez-Perez14}, using the PEGASE.2 SPS model to calculate
for different IMFs the luminosity at different wavelengths of a galaxy
of Solar metallicity and age 100~$\Myr$ forming stars at a constant
rate. For deriving the conversion factors between SFR and the
luminosity of a particular tracer, the H$\alpha$ luminosity is
calculated from the Lyman continuum luminosity assuming Case~B
recombination, the FIR luminosity is assumed to equal the bolometric
stellar luminosity, and the 1.4$\GHz$ radio luminosity is assumed to
be proportional to the rate of Type~II supernovae.


\section{Effects of varying parameters on key observables}
\label{sec:param_var_plots}

In this appendix, we present plots showing the effects of varying
different \galform parameters on the key observational constraints
(both primary and secondary) described in \S\ref{sec:fiducial}.  The
results shown in these plots are discussed in
\S\ref{sec:param-variations}. In each panel of each plot, we vary only
a single parameter around its standard value (indicated by the red
curve in all cases), while keeping all other parameters fixed at their
standard values as given in Table~\ref{table:param}. We group the
plots according to the observational constraint being compared to, to
make it easier to see to which \galform parameters a particular
observational constraint is most sensitive. For brevity, we do not
show here how all observational constraints respond to changes in all
\galform parameters, but instead focus on those combinations which
show some interesting dependence. A more complete set of plots will be
made available online at http://icc.dur.ac.uk/data/.

For convenience, we summarize here the physical meanings of the
\galform parameters which are varied in the following plots:
\begin{itemize}
\item $\gammaSN$ and $\VSN$ specify respectively the slope and
  normalization of the mass-loading factor for SN feedback
  (eqn.(\ref{eq:Meject})).
\item $\alpharet$ controls the timescale for gas ejected by SN
  feedback to return to the halo (eqn.(\ref{eq:Mreturn})).
\item $\alphacool$ controls which halos are subject to AGN feedback
  through a hydrostatic cooling criterion (eqn.(\ref{eq:alpha_cool})).
\item $\Fstab$ sets the threshold for disks to become unstable to bar
  formation (eqn.(\ref{eq:disk_stability})).
\item $\nuSF$ is the normalization of the molecular SFR law in
  disks(eqn.(\ref{eq:SFR_mol})).
\item $\fdyn$ and $\tauburstmin$ respectively control the scaling of
  the SFR timescale in bursts with bulge dynamical timescale and the
  floor value of this SFR timescale (eqn.(\ref{eq:taustar_burst})).
\item $x$ is the slope of the IMF in starbursts (eqn.(\ref{eq:IMF_slope})).
\end{itemize}


\subsection{$B$- and $K$-band galaxy LFs at $z=0$}

\begin{figure*}

\begin{center}
\begin{minipage}{5.4cm}
\includegraphics[width=5.5cm, bb= 20 55 275 510]{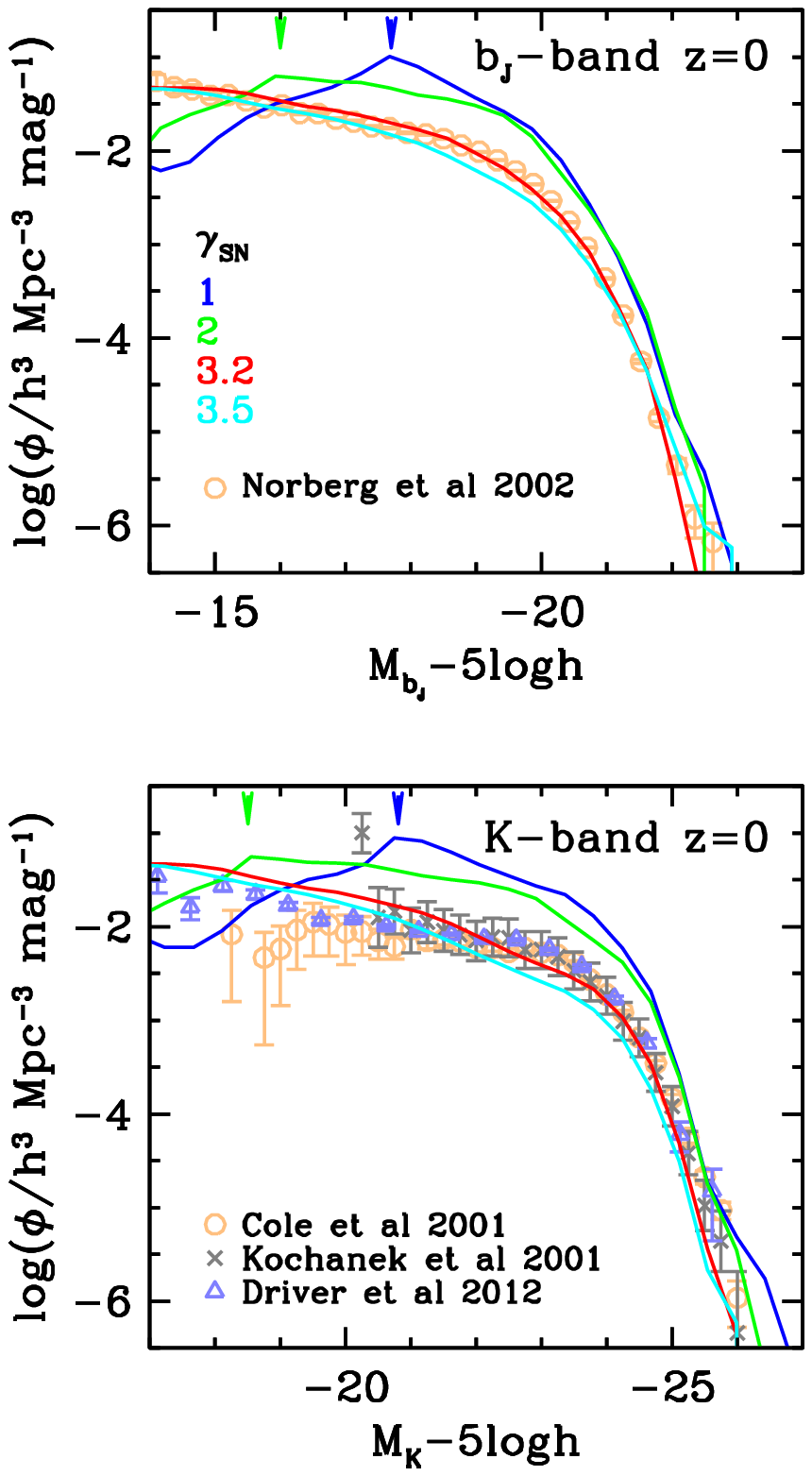}
\end{minipage}
\hspace{0.4cm}
\begin{minipage}{5.4cm}
\includegraphics[width=5.4cm, bb= 20 55 275 510]{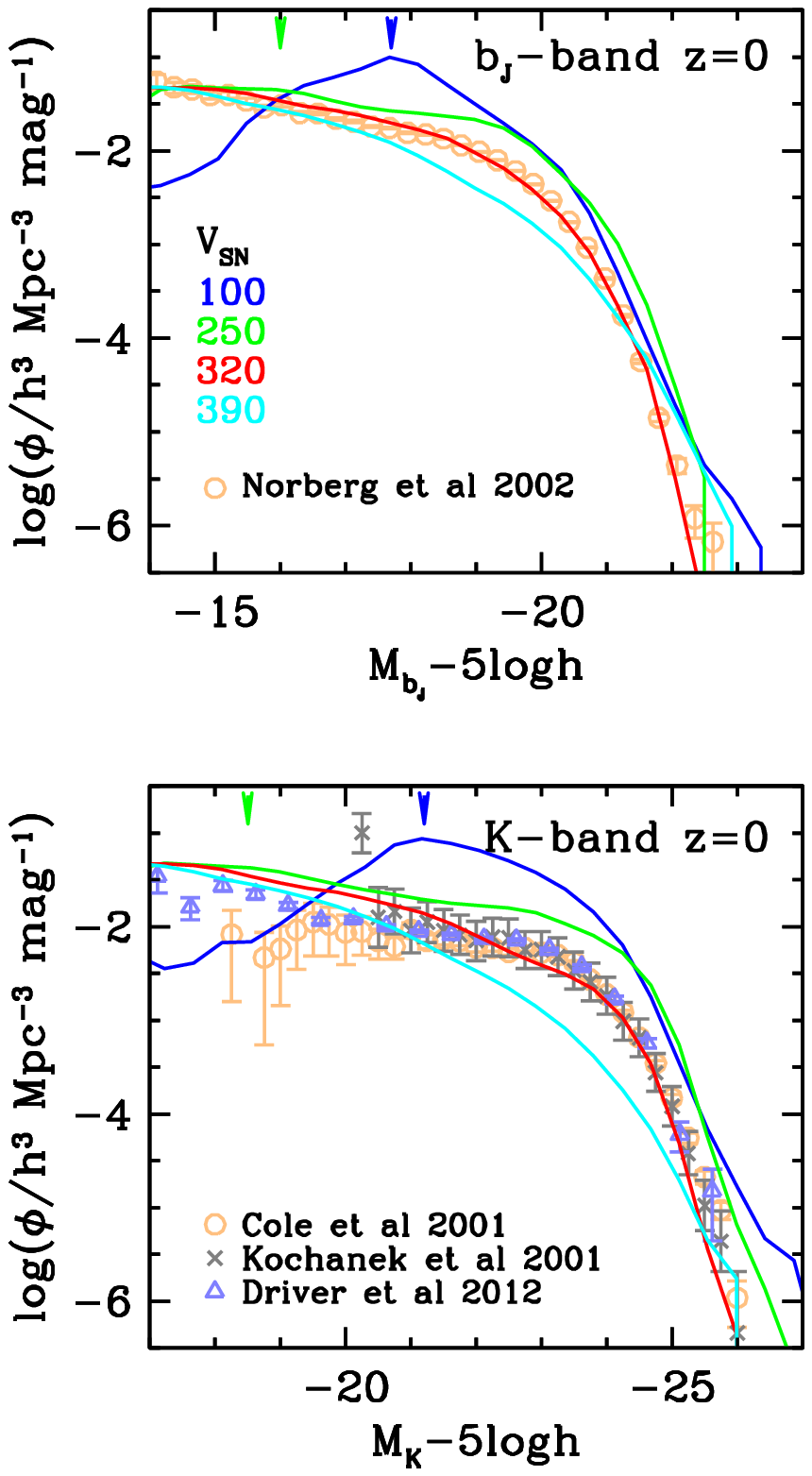}
\end{minipage}
\hspace{0.4cm}
\begin{minipage}{5.4cm}
\includegraphics[width=5.4cm, bb= 20 55 275 510]{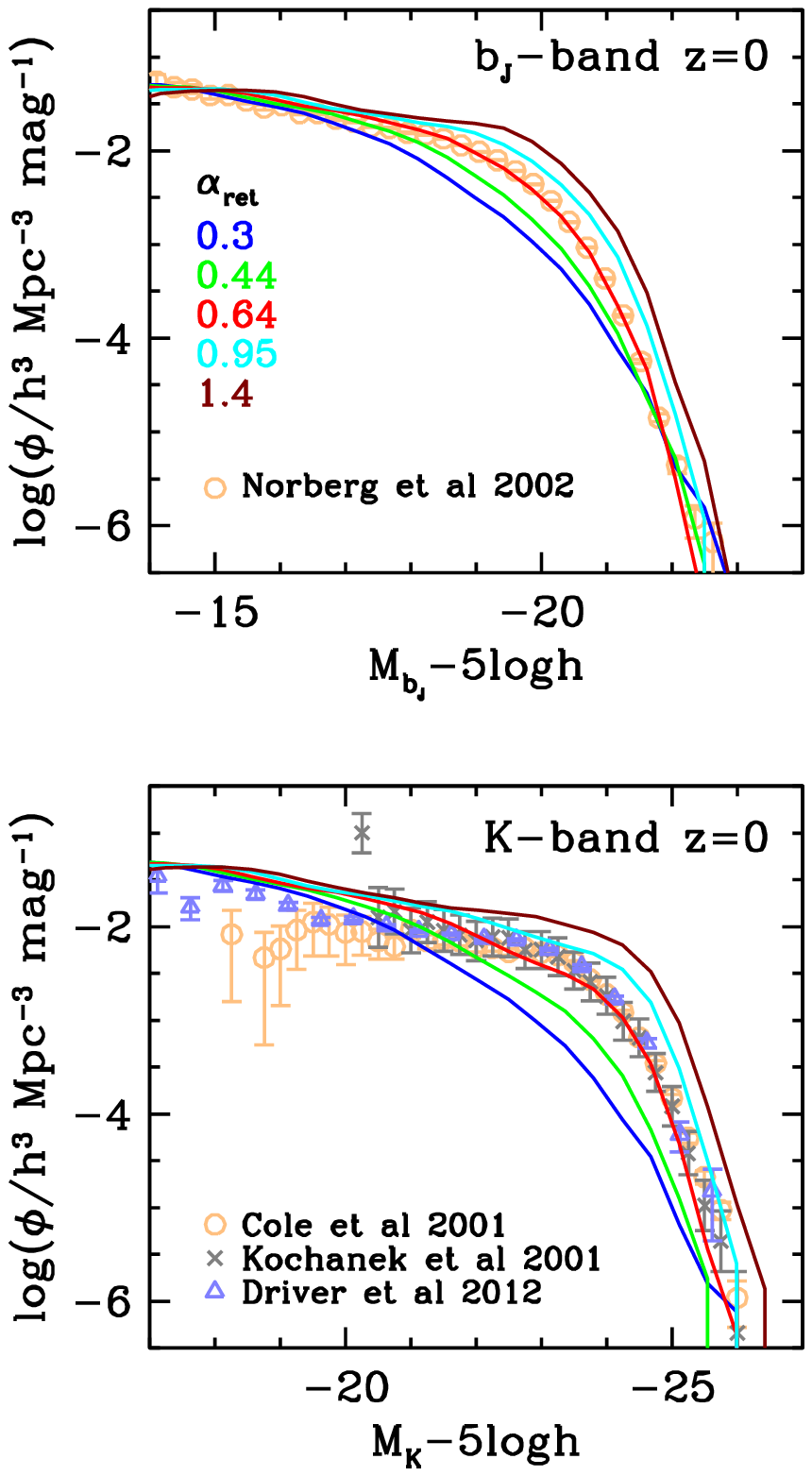}
\end{minipage}
\end{center}

\caption{Effects on the $b_J$- and $K$-band LFs at $z=0$ of varying
  the supernova feedback parameters $\gammaSN$ and $\VSN$ (left and
  middle columns) and the gas return parameter $\alpharet$ (right
  column). Only one parameter is varied in each column, and the values
  are given in the key in each panel, with the red curve showing the
  standard model in all cases. { The vertical arrows at the top of
    each panel indicate the luminosity below which the results for the
    corresponding model are affected by the halo mass resolution.} The
  observational data plotted are the same as in
  Fig.~\ref{fig:lf_default}.}

\label{fig:lf_SNfeedback}
\end{figure*}

\begin{figure*}

\begin{center}
\begin{minipage}{5.4cm}
\includegraphics[width=5.4cm, bb= 20 55 275 510]{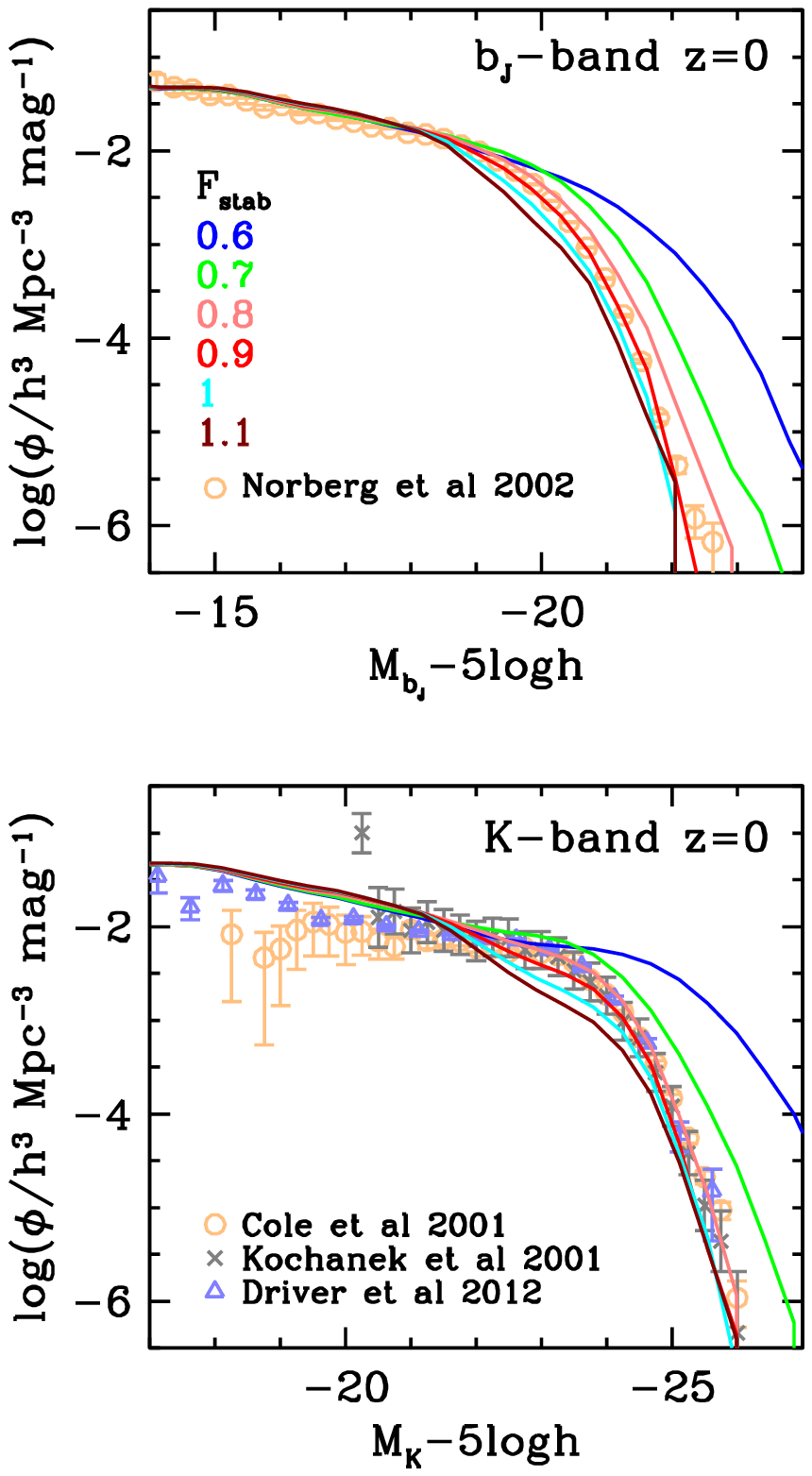}
\end{minipage}
\hspace{0.4cm}
\begin{minipage}{5.4cm}
\includegraphics[width=5.4cm, bb= 20 55 275 510]{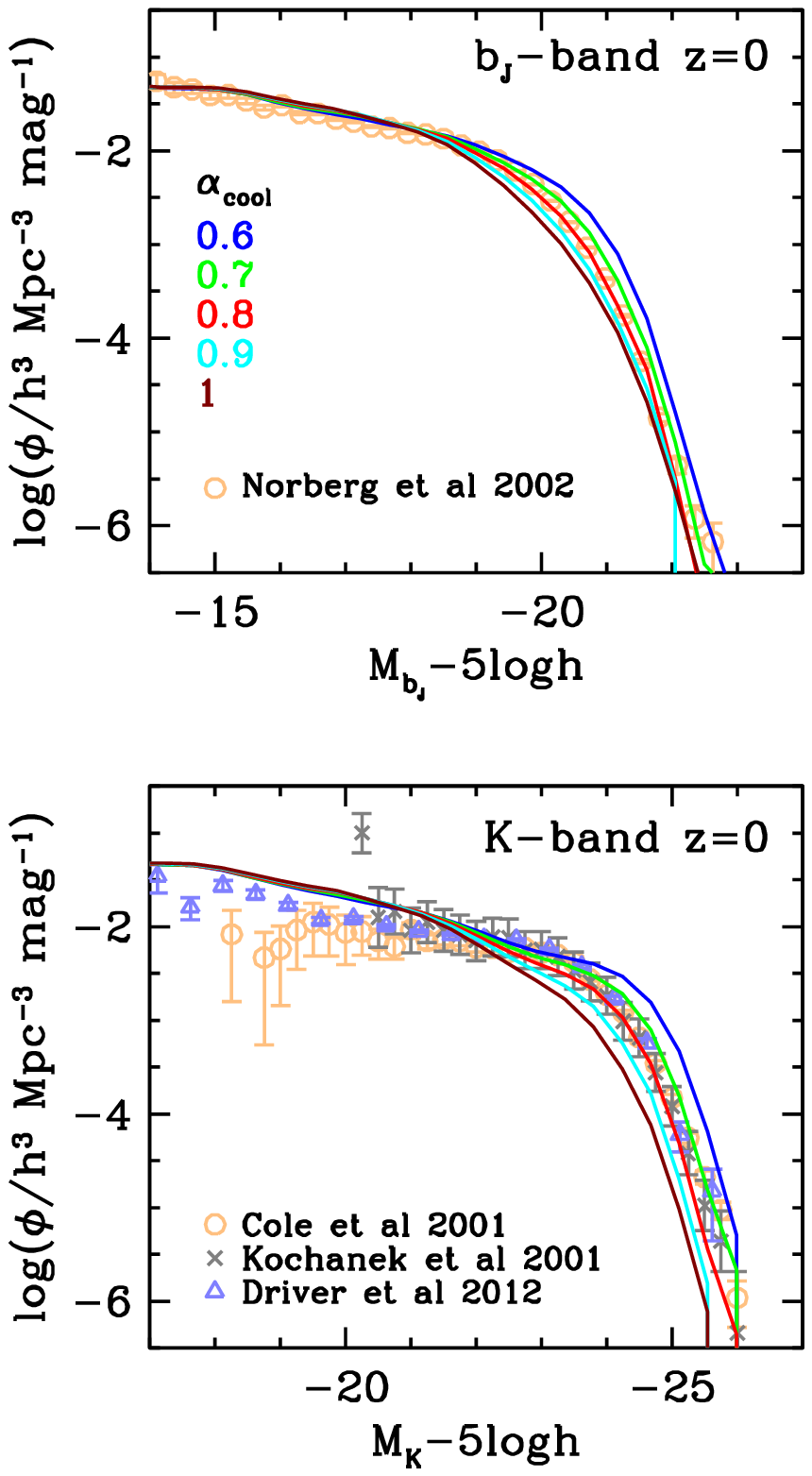}
\end{minipage}
\hspace{0.4cm}
\begin{minipage}{5.4cm}
\includegraphics[width=5.4cm, bb= 20 55 275 510]{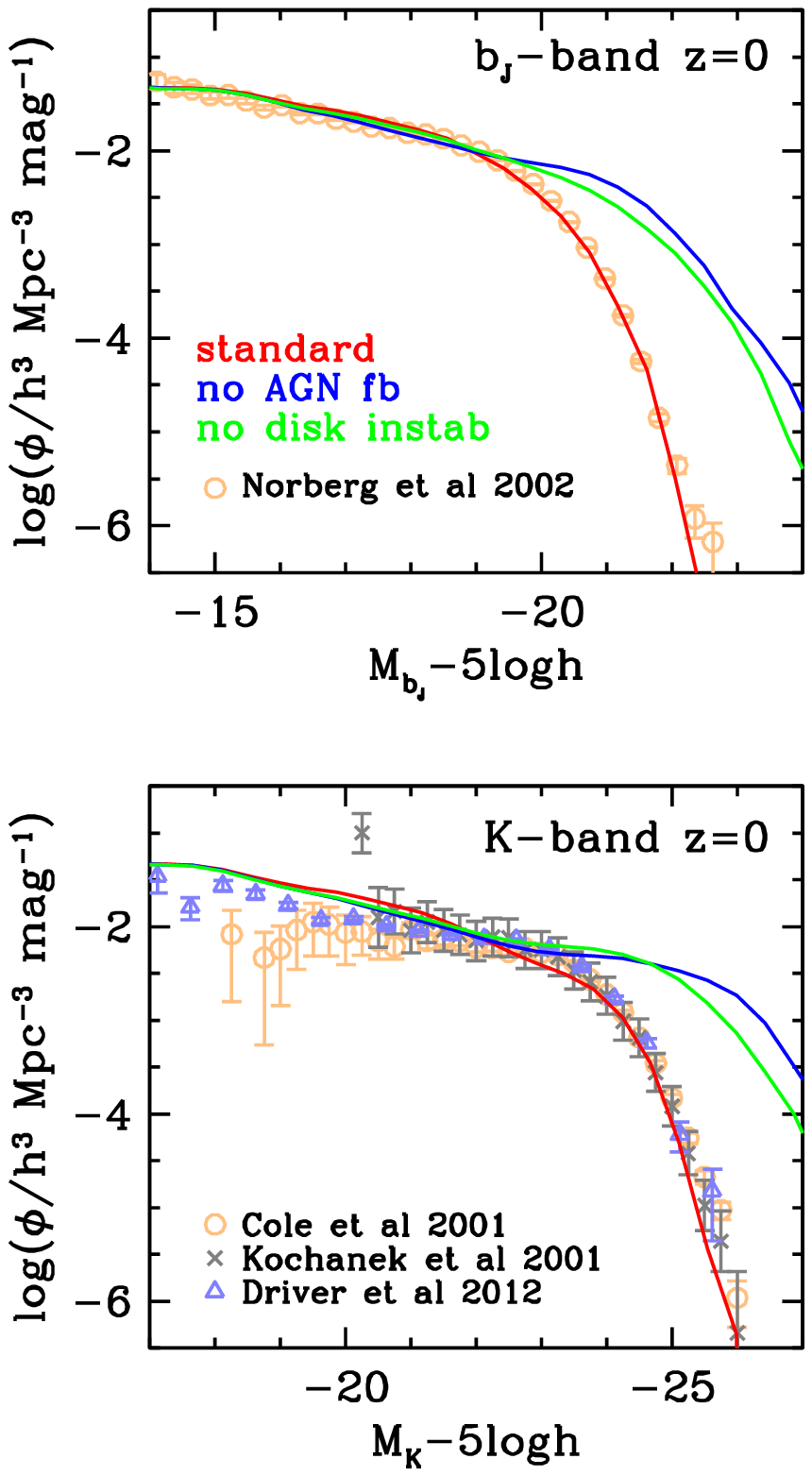}
\end{minipage}
\end{center}

\caption{Effects on the $b_J$- and $K$-band LFs at $z=0$ of varying
  (a) the disk stability parameter $\Fstab$ and (b) the AGN feedback
  parameters $\alphacool$, and (c) of turning off AGN feedback or disk
  instabilities, as shown by the key in each panel. A single parameter
  is varied in each column, with the red curves showing the standard
  model.}

\label{fig:lf_AGNfeedback}
\end{figure*}

\begin{figure*}
\begin{center}
\begin{minipage}{5.4cm}
\includegraphics[width=5.4cm, bb= 20 55 275 510]{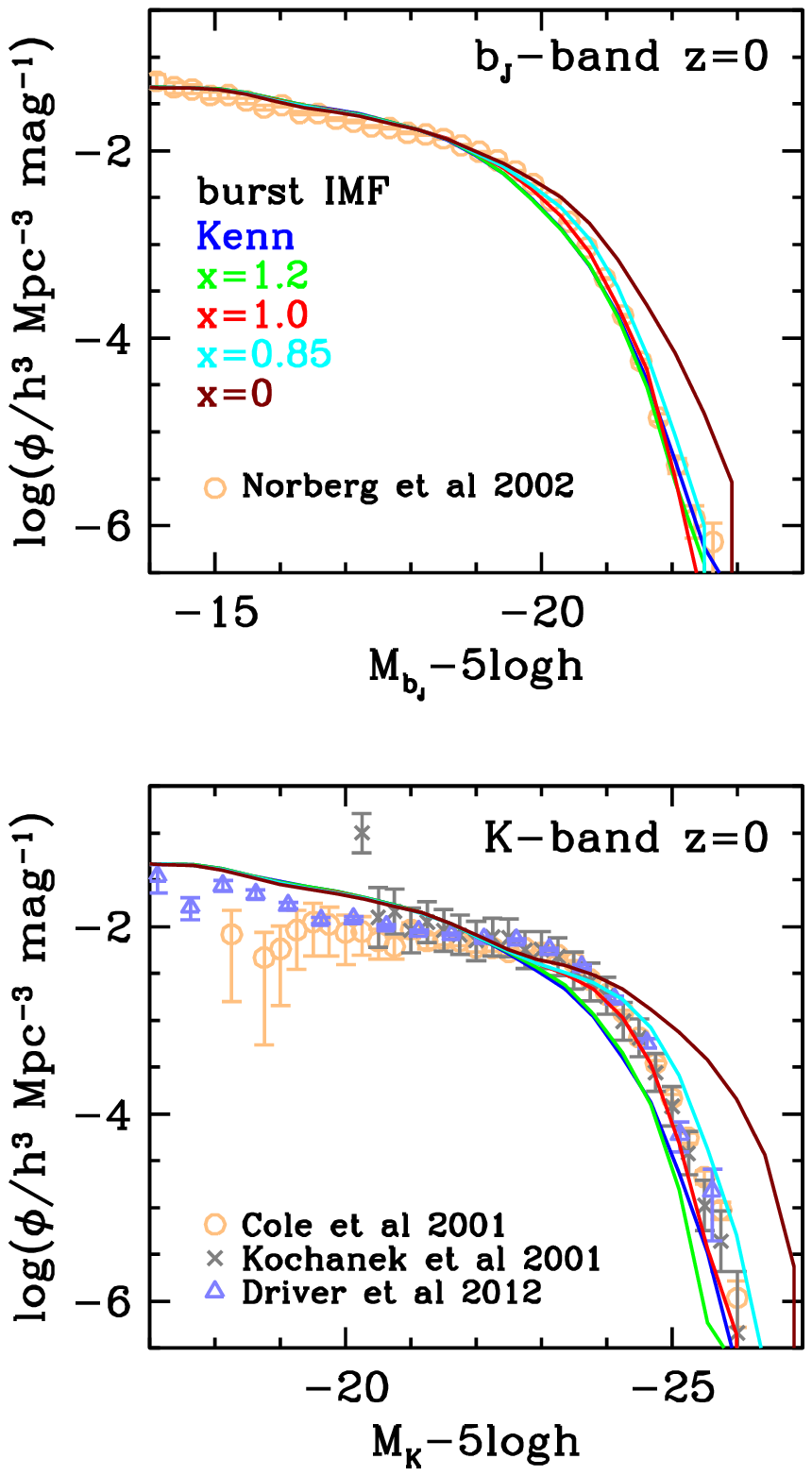}
\end{minipage}
\hspace{0.4cm}
\begin{minipage}{5.4cm}
\includegraphics[width=5.4cm, bb= 20 55 275 510]{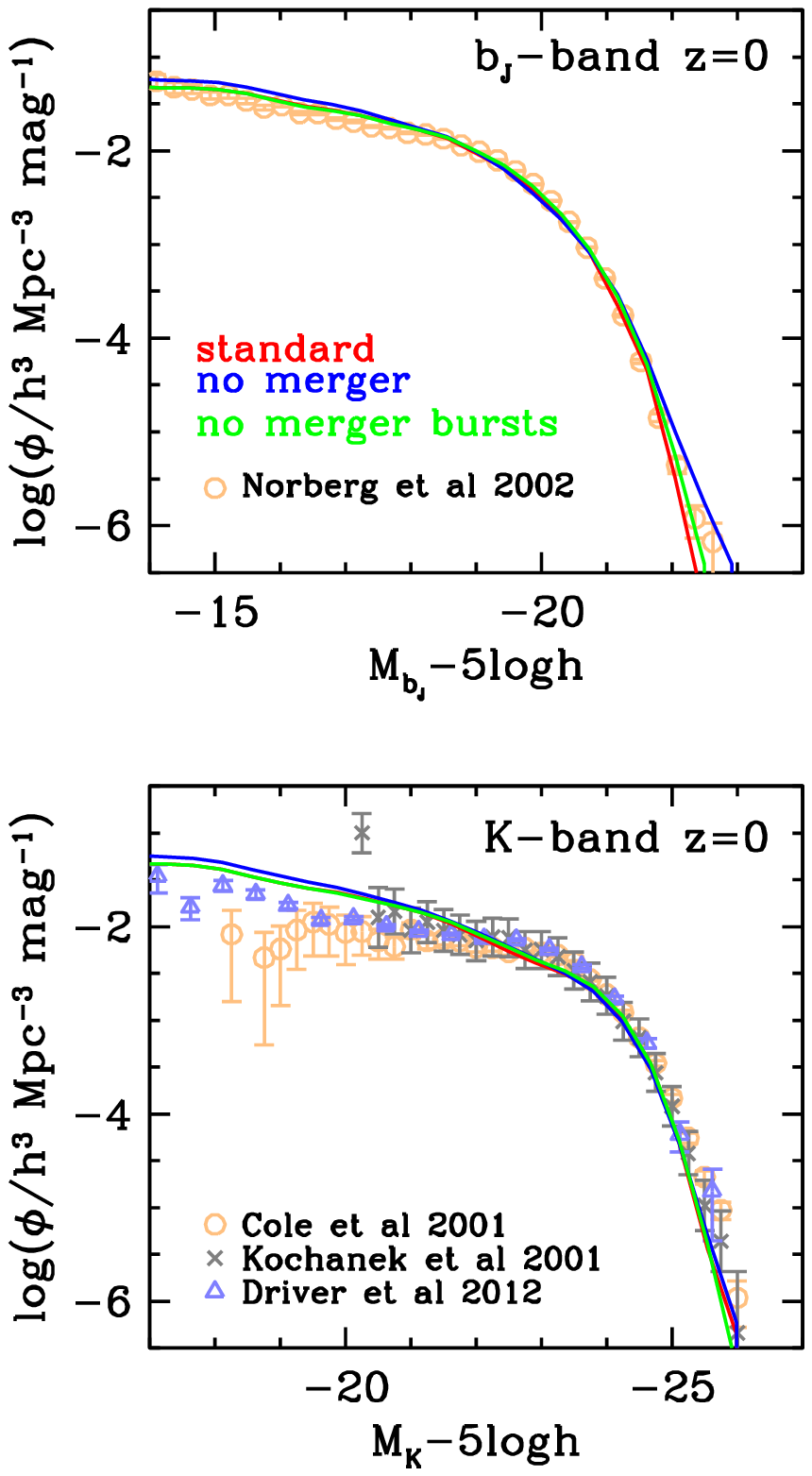}
\end{minipage}
\end{center}

\caption{Effects on the $b_J$- and $K$-band LFs at $z=0$ of (a)
  changing IMF in starbursts, and (b) turning off galaxy mergers or
  starbursts triggered by galaxy mergers, as shown by the key in each
  panel. A single parameter is varied in each column, with the red
  curves showing the standard model.}

\label{fig:lf_IMF_nomerge}
\end{figure*}

We show the effects on the $b_J$- and $K$-band LFs at $z=0$ of varying
the supernova feedback and gas return rate
(Fig.~\ref{fig:lf_SNfeedback}), disk instabilities and AGN feedback
(Fig.~\ref{fig:lf_AGNfeedback}), and the starburst IMF and galaxy
mergers (Fig.~\ref{fig:lf_IMF_nomerge}).

\subsection{$HI$ mass function at $z=0$}

\begin{figure*}

\begin{center}
\begin{minipage}{5.4cm}
\includegraphics[width=5.4cm, bb= 20 520 280 750]{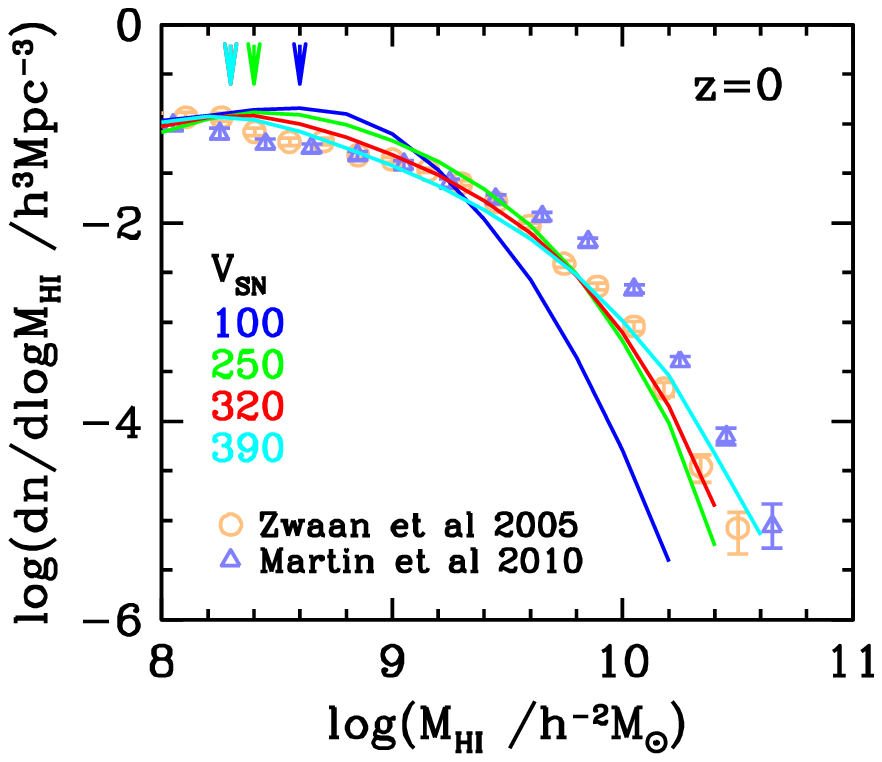}
\end{minipage}
\hspace{0.4cm}
\begin{minipage}{5.4cm}
\includegraphics[width=5.4cm, bb= 20 520 280 750]{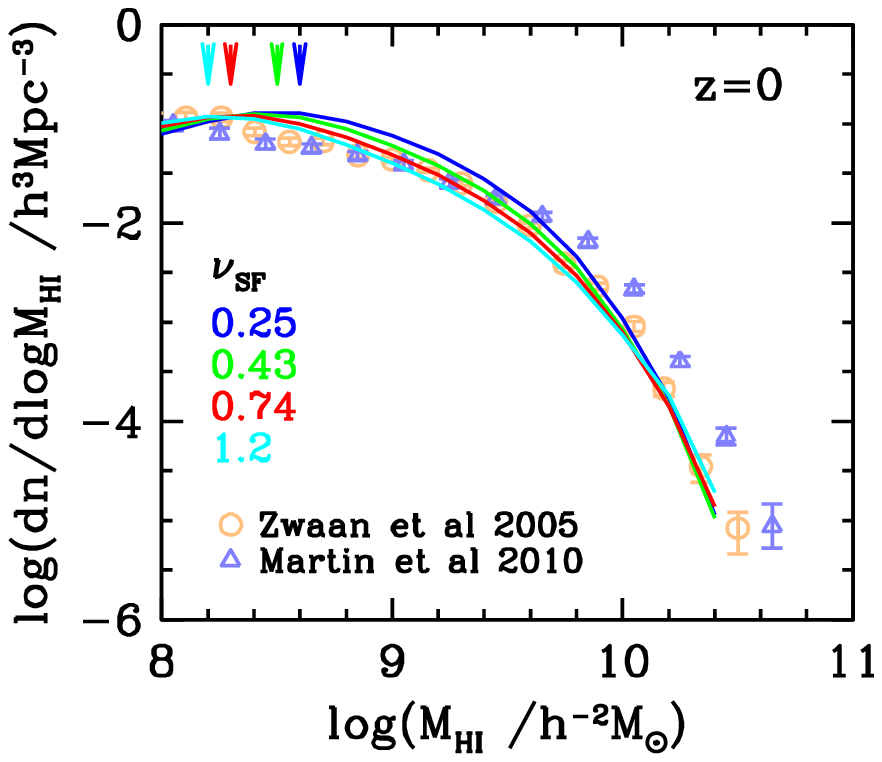}
\end{minipage}
\hspace{0.4cm}
\begin{minipage}{5.4cm}
\includegraphics[width=5.4cm, bb= 20 520 280 750]{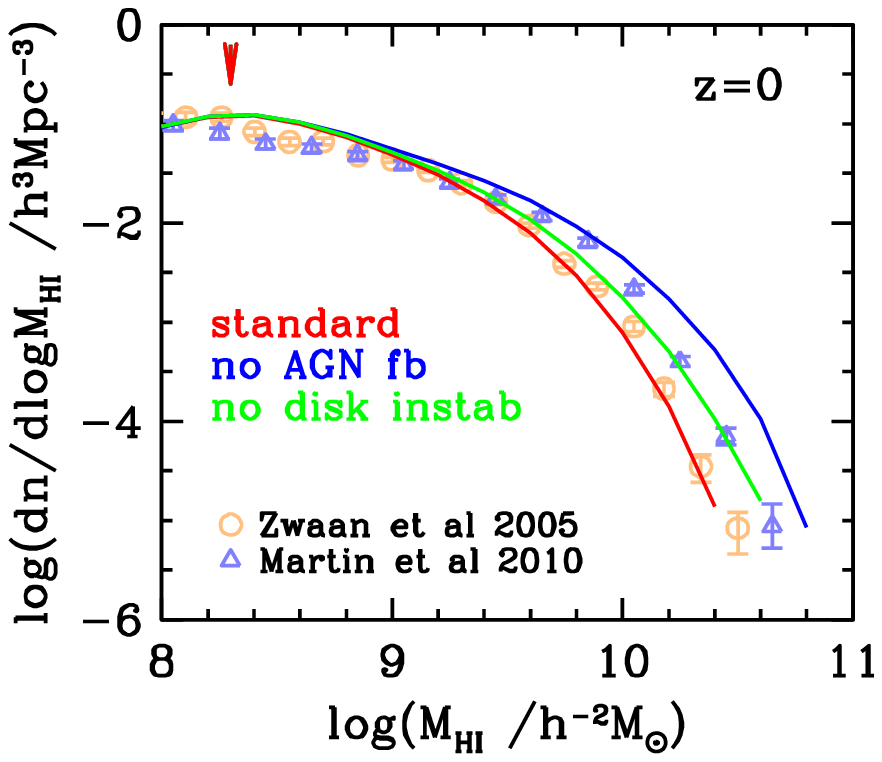}
\end{minipage}
\end{center}

\caption{Effects on the HI mass function at $z=0$ of varying (a) the
  strength of supernova feedback, specified by $\VSN$, and (b) the
  disk star formation rate, specified by $\nuSF$, and (c) of turning
  off AGN feedback or disk instabilities, as shown by the key in each
  panel. The red curves show the standard model. { The vertical
    arrows at the top of each panel indicate the HI mass below which
    the results for the corresponding model are affected by the halo
    mass resolution.}  The observational data plotted are the same as
  in Fig.~\ref{fig:mfHI_default}. }

\label{fig:mfHI_Vhot_SF_noAGN}
\end{figure*}

We show the effects on the $HI$ mass function at $z=0$ of varying the
normalization of the SN feedback, the normalization of the disk SFR
law, and of disk instabilities and AGN feedback
(Fig.~\ref{fig:mfHI_Vhot_SF_noAGN}). 


\subsection{Early vs late type morphological fractions at $z=0$}

\begin{figure*}

\begin{center}
\begin{minipage}{5.4cm}
\includegraphics[width=5.4cm, bb= 20 520 275 750]{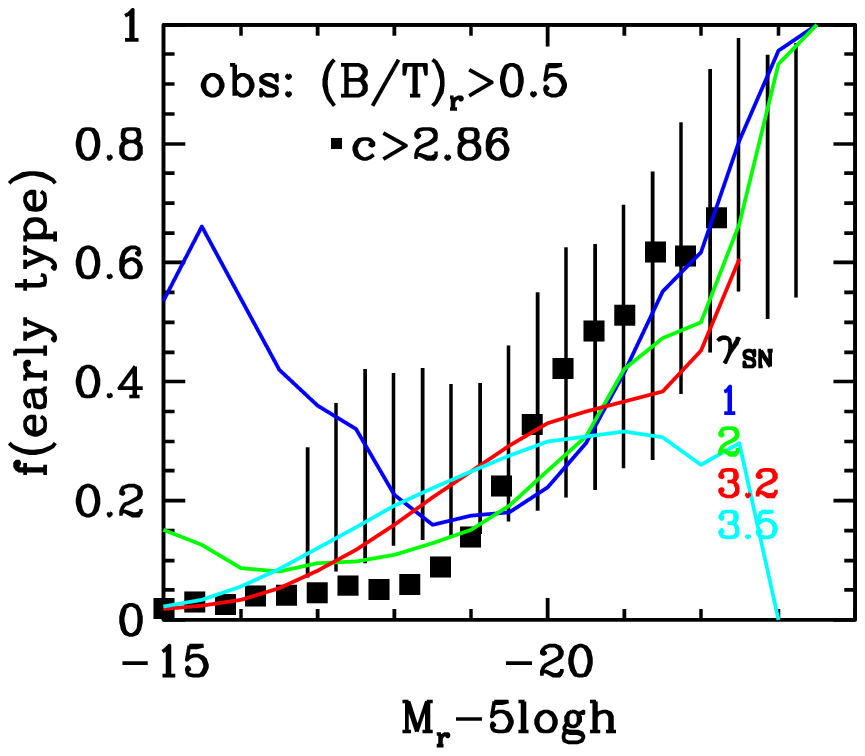}
\end{minipage}
\hspace{0.4cm}
\begin{minipage}{5.4cm}
\includegraphics[width=5.4cm, bb= 20 520 275 750]{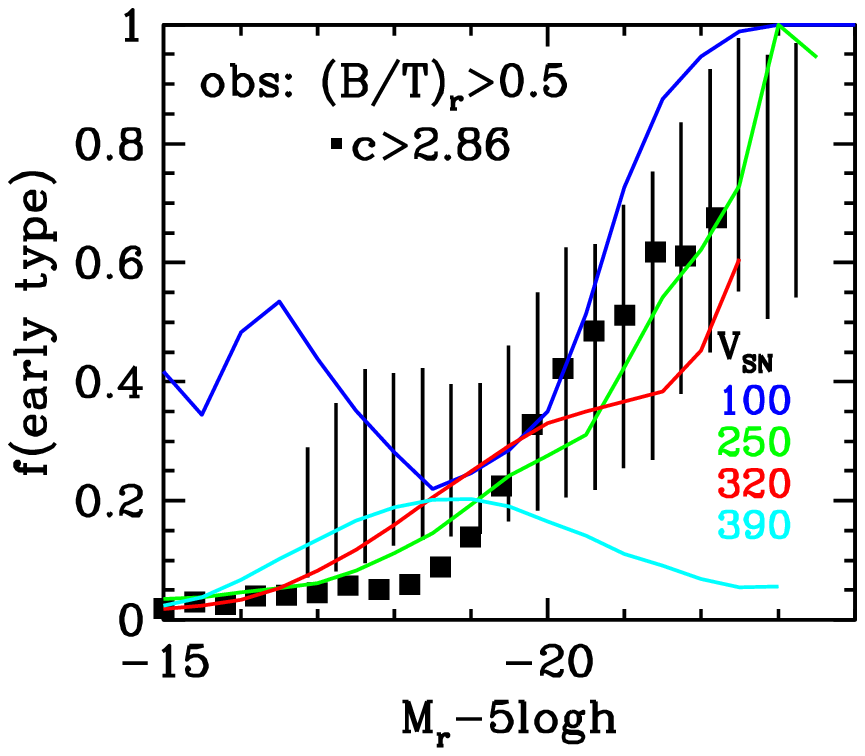}
\end{minipage}
\hspace{0.4cm}
\begin{minipage}{5.4cm}
\includegraphics[width=5.4cm, bb= 20 520 275 750]{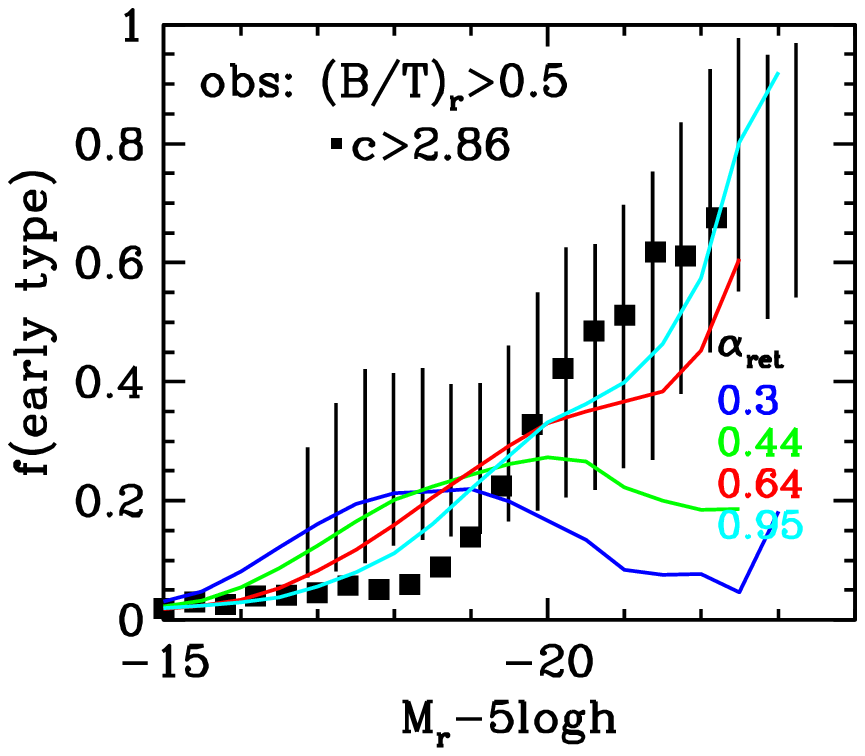}
\end{minipage}
\end{center}

\caption{Effects on the fraction of early-type galaxies at $z=0$ of
  varying the supernova feedback parameters $\gammaSN$ and $\VSN$ and
  the gas return parameter $\alpharet$. The red curves show the
  standard model. The definition of early-type galaxies in the model
  and the observational data plotted are the same as in
  Fig.~\ref{fig:morph_default}.}

\label{fig:morph_SNfeedback}
\end{figure*}

\begin{figure*}

\begin{center}

\begin{minipage}{5.4cm}
\includegraphics[width=5.4cm, bb= 20 520 275 750]{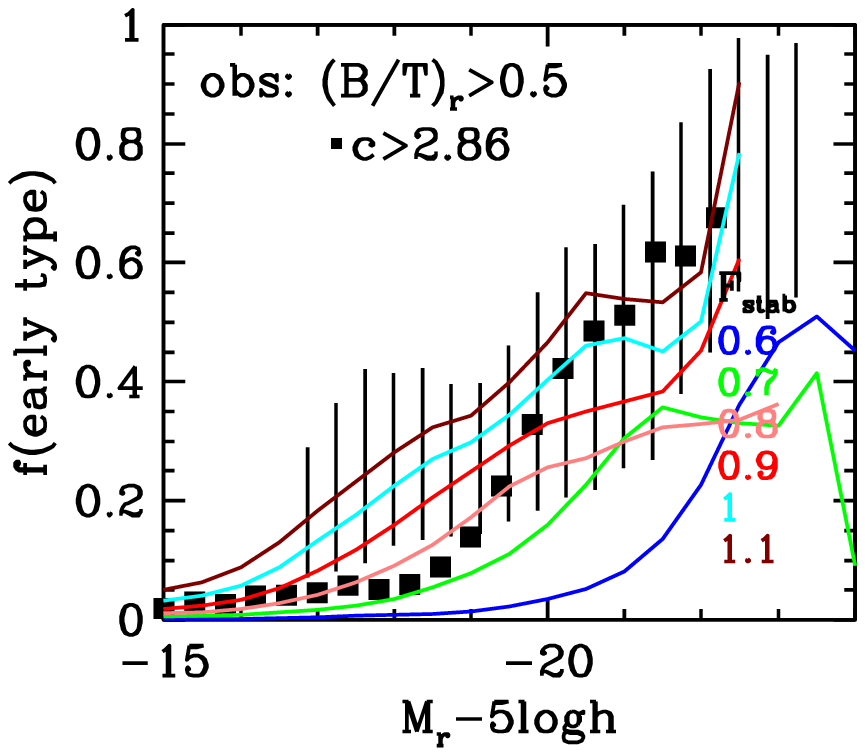}
\end{minipage}
\hspace{0.4cm}
\begin{minipage}{5.4cm}
\includegraphics[width=5.4cm, bb= 20 520 275 750]{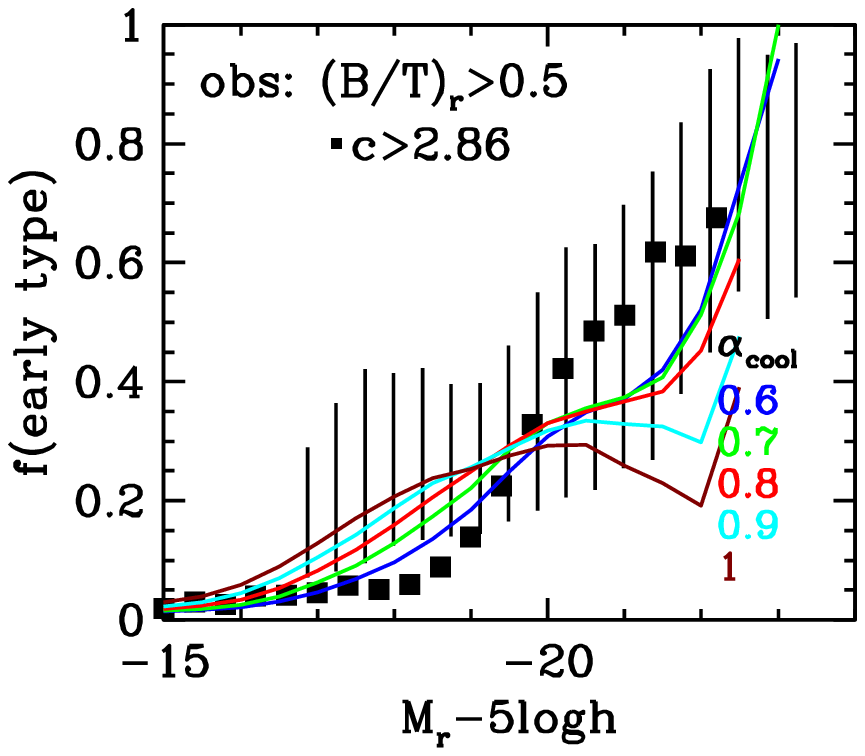}
\end{minipage}
\hspace{0.4cm}
\begin{minipage}{5.4cm}
\includegraphics[width=5.4cm, bb= 20 520 275 750]{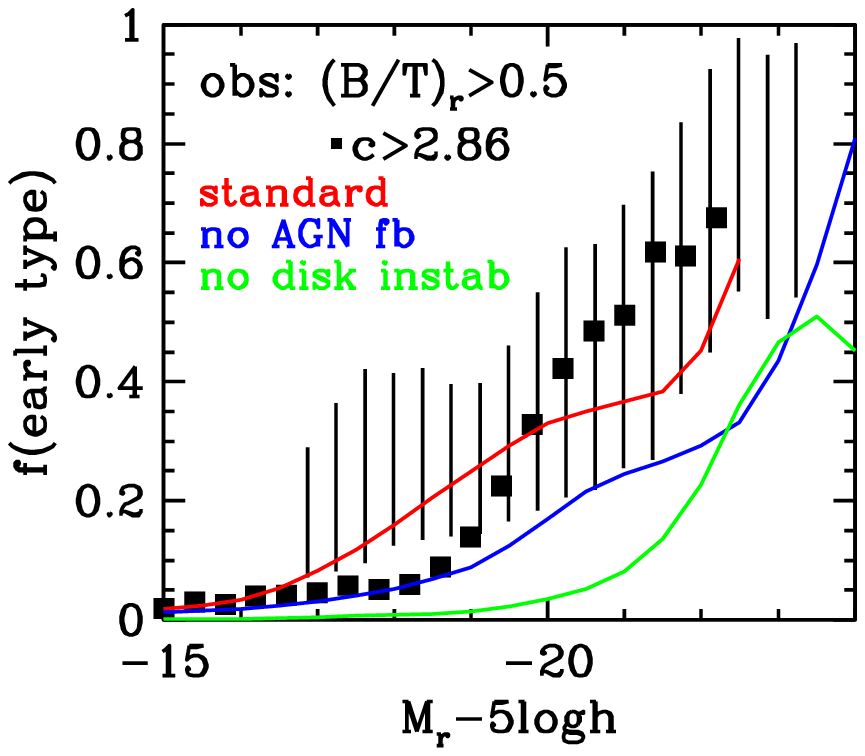}
\end{minipage}

\end{center}

\caption{Effects on the fraction of early-type galaxies at $z=0$ of
  varying (a) the disk stability $\Fstab$ and (b) the AGN feedback
  parameter $\alphacool$, and (c) of turning off AGN feedback or disk
  instabilities. The red curves show the standard model.}

\label{fig:morph_AGNfeedback}
\end{figure*}

\begin{figure*}
\begin{center}
\begin{minipage}{5.4cm}
\includegraphics[width=5.4cm, bb= 20 520 275 750]{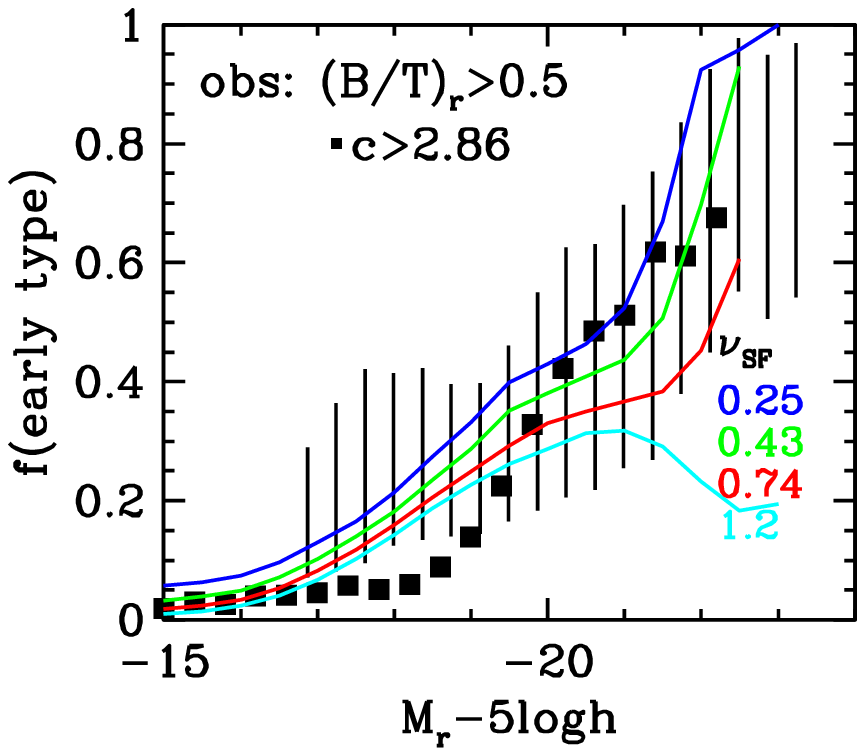}
\end{minipage}
\hspace{0.4cm}
\begin{minipage}{5.4cm}
\includegraphics[width=5.4cm, bb= 20 520 275 750]{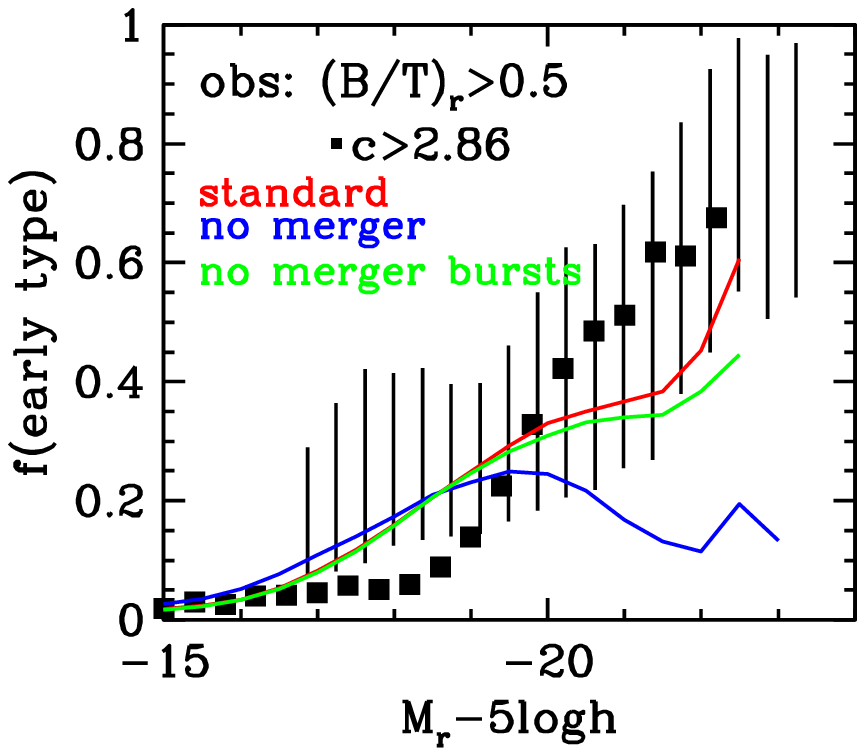}
\end{minipage}
\end{center}

\caption{Effects on the fraction of early-type galaxies at $z=0$ of
  (a) varying the disk star formation rate parameter $\nuSF$ and (b)
  turning off galaxy mergers or starbursts triggered by galaxy
  mergers. The red curves show the standard model.}

\label{fig:morph_SF_nomerge}
\end{figure*}

We show the effects on the fraction of early-type galaxies as a
function of lumiminosity at $z=0$ of varying the supernova feedback
and gas return rate (Fig.~\ref{fig:morph_SNfeedback}), disk instabilities and
AGN feedback (Fig.~\ref{fig:morph_AGNfeedback}), and the normalization of the
disk SFR law and of galaxy mergers (Fig.~\ref{fig:morph_SF_nomerge}).

\subsection{SMBH vs bulge mass relation at $z=0$}

\begin{figure*}

\begin{center}
\begin{minipage}{5.4cm}
\includegraphics[width=5.4cm, bb= 20 520 275 750]{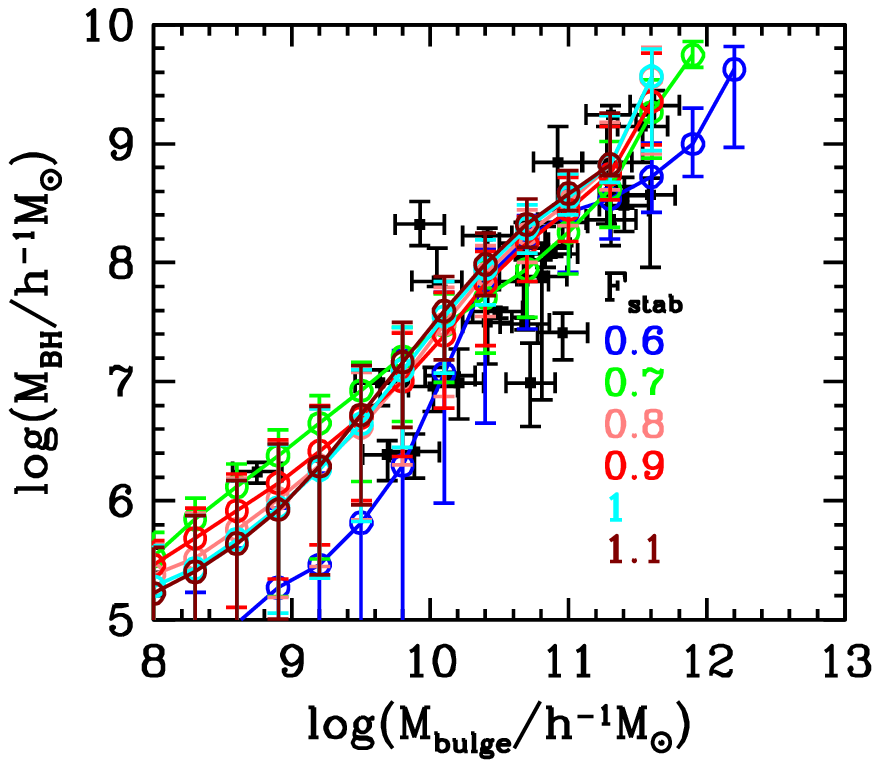}
\end{minipage}
\hspace{0.4cm}
\begin{minipage}{5.4cm}
\includegraphics[width=5.4cm, bb= 20 520 275 750]{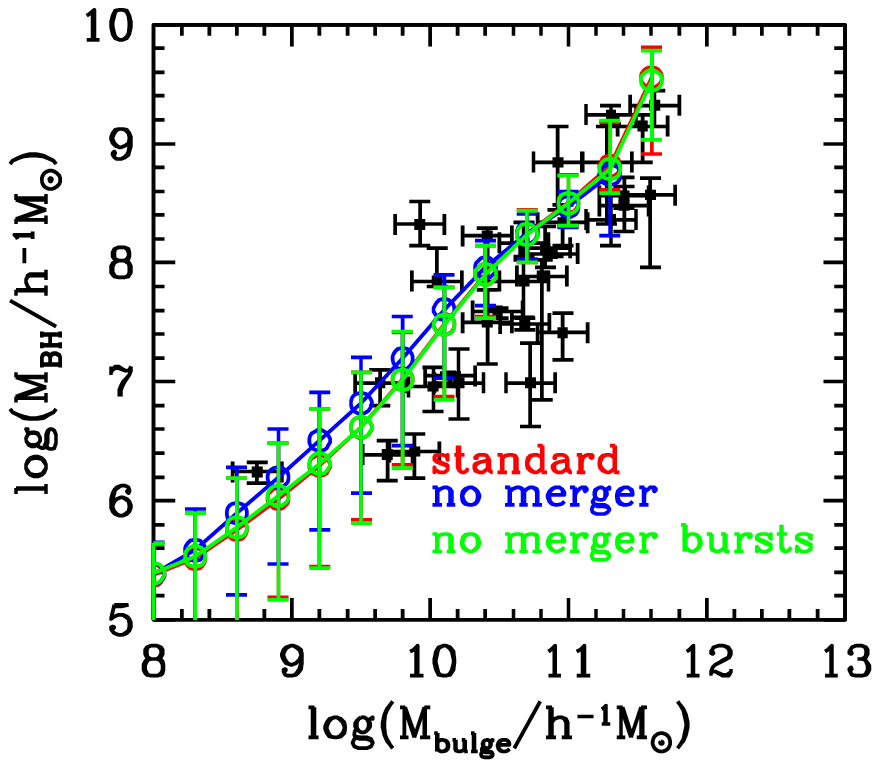}
\end{minipage}
\end{center}

\caption{Effects on the SMBH - bulge relation at $z=0$ of (a) varying
  the disk stability parameter $\Fstab$ and (b) turning off galaxy
  mergers or starbursts triggered by galaxy mergers. The red curves
  show the standard model. The observational data plotted are the same
  as in Fig.~\ref{fig:SMBH_default}.}

\label{fig:SMBH_AGNfeedback_nomerge}
\end{figure*}

We show the effects on the SMBH vs bulge mass relation at $z=0$ of
varying the disk stability threshold and of galaxy mergers
(Fig.~\ref{fig:SMBH_AGNfeedback_nomerge}).

\subsection{Tully-Fisher relation at $z=0$}

\begin{figure*}

\begin{center}
\begin{minipage}{5.4cm}
\includegraphics[width=5.5cm, bb= 270 290 530 510]{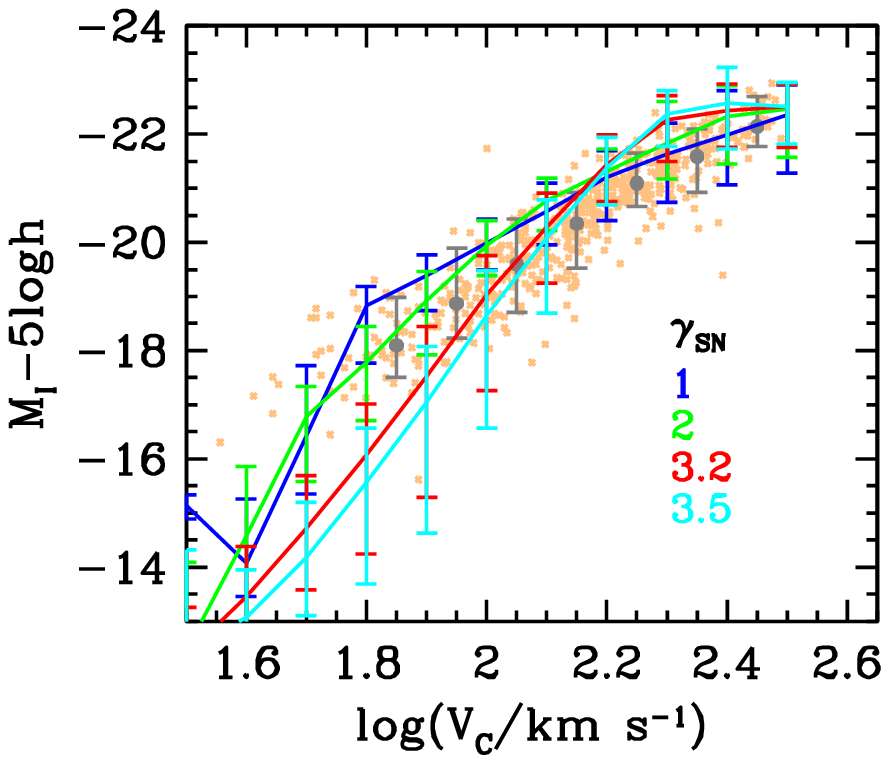}
\end{minipage}
\hspace{0.4cm}
\begin{minipage}{5.4cm}
\includegraphics[width=5.4cm, bb= 270 290 530 510]{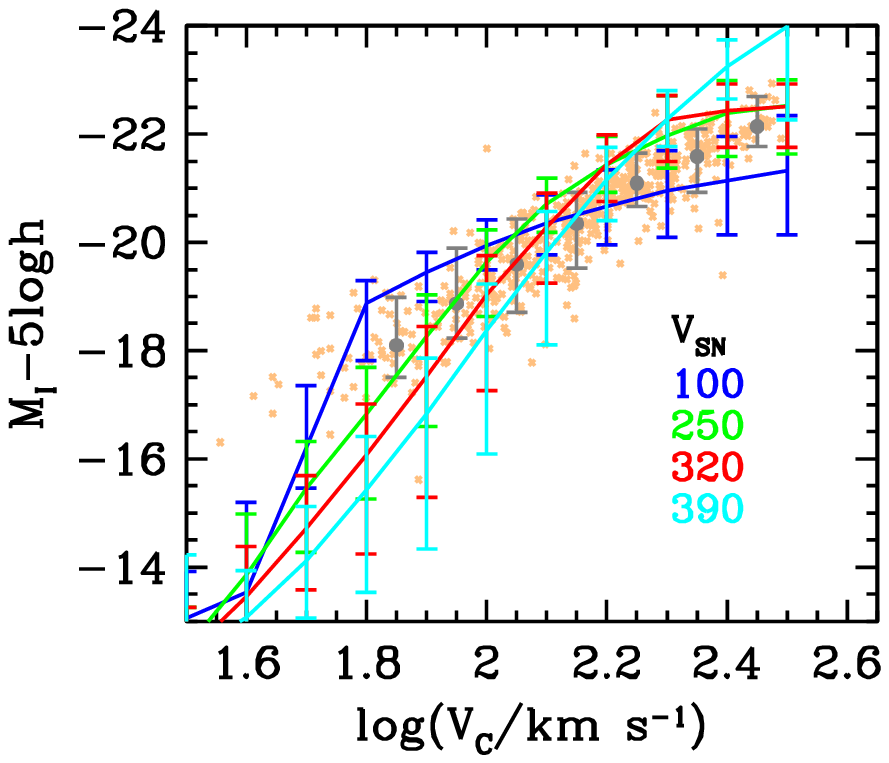}
\end{minipage}
\end{center}

\caption{Effects on the $I$-band Tully-Fisher relation at $z=0$ of
  varying the supernova feedback parameters $\gammaSN$ and $\VSN$.
  The red curves show the standard model. The observational data
  plotted are the same as in Fig.~\ref{fig:TF_default}.}

\label{fig:TF_SNfeedback}
\end{figure*}

We show the effects on the Tully-Fisher relation at $z=0$ of varying
the slope and amplitude of the SN feedback
(Fig.~\ref{fig:TF_SNfeedback}). 

\subsection{Galaxy sizes at $z=0$}

\begin{figure*}

\begin{center}
\begin{minipage}{5.4cm}
\includegraphics[width=5.4cm, bb= 10 285 280 750]{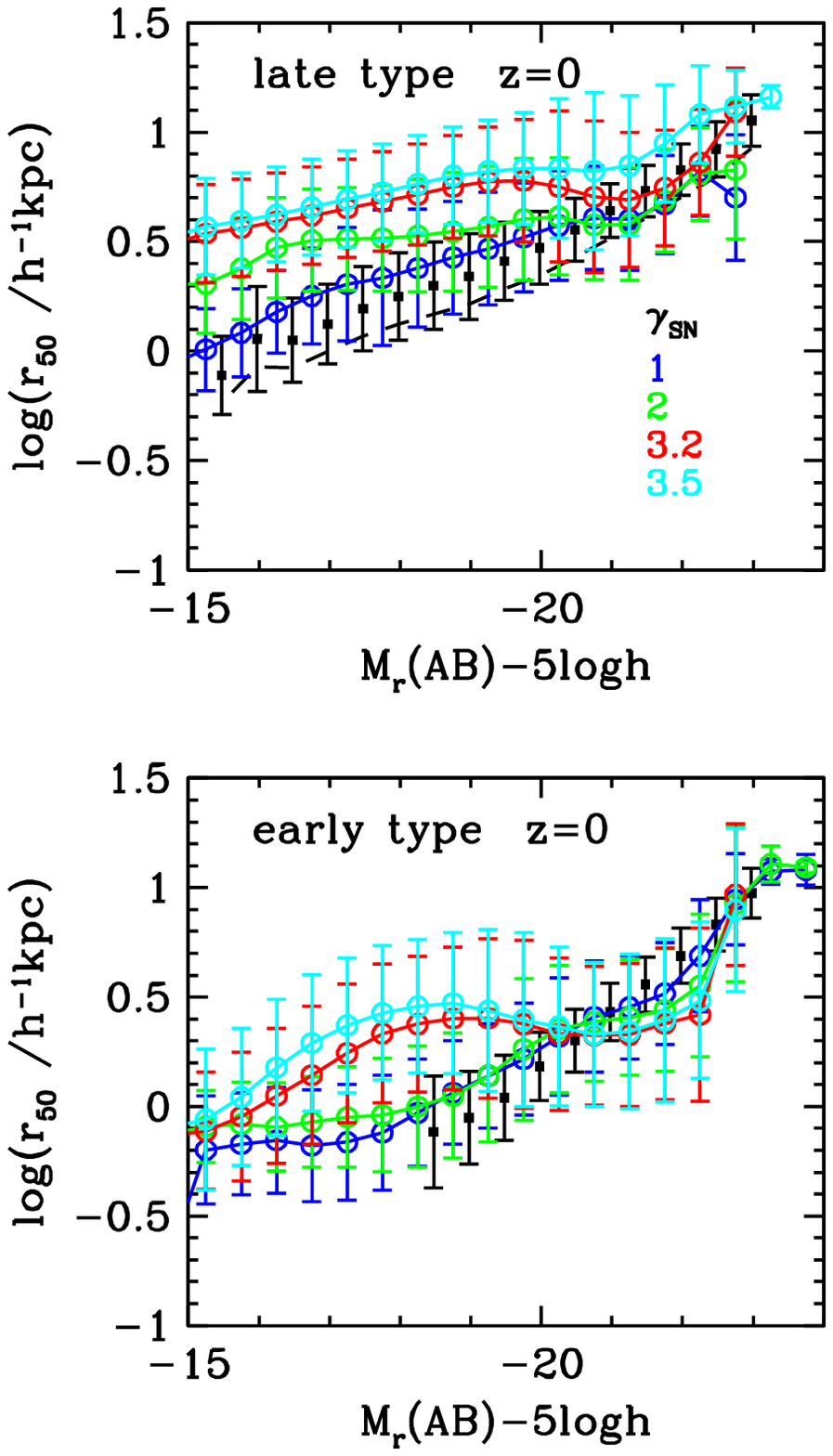}
\end{minipage}
\hspace{0.4cm}
\begin{minipage}{5.4cm}
\includegraphics[width=5.4cm, bb= 10 285 280 750]{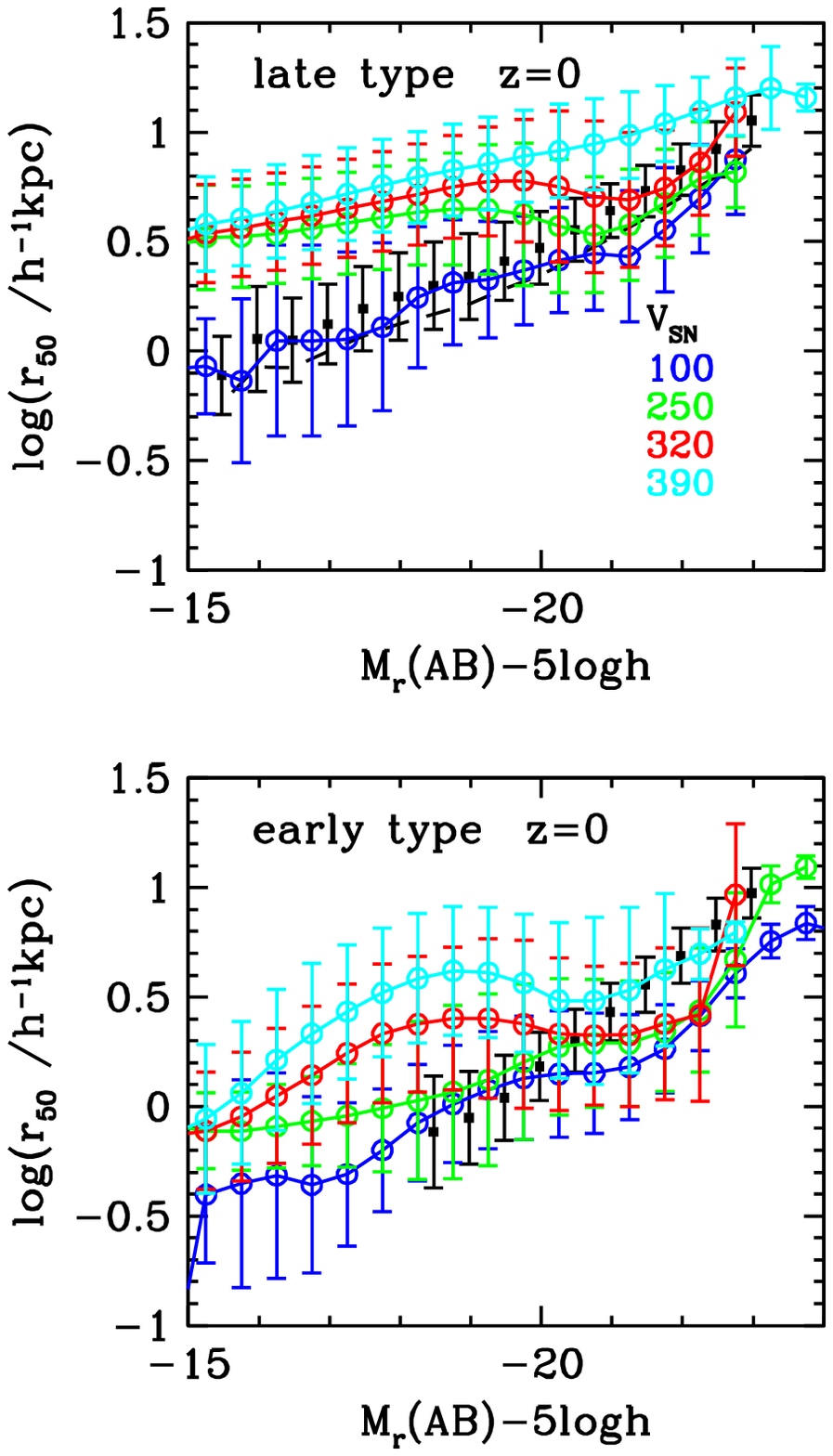}
\end{minipage}
\end{center}

\caption{Effects on the half-light radii of late- and early-type
  galaxies at $z=0$ of varying the supernova feedback parameters
  $\gammaSN$ and $\VSN$. A single parameter is varied in each column,
  with the red curves showing the standard model. The observational
  data plotted are the same as in Fig.~\ref{fig:sizes_default}.}

\label{fig:sizes_SNfeedback}
\end{figure*}

\begin{figure*}
\begin{center}
\begin{minipage}{5.4cm}
\includegraphics[width=5.4cm, bb= 10 285 280 750]{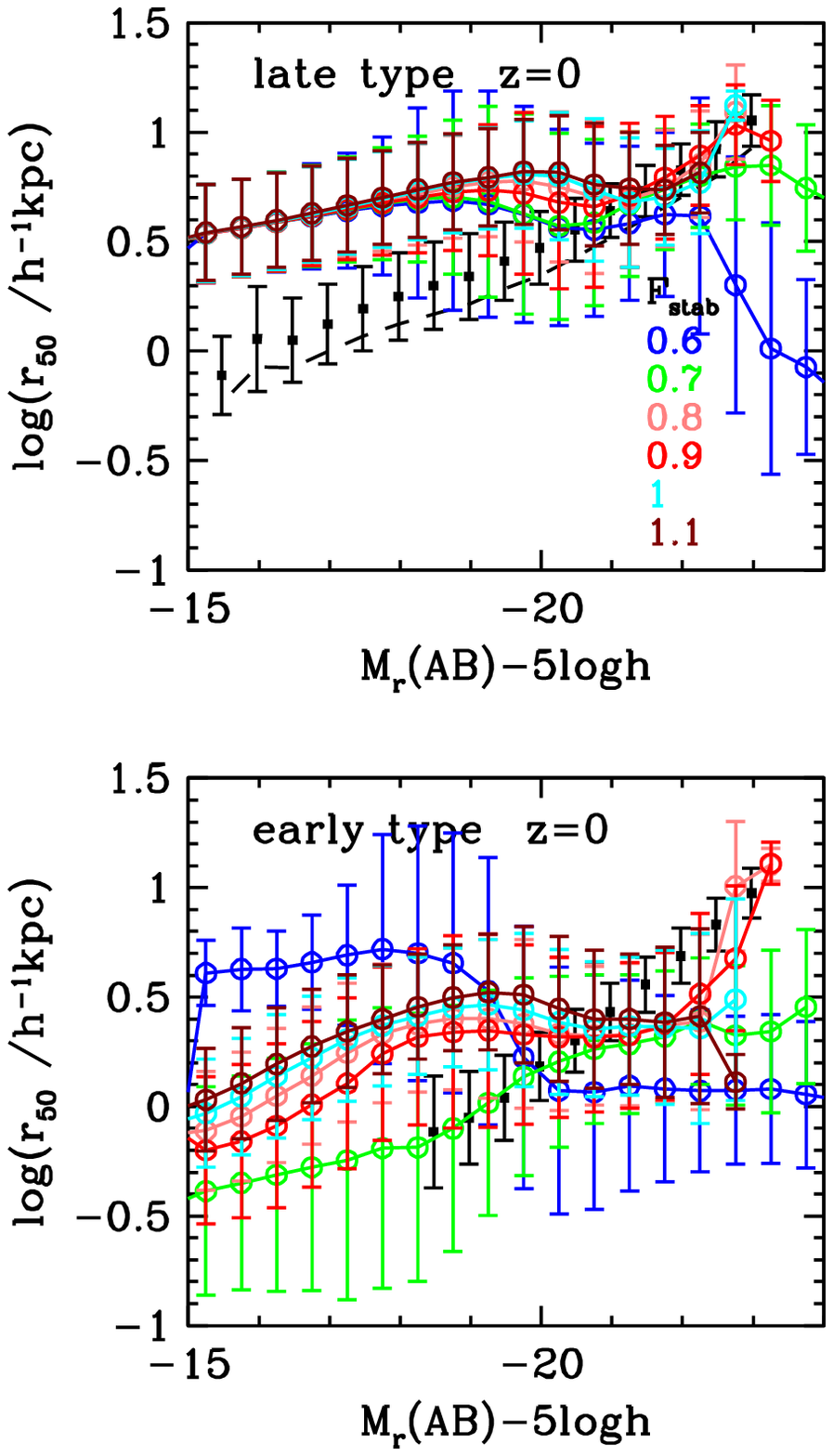}
\end{minipage}
\hspace{0.4cm}
\begin{minipage}{5.4cm}
\includegraphics[width=5.4cm, bb= 10 285 280 750]{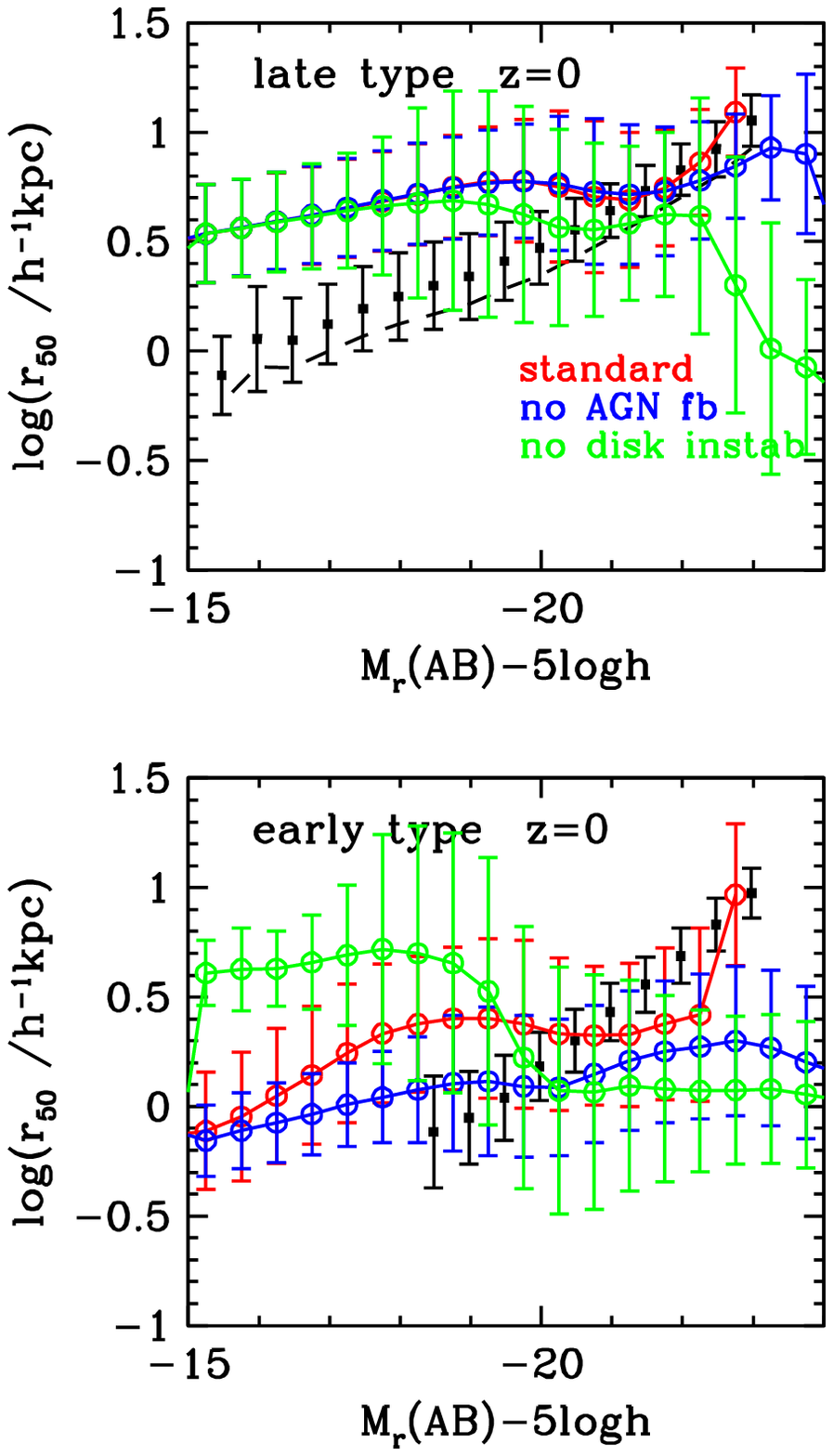}
\end{minipage}
\end{center}

\caption{Effects on the half-light radii of late- and early-type
  galaxies at $z=0$ of (a) varying the disk stability parameter
  $\Fstab$, and (b) of turning off AGN feedback or disk instabilities.
  A single parameter is varied in each column, with the red curves
  showing the standard model.}

\label{fig:sizes_AGNfeedback}
\end{figure*}

We show the effects on the size-luminosity relations of late- and
early-type galaxies at $z=0$ of varying the supernova feedback
(Fig.~\ref{fig:sizes_SNfeedback}) and of disk instabilities and AGN
feedback (Fig.~\ref{fig:sizes_AGNfeedback}).

\subsection{Stellar metallicities at $z=0$}

\begin{figure*}

\begin{center}
\begin{minipage}{5.4cm}
\includegraphics[width=5.4cm, bb= 10 520 275 750]{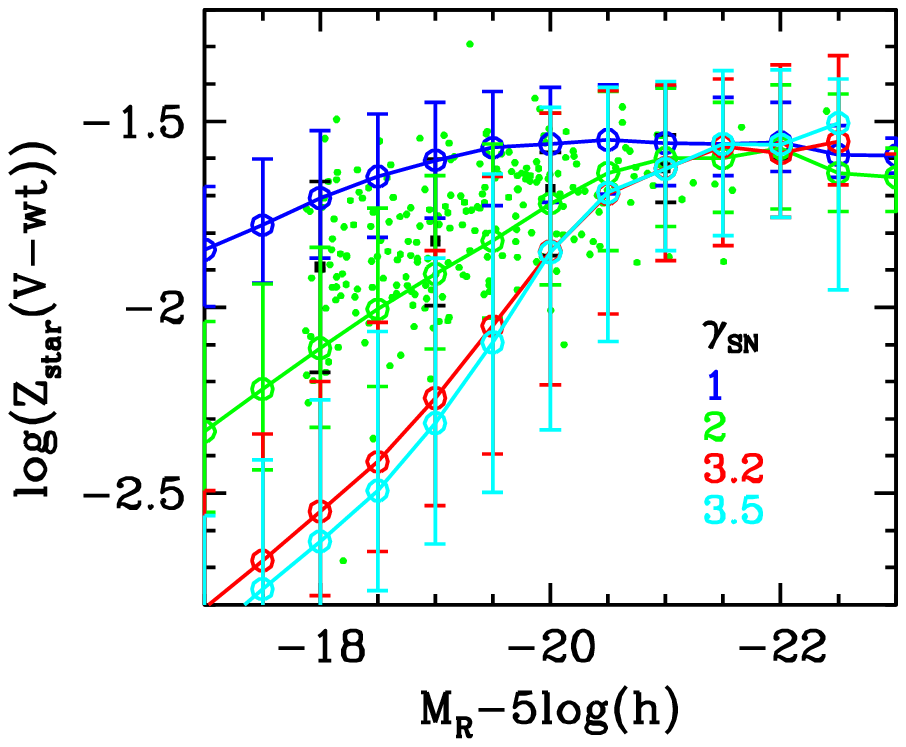}
\end{minipage}
\hspace{0.4cm}
\begin{minipage}{5.4cm}
\includegraphics[width=5.4cm, bb= 10 520 275 750]{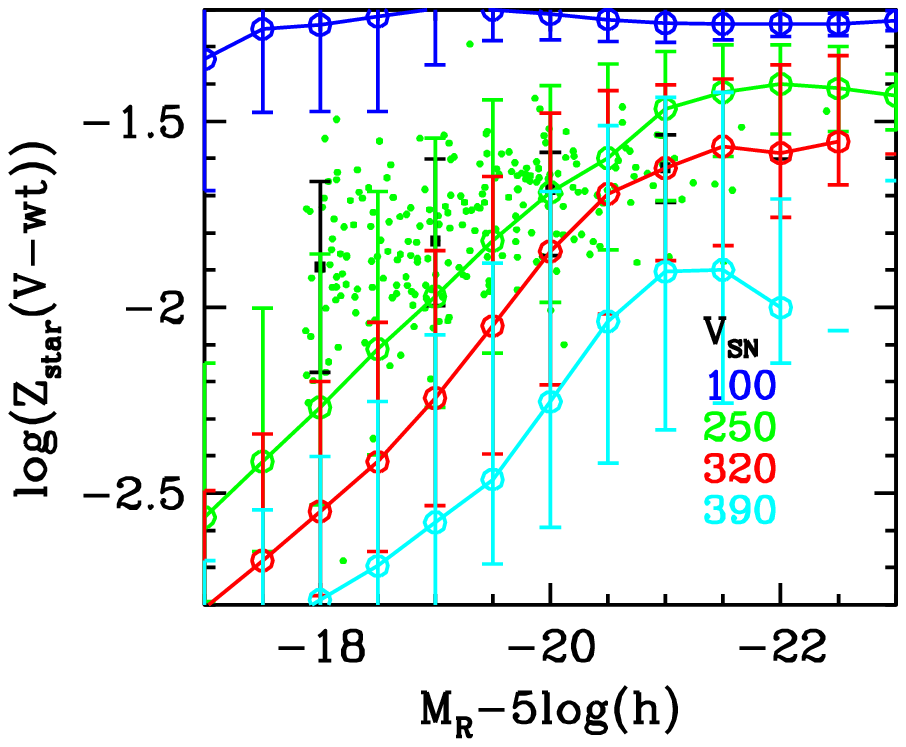}
\end{minipage}
\hspace{0.4cm}
\begin{minipage}{5.4cm}
\includegraphics[width=5.4cm, bb= 10 520 275 750]{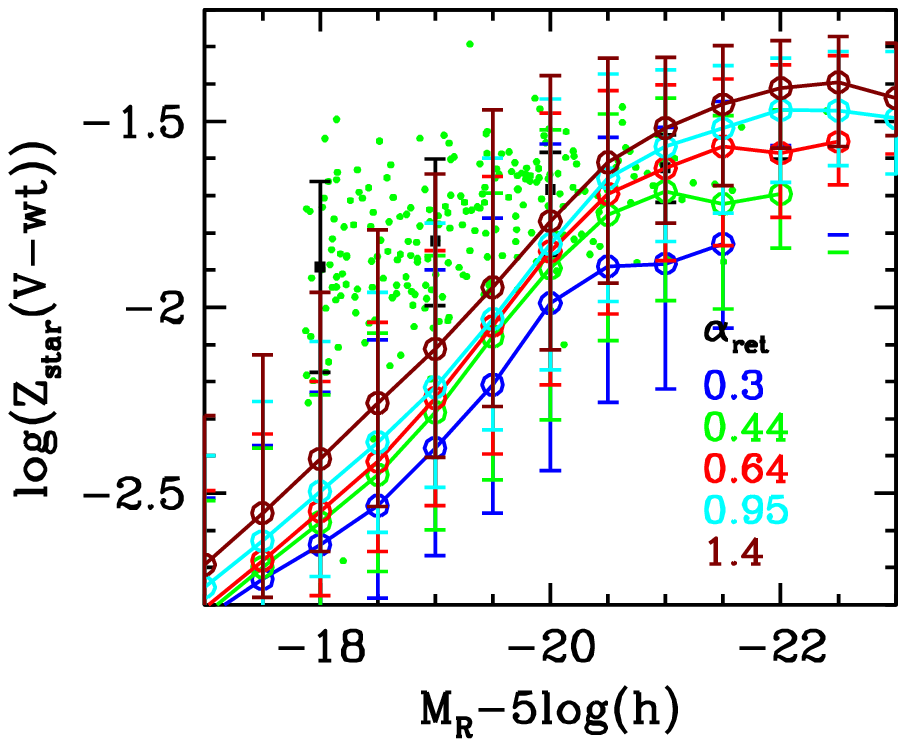}
\end{minipage}
\end{center}

\caption{Effects on the stellar metallicity in early-type galaxies at
  $z=0$ of varying the supernova feedback parameters $\gammaSN$ and
  $\VSN$, and the gas return timescale parameter $\alpharet$. The red
  curves show the standard model. The observational data plotted are
  the same as in Fig.~\ref{fig:Zstar_default}.}

\label{fig:Zstar_SNfeedback}
\end{figure*}

\begin{figure*}
\begin{center}
\begin{minipage}{5.4cm}
\includegraphics[width=5.4cm, bb= 10 520 275 750]{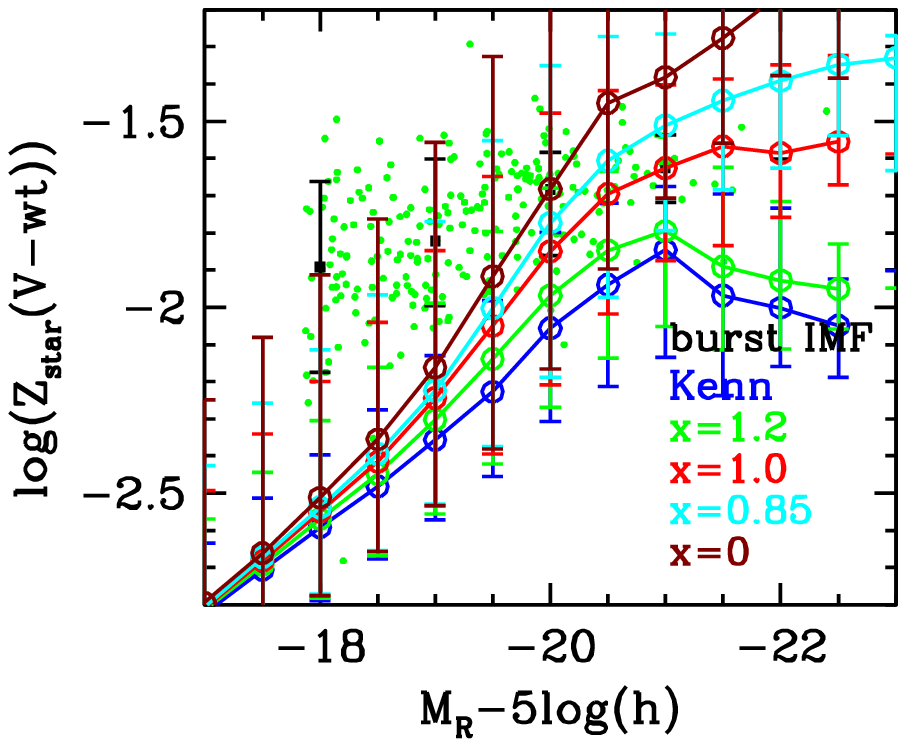}
\end{minipage}
\end{center}

\caption{Effects on the stellar metallicity in early-type galaxies at
  $z=0$ of changing the slope $x$ of the starburst IMF.}

\label{fig:Zstar_IMF}
\end{figure*}

We show the effects on the stellar metallicity vs luminosity relation
in early-type galaxies at $z=0$ of varying the supernova feedback and
gas return rate (Fig.~\ref{fig:Zstar_SNfeedback}) and the starburst
IMF (Fig.~\ref{fig:Zstar_IMF}).

\subsection{Evolution of $K$-band LF}

\begin{figure*}

\begin{center}
\begin{minipage}{5.4cm}
\includegraphics[width=5.4cm, clip=true, bb= 280 525 530 750]{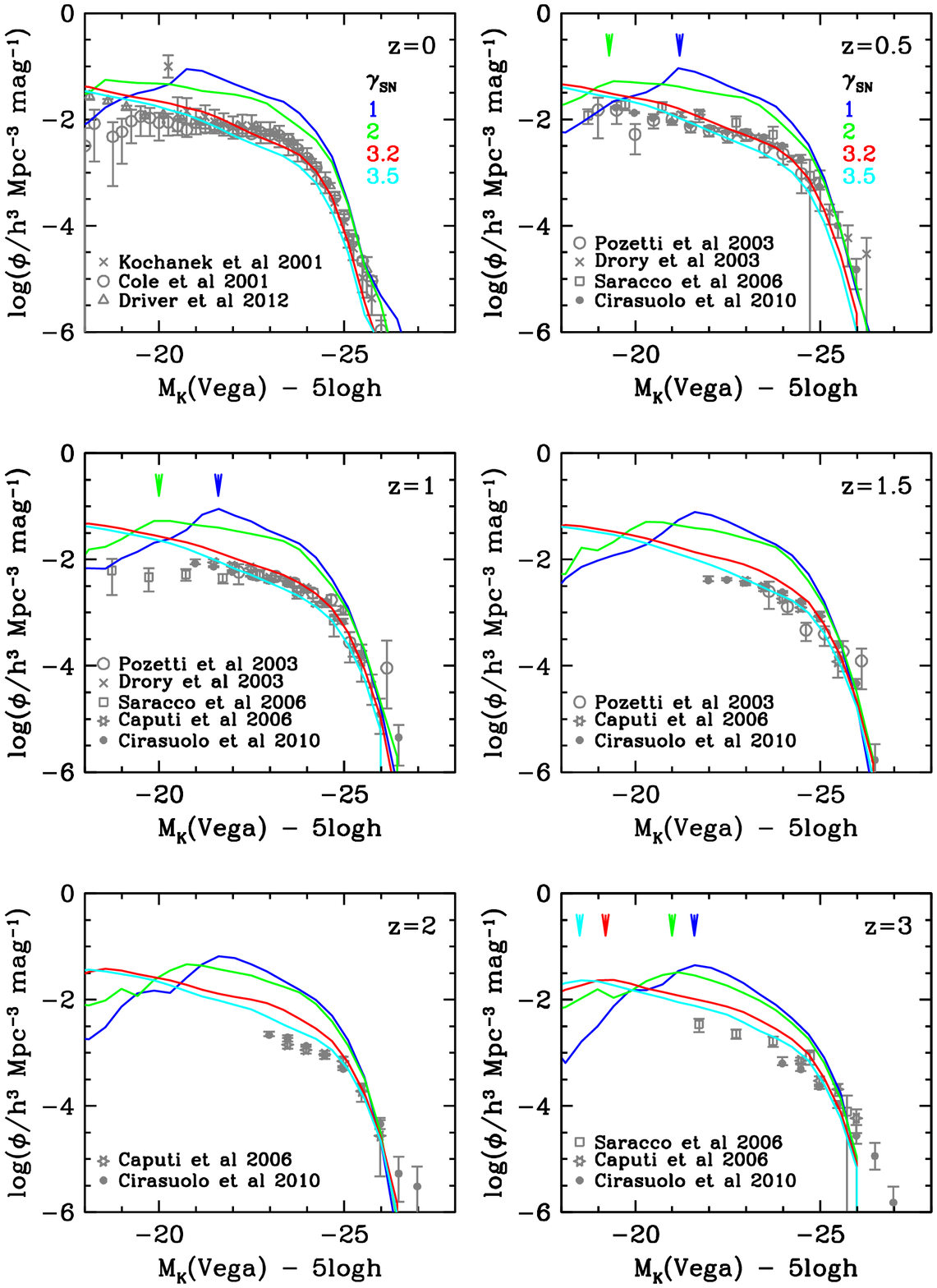}
\end{minipage}
\hspace{0.4cm}
\begin{minipage}{5.4cm}
\includegraphics[width=5.4cm, clip=true, bb= 24 288 275 514]{figs/Klfz_alphahot.ps}
\end{minipage}
\hspace{0.4cm}
\begin{minipage}{5.4cm}
\includegraphics[width=5.4cm, clip=true, bb= 280 51 530 277]{figs/Klfz_alphahot.ps}
\end{minipage}
\end{center}

\begin{center}
\begin{minipage}{5.4cm}
\includegraphics[width=5.4cm, clip=true, bb= 280 525 530 750]{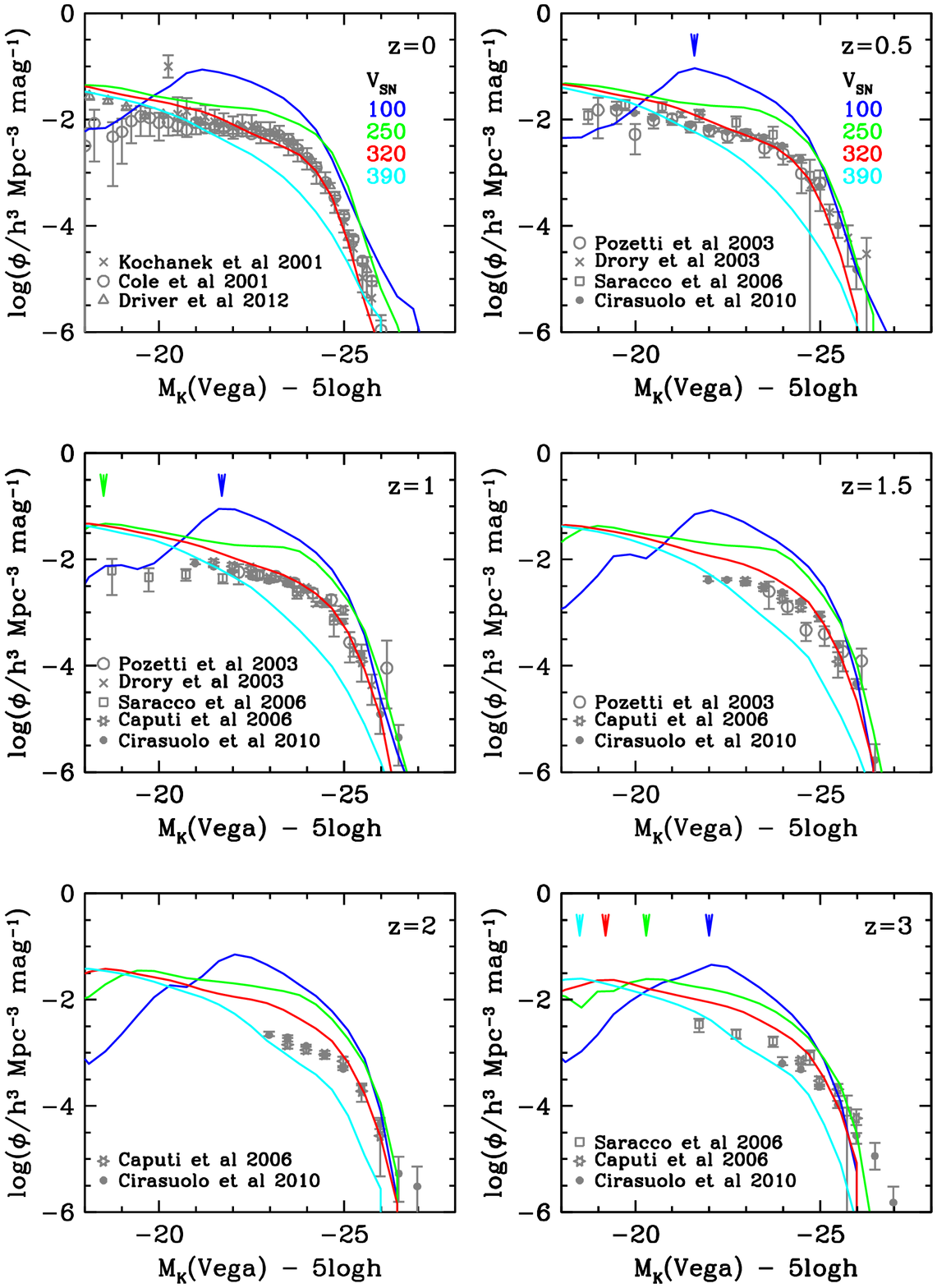}
\end{minipage}
\hspace{0.4cm}
\begin{minipage}{5.4cm}
\includegraphics[width=5.4cm, clip=true, bb= 24 288 275 514]{figs/Klfz_vhot.ps}
\end{minipage}
\hspace{0.4cm}
\begin{minipage}{5.4cm}
\includegraphics[width=5.4cm, clip=true, bb= 280 51 530 277]{figs/Klfz_vhot.ps}
\end{minipage}
\end{center}

\begin{center}
\begin{minipage}{5.4cm}
\includegraphics[width=5.4cm, clip=true, bb= 280 525 530 750]{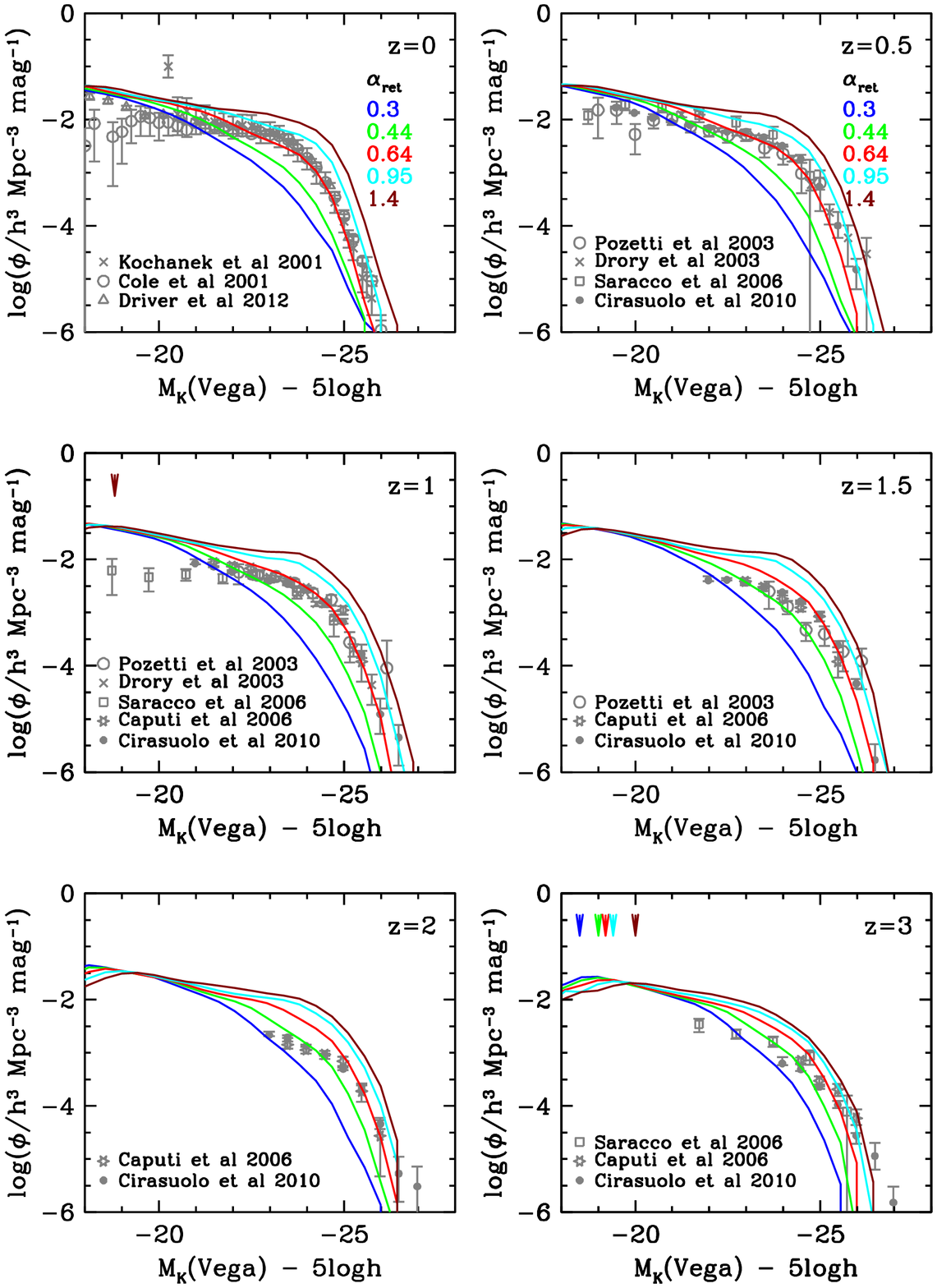}
\end{minipage}
\hspace{0.4cm}
\begin{minipage}{5.4cm}
\includegraphics[width=5.4cm, clip=true, bb= 24 288 275 514]{figs/Klfz_alphareheat.ps}
\end{minipage}
\hspace{0.4cm}
\begin{minipage}{5.4cm}
\includegraphics[width=5.4cm, clip=true, bb= 280 51 530 277]{figs/Klfz_alphareheat.ps}
\end{minipage}
\end{center}

\caption{Effect on the evolution of the K-band luminosity function of
  varying the supernova feedback parameters $\gammaSN$ and $\VSN$ and
  the gas return timescale parameter $\alpharet$. A single parameter
  is varied in each row of panels, with the red curves showing the
  standard model. { The vertical arrows at the top of each panel
    indicate the luminosity below which the results for the
    corresponding model are affected by the halo mass resolution.} The
  observational data plotted are the same as in
  Fig.~\ref{fig:lfKz_default}.}

\label{fig:lfKz_SNfeedback}
\end{figure*}

\begin{figure*}

\begin{center}
\begin{minipage}{5.4cm}
\includegraphics[width=5.4cm, clip=true, bb= 280 525 530 750]{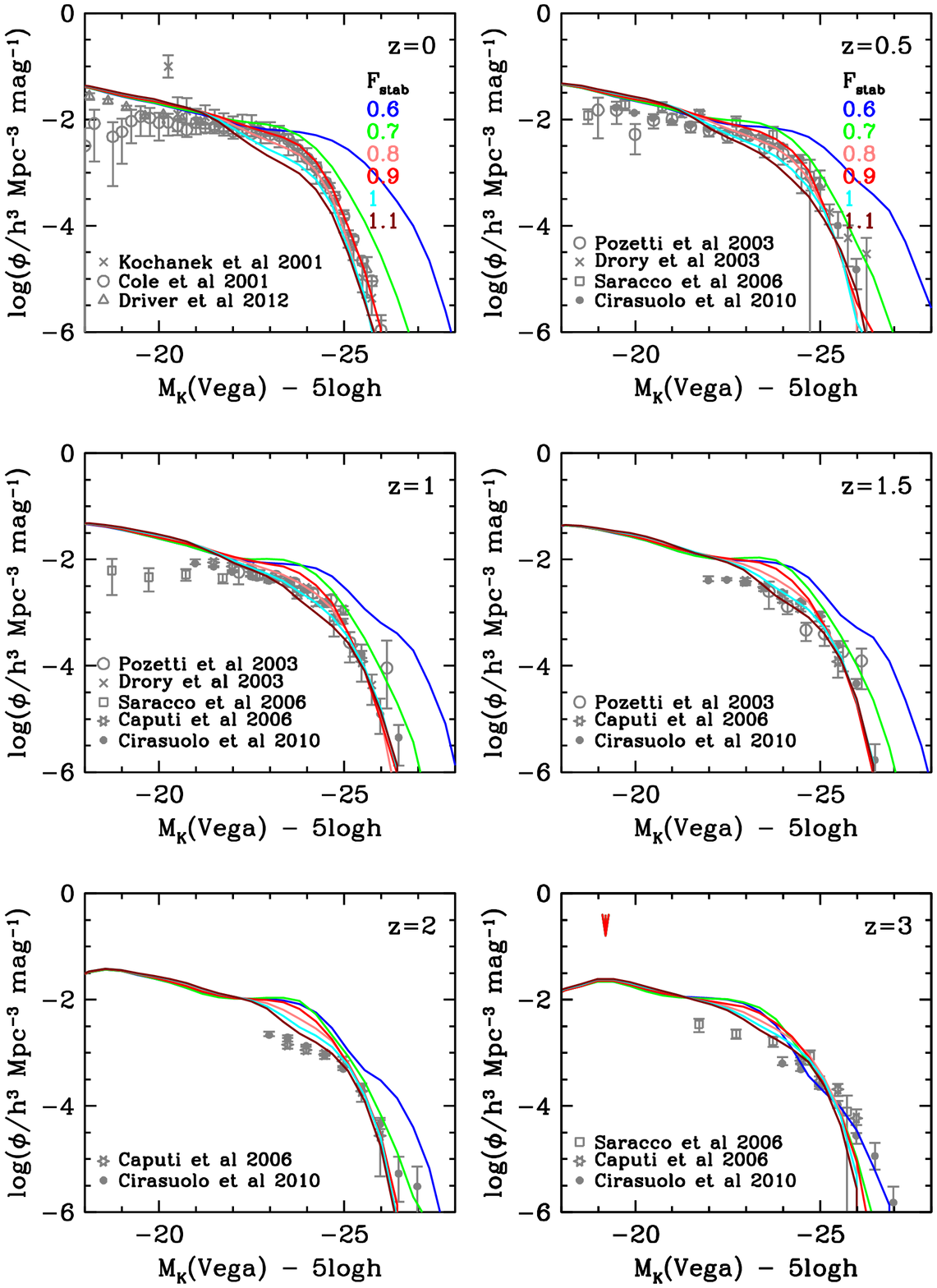}
\end{minipage}
\hspace{0.4cm}
\begin{minipage}{5.4cm}
\includegraphics[width=5.4cm, clip=true, bb= 24 288 275 514]{figs/Klfz_stabledisk.ps}
\end{minipage}
\hspace{0.4cm}
\begin{minipage}{5.4cm}
\includegraphics[width=5.4cm, clip=true, bb= 280 51 530 277]{figs/Klfz_stabledisk.ps}
\end{minipage}
\end{center}

\begin{center}
\begin{minipage}{5.4cm}
\includegraphics[width=5.4cm, clip=true, bb= 280 525 530 750]{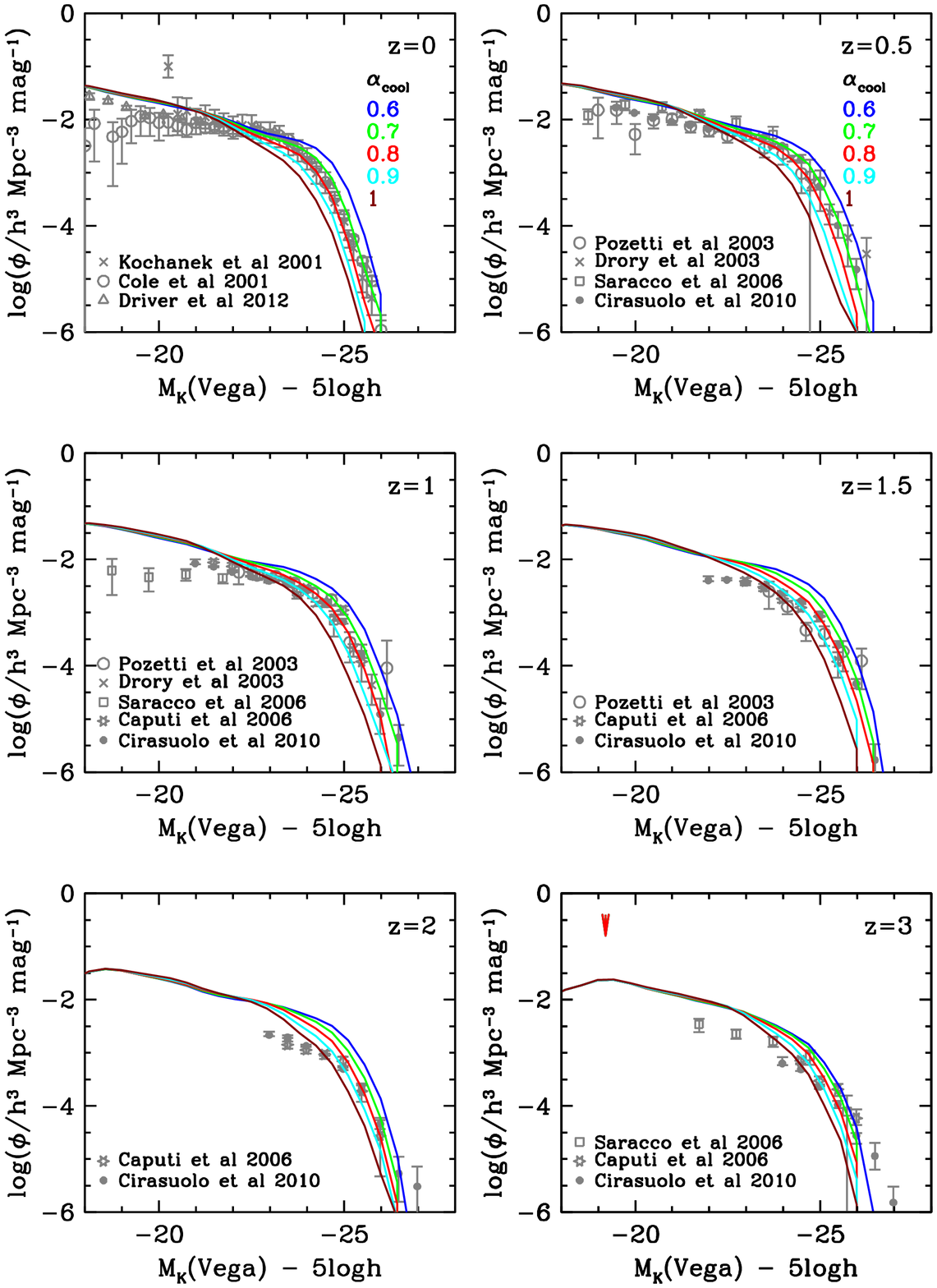}
\end{minipage}
\hspace{0.4cm}
\begin{minipage}{5.4cm}
\includegraphics[width=5.4cm, clip=true, bb= 24 288 275 514]{figs/Klfz_alphacool.ps}
\end{minipage}
\hspace{0.4cm}
\begin{minipage}{5.4cm}
\includegraphics[width=5.4cm, clip=true, bb= 280 51 530 277]{figs/Klfz_alphacool.ps}
\end{minipage}
\end{center}

\begin{center}
\begin{minipage}{5.4cm}
\includegraphics[width=5.4cm, clip=true, bb= 280 525 530 750]{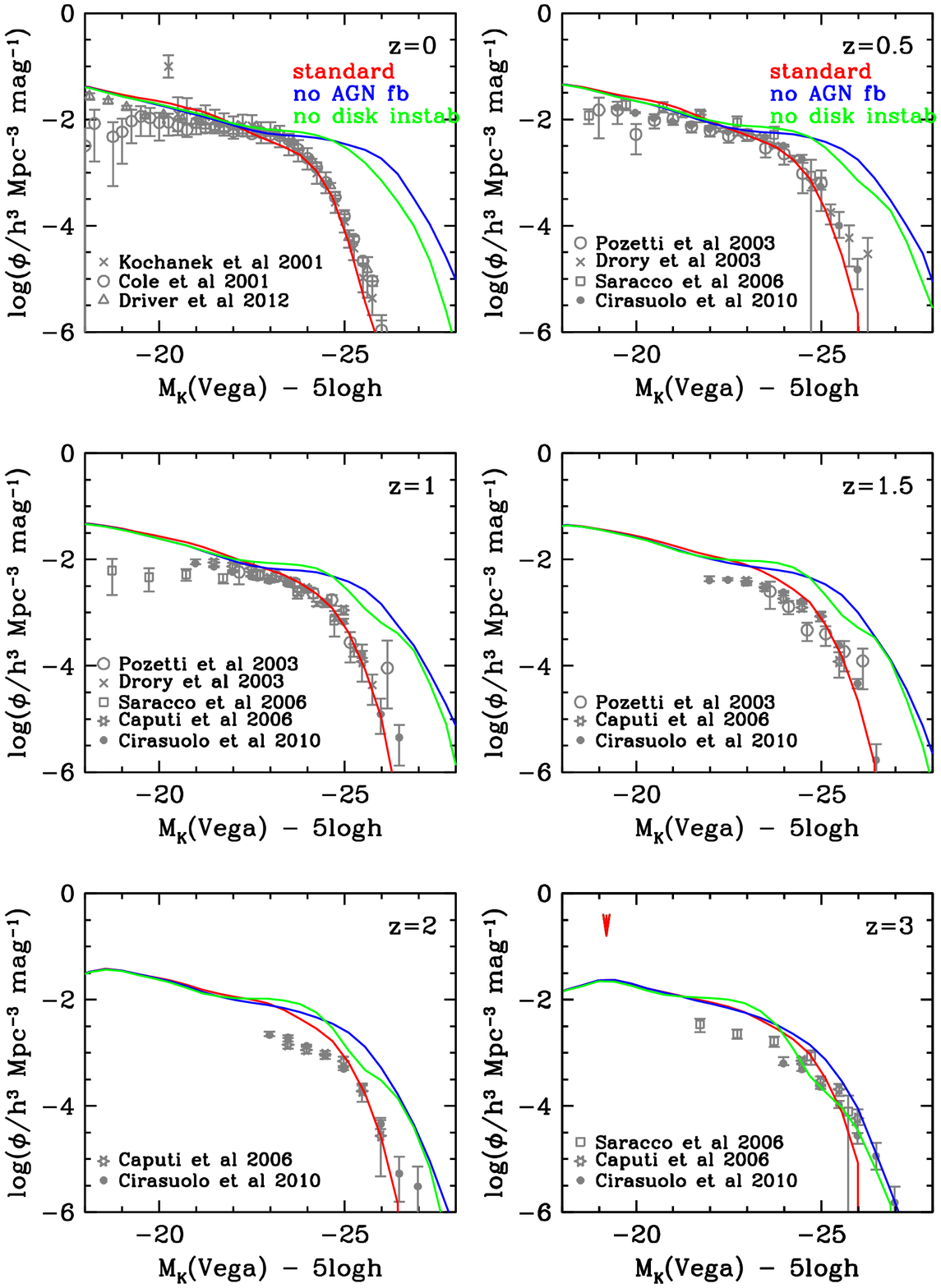}
\end{minipage}
\hspace{0.4cm}
\begin{minipage}{5.4cm}
\includegraphics[width=5.4cm, clip=true, bb= 24 288 275 514]{figs/Klfz_noAGNfb_nodiskinstab.ps}
\end{minipage}
\hspace{0.4cm}
\begin{minipage}{5.4cm}
\includegraphics[width=5.4cm, clip=true, bb= 280 51 530 277]{figs/Klfz_noAGNfb_nodiskinstab.ps}
\end{minipage}
\end{center}

\caption{Effect on the evolution of the K-band luminosity function of
  varying (a) the disk stability parameter $\Fstab$ and (b) the AGN
  feedback parameter $\alphacool$, and (c) of turning off AGN feedback
  or disk instabilities. A single parameter is varied in each row of
  panels, with the red curves showing the standard model.}

\label{fig:lfKz_AGNfeedback}
\end{figure*}

\begin{figure*}

\begin{center}
\begin{minipage}{5.4cm}
\includegraphics[width=5.4cm, clip=true, bb= 280 525 530 750]{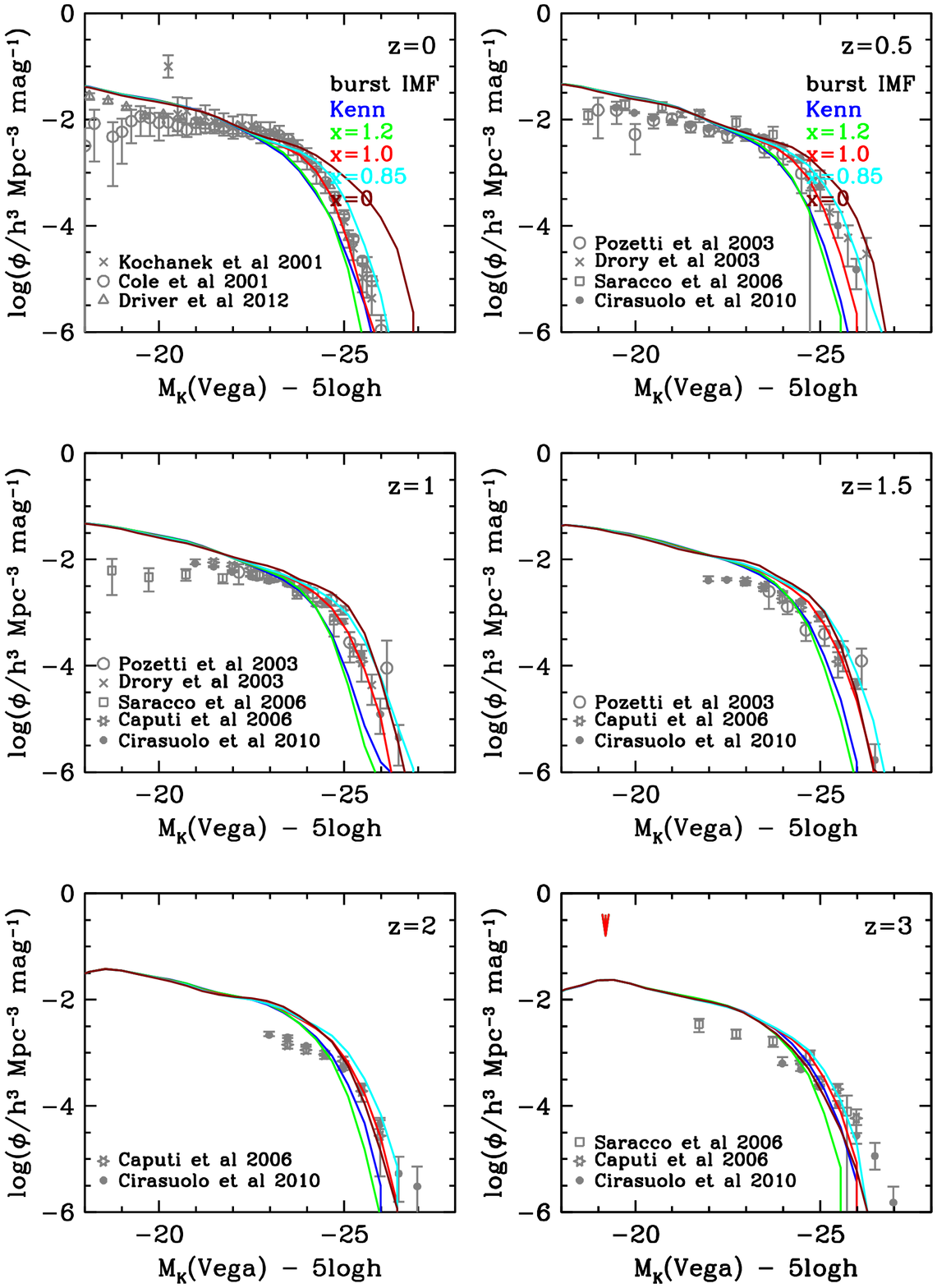}
\end{minipage}
\hspace{0.4cm}
\begin{minipage}{5.4cm}
\includegraphics[width=5.4cm, clip=true, bb= 24 288 275 514]{figs/Klfz_IMF.ps}
\end{minipage}
\hspace{0.4cm}
\begin{minipage}{5.4cm}
\includegraphics[width=5.4cm, clip=true, bb= 280 51 530 277]{figs/Klfz_IMF.ps}
\end{minipage}
\end{center}

\begin{center}
\begin{minipage}{5.4cm}
\includegraphics[width=5.4cm, clip=true, bb= 280 525 530 750]{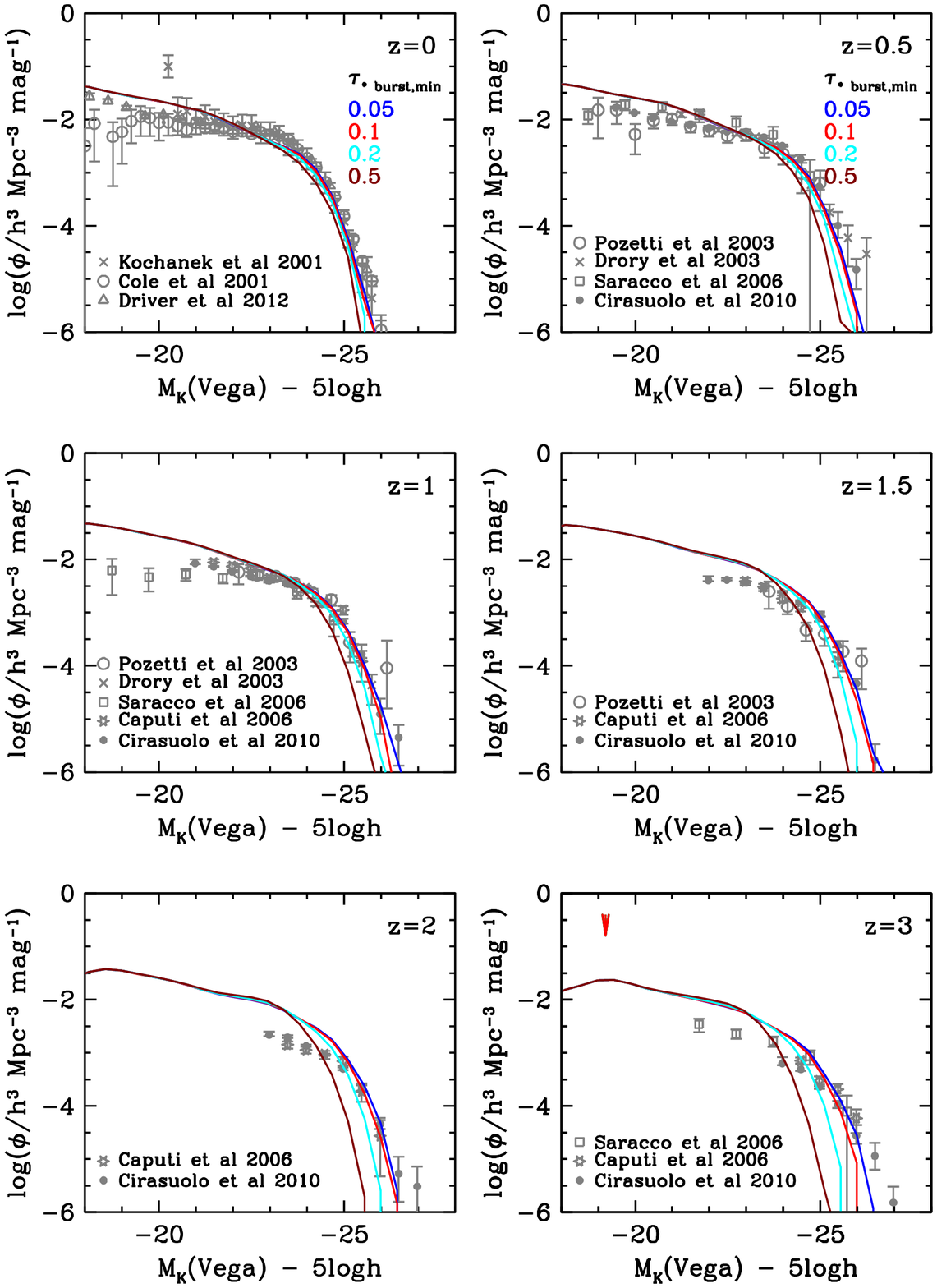}
\end{minipage}
\hspace{0.4cm}
\begin{minipage}{5.4cm}
\includegraphics[width=5.4cm, clip=true, bb= 24 288 275 514]{figs/Klfz_taustarminburst.ps}
\end{minipage}
\hspace{0.4cm}
\begin{minipage}{5.4cm}
\includegraphics[width=5.4cm, clip=true, bb= 280 51 530 277]{figs/Klfz_taustarminburst.ps}
\end{minipage}
\end{center}

\begin{center}
\begin{minipage}{5.4cm}
\includegraphics[width=5.4cm, clip=true, bb= 280 525 530 750]{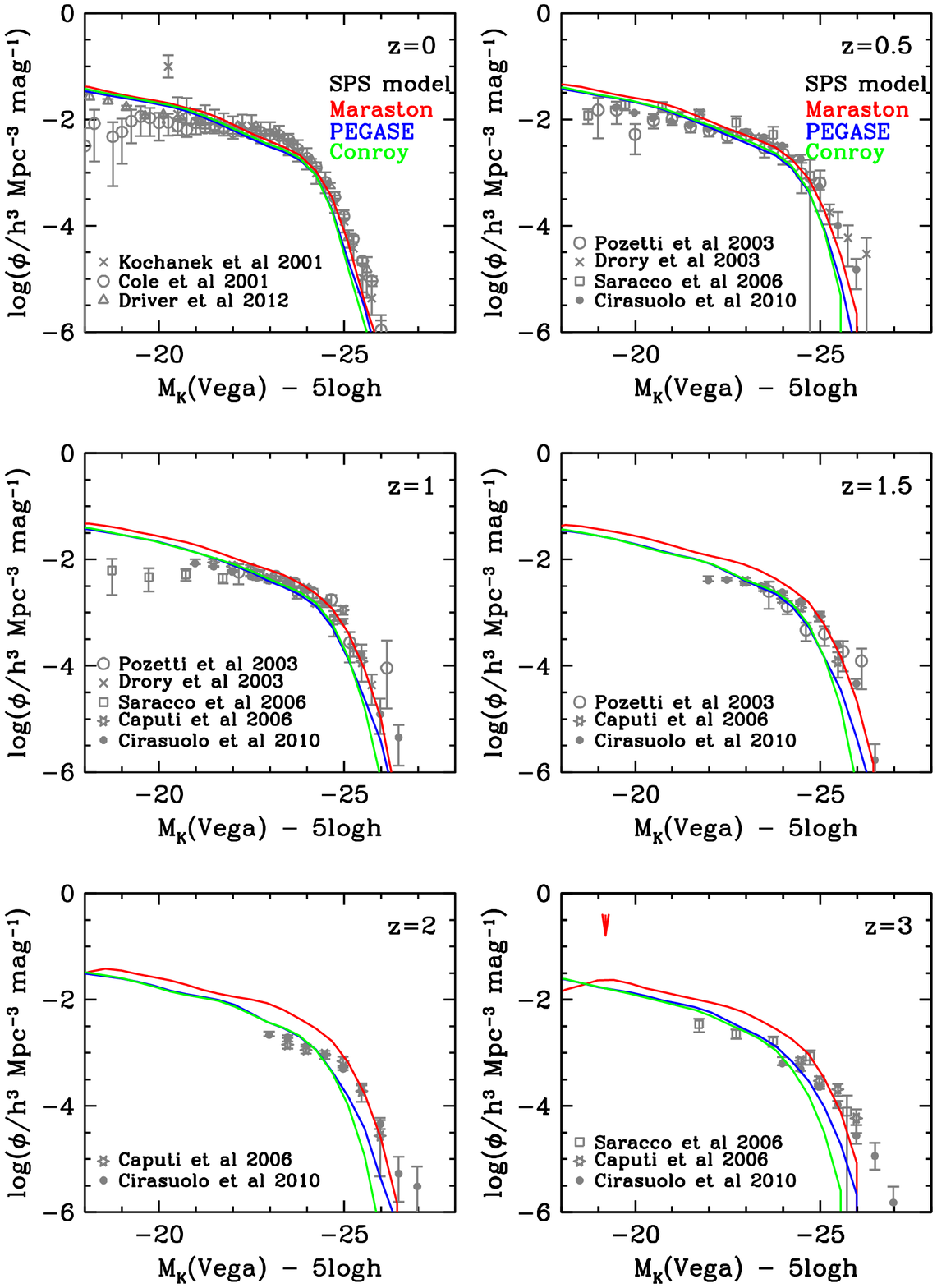}
\end{minipage}
\hspace{0.4cm}
\begin{minipage}{5.4cm}
\includegraphics[width=5.4cm, clip=true, bb= 24 288 275 514]{figs/Klfz_SPS.ps}
\end{minipage}
\hspace{0.4cm}
\begin{minipage}{5.4cm}
\includegraphics[width=5.4cm, clip=true, bb= 280 51 530 277]{figs/Klfz_SPS.ps}
\end{minipage}
\end{center}

\caption{Effect on the evolution of the K-band luminosity function of
  varying (a) the slope $x$ of the starburst IMF, (b) the minimum
  starburst timescale $\tauburstmin$ and (c) the SPS model. A single
  parameter is varied in each row of panels, with the red curves
  showing the standard model.}

\label{fig:lfKz_IMF_SPS}
\end{figure*}

We show the effects on the evolution of the $K$-band LF at $z=0.5-3$
of varying the supernova feedback and gas return rate
(Fig.~\ref{fig:lfKz_SNfeedback}), disk instabilities and AGN feedback
(Fig.~\ref{fig:lfKz_AGNfeedback}), and the starburst IMF, minimum
starburst timescale and SPS model (Fig.~\ref{fig:lfKz_IMF_SPS}).

\cleardoublepage

\subsection{Far-IR number counts}

\begin{figure*}

\begin{center}
\begin{minipage}{5.4cm}
\includegraphics[width=5.4cm, clip=true, bb= 24 525 275 750]{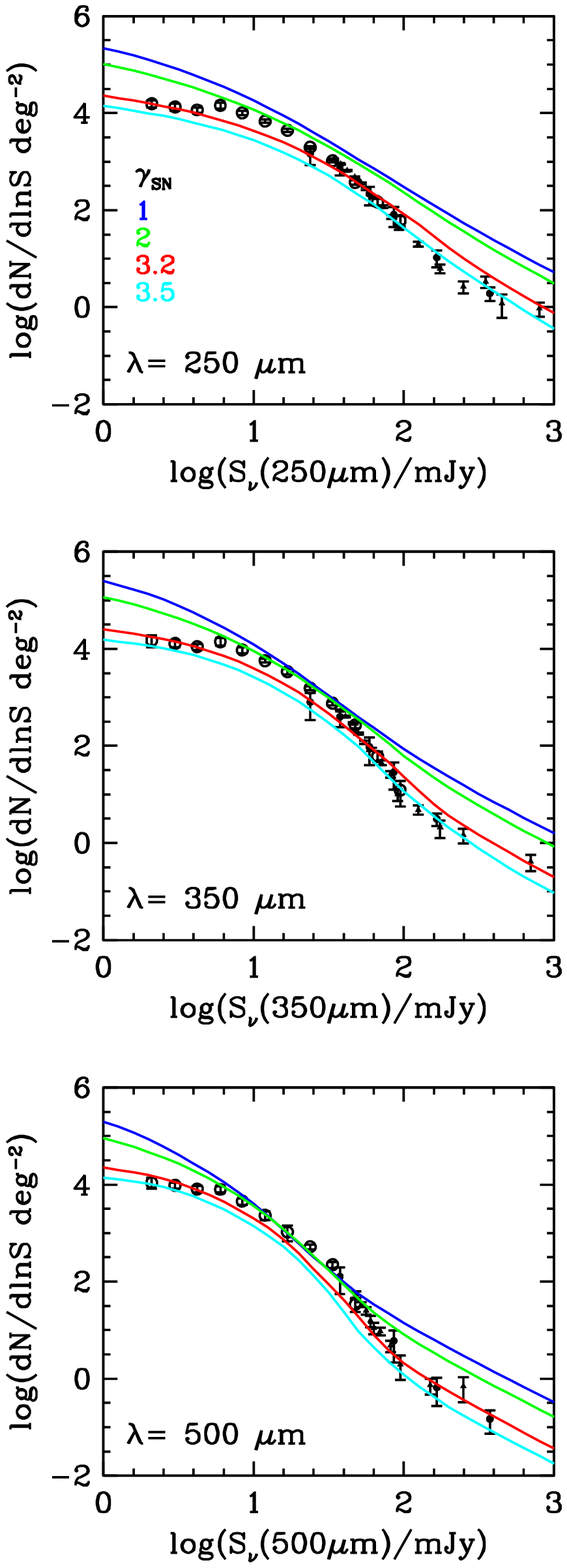}
\end{minipage}
\hspace{0.4cm}
\begin{minipage}{5.4cm}
\includegraphics[width=5.4cm, clip=true, bb= 24 288 275 514]{figs/FIRcounts_alphahot.ps}
\end{minipage}
\hspace{0.4cm}
\begin{minipage}{5.4cm}
\includegraphics[width=5.4cm, clip=true, bb= 24 51 275 277]{figs/FIRcounts_alphahot.ps}
\end{minipage}
\end{center}

\begin{center}
\begin{minipage}{5.4cm}
\includegraphics[width=5.4cm, clip=true, bb= 24 525 275 750]{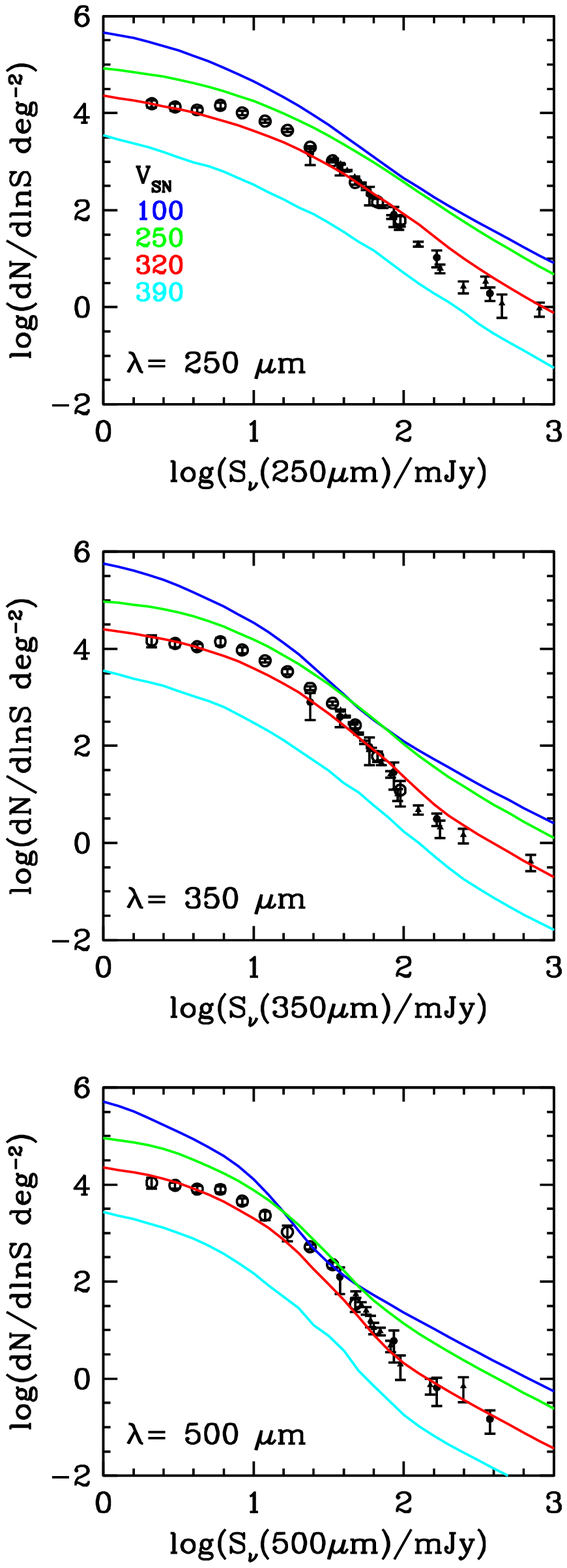}
\end{minipage}
\hspace{0.4cm}
\begin{minipage}{5.4cm}
\includegraphics[width=5.4cm, clip=true, bb= 24 288 275 514]{figs/FIRcounts_vhot.ps}
\end{minipage}
\hspace{0.4cm}
\begin{minipage}{5.4cm}
\includegraphics[width=5.4cm, clip=true, bb= 24 51 275 277]{figs/FIRcounts_vhot.ps}
\end{minipage}
\end{center}

\begin{center}
\begin{minipage}{5.4cm}
\includegraphics[width=5.4cm, clip=true, bb= 24 525 275 750]{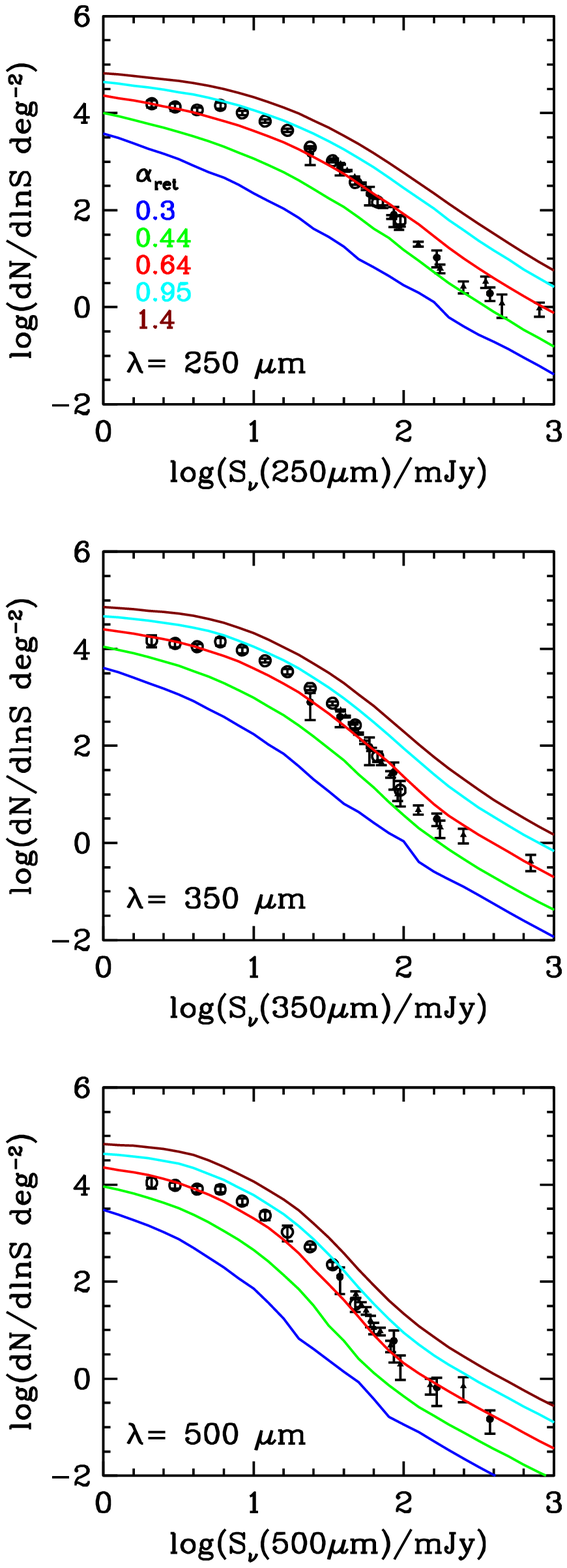}
\end{minipage}
\hspace{0.4cm}
\begin{minipage}{5.4cm}
\includegraphics[width=5.4cm, clip=true, bb= 24 288 275 514]{figs/FIRcounts_alphareheat.ps}
\end{minipage}
\hspace{0.4cm}
\begin{minipage}{5.4cm}
\includegraphics[width=5.4cm, clip=true, bb= 24 51 275 277]{figs/FIRcounts_alphareheat.ps}
\end{minipage}
\end{center}

\begin{center}
\begin{minipage}{5.4cm}
\includegraphics[width=5.4cm, clip=true, bb= 24 525 275 750]{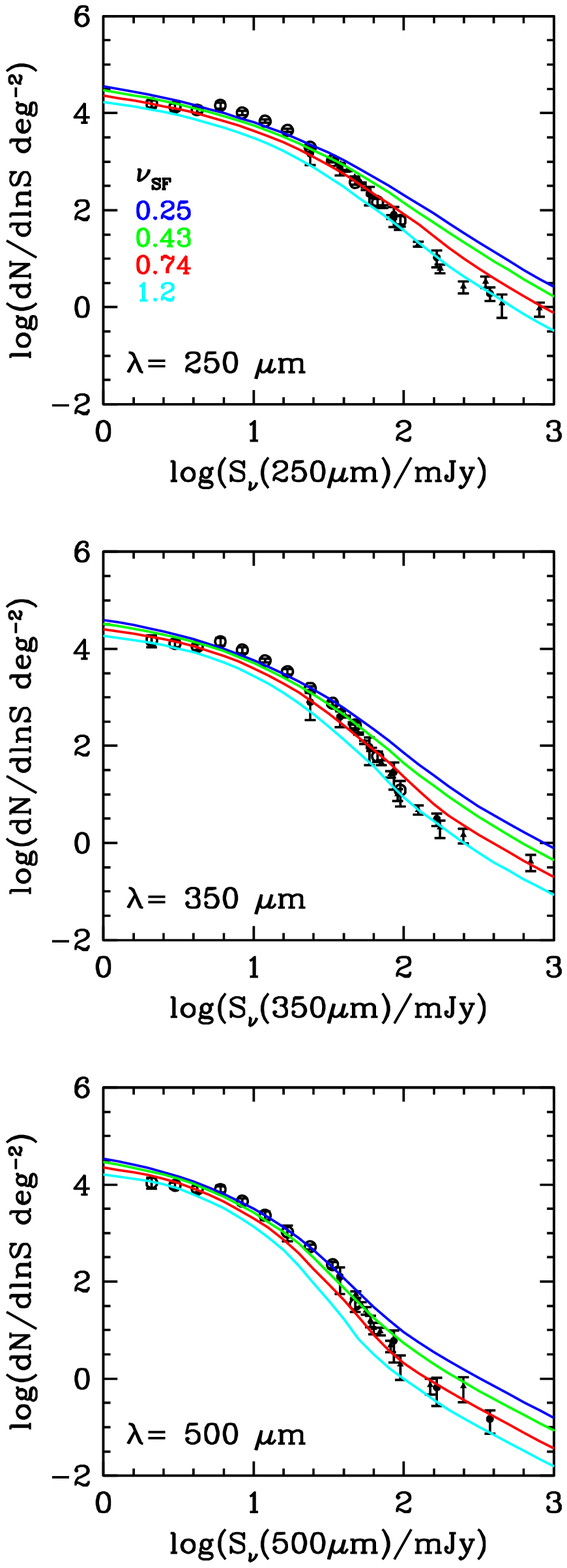}
\end{minipage}
\hspace{0.4cm}
\begin{minipage}{5.4cm}
\includegraphics[width=5.4cm, clip=true, bb= 24 288 275 514]{figs/FIRcounts_nusf.ps}
\end{minipage}
\hspace{0.4cm}
\begin{minipage}{5.4cm}
\includegraphics[width=5.4cm, clip=true, bb= 24 51 275 277]{figs/FIRcounts_nusf.ps}
\end{minipage}
\end{center}

\caption{Effect on the far-IR number counts at 250, 350 and 500$\mum$
  of varying the supernova feedback parameters $\gammaSN$ and $\VSN$,
  the gas return parameter $\alpharet$, and the disk star formation
  parameter $\nuSF$. A single parameter is varied in each row of
  panels, with the red curves showing the standard model. The
  observational data plotted are the same as in
  Fig.~\ref{fig:FIRcounts_default}.}

\label{fig:FIRcounts_SNfeedback}
\end{figure*}

\begin{figure*}

\begin{center}
\begin{minipage}{5.4cm}
\includegraphics[width=5.4cm, clip=true, bb= 24 525 275 750]{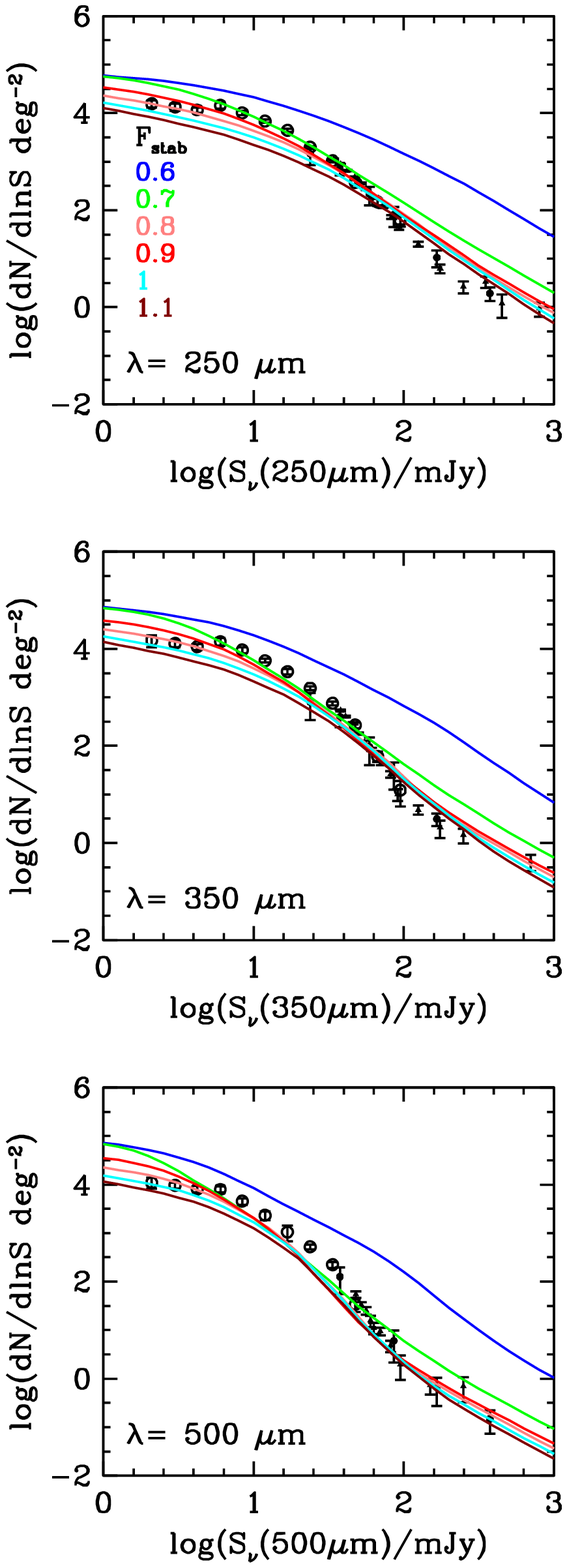}
\end{minipage}
\hspace{0.4cm}
\begin{minipage}{5.4cm}
\includegraphics[width=5.4cm, clip=true, bb= 24 288 275 514]{figs/FIRcounts_stabledisk.ps}
\end{minipage}
\hspace{0.4cm}
\begin{minipage}{5.4cm}
\includegraphics[width=5.4cm, clip=true, bb= 24 51 275 277]{figs/FIRcounts_stabledisk.ps}
\end{minipage}
\end{center}

\begin{center}
\begin{minipage}{5.4cm}
\includegraphics[width=5.4cm, clip=true, bb= 24 525 275 750]{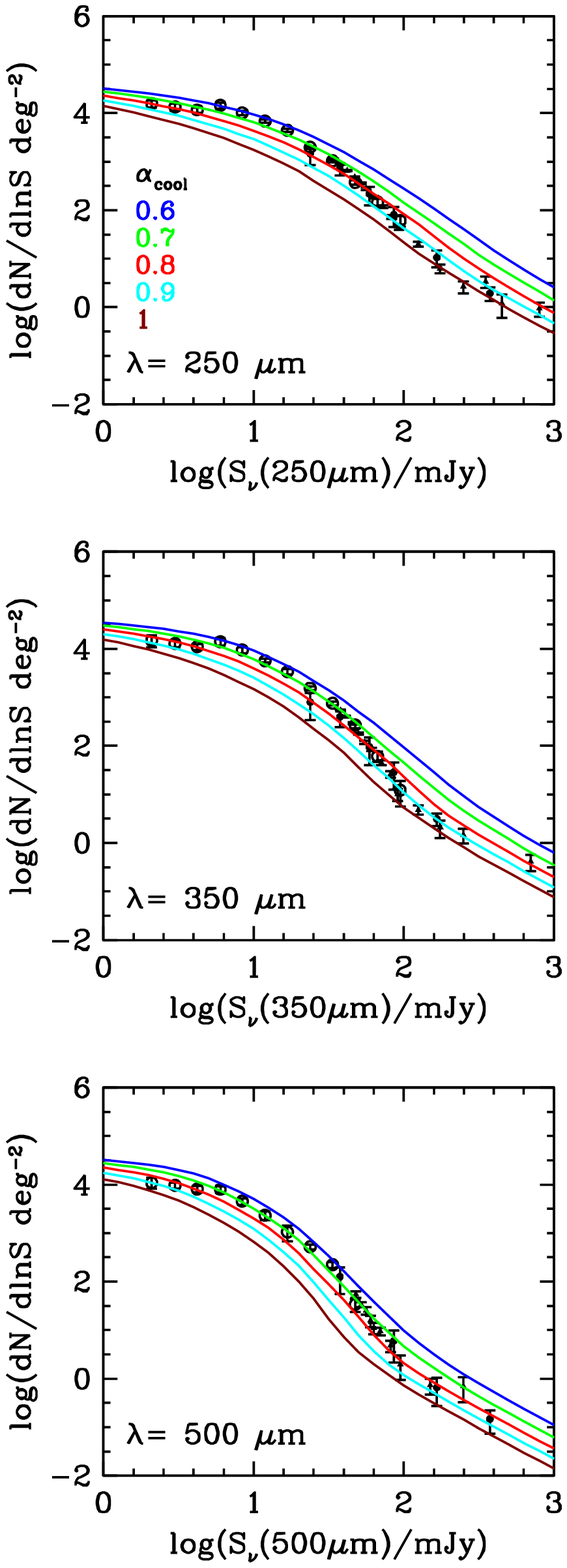}
\end{minipage}
\hspace{0.4cm}
\begin{minipage}{5.4cm}
\includegraphics[width=5.4cm, clip=true, bb= 24 288 275 514]{figs/FIRcounts_alphacool.ps}
\end{minipage}
\hspace{0.4cm}
\begin{minipage}{5.4cm}
\includegraphics[width=5.4cm, clip=true, bb= 24 51 275 277]{figs/FIRcounts_alphacool.ps}
\end{minipage}
\end{center}

\begin{center}
\begin{minipage}{5.4cm}
\includegraphics[width=5.4cm, clip=true, bb= 24 525 275 750]{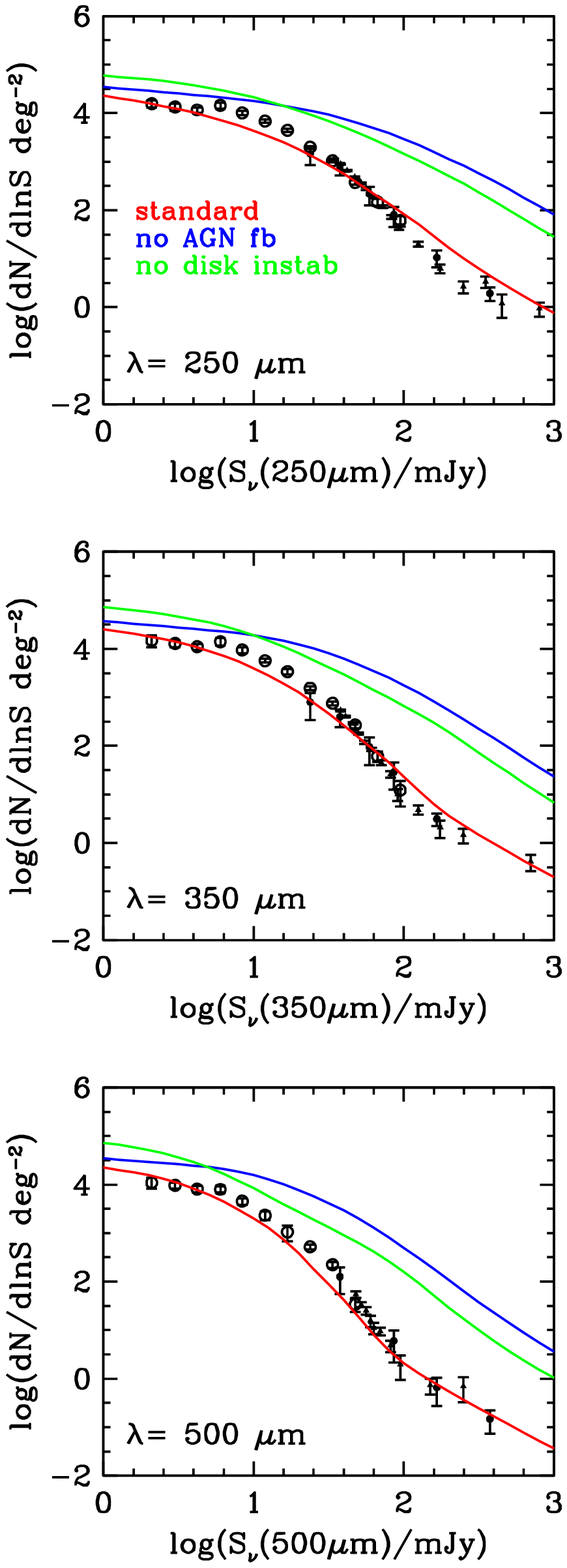}
\end{minipage}
\hspace{0.4cm}
\begin{minipage}{5.4cm}
\includegraphics[width=5.4cm, clip=true, bb= 24 288 275 514]{figs/FIRcounts_noAGNfb_nodiskinstab.ps}
\end{minipage}
\hspace{0.4cm}
\begin{minipage}{5.4cm}
\includegraphics[width=5.4cm, clip=true, bb= 24 51 275 277]{figs/FIRcounts_noAGNfb_nodiskinstab.ps}
\end{minipage}
\end{center}

\begin{center}
\begin{minipage}{5.4cm}
\includegraphics[width=5.4cm, clip=true, bb= 24 525 275 750]{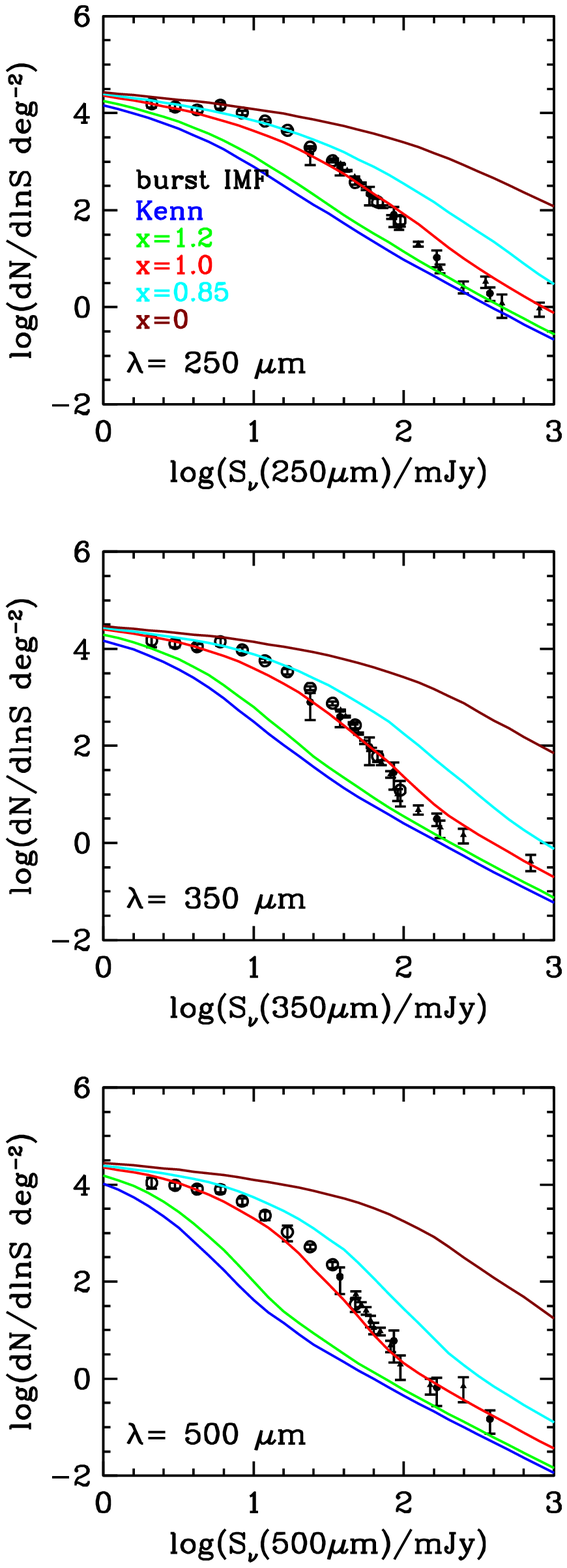}
\end{minipage}
\hspace{0.4cm}
\begin{minipage}{5.4cm}
\includegraphics[width=5.4cm, clip=true, bb= 24 288 275 514]{figs/FIRcounts_IMF.ps}
\end{minipage}
\hspace{0.4cm}
\begin{minipage}{5.4cm}
\includegraphics[width=5.4cm, clip=true, bb= 24 51 275 277]{figs/FIRcounts_IMF.ps}
\end{minipage}
\end{center}

\caption{Effect on the far-IR number counts at 250, 350 and 500$\mum$
  of varying (a) the disk stability parameter $\Fstab$ and (b) AGN
  feedback parameter $\alphacool$, (c) of turning off AGN feedback or
  disk instabilities, and (d) of varying the slope $x$ of the
  starburst IMF. A single parameter is varied in each row of panels,
  with the red curves showing the standard model.}

\label{fig:FIRcounts_AGNfeedback_IMF}
\end{figure*}

We show the effects on the far-IR number counts of galaxies at
wavelengths 250--500~$\mum$ of varying the supernova feedback, gas
return rate and normalization of the disk SFR law
(Fig.~\ref{fig:FIRcounts_SNfeedback}), and disk instabilities, AGN
feedback and the starburst IMF
(Fig.~\ref{fig:FIRcounts_AGNfeedback_IMF}).

\subsection{Number counts and redshifts of sub-mm galaxies}

\begin{figure*}

\begin{center}
\begin{minipage}{5.4cm}
\includegraphics[width=5.5cm, bb= 20 295 275 750]{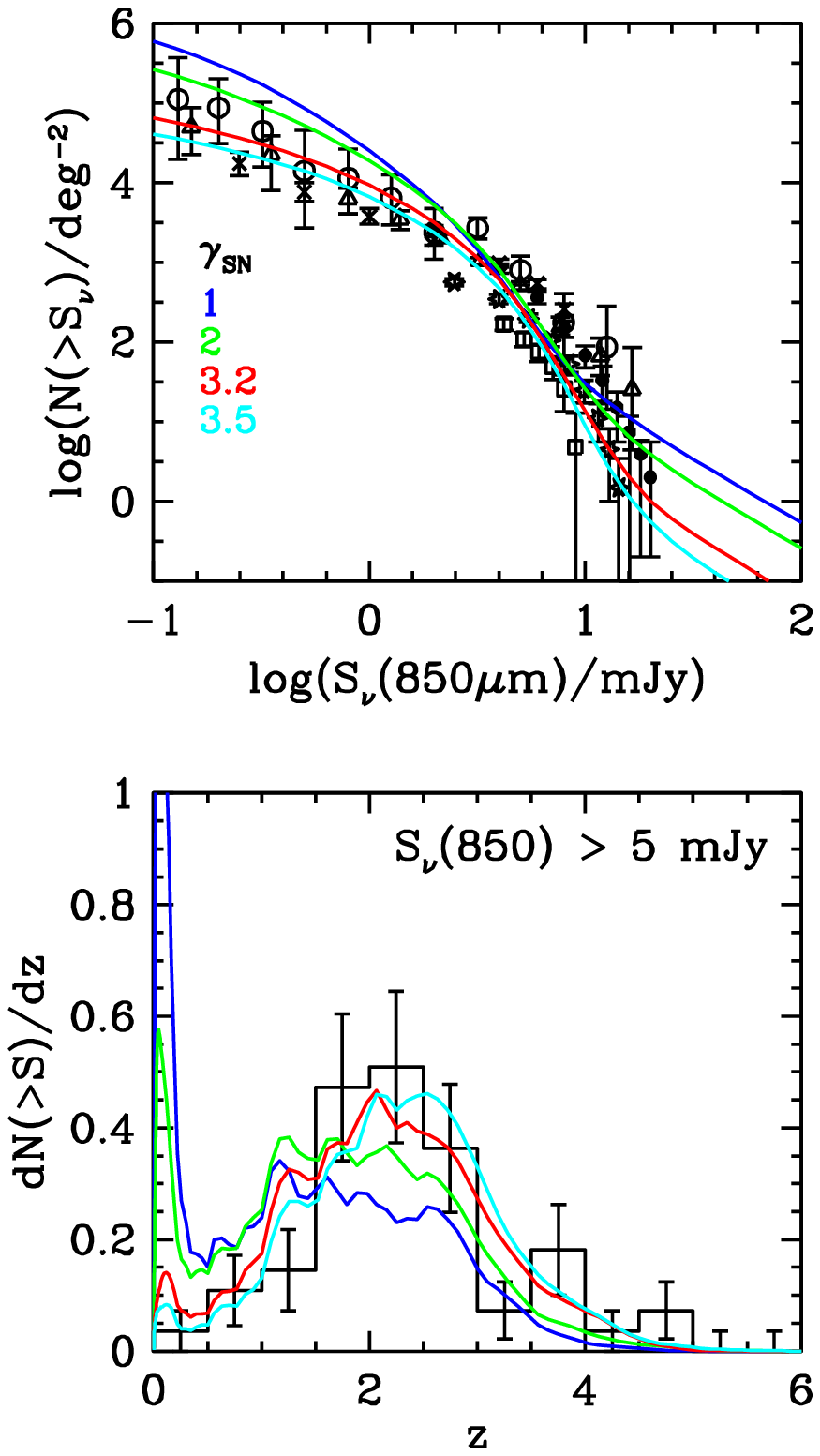}
\end{minipage}
\hspace{0.4cm}
\begin{minipage}{5.4cm}
\includegraphics[width=5.4cm, bb= 20 295 275 750]{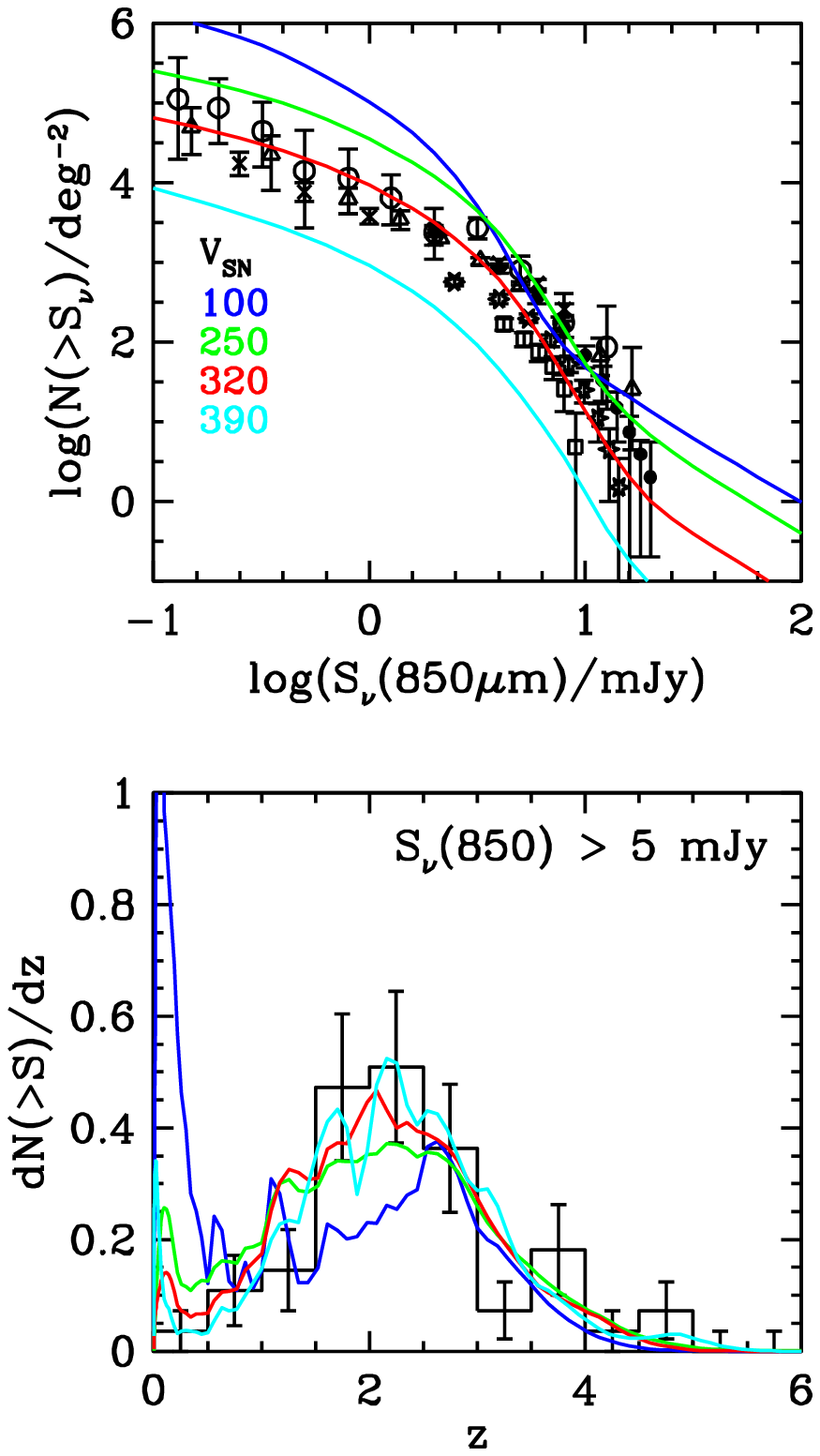}
\end{minipage}
\hspace{0.4cm}
\begin{minipage}{5.4cm}
\includegraphics[width=5.4cm, bb= 20 295 275 750]{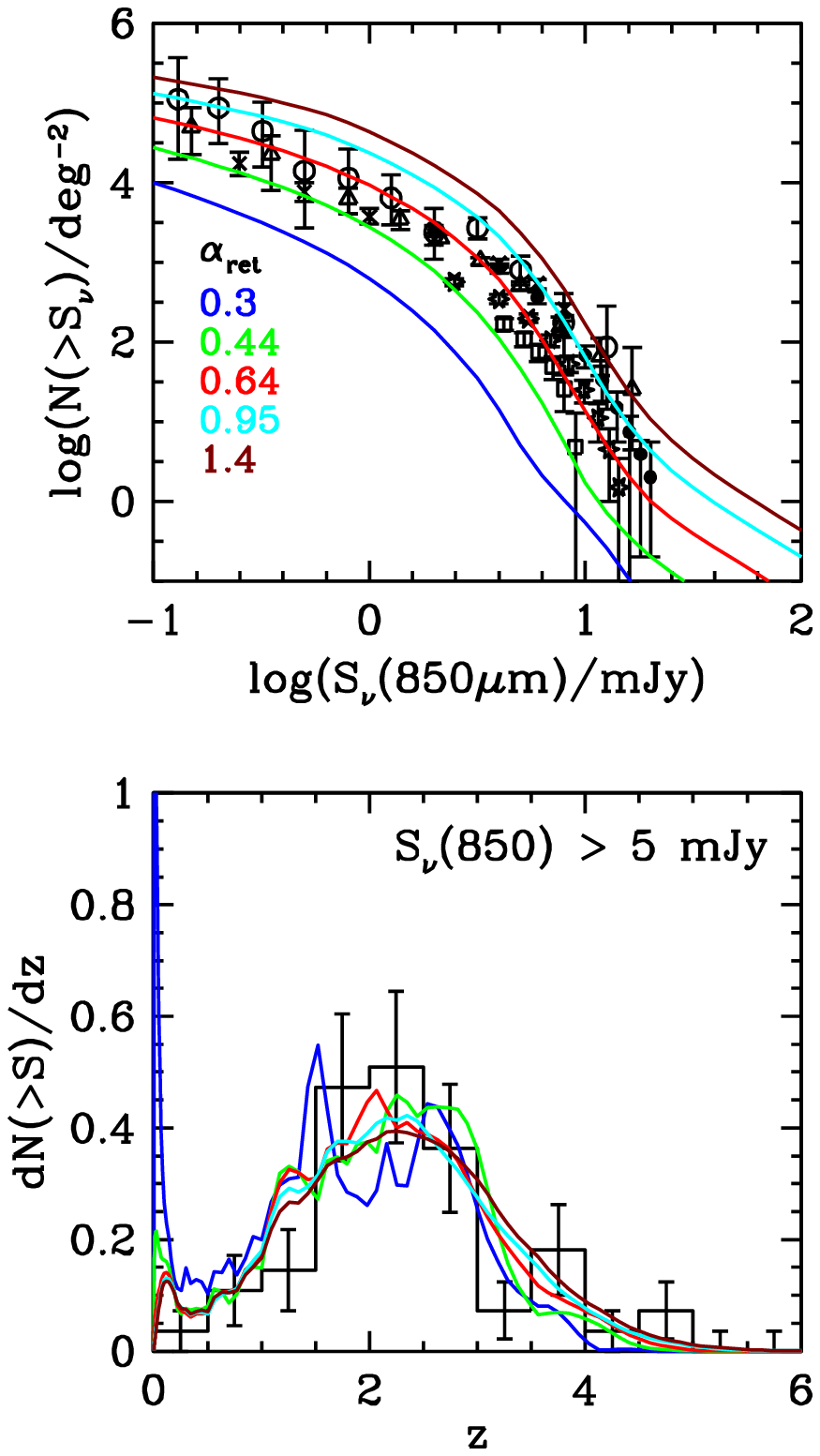}
\end{minipage}
\end{center}

\caption{Effects on the $850\mum$ number counts and redshift
  distribution of varying the supernova feedback parameters $\gammaSN$
  and $\VSN$, and the gas return parameter $\alpharet$. A single
  parameter is varied in each column, with the red curves showing the
  standard model. The observational data plotted are the same as in
  Fig.~\ref{fig:SMGs_default}.}

\label{fig:SMGs_SNfeedback}
\end{figure*}

\begin{figure*}

\begin{center}
\begin{minipage}{5.4cm}
\includegraphics[width=5.4cm, bb= 20 295 275 750]{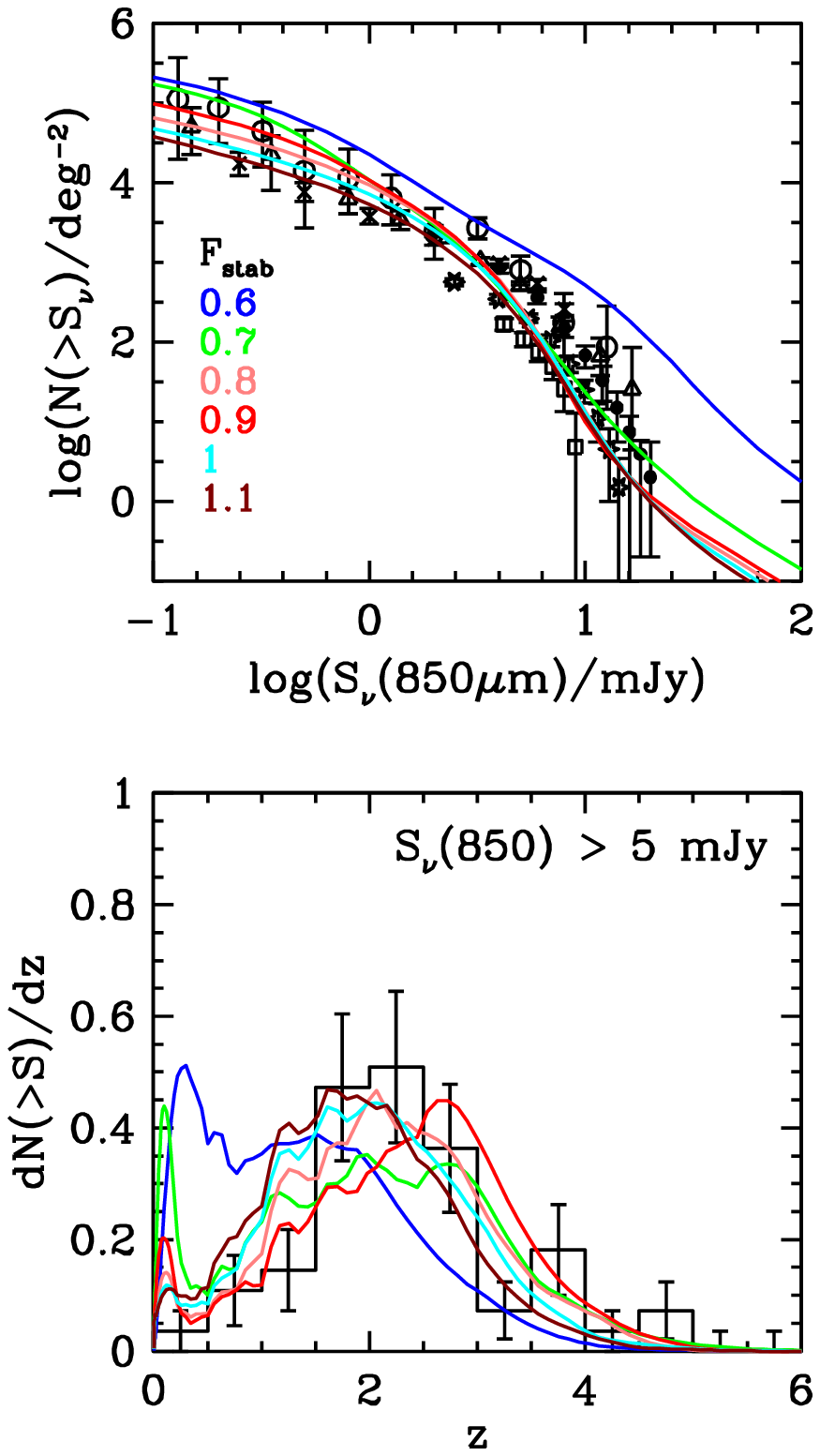}
\end{minipage}
\hspace{0.4cm}
\begin{minipage}{5.4cm}
\includegraphics[width=5.4cm, bb= 20 295 275 750]{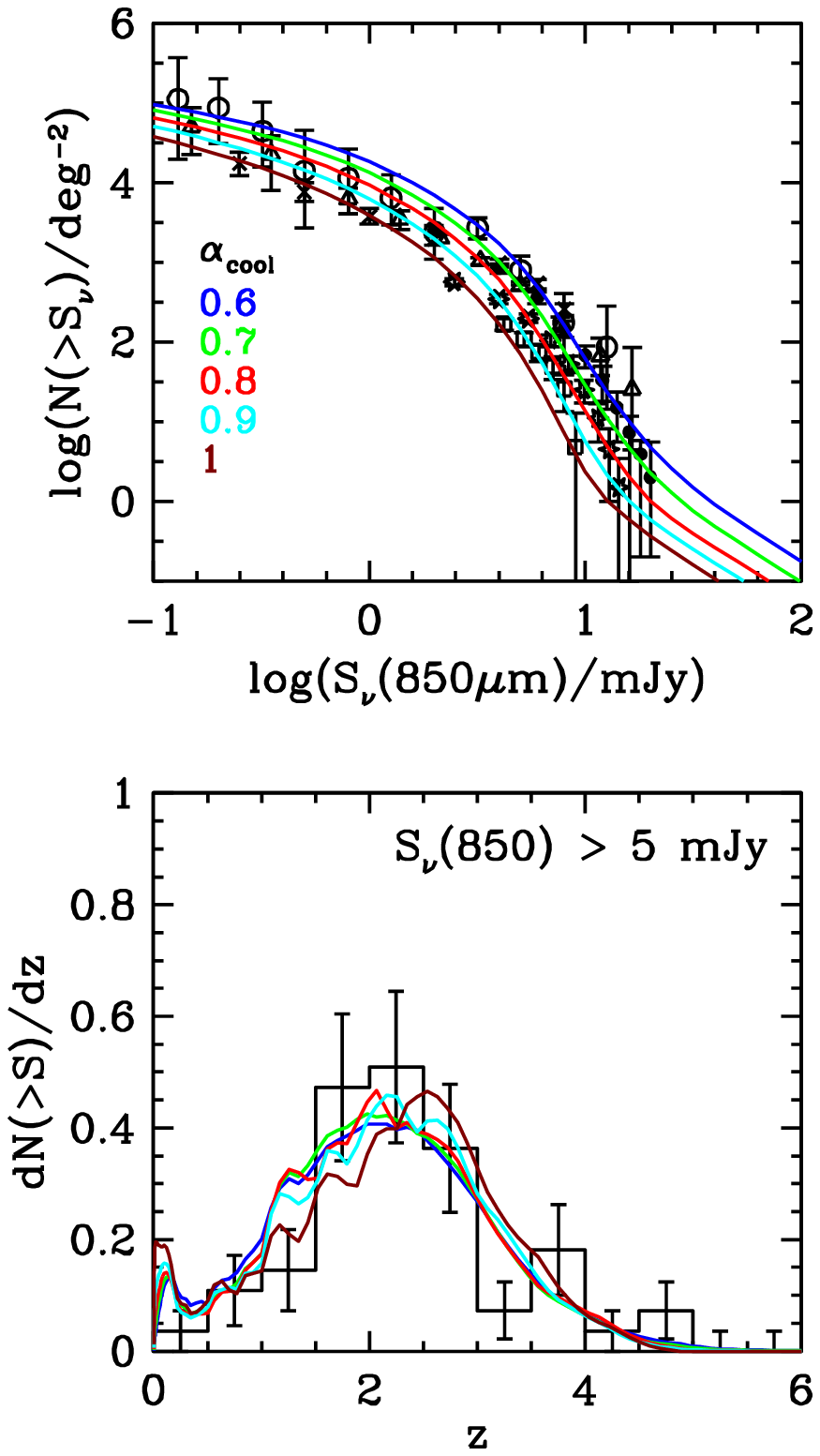}
\end{minipage}
\hspace{0.4cm}
\begin{minipage}{5.4cm}
\includegraphics[width=5.5cm, bb= 20 295 275 750]{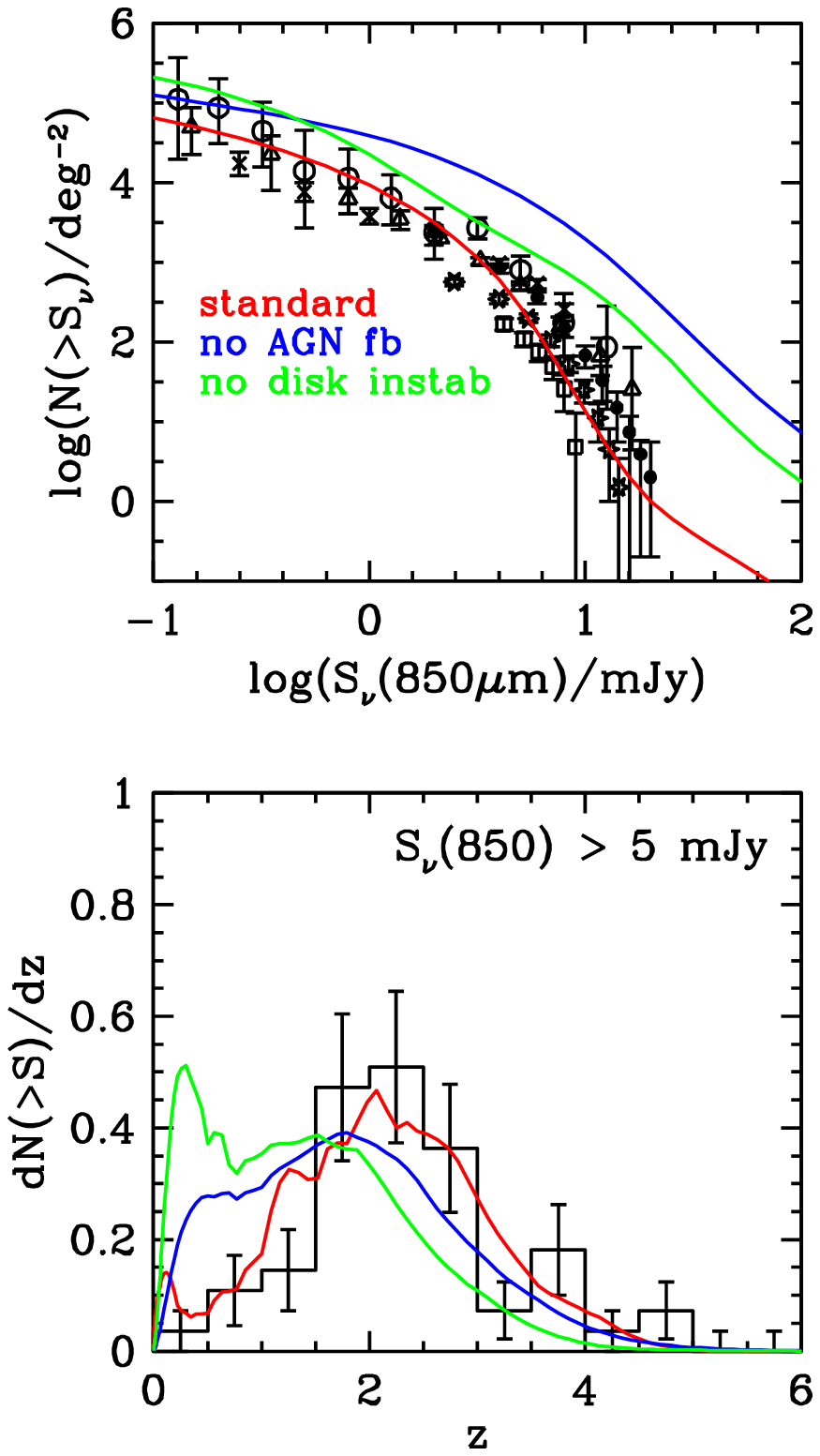}
\end{minipage}
\end{center}

\caption{Effects on the $850\mum$ number counts and redshift
  distribution of varying (a) the disk stability parameter $\Fstab$
  and (b) AGN feedback parameter $\alphacool$, and (c) of turning off
  AGN feedback or disk instabilities. A single parameter is varied in
  each column, with the red curves showing the standard model.}

\label{fig:SMGs_AGNfeedback}
\end{figure*}

\begin{figure*}
\begin{center}
\begin{minipage}{5.4cm}
\includegraphics[width=5.4cm, bb= 20 295 275 750]{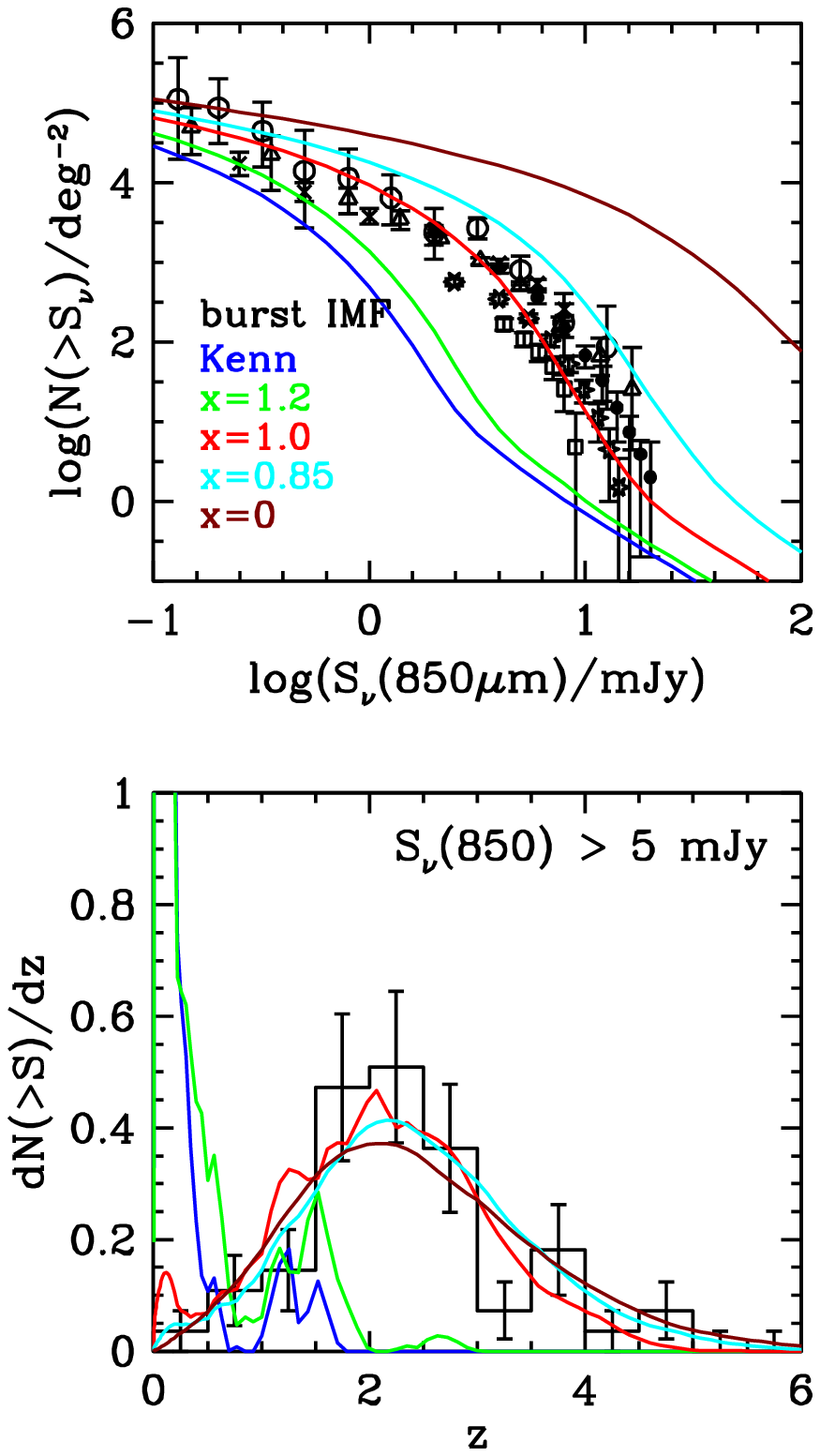}
\end{minipage}
\hspace{0.4cm}
\begin{minipage}{5.4cm}
\includegraphics[width=5.4cm, bb= 20 295 275 750]{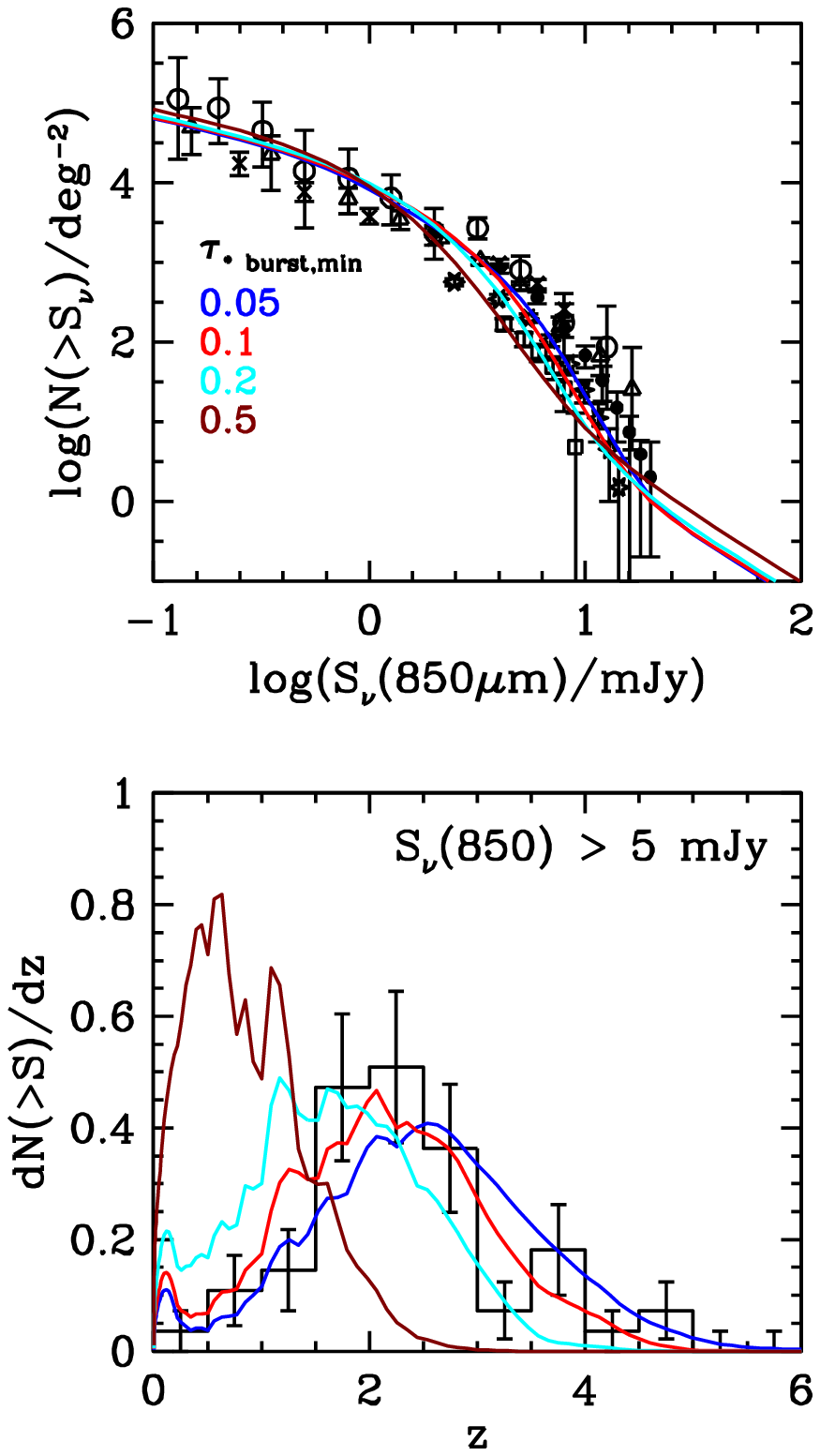}
\end{minipage}
\hspace{0.4cm}
\begin{minipage}{5.4cm}
\includegraphics[width=5.4cm, bb= 20 295 275 750]{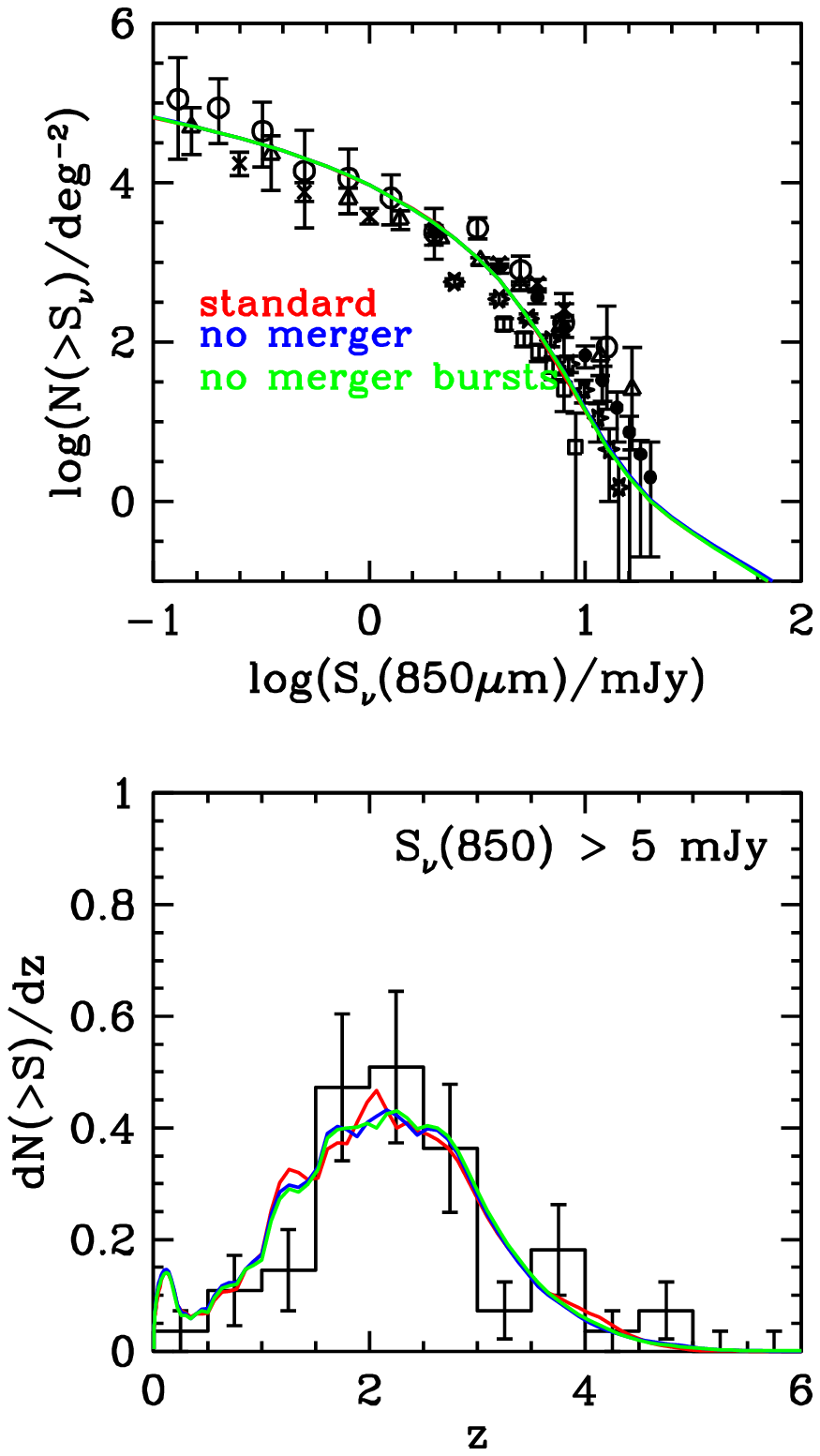}
\end{minipage}
\end{center}

\caption{Effects on the $850\mum$ number counts and redshift
  distribution of varying (a) the slope $x$ of the starburst IMF and
  (b) the minimum star formation timescale in bursts $\tauburstmin$,
  and (c) of turning off galaxy mergers or starbursts triggered by
  galaxy mergers. A single parameter is varied in each column, with
  the red curves showing the standard model.}

\label{fig:SMGs_IMF_tauburst_nomerge}
\end{figure*}

We show the effects on the 850~$\mum$ number counts and redshift
distribution of varying the supernova feedback and gas return rate
(Fig.~\ref{fig:SMGs_SNfeedback}), disk instabilities and AGN feedback
(Fig.~\ref{fig:SMGs_AGNfeedback}), and the starburst IMF, minimum
starburst timescale of of galaxy mergers
(Fig.~\ref{fig:SMGs_IMF_tauburst_nomerge}).

\subsection{Far-UV LFs of Lyman-break galaxies}

\begin{figure*}

\begin{center}
\begin{minipage}{5.4cm}
\includegraphics[width=5.5cm, bb= 20 295 275 750]{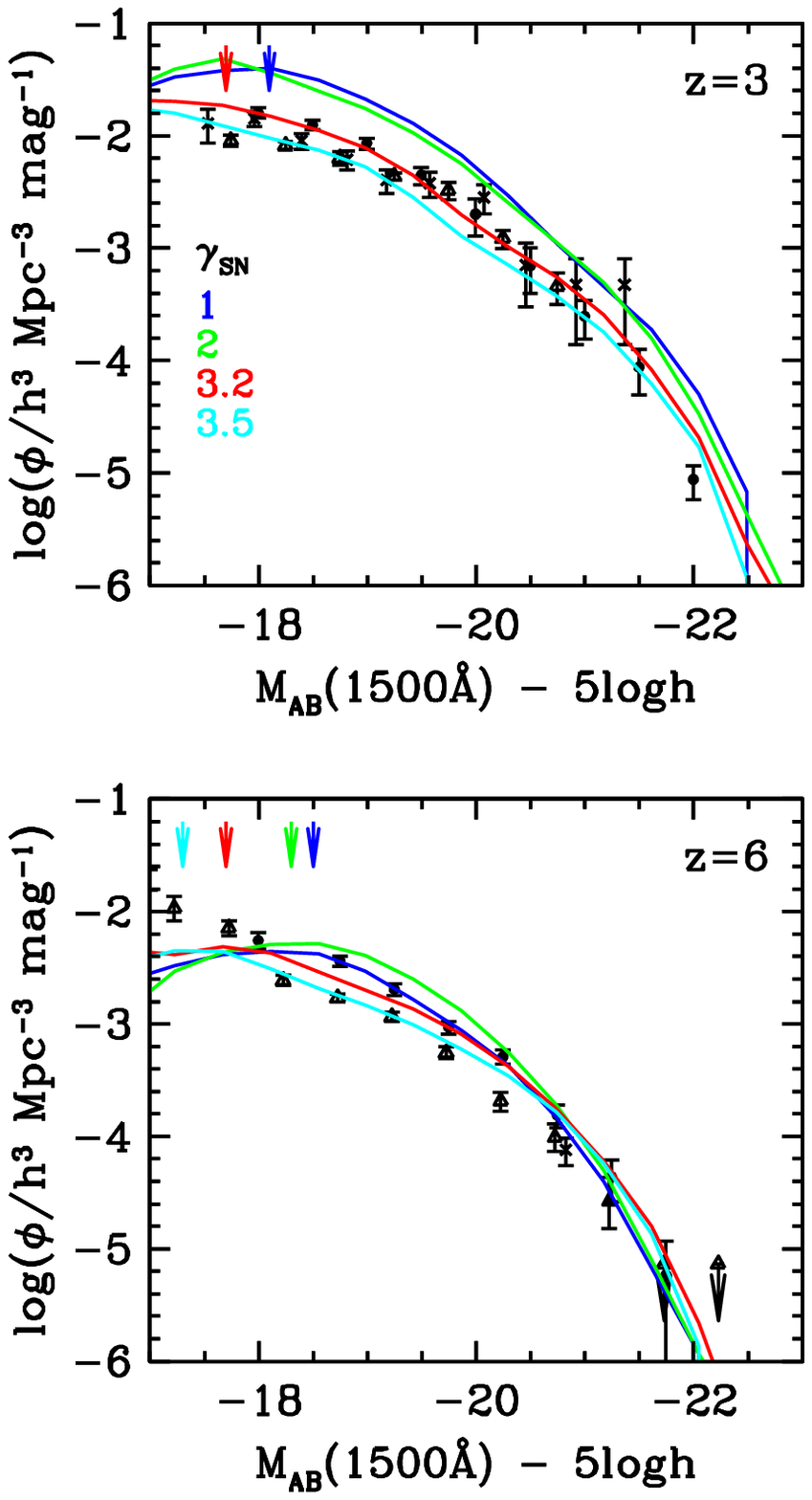}
\end{minipage}
\hspace{0.4cm}
\begin{minipage}{5.4cm}
\includegraphics[width=5.4cm, bb= 20 295 275 750]{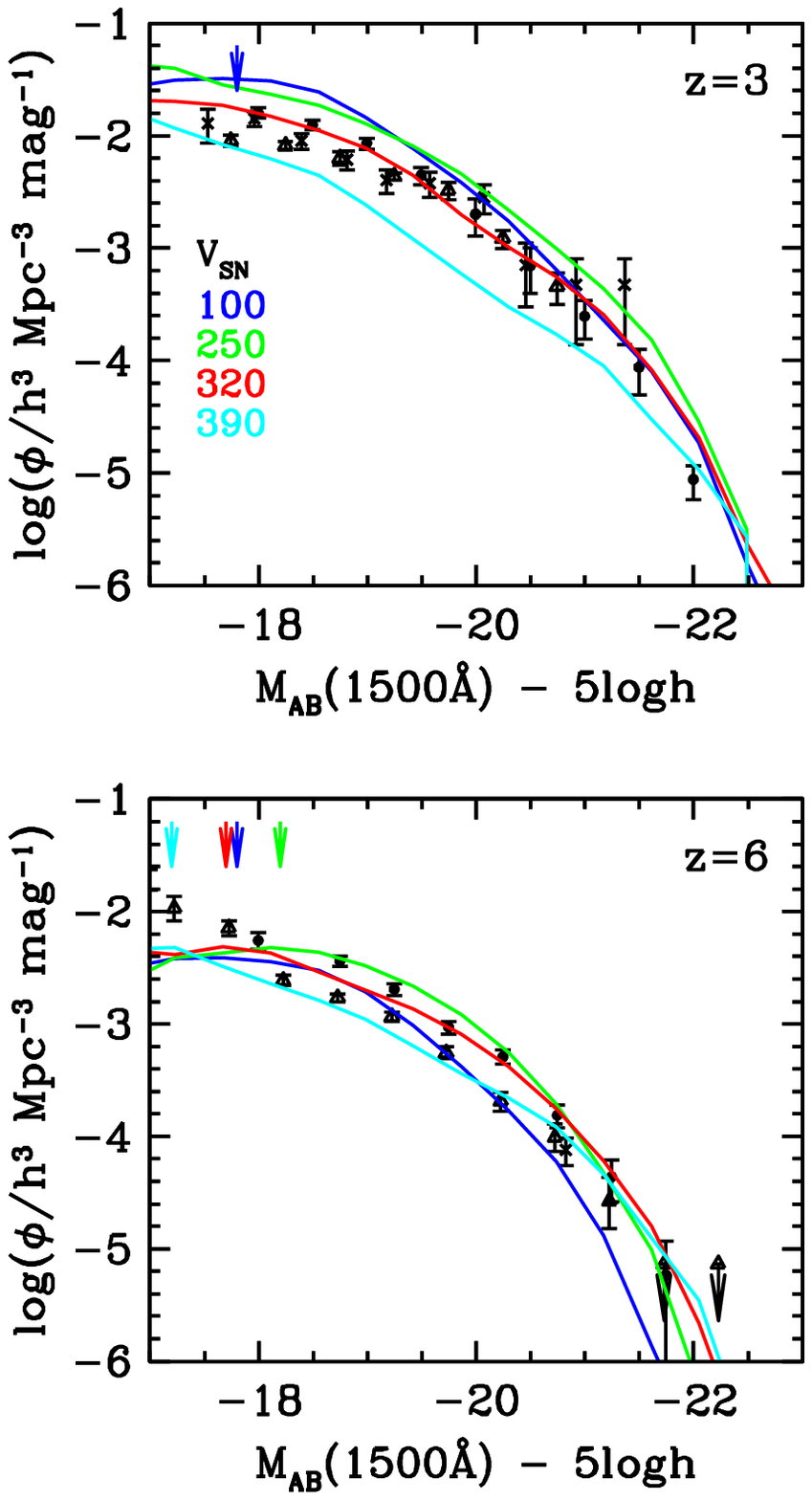}
\end{minipage}
\hspace{0.4cm}
\begin{minipage}{5.4cm}
\includegraphics[width=5.4cm, bb= 20 295 275 750]{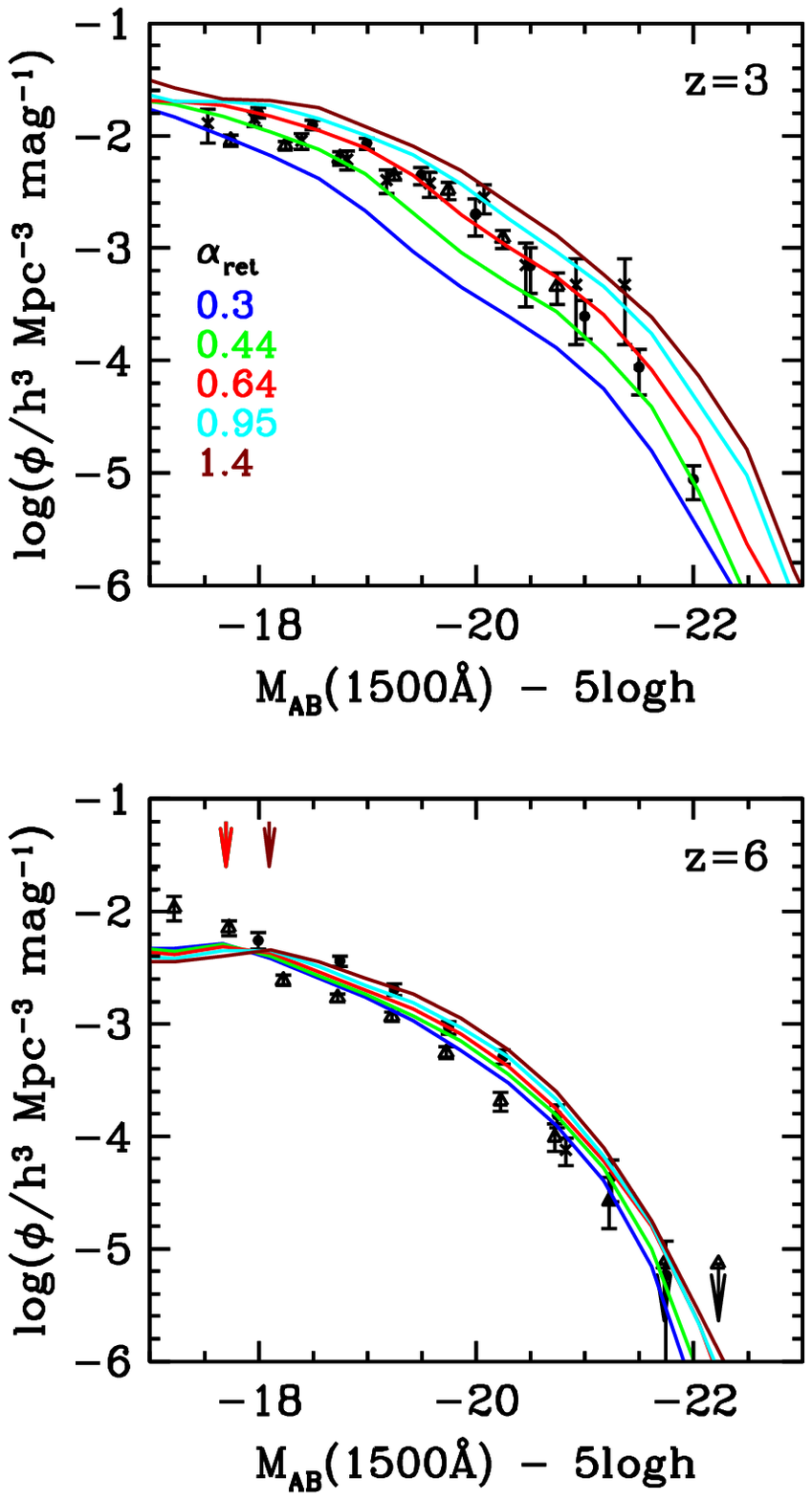}
\end{minipage}
\end{center}

\caption{Effects on the rest-frame far-UV (1500\AA) LF at $z=3$ and
  $z=6$ of varying the supernova feedback parameters $\gammaSN$ and
  $\VSN$ and gas return parameter $\alpharet$. A single parameter is
  varied in each column, with the red curves showing the standard
  model. { The vertical arrows at the top of each panel indicate
    the luminosity below which the results for the corresponding model
    are affected by the halo mass resolution.} The observational data
  plotted are the same as in Fig.~\ref{fig:LBGs_default}.}

\label{fig:LBGs_SNfeedback}
\end{figure*}

\begin{figure*}

\begin{center}
\begin{minipage}{5.4cm}
\includegraphics[width=5.4cm, bb= 20 295 275 750]{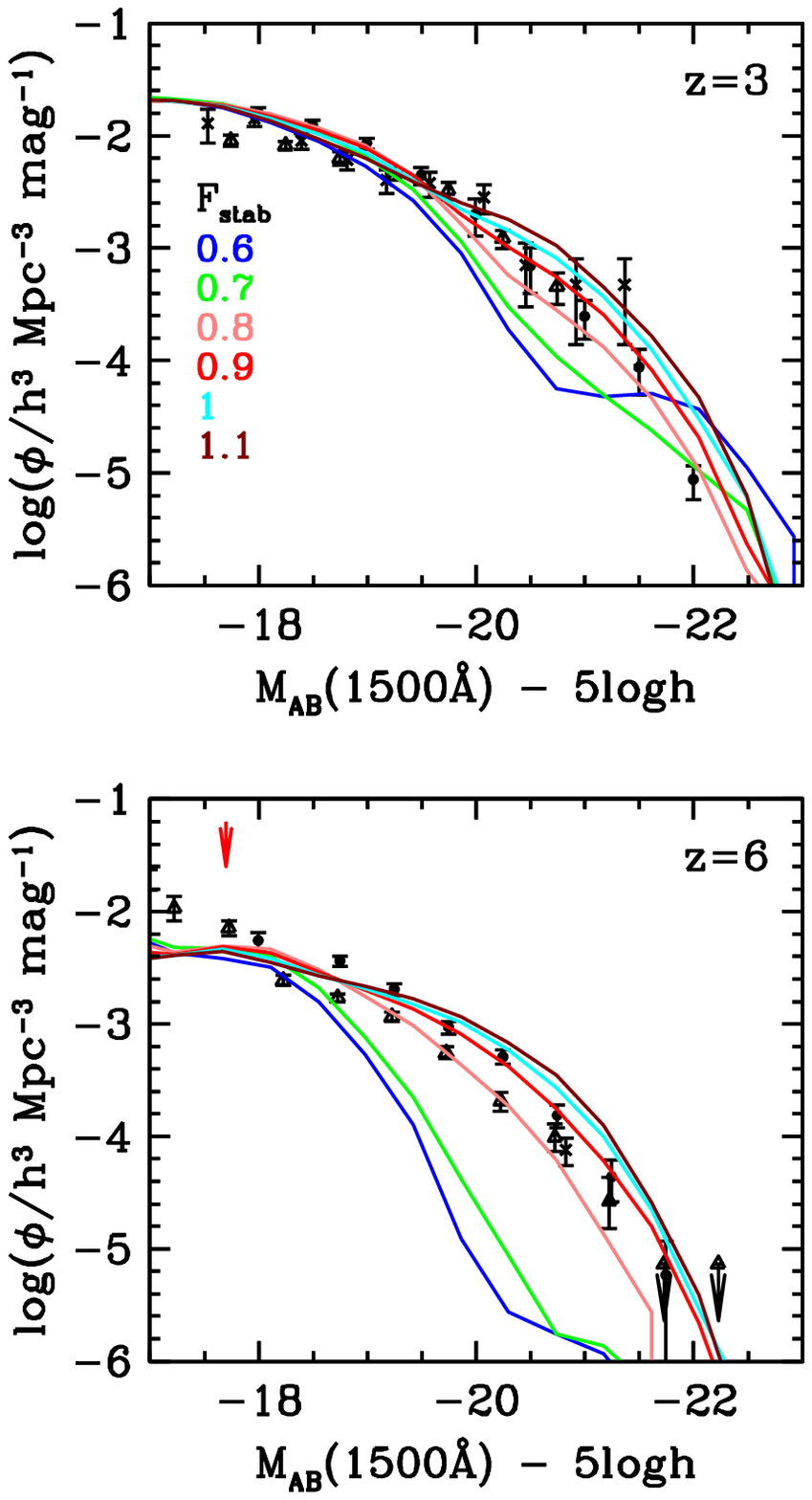}
\end{minipage}
\hspace{0.4cm}
\begin{minipage}{5.4cm}
\includegraphics[width=5.4cm, bb= 20 295 275 750]{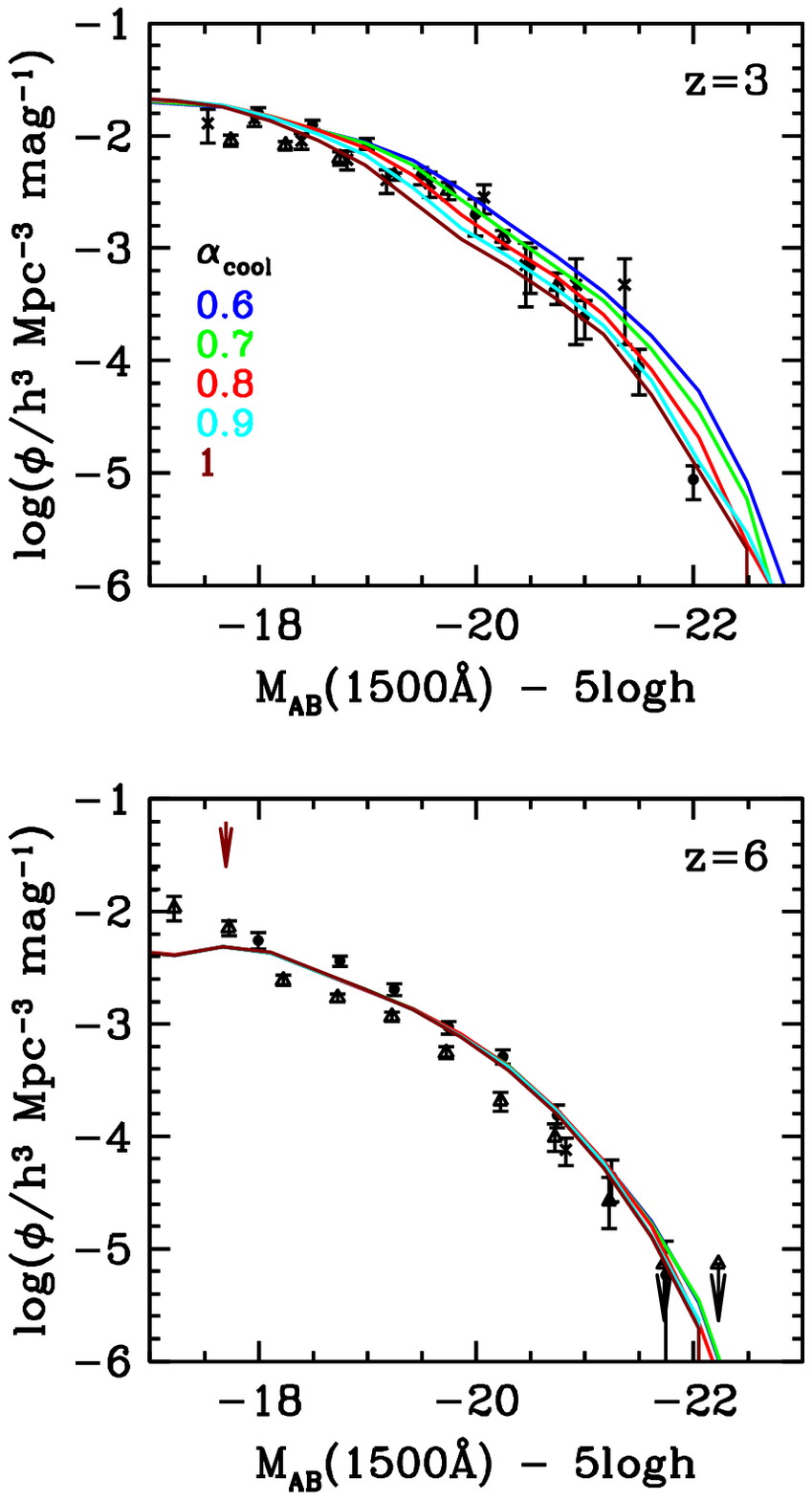}
\end{minipage}
\hspace{0.4cm}
\begin{minipage}{5.4cm}
\includegraphics[width=5.5cm, bb= 20 295 275 750]{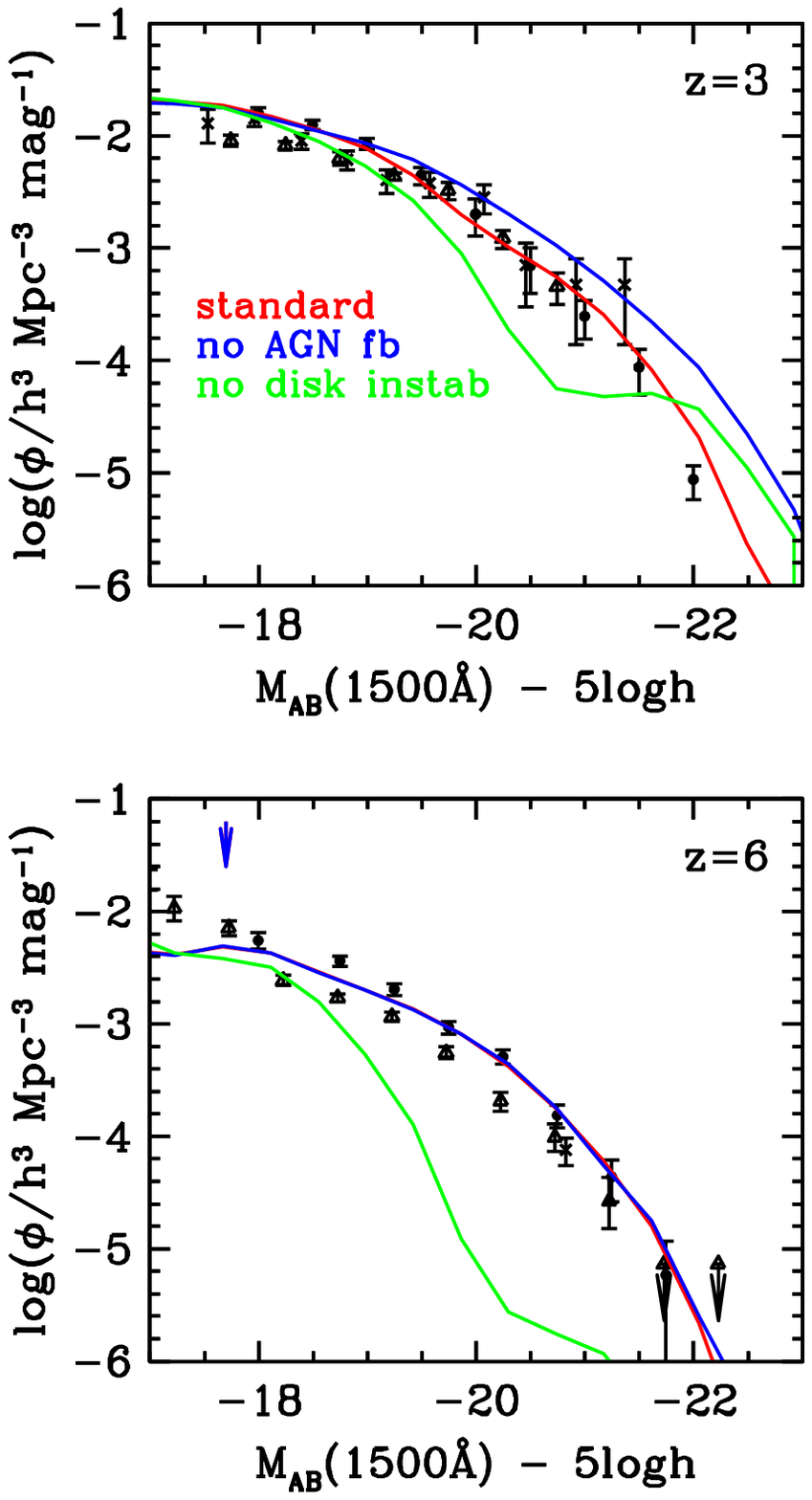}
\end{minipage}
\end{center}

\caption{Effects on the rest-frame far-UV (1500\AA) LF at $z=3$ and
  $z=6$ of varying (a) the disk stability parameter $\Fstab$ and (b)
  the AGN feedback parameter $\alphacool$, and (c) of turning off AGN
  feedback or disk instabilities. A single parameter is varied in each
  column, with the red curves showing the standard model.}

\label{fig:LBGs_AGNfeedback}
\end{figure*}

\begin{figure*}

\begin{center}
\begin{minipage}{5.4cm}
\includegraphics[width=5.4cm, bb= 20 295 275 750]{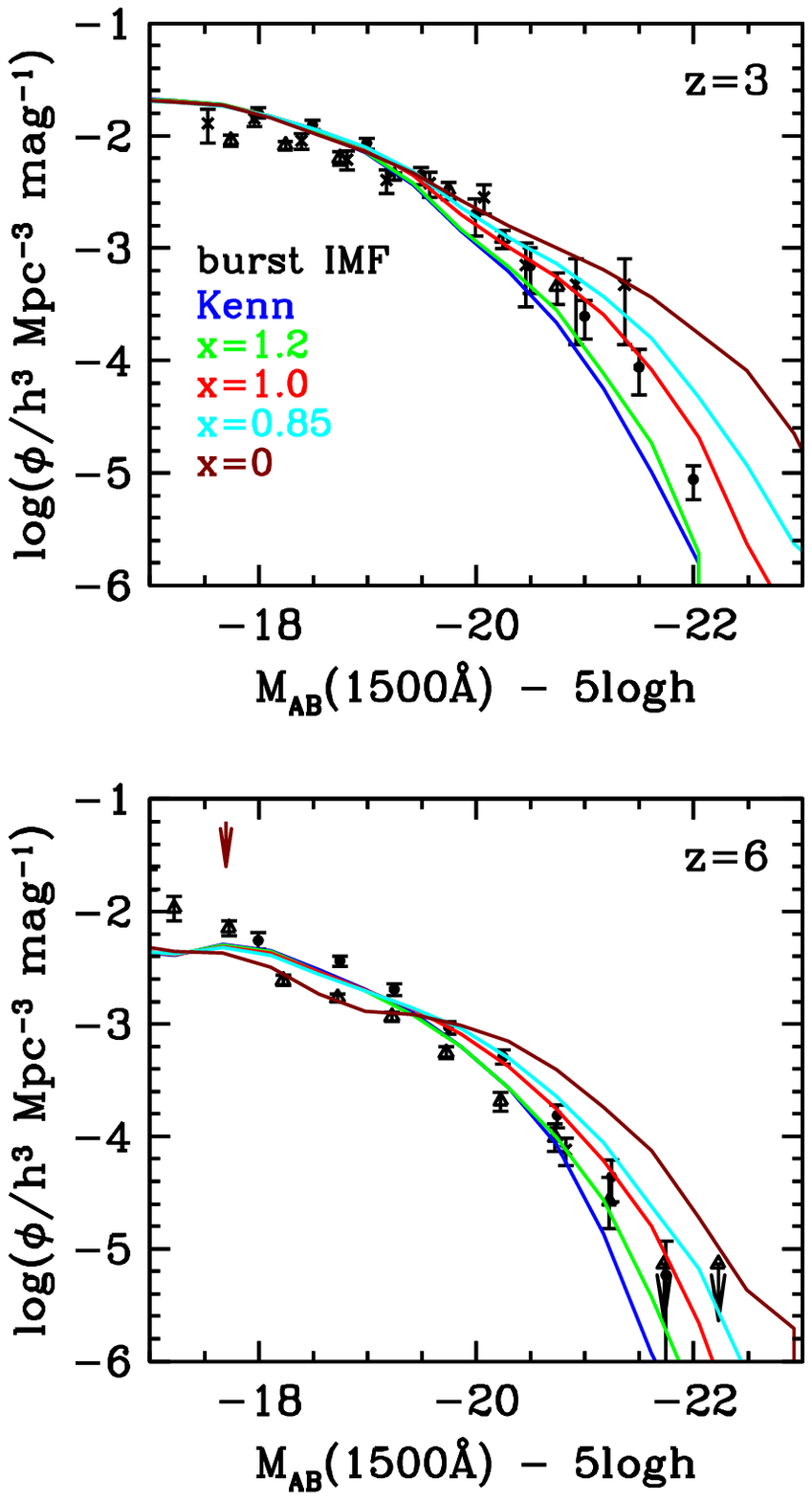}
\end{minipage}
\hspace{0.4cm}
\begin{minipage}{5.4cm}
\includegraphics[width=5.4cm, bb= 20 295 275 750]{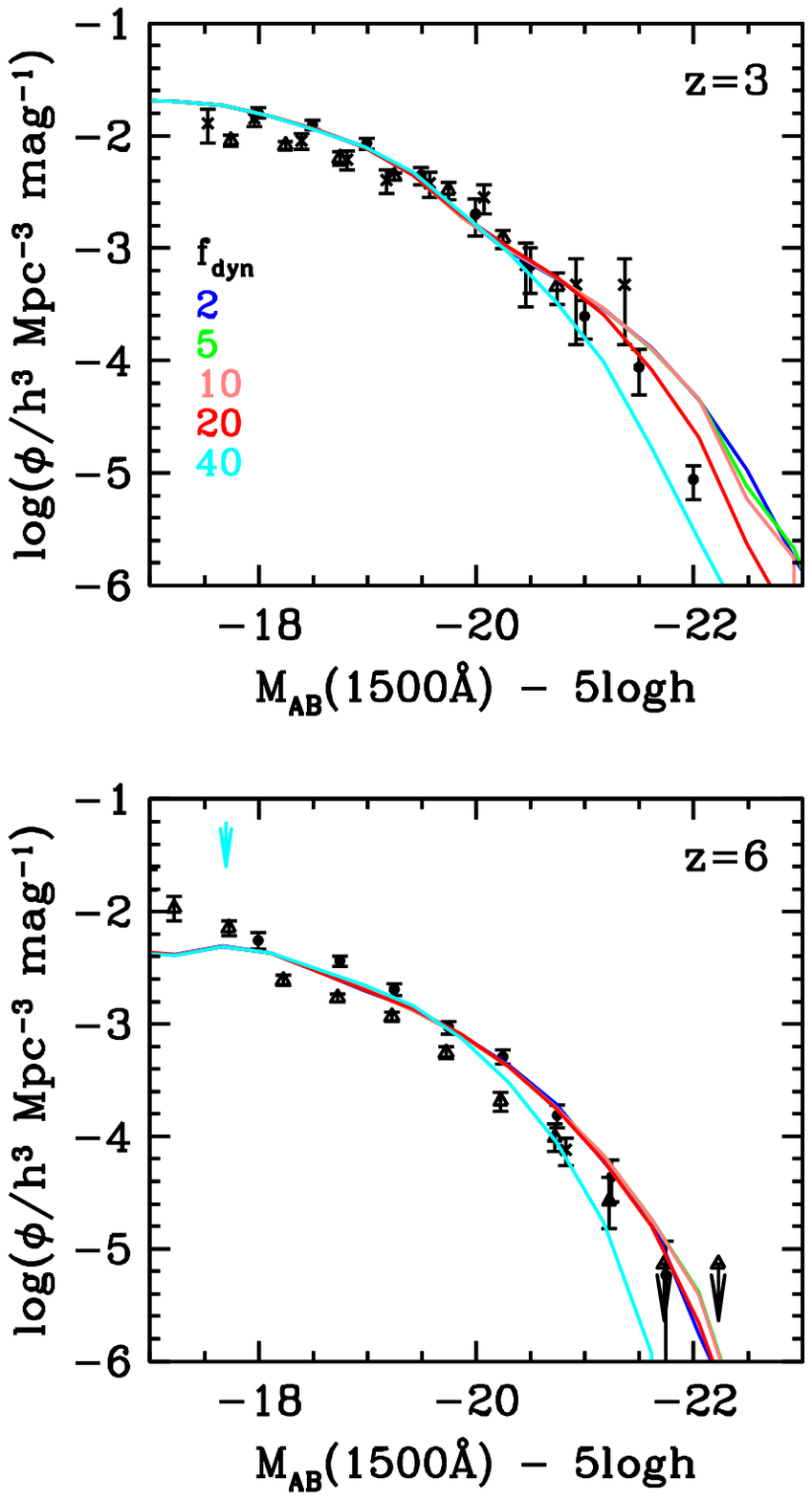}
\end{minipage}
\hspace{0.4cm}
\begin{minipage}{5.4cm}
\includegraphics[width=5.4cm, bb= 20 295 275 750]{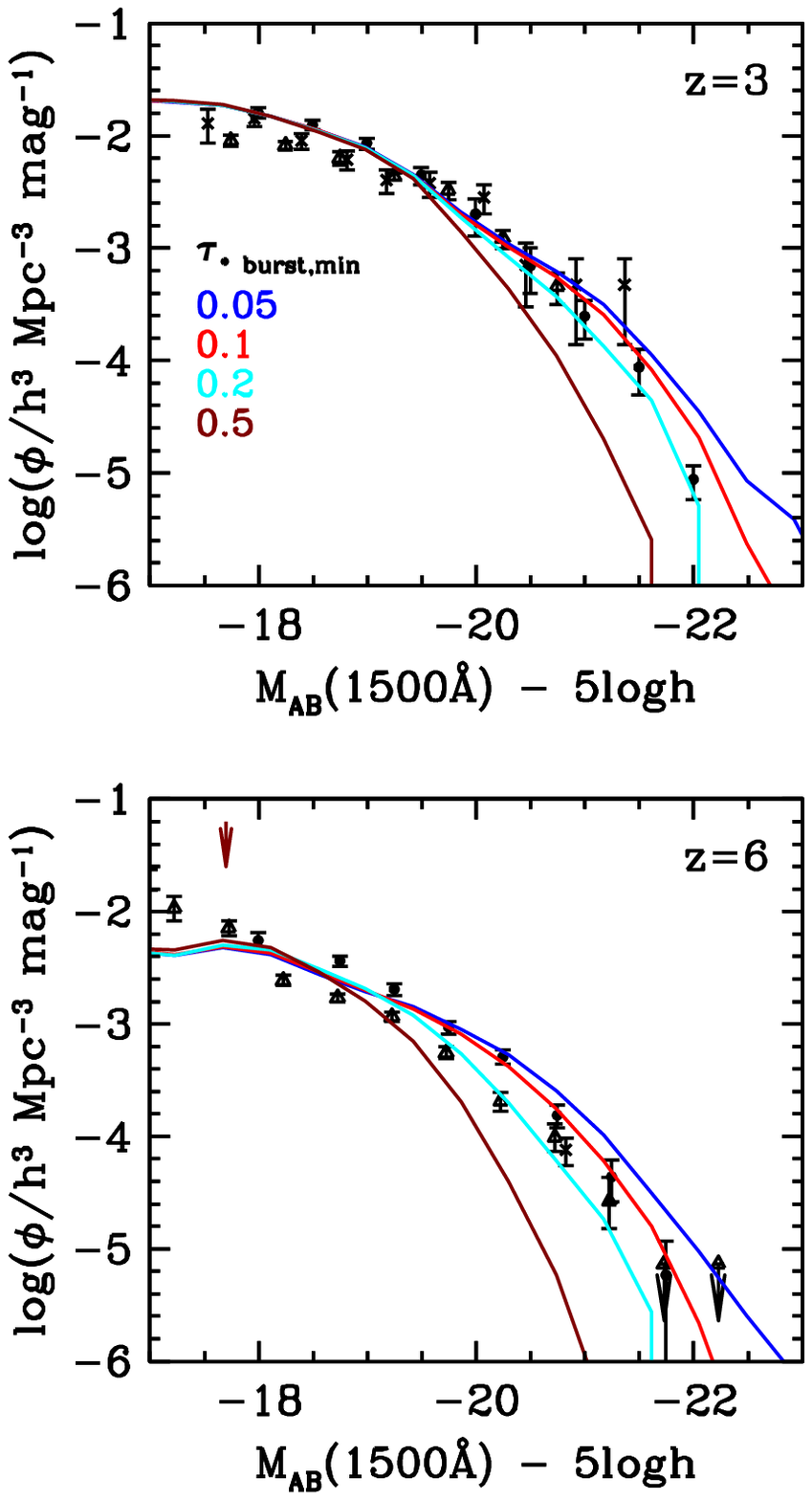}
\end{minipage}
\end{center}

\caption{Effects on the rest-frame far-UV (1500\AA) LF at $z=3$ and
  $z=6$ of varying the slope $x$ of the starburst IMF and the burst
  timescale parameters $\fdyn$ and $\tauburstmin$. A single parameter
  is varied in each column, with the red curves showing the standard
  model.}

\label{fig:LBGs_IMF_tauburst}
\end{figure*}

We show the effects on the far-UV (1500~\AA) LF at $z=3$ and $z=6$ of
varying the supernova feedback and gas return rate
(Fig.~\ref{fig:LBGs_SNfeedback}), disk instabilities and AGN feedback
(Fig.~\ref{fig:LBGs_AGNfeedback}), and the starburst IMF and starburst
timescale (Fig.~\ref{fig:LBGs_IMF_tauburst}).

\begin{figure*}
\begin{center}
\begin{minipage}{5.4cm}
\includegraphics[width=5.4cm, bb= 20 295 275 750]{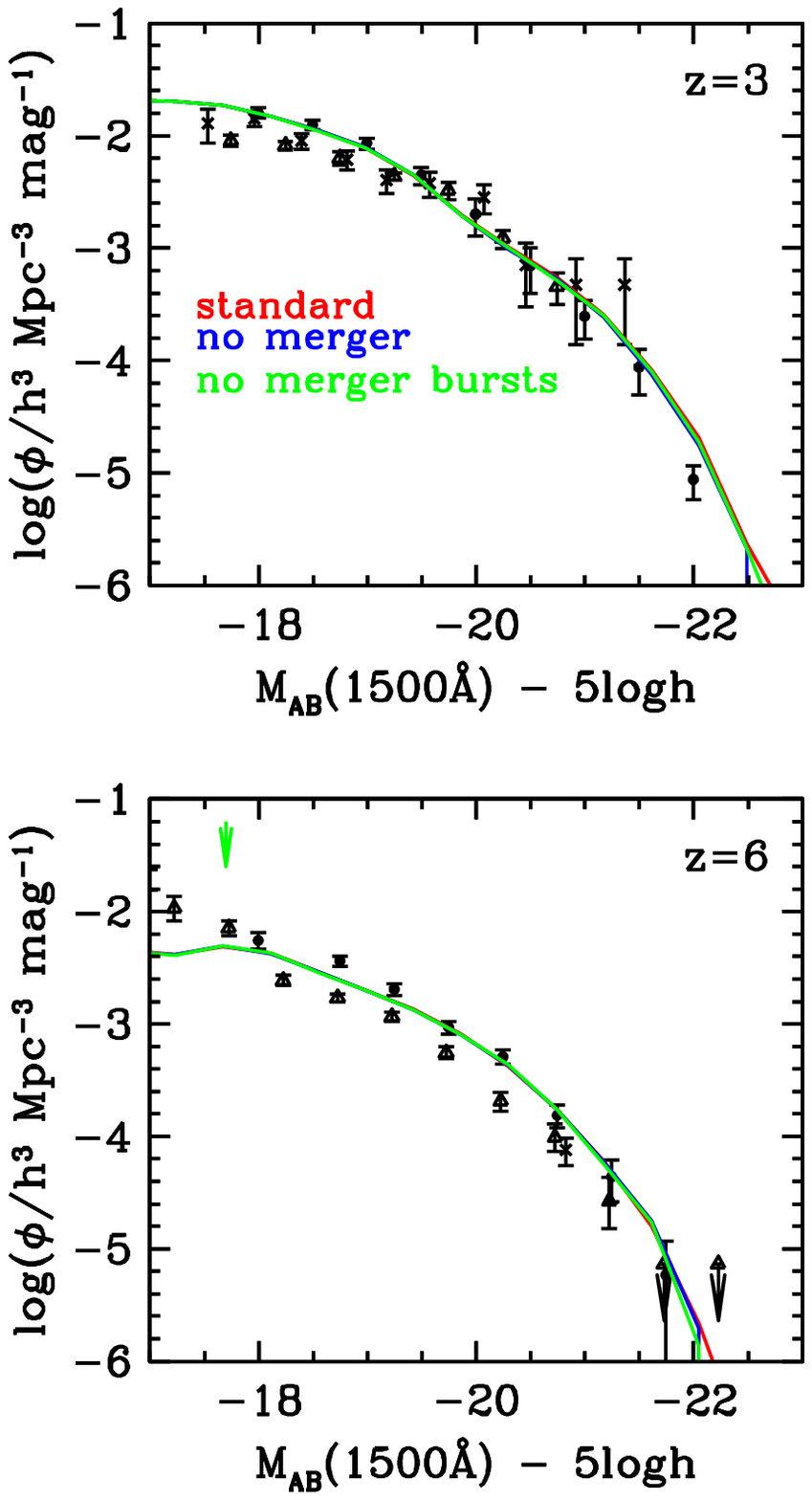}
\end{minipage}
\end{center}

\caption{Effects on the rest-frame far-UV (1500\AA) LF at $z=3$ and
  $z=6$ of turning off galaxy mergers or starbursts triggered by
  galaxy mergers. The red curves show the standard model.}

\label{fig:LBGs_nomerge}
\end{figure*}

\end{document}